\documentclass{aa}  

\usepackage{graphicx}
\usepackage{txfonts}
\usepackage[utf8]{inputenc}
\usepackage{dcolumn}
\usepackage{supertabular}
\usepackage{rotating}
\usepackage{longtable}
\usepackage{lscape}
\usepackage{float}
\usepackage{subfig}
\usepackage{url}
\usepackage{color}
\usepackage{natbib}
\usepackage[dvipsnames]{xcolor}
\usepackage{ulem}
\usepackage{soul}
\usepackage[colorinlistoftodos]{todonotes}
\usepackage[version=4]{mhchem} 
\usepackage{supertabular}  
\usepackage[colorlinks=true,
            linkcolor=blue,
            urlcolor=blue,
            citecolor=blue]{hyperref}
\newcommand{\kms}{\mbox{km s$^{-1}$}}
\newcolumntype{d}[1]{D{.}{\cdot}{#1}}
\newcolumntype{.}{D{.}{.}{-1}}

\newcommand{\hi}{H{\sc i}}
\newcommand{\hii}{H{\sc ii}}

\begin{document}

   \title{HyGAL: Characterizing the Galactic ISM with observations of hydrides and other small molecules}

   \subtitle{II. The absorption line survey with the IRAM 30~m telescope }

   \author{W.-J. Kim\inst{1} \and 
           P. Schilke\inst{1} \and 
           D. A. Neufeld\inst{2} \and 
           A. M. Jacob\inst{2} \and 
           \'A. S\'anchez-Monge\inst{1} \and
           D. Seifried\inst{1} \and 
           B. Godard \inst{3} \and 
           K. M. Menten\inst{4} \and
           S. Walch\inst{1} \and
           E. Falgarone\inst{5} \and 
           V.S. Veena\inst{4} \and
           S. Bialy\inst{6} \and
           T. M\"{o}ller\inst{1} \and 
           F. Wyrowski\inst{4} 
           }

   \institute{I. Physikalisches Institut, Universit\"at zu Köln, Z\"ulpicher Str. 77, 50937 K\"oln, Germany\\
        \email{wonjukim@ph1.uni-koeln.de}
        \and
        William H. Miller III Department of Physics \& Astronomy, Johns Hopkins University, Baltimore, MD 21218, USA 
        \and 
        LERMA, Observatoire de Paris, PSL Research University, CNRS, Sorbonne Universit\'{e}s, F-75014 Paris, France
        \and 
        Max-Planck-Institut f\"{u}r Radioastronomie, Auf dem H\"{u}gel 69, 53121 Bonn, Germany
        \and 
        Laboratoire de Physique de l'ENS, ENS, Universit\'{e} PSL, CNRS, Sorbonne Universit\'{e}, Universit\'{e} de Paris, 24 rue Lhomond, 75005 Paris, France
        \and 
        Department of Astronomy, University of Maryland, College Park, MD 20742-2421, USA}

   \date{Received ; accepted }

 \abstract{As a complement to the HyGAL Stratospheric Observatory for Infrared Astronomy Legacy Program, we report the results of a ground-based absorption line survey of simple molecules in diffuse and translucent Galactic clouds. Using the Institut de Radioastronomie Millim\'etrique (IRAM) 30\,m telescope, we surveyed molecular lines in the 2\,mm and 3\,mm wavelength ranges toward 15 millimeter continuum sources. These sources, which are all massive star-forming regions located mainly in the first and second quadrants of the Milky Way, form the subset of the HyGAL sample that can be observed by the IRAM 30\,m telescope. We detected \ce{HCO+} absorption lines toward 14 sightlines, toward which we identified 78 foreground cloud components, as well as lines from HCN, HNC, \ce{C2H}, and \ce{c-C3H2} toward most sightlines. In addition, CS and \ce{H2S} absorption lines are found toward at least half of the continuum sources. The spectral line data obtained were analyzed to characterize the chemical and physical properties of the absorbing interstellar medium statistically. The column density ratios of the seven molecular species observed are very similar to values found in previous absorption line studies carried out toward diffuse clouds at high latitudes. As expected, the \ce{C2H} and \ce{c-C3H2} column densities show a tight correlation with that of $N$(\ce{HCO+}), because of these all these molecules are considered to be proxies for the \ce{H2} column density toward diffuse and translucent clouds. The HCN and HNC column densities, by contrast, exhibit nonlinear correlations with those of \ce{C2H}, \ce{c-C3H2}, and \ce{HCO+}, increasing rapidly at $A_{\rm v}\approx 1$ in translucent clouds. Static Meudon photodissociation region (PDR) isobaric models that consider ultraviolet-dominated chemistry were unable to reproduce the column densities of all seven molecular species by just a factor of a few, except for \ce{H2S}. The inclusion of other formation routes driven by turbulent dissipation could possibly explain the observed high column densities of these species in diffuse clouds. There is a tentative trend for \ce{H2S} and CS abundances relative to \ce{H2} to be larger in diffuse clouds ($X$(\ce{H2S}) and $X$(CS) $\sim 10^{-8} - 10^{-7}$) than in translucent clouds ($X$(\ce{H2S}) and $X$(CS) $\sim 10^{-9} - 10^{-8}$) toward a small sample; however, a larger sample is required in order to confirm this trend. The derived \ce{H2S} column densities are higher than the values predicted from the isobaric PDR models, suggesting that chemical desorption of \ce{H2S} from sulfur-containing ice mantles may play a role in increasing the \ce{H2S} abundance.
 }

   \keywords{astrochemistry -- techniques: spectroscopic -- ISM: molecules}

   \maketitle
%

\section{Introduction}\label{sec:introduction}
Our understanding of the interstellar medium (ISM) has significantly improved in the past decade, through extensive observational efforts in absorption line spectroscopy performed at millimeter (mm) and submillimeter (submm) wavelengths toward background continuum sources. These observations have revealed both physical complexity and chemical richness as evidenced by the detection of various diatomic (e.g., CN, CO; \citealt{liszt2001cn-bearing, liszt2019co_dark_hcop}), triatomic, and even polyatomic molecules (e.g., \ce{C3H+}, \ce{CH2}, \ce{H2O}, \ce{HCO+}, \ce{C2H}, \ce{c-C3H}, \ce{l-C3H}, HCN, HNC, \ce{CH3CN}, \ce{NH3}, \ce{H2CO}, and \ce{c-C3H2}; \citealt{polehampton2005,flagey2013,gerin2011_smallhydro, gerin2019_ions,godard2010_3mmabs, lisztlucas2006, liszt2014_c-c3h, liszt2018}). The ISM has been shown to consist of multiphase gas combining a cold neutral medium (CNM) and warm neutral medium (WNM) \citep{bialy2019,bellomi2020_multi_phase} in diffuse clouds where chemistry begins with the formation of simple molecules. These diffuse clouds contain CO-dark \ce{H2} gas, in which CO is not detected in emission, but it can still be detected at a low abundance in absorption. Thus, diffuse clouds (defined as those in which the local fraction of gas-phase carbon in \ce{C+}, $f^n_{\rm \ce{C+}}$, exceeds 0.5) and translucent clouds (defined as those in which $f^n_{\rm \ce{C+}}$ < 0.5 but the local fraction of gas-phase carbon in CO, $f^n_{\rm CO}$, remains less than 0.9) in the Milky Way are excellent laboratories to study the formation and destruction of molecular clouds. The very different conditions in dense molecular clouds, where eventually star formation occurs, lead to a substantially different molecular composition. Significant efforts have been undertaken to determine the physical parameters that characterize the gas properties of high latitude diffuse clouds and solar neighborhood clouds (e.g., \citealt{liszt2001cn-bearing, liszt2004_hcop_hcn_co,lucas2000_small_hydrocarbons,lucas2002_sulfur}) as well as for a few specific sightlines in the Galactic plane (e.g., \citealt{godard2010_3mmabs,gerin2011_smallhydro,neufeld2015}). In contrast to high latitude clouds, the ISM in the disk of the Milky Way has many more complicated structures of atomic and molecular gas. Differing in their physical and chemical properties, diffuse and translucent clouds have, for example, different molecular gas fractions and different abundances of carbon-bearing species \citep[see][and references therein]{snow2006}. In particular, if only the total visual extinction ($A_{\rm v}$) is known, distinguishing translucent clouds from the superposition of multiple diffuse clouds yielding similar fractions of integrated column densities for atomic ($f^N_{\rm H}$) or molecular ($f^N_{\rm \ce{H2}}$) hydrogen is not straightforward. Nonetheless, the chemistry in the two types of clouds is certainly different as the primary reservoirs of gas-phase carbon vary from diffuse clouds to translucent clouds.

Absorption lines of \ce{HCO+} sample low-density regions, which can neither be observed in molecular emission nor in CO millimeter absorption because of both low excitation and the absence of CO \citep{hogerheijde1995,liszt2018,liszt2019co_dark_hcop}, making them an excellent tracer of \ce{H2} \citep{gerin2019_ions} that is not directly detectable except through ultraviolet (UV) observations toward nearby hot stars in very diffuse clouds. At millimeter and submillimeter wavelengths, where observations are unaffected by dust absorption, background sources (and foreground clouds) can be observed at large distances within the Galactic plane, and at higher visual extinctions, that are higher column densities. As carbon is a key element in interstellar chemistry, we also explored absorption by carbon-bearing molecules (HCN, HNC, \ce{C2H}, and \ce{c-C3H2}) that have been observed to be surprisingly as abundant (e.g., \citealt{godard2010_3mmabs,gerin2011_smallhydro}) as \ce{HCO+} in diffuse and translucent clouds. Their observed abundances are much higher than predicted from simple photodissociation region (PDR) models (e.g., \citealt{godard2010_3mmabs,gerin2011_smallhydro}).

The unexpectedly large column densities and broad line widths of the \ce{HCO+} and HCN molecules observed at millimeter wavelengths can be explained by their formation being enhanced by the dissipation of interstellar turbulence; this creates small regions of elevated temperature and large ion-neutral drift that enables endothermic reactions not possible under the standard conditions typical of the CNM (e.g., \citealt{hogerheijde1995,godard2010_3mmabs,godard2014_tdr}). For similar reasons, the observed abundances of sulfur-bearing molecules (such as \ce{H2S}, CS, SO, and SH) greatly exceed values predicted by standard models of cold diffuse molecular clouds, providing further evidence for the enhancement of endothermic reaction rates by elevated temperatures or ion-neutral drift \citep{neufeld2015}. Therefore, observations of these molecules provide important constraints on shock and turbulent dissipation region (TDR) models. 

By combining observations of multiple species, examining correlations and measuring abundance ratios, it has been possible to determine the distribution function for the \ce{H2} fraction within the diffuse and translucent ISM. This provides an important constraint on global models for the formation and destruction of molecular hydrogen in a turbulent medium (e.g., \citealt{bialy2017,bialy2019_turbulent,bialy2020}), as the differing abundances and kinematics of these molecules studied here demonstrate the importance of considering gas dynamical processes that enhance the formation rate of some of their progenitor species.

\begin{figure}[h!]
    \centering
    \includegraphics[width=0.48\textwidth]{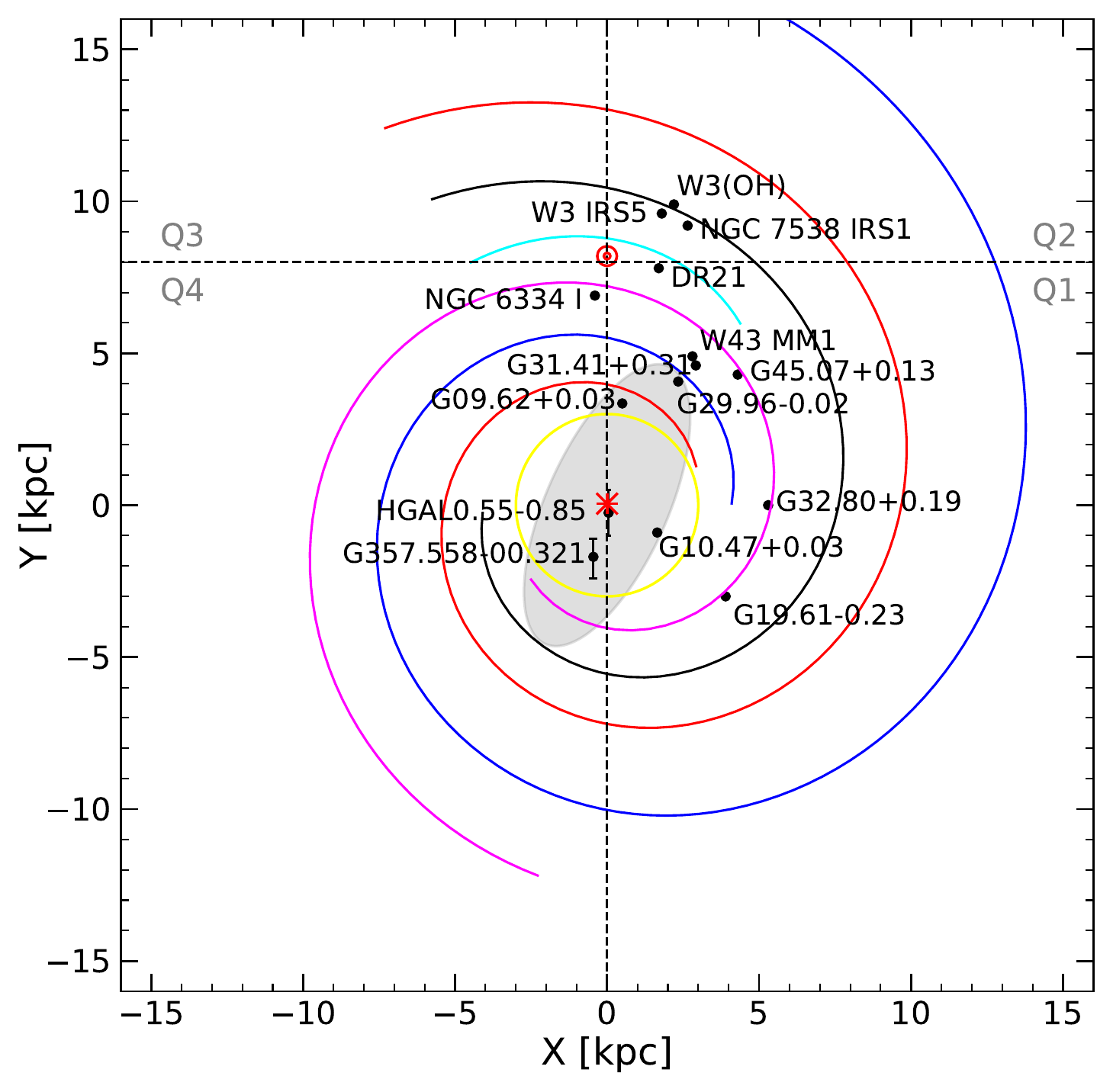}
    \caption{Positions of fifteen sightlines (black dots with names) from the HyGAL program in the Galactic plane. This distribution plot is reproduced based on the Figure 3 of Paper I. Colored curves represent different Milky Way spiral arms; the 3-kpc arm is in yellow, the Norma and outer arms  are in red, the Scutum-Centaurus arm is in blue, the Sagittarius-Carina arm is in purple, the Perseus arm is in black, and the local arm is in cyan. The Sun and the Galactic center (GC) are marked with the red Sun and asterisk symbols, respectively. The gray-shaded ellipse around the GC represents the long bar. }
    \label{fig:sources}
\end{figure}

\begin{table*}[h!]
    \centering
    \caption{Source summary.}
    \label{tab:conti_sou}
    \begin{tabular}{l c c l l c c .}
    \hline \hline
     Source & R.A. (J2000)  & Dec. (J2000)  & \multicolumn{1}{c}{Gal. Long.} & \multicolumn{1}{c}{Gal. Lat.} & $d$   & $R_{\rm GAL}$ & \multicolumn{1}{c}{$\varv_{\rm sys}$} \\
            & (hh:mm:ss)    & (dd:mm:ss) &  \multicolumn{1}{c}{(deg)} & \multicolumn{1}{c}{(deg)} & (kpc) & (kpc)         & \multicolumn{1}{c}{(\kms)}        \\
    \hline
    W3 IRS5         & 02:25:40.5  & $+$62:05:51.4  & 133.715 & $+$1.215 & 2.3 & 9.9 & -39.0 \\
    W3(OH)          & 02:27:04.1  & $+$61:52:22.1  & 133.948 & $+$1.064 & 2.0 & 9.6 & -47.0 \\
    NGC 6334 I      & 17:20:53.4  & $-$35:47:01.5  & 351.417 & $+$0.645 & 1.3 & 7.0 & -7.6 \\ 
    HGAL0.55$-$0.85 & 17:50:14.5  & $-$28:54:30.7  & 0.546   & $-$0.851 & 7.7$-$9.2 & 0.4$-$1.0 & 17.5 \\ 
    G09.62$+$0.19   & 18:06:14.9  & $-$20:31:37.0  & 9.620   & $+$0.194 & 5.2 & 3.1 & 4.4 \\ 
    G10.47$+$0.03   & 18:08:38.4  & $-$19:51:52.0  & 10.472  & $+$0.026 & 8.6 & 1.6 & 66.9 \\ 
    G19.61$-$0.23   & 18:27:38.0  & $-$11:56:39.5  & 19.608  & $-$0.234 & 12.6 & 4.4 & 41.8 \\
    *G29.96$-$0.02   & 18:46:03.7  & $-$02:39:21.2  & 29.954  & $-$0.016 & 4.8 & 4.7 & 97.0 \\
    *W43 MM1         & 18:47:47.0  & $-$01:54:28.0  & 30.817  & $-$0.057 & 3.1 & 5.7 & 98.0 \\    
    *G31.41$+$0.31   & 18:47:34.1  & $-$01:12:49.0  & 31.411  & $+$0.307 & 3.7 & 5.4 & 97.0 \\
    *G32.80$+$0.19   & 18:50:30.6  & $-$00:02:00.0  & 32.796  & $+$0.191 & 9.7 & 5.3 & 15.0 \\
    *G45.07$+$0.13   & 19:13:22.0  & $+$10:50:54.0  & 45.071  & $+$0.133 & 7.8 & 6.1 & 60.0 \\
    DR21            & 20:39:01.6  & $+$42:19:37.9  & 81.681  & $+$0.537 & 1.5 & 7.4 & -3.0 \\
    NGC 7538 IRS1   & 23:13:45.3  & $+$61:28:11.7  & 111.542 & $+$0.777 & 2.6 & 9.8 & -59.0 \\
    G357.558$-$00.321 &  17:40:57.2 & $-$31:10:59.3 & 357.557 & $-$0.321 & 9.0$-$11.8 & 1.0$-$3.6 & 5.3 \\
    \hline
    \end{tabular}
    \tablefoot{Except for the sources whose names are preceded by an asterisk, heliocentric and galactocentric distances are the values listed by Paper I. For those five sources, revised distances are taken from \citet{reid2019}, who cite the primary references. The systemic velocity, $\varv_{\rm sys}$, refers to the velocity at which the \ce{H^{13}CO+} emission line peaks, for each source in our sample as per the observations presented here.}
\end{table*}

This paper presents an absorption line survey, in the 2~mm and 3~mm wavelength regions, carried out using the Institut de Radioastronomie Millim\'etrique (IRAM) 30\,m telescope, as a complement to the HyGAL Stratospheric Observatory for Infrared Astronomy (SOFIA) Legacy Program (see \citealt{jacob2022_hygal}, hereafter Paper I, for an overview of the HyGAL program) that surveys six hydride species (i.e., \ce{ArH+}, \ce{OH+}, \ce{H2O+}, SH, OH, and CH), \ce{C+}, and O toward 25 bright Galactic continuum background sources at submillimeter wavelengths. This millimeter absorption line program surveys 15 sightlines targeted by the HyGAL program. Here we will explore the absorption line data and investigate associations between simple molecules and the hydrides studied in Paper I. Section\,\ref{sec:observation} describes the acquisition of the spectroscopic data, and the calibration and data reduction procedures employed. In Sect.\,\ref{sec:results}, we present the spectra of lines from \ce{HCO+}, HCN, HNC, CS, \ce{H2S}, \ce{C2H}, and \ce{c-C3H2} that are detected in absorption toward the different sightlines; here we discuss both their general properties and the CO emission counterparts that have previously been observed along some of these sightlines. The determination of the column densities of these molecular species will be explained in Sect.\,\ref{sec:analysis}, together with a principal component analysis for all species reported both in this paper and in Paper I toward W3(OH) and W3 IRS5. In Sect.\,\ref{sec:discussion}, we discuss the results of the principal component analysis and the derived observational quantities (column densities and abundances), and we compare these measurements with the results of isobaric PDR models, trying to understand the chemical and physical properties of diffuse and translucent clouds across the Milky Way. Finally, we conclude and summarize all the results from this study in Sect.\,\ref{sec:summary}.

\begin{figure}[h!]
    \centering
    \includegraphics[width=0.48\textwidth]{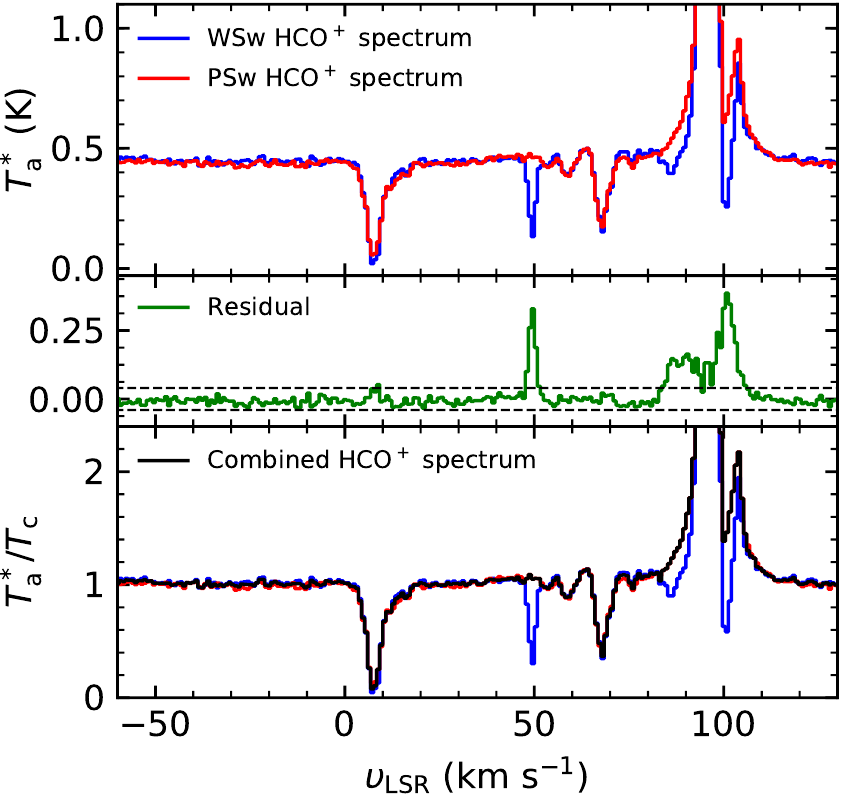}
    \caption{Comparison of the \ce{HCO+} spectra from observations with wobbler switching and position switching modes. The top panel shows the \ce{HCO+} spectra (velocities relative to the local standard of rest, LSR) taken with WSw (blue) and PSw (red) modes toward G29.96$-$0.02. The middle panel shows the residual spectrum (green) after fitting the WSw and PSw spectra. Black horizontal dashed lines mark the 3\,rms level. The bottom panel presents the combined spectrum (black) of the WSw and rescaled PSw (A$T_{\rm a}^*$(PSw)~$+$~B, where A and B are the slope and y-intercept of a linear least-squares regression fit, respectively), with the individual WSw and PSw spectra in the same colors as in the top panel. }
    \label{fig:wsw_psw}
\end{figure}

\begin{table*}[h!]
	\centering
	\small
	\caption{\label{tab:mol_info}Observed molecular transitions.}
	\begin{tabular}{l c . . c r c}
	\hline \hline
    Molecule & Transition & \multicolumn{1}{c}{Rest Frequency} & \multicolumn{1}{c}{Beam Size} & \multicolumn{1}{c}{$E_{\rm u}/k$} & \multicolumn{1}{c}{$E_{\rm l}/k$} & \multicolumn{1}{c}{$A_{\rm ul}$} \\
        &     & \multicolumn{1}{c}{(GHz)} & \multicolumn{1}{c}{($''$)} & \multicolumn{1}{c}{(K)} & \multicolumn{1}{c}{(K)} & \multicolumn{1}{c}{(s$^{-1}$)} \\
    \hline 
    \ce{HCO+} & $J=1\leftarrow0$                                    & 89.1885  & 27.6 & 4.28  & 0.0   & $4.19\times 10^{-5}$  \\
    HCN     & $J=1\leftarrow0, F=1\leftarrow1$                      & 88.6304  & 27.8 & 4.25  & 0.0   & $2.41\times 10^{-5}$ \\
            & $\,\,\,J=1\leftarrow0, F=2\leftarrow1 ^{\dagger}$                      & 88.6318  & 27.8 & 4.25  & 0.0   & $2.41\times 10^{-5}$ \\
            & $J=1\leftarrow0, F=0\leftarrow1$                      & 88.6339  & 27.8 & 4.25  & 0.0   & $2.41\times 10^{-5}$ \\
    HNC     & $J=1\leftarrow0$                                      & 90.6635  & 27.1 & 4.35  & 0.0   & $2.69\times 10^{-5}$  \\
    CS      & $J=2\leftarrow1$                                      & 97.9809  & 25.1 & 7.05  & 2.35  & $1.68\times 10^{-5}$ \\
    \ce{H2S}& $J_{\rm K_a,K_c}=1_{1,0} \leftarrow 1_{0,1}$          & 168.7627 & 14.6 & 8.10 & 0.0 & $2.68\times 10^{-5}$ \\
    \ce{C2H}& $N=1\leftarrow0, J=3/2\leftarrow1/2, F=1\leftarrow1$  & 87.2841  & 28.2 & 4.19  & 0.0   & $2.60\times 10^{-7}$\\
            & $\,\,N=1\leftarrow0, J=3/2\leftarrow1/2, F=2\leftarrow1 ^{\dagger}$  & 87.3168  & 28.2 & 4.19  & 0.0   & $1.53\times 10^{-6}$ \\
            & $N=1\leftarrow0, J=3/2\leftarrow1/2, F=1\leftarrow0$  & 87.3285  & 28.2 & 4.19  & 0.0   & $1.27\times 10^{-6}$ \\
            & $N=1\leftarrow0, J=1/2\leftarrow1/2, F=1\leftarrow1$  & 87.4019  & 28.1 & 4.20  & 0.0   & $1.27\times 10^{-6}$ \\
            & $N=1\leftarrow0, J=1/2\leftarrow1/2, F=0\leftarrow1$  & 87.4071  & 28.1 & 4.20  & 0.0   & $1.54\times 10^{-6}$ \\
            & $N=1\leftarrow0, J=1/2\leftarrow1/2, F=1\leftarrow0$  & 87.4464  & 28.1 & 4.20  & 0.0   & $2.61\times 10^{-7}$ \\
    \ce{c-C3H2}    & $J_{\rm K_{a},K_{c}}=2_{1,2}\leftarrow1_{0,1}$ & 85.3388  & 28.8 & 4.10  & 0.0  & $2.32\times 10^{-5}$ \\
    \hline
    \end{tabular}
    \tablefoot{The dagger ($\dagger$) marks the reference frequency corresponding to the reference velocity ($\varv_{\rm sys}$) for a given sightline. The spectroscopic information on these transitions is taken from Jet Propulsion Laboratory (JPL)\footnote{\url{https://spec.jpl.nasa.gov}} and the Cologne Database for Molecular Spectroscopy (CDMS)\footnote{\url{https://cdms.astro.uni-koeln.de}} catalogs. The frequencies of \ce{H2S} and \ce{c-C3H2} transitions correspond to that of their ortho spin state, and so the subsequently derived column densities and abundances for these species refer to ortho-\ce{H2S} and ortho-\ce{c-C3H2}. The upper energy levels of ortho-\ce{H2S} and ortho-\ce{c-C3H2} listed in the table are relative to the ground state of the orhto species; without that, for ortho-\ce{H2S}, $E_{\rm u}/k$ = 27.88~K and $E_{\rm l}/k$ = 19.78~K, and for ortho-\ce{c-C3H2}, $E_{\rm u}/k$ = 6.45~K and $E_{\rm l}/k$ = 2.35~K.
    } 
\end{table*}

\begin{table*}[h!]
\tiny
\centering
\caption{Coefficients from linear regression fits.}\label{tab:constant}
\begin{tabular}{l c c c c c c c c c c c c c c  }
\hline \hline
Source   & \multicolumn{2}{c}{\ce{HCO+}} & \multicolumn{2}{c}{HCN} & \multicolumn{2}{c}{HNC} & \multicolumn{2}{c}{CS} & \multicolumn{2}{c}{\ce{H2S}} & \multicolumn{2}{c}{\ce{C2H}} & \multicolumn{2}{c}{\ce{c-C3H2}} \\
        & A& B   & A & B & A & B & A & B & A & B & A & B & A & B \\
\hline 
W3 IRS5         &0.977&	$-$0.225&	0.963&	$-$0.199&	0.942&	$-$0.218&	&	$\cdots$ &$\cdots$	&$\cdots$	&	0.944&	$-$0.217&	0.800&	$-$0.128\\
W3(OH)          &1.163&	$-$0.441&	0.960&	$-$0.281&	0.856&	$-$0.280&	&$\cdots$	&$\cdots$	&$\cdots$	&	0.869&	$-$0.168&	0.620&	0.005	\\       
NGC 6334 I      &0.605&	$-$2.148&	0.716&	$-$2.538&	0.670&	$-$2.438&$\cdots$		&	$\cdots$	&	$\cdots$	&	$\cdots$	&	0.454	&	$-$1.577&	0.520	&	$-$1.904\\ 
HGAL0.55$-$0.85 &$\cdots$	&$\cdots$		&	$\cdots$	&	$\cdots$	&	$\cdots$	&	$\cdots$	&	$\cdots$	&$\cdots$	&	$\cdots$	&	$\cdots$	&	$\cdots$	&$\cdots$	&$\cdots$		&$\cdots$		\\ 
G09.62$+$0.19  	&0.987&	$-$1.133&	1.033&	$-$1.188&	1.015&	$-$1.145&$\cdots$	&	$\cdots$	&	$\cdots$	&	$\cdots$	&	0.486&	$-$0.539&	0.156	&	$-$0.126	\\
G10.47$+$0.03   &1.013&	$-$0.047&	1.037&	$-$0.043&	0.896&	$-$0.014&$\cdots$	&	$\cdots$	&	$\cdots$	&	$\cdots$	&	0.907&	$-$0.033&	0.907	&	$-$0.021	\\
G19.61$-$0.23  	&1.049&	0.867	&	1.050&	0.862	&	1.024&	0.887	&$\cdots$	&	$\cdots$	&	$\cdots$	&	$\cdots$	&	0.889&	0.836	&	0.819	&	0.807	\\
G29.96$-$0.02   &0.976&	$-$0.681&	0.969&	$-$0.663&	0.942&	$-$0.644&	0.580&	$-$0.235&	0.112&	$-$0.028&	0.922	&	$-$0.633&	0.928	&	$-$0.655\\
W43 MM1      	&0.626&	$-$0.185&	0.749&	$-$0.229&	0.628&	$-$0.185&	0.763&	$-$0.038&	0.700&	0.184	&	0.809	&	$-$0.264&	0.891	&	$-$0.288\\   
G31.41$+$0.31  	&1.018&	0.098	&	1.042&	0.089	&	0.902&	0.099	&	0.977&	$-$0.117&	0.942&	$-$0.228&	1.029	&	0.080	&	1.074	&	0.073	\\
G32.80$+$0.19  	&1.075&	$-$0.015&	1.072&	$-$0.005&	1.100&	$-$0.323&	0.931&	0.212&	0.592&	0.421&	1.080&	$-$0.287&	1.031	&	$-$0.238\\
G45.07$+$0.13   &0.946&	0.193	&	0.906&	0.184	&	0.260&	0.199	&	$\cdots$	&	$\cdots$	&	$\cdots$	&	$\cdots$	&	0.930	&	0.212	&	0.625	&	0.220	\\
DR21        	&0.847&	$-$0.322&	0.851&	0.293	&	0.919&	$-$0.505&	$\cdots$	&	$\cdots$	&	$\cdots$	&	$\cdots$	&	0.906	&	0.220	&	1.010	&	$-$0.306\\
NGC 7538 IRS1  	&0.761&	0.224	&	0.932&	0.238	&	0.536&	0.283	&	$\cdots$	&	$\cdots$	&	$\cdots$	&	$\cdots$	&	0.761	&	0.225	&	0.278	&	0.342	\\
\hline  
\end{tabular}

\tablefoot{The coefficients A and B refer to the slope and y-intercept of a linear least-square regression fit. The molecular transitions with blank A and B values correspond to WSw mode-only observations with noncontaminated WSw off positions.} 
\end{table*}

\section{Observations and data reduction}\label{sec:observation}
We carried out observations of a molecular absorption line survey (Project ID: 003--20) in the 2~mm and 3~mm wavelength regions, during the period of 2020 June 23$-$30, with the Eight MIxer Receiver (EMIR)\footnote{\url{http://www.iram.es/IRAMES/mainWiki/EmirforAstronomers}} on the IRAM 30\,m telescope, toward the 15 HyGAL continuum sources listed in Table\,\ref{tab:conti_sou}. The continuum sources observed here cover slightly more than half of the entire HyGAL source sample (see Paper I) and are mostly located in the first and second quadrants of the Milky Way as shown in Fig.\,\ref{fig:sources}, except for two sources, NGC6334 I and G357.558$-$00.321, which are located in the fourth quadrant. For a given column density of a target species, the signal-to-noise ratio for absorption line features is proportional to the continuum flux at the relevant wavelength. The Herschel InfraRed Galactic Plane Survey (Hi-GAL, \citealt{molinari2016_HiGAL}) provides a compact source catalog \citep{elia2017_higal_compact, Elia2021} with continuum fluxes at 70, 160, 250, 350, and 500\,$\mu$m coinciding with the wavelengths of transitions targeted with HyGAL (approximately spanning 149--494\,$\mu$m). The HyGAL program selected 25 submillimeter sources with flux limits over 2000\,Jy at 160\,$\mu$m for sources in the inner Galaxy and ${>~1000}$\,Jy for sources in the outer Galaxy (see Paper I for more details on the source selection). 

The observations were done with two different receiver setups: the first configuration combines the E090 and E150 receivers (covering 92$-$99.7\,GHz and 166.5$-$174.2\,GHz) and the other uses the E090 receiver alone (covering 83.2$-$90.9\,GHz and 98.9--106.6\,GHz). These setups allowed us to target a number of molecular transitions with a total bandwidth coverage of 15~GHz (including a 1~GHz bandwidth overlap between the two setups); they proved ideal for obtaining widespread absorption line features arising from foreground clouds at different velocities. Both receiver setups are connected to the FTS200 backend offering a frequency channel resolution of 200\,kHz (equivalent to $\Delta\varv$ $\sim$ 0.35\,\kms\ for \ce{H2S} and 0.62$-$0.66\,\kms\ for the other transitions). For our analysis, we resampled all the spectral line data to a fixed velocity resolution of 0.8\,\kms, except \ce{H2S} for which we use a better resampled resolution of 0.36\,\kms. We utilized the Continuum and Line Analysis Single-dish Software (CLASS) software\footnote{\url{https://www.iram.fr/IRAMFR/GILDAS/doc/html/class-html/class.html}} of the Grenoble Image and Line Data Analysis Software (GILDAS) package \citep{pety2005_gildas} for the initial data reduction of the molecular line data, and further post-processed the data using Python (e.g., when combining the observations made in two different switching modes).

\subsection{Calibration for PSw and WSw mode observations}
The observed data sets consist of wobbler switching (WSw) and position switching (PSw) On-Off mode observations. For WSw mode observations, all sources were observed with throw angles of $\pm$120$''$. In active star-forming regions, which often exhibit extended emission structures, these relatively small throw angles can result in reference positions that are significantly contaminated by emission in the stronger lines such as \ce{HCO+} (1$\rightarrow$0). This causes artificial absorption features in the final calibrated spectra. On the other hand, standing waves are often generated in the PSw mode as the telescope slews to distant reference positions while the WSw mode provides reliable continuum level measurements and flat baselines. Since standing waves distort the emission and absorption line profiles, we carefully investigated all the PSw spectra and excluded those scans that are severely affected by standing wave effects before averaging all observational scans.

\begin{table*}[h!]
\tiny
\centering
\caption{Continuum levels and rms noise of the spectra.}\label{tab:tcon_rms}
\begin{tabular}{l c c c c c c c c c c c c c c  }
\hline \hline
Source   & \multicolumn{2}{c}{\ce{HCO+}} & \multicolumn{2}{c}{HCN} & \multicolumn{2}{c}{HNC} & \multicolumn{2}{c}{CS} & \multicolumn{2}{c}{\ce{H2S}} & \multicolumn{2}{c}{\ce{C2H}} & \multicolumn{2}{c}{\ce{c-C3H2}} \\
        & $T_{\rm c}$& $T_{\rm rms}$   & $T_{\rm c}$& $T_{\rm rms}$& $T_{\rm c}$& $T_{\rm rms}$& $T_{\rm c}$& $T_{\rm rms}$& $T_{\rm c}$& $T_{\rm rms}$& $T_{\rm c}$& $T_{\rm rms}$&$T_{\rm c}$& $T_{\rm rms}$ \\
        & (K) & (K) & (K) & (K) & (K) & (K) & (K) & (K) & (K) & (K) & (K) & (K) & (K) & (K) \\
\hline 
W3 IRS5         & 0.525 & 0.007 & 0.514 & 0.008 & 0.509 & 0.007 & 0.418 & 0.005 & 0.181 & 0.012 & 0.566 & 0.010 & 0.606 & 0.008 \\
W3(OH)          & 0.626 & 0.006 & 0.626 & 0.009 & 0.647 & 0.008 & 0.589 & 0.005 & 0.551 & 0.011 & 0.638 & 0.009 & 0.629 & 0.009 \\
NGC 6334 I      & 0.427 & 0.010 & 0.531 & 0.013 & 0.464 & 0.014 & 0.677 & 0.017 & 1.031 & 0.048 & 0.455 & 0.014 & 0.438 & 0.013 \\
HGAL0.55$-$0.85 & 0.116 & 0.008 & 0.111 & 0.008 & 0.116 & 0.009 & 0.141 & 0.006 & 0.257 & 0.014 & 0.112 & 0.009 & 0.109 & 0.009 \\
G09.62$+$0.19   & 0.075 & 0.008 & 0.073 & 0.008 & 0.077 & 0.009 & 0.075 & 0.005 & 0.098 & 0.013 & 0.076 & 0.009 & 0.076 & 0.009 \\
G10.47$+$0.03   & 0.117 & 0.006 & 0.122 & 0.009 & 0.121 & 0.008 & 0.159 & 0.005 & 0.367 & 0.015 & 0.115 & 0.011 & 0.115 & 0.007 \\
G19.61$-$0.23   & 0.353 & 0.004 & 0.353 & 0.007 & 0.366 & 0.006 & 0.384 & 0.005 & 0.325 & 0.011 & 0.370 & 0.006 & 0.379 & 0.006 \\
G29.96$-$0.02   & 0.439 & 0.004 & 0.439 & 0.008 & 0.458 & 0.007 & 0.390 & 0.008 & 0.182 & 0.024 & 0.456 & 0.007 & 0.472 & 0.007 \\
W43 MM1         & 0.059 & 0.004 & 0.059 & 0.009 & 0.069 & 0.005 & 0.116 & 0.010 & 0.256 & 0.024 & 0.065 & 0.006 & 0.060 & 0.005 \\
G31.41$+$0.31   & 0.155 & 0.006 & 0.155 & 0.009 & 0.158 & 0.006 & 0.167 & 0.006 & 0.271 & 0.013 & 0.167 & 0.008 & 0.161 & 0.007 \\
G32.80$+$0.19   & 0.497 & 0.004 & 0.486 & 0.007 & 0.506 & 0.008 & 0.480 & 0.003 & 0.330 & 0.008 & 0.516 & 0.008 & 0.523 & 0.007 \\
G45.07$+$0.13   & 0.194 & 0.004 & 0.194 & 0.005 & 0.201 & 0.006 & 0.191 & 0.004 & 0.183 & 0.011 & 0.199 & 0.005 & 0.198 & 0.005 \\
DR21            & 1.500 & 0.005 & 1.490 & 0.009 & 1.496 & 0.009 & 1.540 & 0.006 & 0.751 & 0.012 & 1.563 & 0.008 & 1.590 & 0.008 \\
NGC 7538 IRS1   & 0.408 & 0.005 & 0.401 & 0.008 & 0.415 & 0.008 & 0.387 & 0.005 & 0.369 & 0.011 & 0.406 & 0.009 & 0.402 & 0.008 \\
\hline  
\end{tabular}

\tablefoot{The continuum and rms noise levels are on the $T_{\rm a}^*$ scale, with a spectral resolution ($\Delta \varv$) of 0.8\,\kms\ for all species except for \ce{H2S} for which $\Delta \varv=$0.36\,\kms. }
\end{table*}

\begin{table*}[h!]
\centering
\caption{Summary of detected species in absorption.}\label{tab:detections}
\begin{tabular}{l c c c c c c c  }
\hline \hline
Source   & \ce{HCO+} & HCN & HNC & CS & \ce{H2S} & \ce{C2H} & \ce{c-C3H2} \\
\hline 
W3 IRS5         & Y & Y & N & N & N & Y & Y \\
W3(OH)          & Y & Y & Y & Y & Y & Y & Y \\
NGC 6334 I      & Y & Y & Y & N & N & Y & Y \\
HGAL0.55$-$0.85 & Y & N & N & N & N & N & N \\
G09.62$+$0.19   & Y & Y & Y & N & Y & Y & N \\
G10.47$+$0.03   & Y & Y & Y & Y & C & Y & C \\
G19.61$-$0.23   & Y & Y & Y & Y & Y & Y & Y \\
G29.96$-$0.02   & Y & Y & Y & Y & C & Y & Y \\
W43 MM1         & Y & Y & Y & C & C & Y & N \\
G31.41$+$0.31   & Y & Y & Y & Y & C & Y & Y \\
G32.80$+$0.19   & Y & Y & Y & Y & Y & Y & Y \\
G45.07$+$0.13   & Y & Y & Y & Y & Y & Y & Y \\
DR21            & Y & Y & Y & Y & Y & Y & Y \\
NGC 7538 IRS1   & Y & N & N & N & N & N & N \\
\hline  
\end{tabular}

\tablefoot{Clear detection of absorption features, which are either widely separated from the emission lines associated with background continuum sources or superposed on emission features, are marked with ``Y''. Less secure detections, due to either poor S/N or severe contamination from unidentified or identified transitions, are marked with ``C'' and nondetections with ``N''.}
\end{table*}

Figure\,\ref{fig:wsw_psw} shows an example of the \ce{HCO+} spectra obtained with WSw and PSw mode observations toward G29.96$-$0.02. In the top panel of Fig.\,\ref{fig:wsw_psw}, the WSw spectrum shows additional apparent absorption features; these are found, for example, at LSR velocities of 50\,\kms\ and 85\,\kms, the latter lying close to the velocity of the background continuum source (97\,\kms). These components are not detected in the PSw spectrum, implying that they may be dismissed as artifacts arising from contamination in the reference position. To identify such artifacts, we rescaled the PSw spectrum to the WSw spectrum, allowing for both multiplicative and additive offsets, by using the relationship of A$T_{\rm a}^*$(PSw)~$+$~B, where A and B are constants, representing the slope and y-intercept, respectively, of the linear least-squares regression on both the spectra. We note that there are some values of A that are lower than 1, which may be caused by the significant contamination from the WSw reference positions or the less flat baselines from the PSw modes. The coefficients used for rescaling the PSw spectra are listed in Table\,\ref{tab:constant}. We have examined if any excessive residual spectral features remain after subtracting a rescaled PSw spectrum from a WSw spectrum, as shown in the green profile in the middle panel of Fig.\,\ref{fig:wsw_psw}. If the residual value in any velocity channel exceeds three times the rms value (marked by black horizontal dashed lines), the WSw value at the selected velocity channel is replaced by the PSw value at the same velocity because the high residual components are regarded not to be real but due to the reference position contamination. The continuum levels and rms noise of the spectra for the seven species toward each source are listed in Table\,\ref{tab:tcon_rms}. Subsequently, we combined the rescaled PSw and the WSw spectra corrected to exclude the WSw spectral range with the counterfeit absorption features. Since the WSw mode was used as the primary observing mode, observations made in this mode have a longer integration time (about 1.5\,hours per source) and in turn a larger number of scans than observations taken in the PSw mode. Thus, before combining both data sets, the spectra are weighted by the inverse squares of their measured rms noise levels. The plot in the bottom panel shows the combined spectrum in black along with the WSw and PSw spectra in blue and red, respectively. The combined spectrum appears to be well-calibrated, with a higher signal-to-noise ratio than spectra obtained in the individual modes, and without any false absorption features (like the 50\,\kms\ component). We used the combined spectra of all observed transitions in all following analyses.

\begin{table*}[h!]
    \centering
    \tiny
    \caption{Summary of identified spiral arms and their velocity ranges along to the observed sightlines.}
    \label{tab:arm_info}
    \begin{tabular}{l c c  c c c c c c c}
    \hline \hline
    Source  &  Local & Aquila-Rift  & Perseus & Sagittarius-Carina  &  Scutum-Centaurus &3-kpc & Inter-arm & Unidentified clouds \\
            &        &              &         &          &          & (Galactic bar/center) & & or envelope \\
            & (\kms) & (\kms)       &  (\kms) &  (\kms)  & (\kms)   &  (\kms) &  (\kms) &  (\kms) \\
    \hline
W3 IRS5        &$-$11.8, 3.6&$\cdots$       &$-$35.8, $-$16.1 &$\cdots$  &$\cdots$    &$\cdots$    &$\cdots$      &$\cdots$  \\      
W3(OH)         &$-$16.0, 3.6&$\cdots$       &$-$35.6, $-$16.0 &$\cdots$  &$\cdots$    &$\cdots$    &$\cdots$      &$\cdots$  \\
NGC 6334 I     &$\cdots$    &$\cdots$       &$\cdots$         &$\cdots$  &$\cdots$    &$\cdots$    &$\cdots$      &4.0, 10.0 \\
HGAL0.55$-$0.85&$\cdots$    &$\cdots$       &$\cdots$         &$\cdots$  &$\cdots$    &$\cdots$    &$\cdots$      &$-$10.0, $-$4.6 \\      
G09.62$+$0.19  &$\cdots$    &$\cdots$       &$\cdots$         &$\cdots$  &10.0, 40.0  &$\cdots$    &$\cdots$      &60.0, 75.0  \\      
G10.47$+$0.03  &$\cdots$    &$\cdots$       &$\cdots$         &3.2, 20.0 &20.0, 50.0  &80.0, 180$^{a}$   & $-$14.4, 2.4 &$\cdots$ \\      
G19.61$-$0.23  &$\cdots$    &$\cdots$       &$\cdots$         &35.0, 93.0 & 105.8, 113.9  & 93.0, 127.0 & $-$10.0, 20.0 & 28.0, 40.0\\      
G29.96$-$0.02  &$\cdots$    &1.6, 18.4      &$\cdots$         &37.0, 85.0&$\cdots$    &$\cdots$    &$\cdots$   &$\cdots$ \\      
W43 MM1        &$-$5.0, 5.0 & 5.0, 21.4     &$\cdots$         &21.4, 50.0&$\cdots$    &$\cdots$    &$\cdots$   &50.0, 100.0  \\      
G31.41$+$0.31  &0.0, 20.0   &$\cdots$       &$\cdots$         &30.0, 58.0&$\cdots$    &$\cdots$    &$\cdots$   &$\cdots$ \\      
*G32.80$+$0.19  &$\cdots$    & 5.0, 6.0   &$\cdots$          &15.0, 57.0& 58.0, 120.0&$\cdots$    &$\cdots$   &$\cdots$ \\
G45.07$+$0.13  &$-$2.0, 20.0&$\cdots$       &$\cdots$         &20.0, 60.0&$\cdots$    &$\cdots$    &$\cdots$   &60.0, 79.0  \\   
DR21           &$\cdots$    &$\cdots$       &$\cdots$         &$\cdots$  &$\cdots$    &$\cdots$    &$\cdots$   &5.0, 20.0 \\
NGC 7538 IRS1  &$-$12.2, 4.4&$\cdots$       &$\cdots$         &$\cdots$  &$\cdots$    &$\cdots$    &$\cdots$   &$\cdots$ \\      
    \hline
    \end{tabular}
    \tablefoot{In the last column, ``envelope'' refers to the diffuse and translucent molecular gas layers surrounding dense molecular gas regions associated with the background continuum sources. For the source (G32.80$+$0.19) marked by an asterisk, its velocity intervals of crossing spiral arms are based on \citealt{reid2019}.\\ $^{a}$ The velocities from 120 -- 180\,\kms\ belong to the 135\,\kms\ arm situated beyond the GC.}
\end{table*}
\begin{figure*}[h!]
    \centering
    \includegraphics[width=0.316\textwidth]{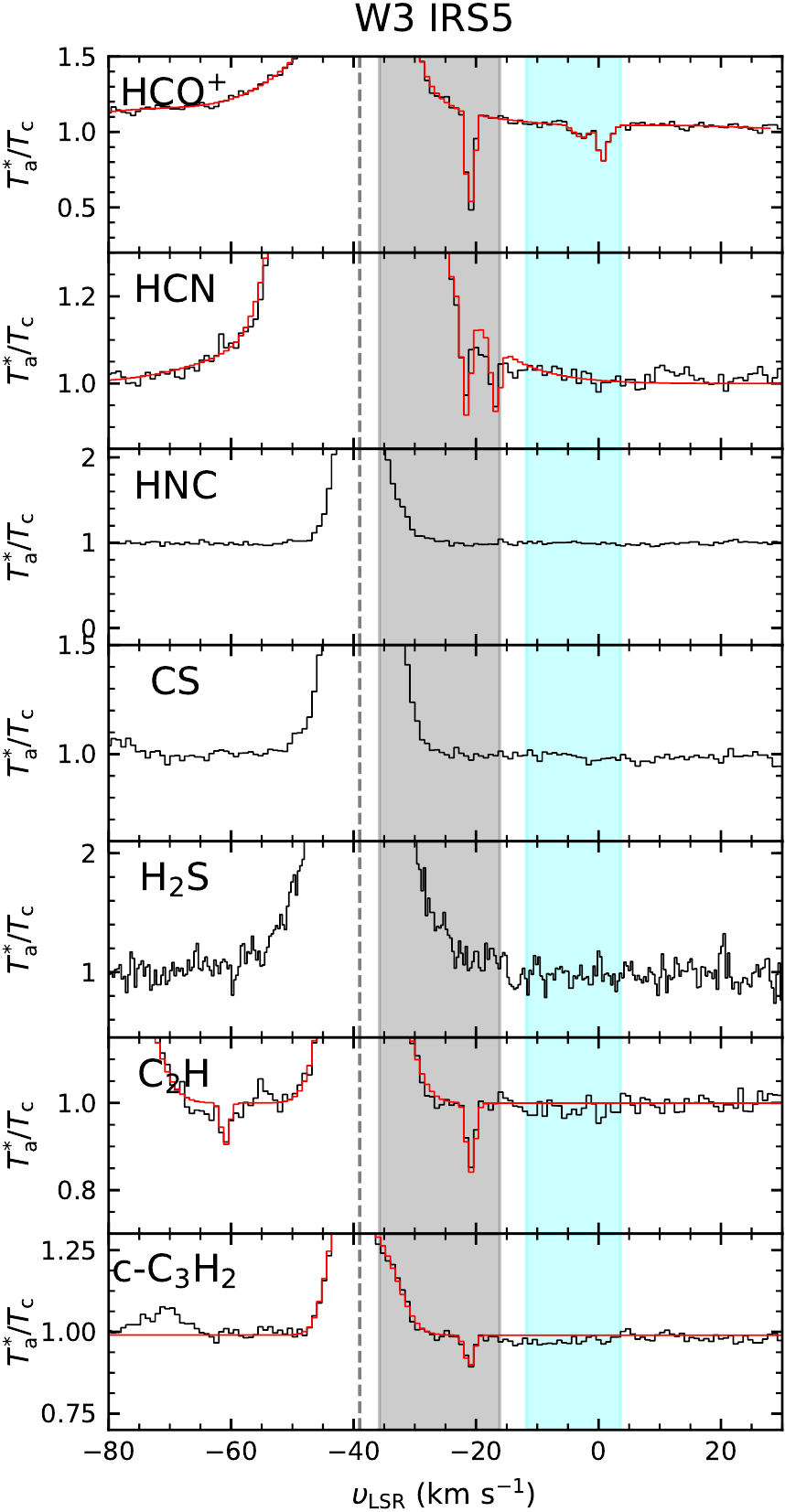} 
    \includegraphics[width=0.316\textwidth]{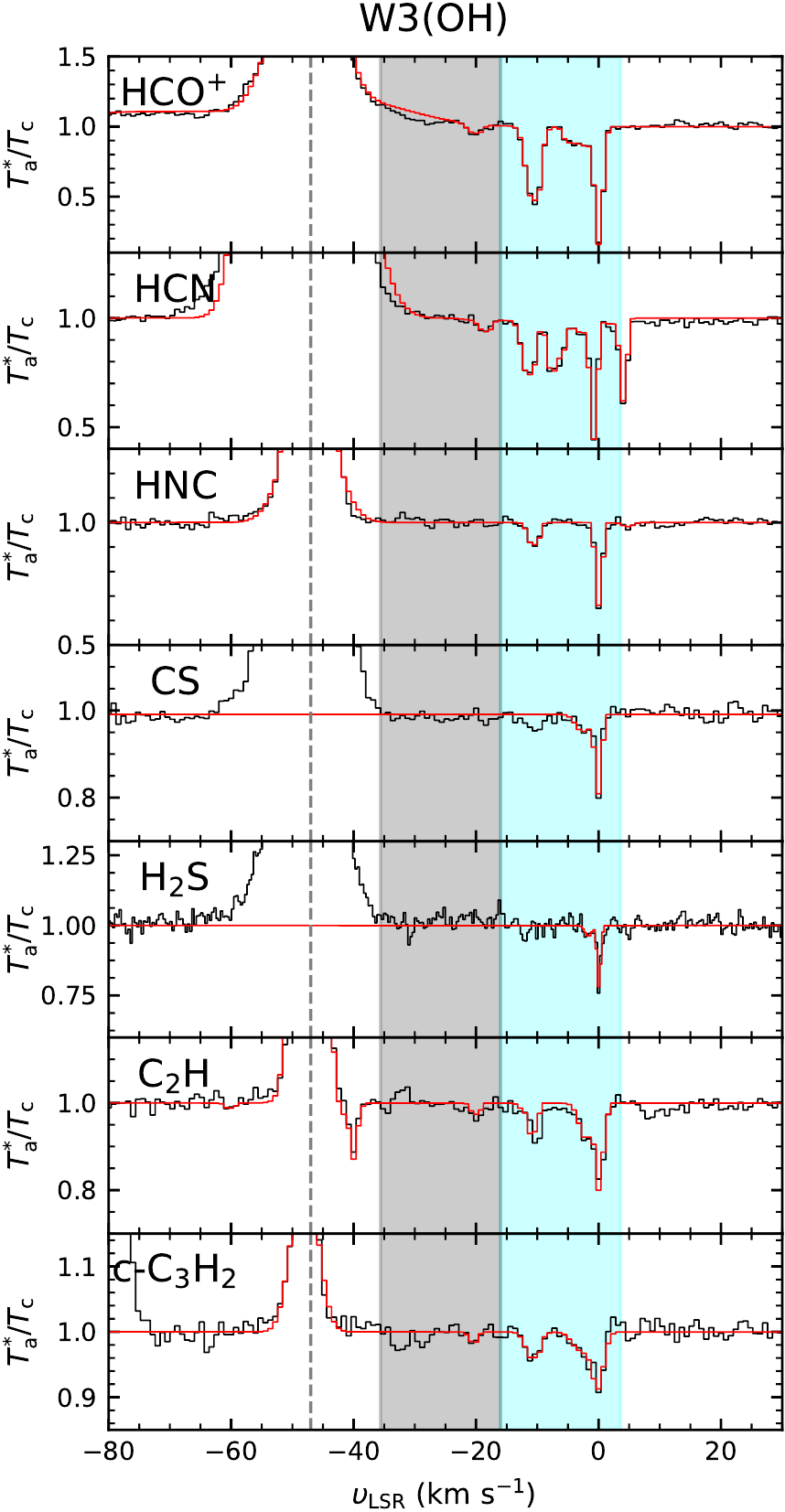}
    \includegraphics[width=0.316\textwidth]{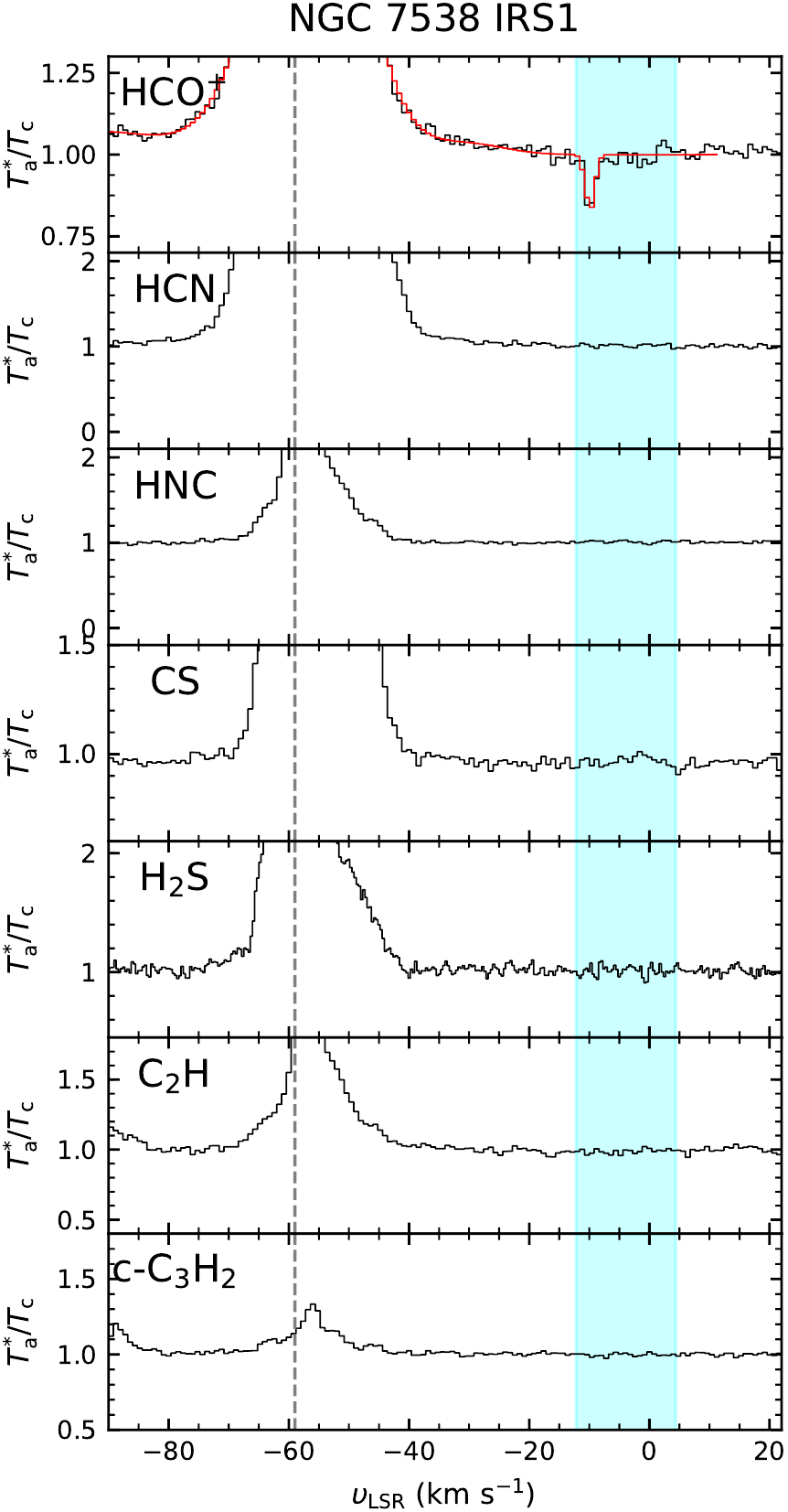} 
     \caption{Observed spectra (black) and the XCLASS modeled spectra (red) in a $T_{\rm a}^*/T_{\rm c}$ scale, toward W3 IRS5, W3(OH), and NGC 7538 IRS1 from left to right. The spectra in each panel correspond to \ce{HCO+}, HCN, HNC, CS, \ce{H2S}, \ce{C2H}, and \ce{c-C3H2}, from top to bottom. The shaded areas indicate velocity intervals of the local arm in cyan and the Perseus arm in gray. The $\varv_{\rm sys}$ of the continuum sources are indicated using vertical dashed lines. }
    \label{fig:abs_spec_w3}   
    \end{figure*}

\section{Results}\label{sec:results}
\subsection{Detected absorption lines}
Table\,\ref{tab:detections} lists the molecular species detected in absorption observed toward 14 out of the 15 millimeter continuum sources. The spectra obtained toward the last target listed in Table~\ref{tab:conti_sou}, G357.558$-$00.321, are not shown because the observations were not sensitive enough to detect the millimeter wave continuum from this weakest of the 15 sources. Some continuum sources have broad emission features of \ce{HCO+}, HCN, HNC, CS, and \ce{H2S} at their $\varv_{\rm sys}$, spanning a few tens of \kms. Several absorption lines are detected at the LSR velocities of such emission line wings and/or emission lines at $\varv_{\rm sys}$. It is not clear if the absorption lines arise from physically unrelated foreground clouds or if they are associated with the molecular gas of the star forming regions in which the continuum sources reside, in particular their dense envelopes. In some cases (see for e.g.,\ Fig.\,\ref{fig:abs_spec_g19_g32} toward G19.61$-$0.23 and G32.80$+$0.19) the absorption line components have distinct offsets in velocity from the emission lines and these components unambiguously trace foreground diffuse or translucent clouds. 

The continuum sources observed here are \hii\ regions and massive star-forming regions (see Paper I for the details of the HyGAL program). Some of them (e.g., G31.41$+$0.31) are chemically rich environments whose submillimeter and millimeter spectra exhibit a multitude of molecular emission lines associated with star-formation (e.g., outflows and hot molecular cores, \citealt{nony2018_w43mm1_hotcore, gorai2021}). The CS, \ce{H2S}, and \ce{c-C3H2} spectra are severely contaminated molecular lines emitted by the background continuum source, various of them unidentified. As a result, the detection rates of absorption features for these transitions are lower than those of the other species. Moreover, the \ce{c-C3H2} transition we targeted lies close to the \ce{HCS+} ($J=2\rightarrow1$) transition at 85.3479\,GHz \citep{Margules2003}, with a velocity separation of only $-$31.6\,\kms, and the H$_2$S transition is relatively close to the \ce{^{34}SO} ($J=4_4\rightarrow3_3$) transition at 168.8151\,GHz. At negative velocities with respect to the source systemic velocity, some \ce{c-C3H2} and \ce{H2S} absorption components -- the presence of which is suggested by corresponding absorption lines of the other molecular species (mainly \ce{HCO+}) -- are blended with the emission lines of \ce{HCS+} and \ce{^{34}SO}. For these cases, we simultaneously fit the \ce{HCS+} and \ce{^{34}SO} emission lines with the \ce{c-C3H2} and \ce{H2S} absorption lines, respectively. 
In total, we have detected \ce{HCO+} absorption features in the spectra of all 14 sightlines listed in Table\,\ref{tab:detections} and so velocities of \ce{HCO+} absorption components were used as a reference for identifying absorption line features in the six other molecular line transitions studied. A significant number of sightlines also have detectable absorption lines of HCN (12 sightlines), HNC (11 sightlines), \ce{C2H} (12 sightlines), and \ce{c-C3H2} (nine sightlines) transitions. CS and \ce{H2S} are only detected toward eight and six sources, respectively, because of the poorer S/N at the higher H$_2$S transition frequency or significant emission line contamination.

\begin{figure}[]
    \centering
    \includegraphics[width=0.316\textwidth]{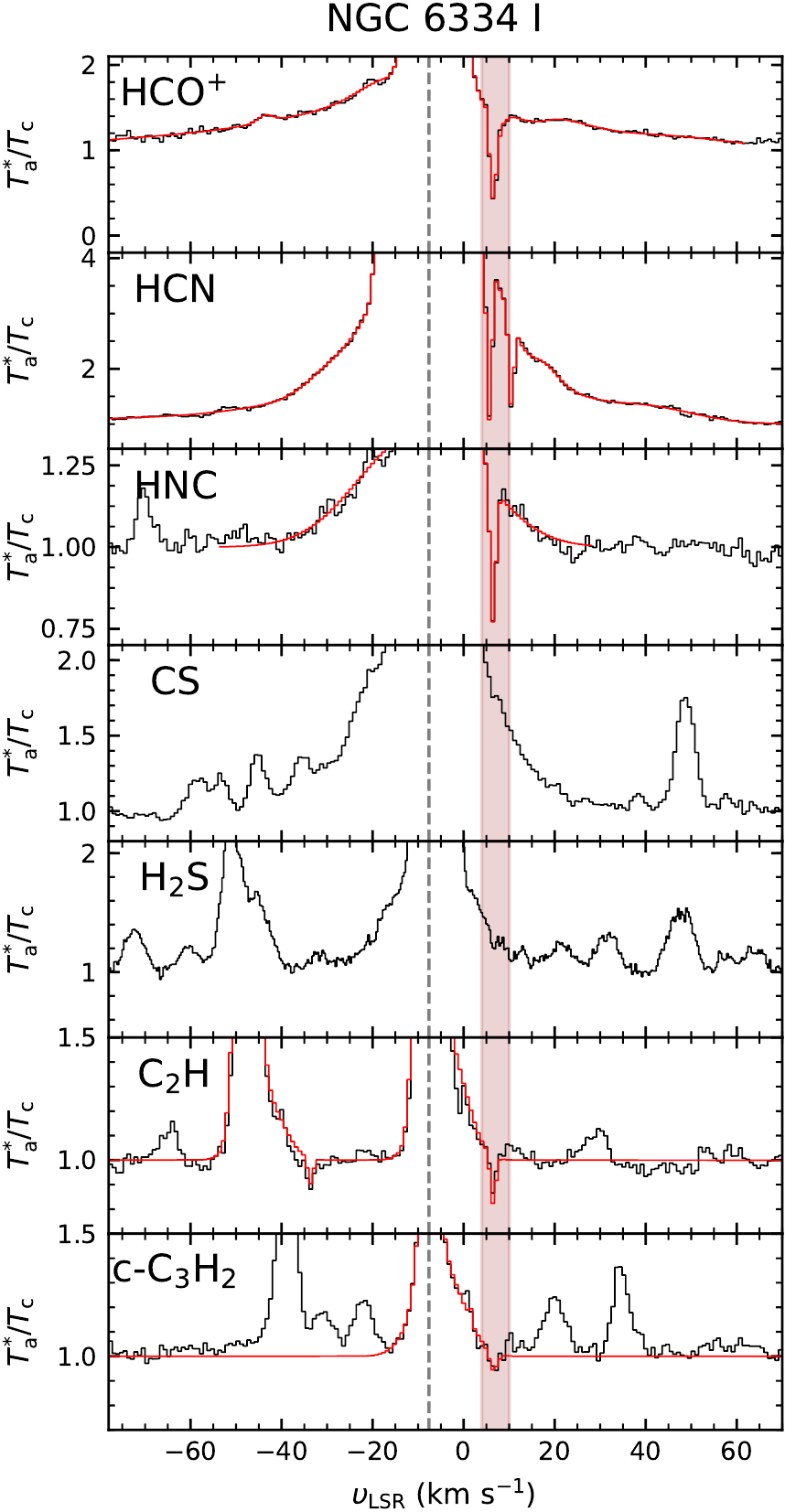}
    \includegraphics[width=0.316\textwidth]{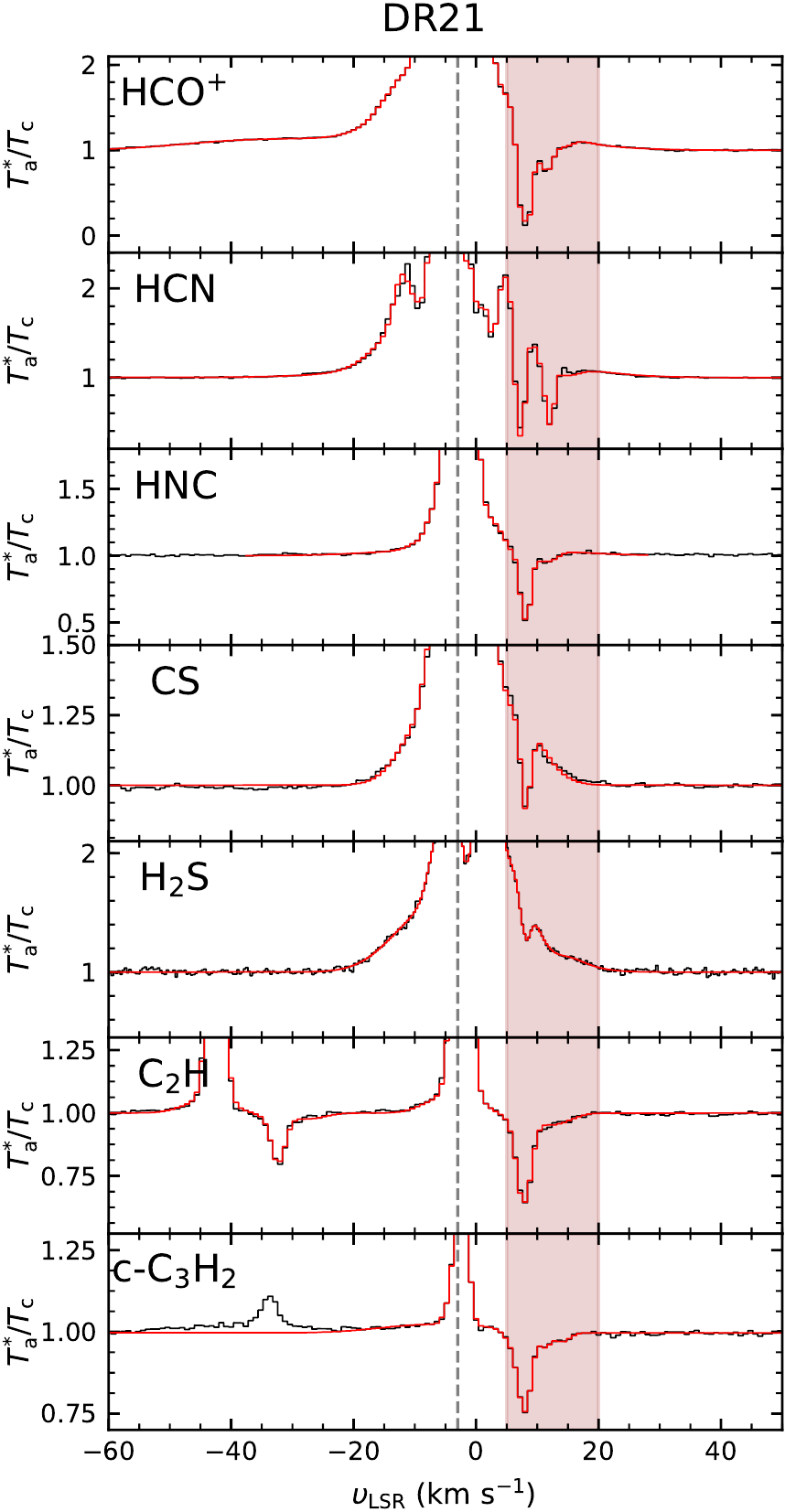}
     \caption{Same as Fig.\,\ref{fig:abs_spec_w3} but toward NGC 6334 I and DR21. The brown shaded areas indicate velocity intervals of the molecular cloud envelope layers or unidentified foreground clouds toward these sightlines.}
    \label{fig:abs_spec_ngc6334_dr21}   
    \end{figure}

Figures\,\ref{fig:abs_spec_w3} - \ref{fig:abs_spec_g19_g32} present the observed spectra, normalized with respect to the continuum flux and displayed as a function of $\varv_{\rm LSR}$. Fits to the spectra, obtained with the eXtended CASA Line Analysis Software Suite (XCLASS, \citealt{moeller2017}), are overlaid in red. The shaded areas designate the velocity intervals of specific spiral arms, inter-arm regions, and unidentified foreground clouds or the envelope layers of dense molecular cloud associated with the continuum sources; their velocity intervals, defined by previous observational studies (see Paper I and references therein), are tabulated in Table\,\ref{tab:arm_info}. In emission, HCN ($J=1\leftarrow0$) has three hyperfine components ($F=1\leftarrow1$, $F=2\leftarrow1$, and $F=0\leftarrow1$) that are completely blended, leading to even broader observed line widths. In absorption, however, the narrower intrinsic line widths in diffuse and translucent clouds often permit all three hyperfine structure (hfs) components to be individually discerned. We see blended HCN hfs profiles in absorption toward some sightlines because several superposed foreground clouds have similar velocity ranges. \ce{C2H} ($J=1\leftarrow0$) is another molecular transition with multiple hyperfine components. This line has six hfs components, with the two weakest components being well separated and each of the two brightest components being blended to the remaining two components. We defined the frequency of the brightest component ($J=3/2\leftarrow1/2$, $F=2\leftarrow1$, 87.3168\,GHz) of \ce{C2H} as the rest frequency for purposes of computing the $\varv_{\rm LSR}$. \ce{C2H} is found to be ubiquitous in the Milky Way, and thus it often traces multiple velocity components. \ce{HCO+} ($\mu_{\rm A}=3.90$), HCN ($\mu_{\rm A}=2.984$) and HNC ($\mu_{\rm A} =3.04)$ have much larger dipole moments than CO ($\mu_{\rm A} = 0.110$) and can therefore be detected in absorption far more easily than CO in low-density clouds.

\subsection{Detected absorption line profiles}

\subsubsection{W3 IRS5, W3(OH), and NGC 7538 IRS1}
The second quadrant sources W3 IRS5, W3(OH), and NGC 7538 IRS1 are located in the Perseus arm in the outer Galaxy. W3 IRS5 and W3(OH) harbor high mass young stellar objects in different evolutionary stages that belong to the W3 molecular clouds and \hii\ region complex, while NGC 7538 IRS1 is the most prominent of several such sources in the NGC 7538 region. As in the host arm of the background sources, the Perseus arm, the sightlines toward these three sources are also aligned along the local arm where they cover a velocity range from $\sim-16$\,\kms\ to 7\,\kms. In the spectra toward W3 IRS5 shown in the left panel of Fig.\,\ref{fig:abs_spec_w3}, three components in \ce{HCO+} absorption are detected at velocities of $-20$, $-3$, and $0.7$\,\kms, respectively, where the $-20$\,\kms\ component is the deepest one. Only the $-20$\,\kms\ component appears in HCN, \ce{C2H}, and \ce{c-C3H2}. Toward W3(OH) (the middle panel of Fig.\,\ref{fig:abs_spec_w3}), we find three \ce{HCO+} components at velocities ($-20$, $-2$ and 0\,\kms) similar to those seen toward W3 IRS5 and also notice an additional component at $-10$\,\kms, not detected in the sightline to W3 IRS5. The deepest feature along this sightline is at 0\,\kms, with that at $-10$\,\kms\ being the next deepest. These velocity components are also detected in the spectra of the six other species except for the $-20$\,\kms\ feature, which is not observed in the HNC and \ce{H2S} spectra. 

Toward NGC 7538 IRS1, we only detect a single \ce{HCO+} component at $-10$\,\kms, as shown in the right panel of Fig.\,\ref{fig:abs_spec_w3}. All the detected absorption components toward these three sightlines are associated with the local arm, except for the $-20$\,\kms\ component toward W3 IRS5. This $-20$\,\kms\ feature likely arises from diffuse clouds belonging to the Perseus arm rather than the envelope of the molecular cloud as \ce{C2H} and \ce{c-C3H2} absorption lines show a clear separation from the background emission. In addition, the spectra of CH, OH, \ce{OH+}, O, \ce{C+}, and \ce{ArH+} from the SOFIA HyGAL observations in Paper I all show absorption line dips around this velocity. The hydride, O, and \ce{C+} absorption lines show variations in absorption line profiles for W3(OH) and W3 IRS5, also seen here in the millimeter spectra.

\subsubsection{NGC 6334 I and DR21}
NGC 6334 I and DR21 are in the vicinity of the solar neighborhood with heliocentric distances of 1.3 and 1.5\,kpc, respectively. However, these sources reside in different spiral arms, namely the Sagittarius arm and the local arm, respectively. Toward NGC 6334 I (upper panel of Fig.\,\ref{fig:abs_spec_ngc6334_dr21}), we find two distinct \ce{HCO+} absorption components at $+$7\,\kms\ and $+$14\,\kms, respectively, of which the first is the most prominent one, while the latter has a significantly broader profile ($\Delta\varv \sim 8$ \kms). HF (an excellent tracer of molecular gas, including CO-dark gas) absorption at 243~$\mu$m \citep{vanderwiel2016_ngc6334i_hf}, shows foreground clouds at $+$6 and $+$8\,\kms\ that are not considered to be physically associated to the NGC 6334 molecular cloud. The \ce{HCO+} absorption component at $+$7\,\kms\ is believed to arise from the same clouds in which HF is detected. The richness of molecular emission lines prohibits a convincing absorption line detection in the CS and \ce{H2S} spectral windows while the other species show an absorption feature close to the $+$7\,\kms\ velocity of \ce{HCO+}. 

Toward DR21 (lower panel of Fig.\,\ref{fig:abs_spec_ngc6334_dr21}), we detected absorption features for all seven species, of which the two small hydrocarbons (\ce{C2H}, \ce{c-C3H2}) show a broader line shape than the other species without a red-shifted emission wing indicating an outflow as traced by HCN, CS, and \ce{H2S}. The deepest absorption lines seen at $+$8.4\,\kms\ for all species originate from diffuse, extended gas surrounding the W75N cloud for which the majority of the emission is seen at $+$9\,\kms\ \citep{schneider2010_hcop_dr21}. 

\begin{figure}[]
    \centering
    \includegraphics[width=0.315\textwidth]{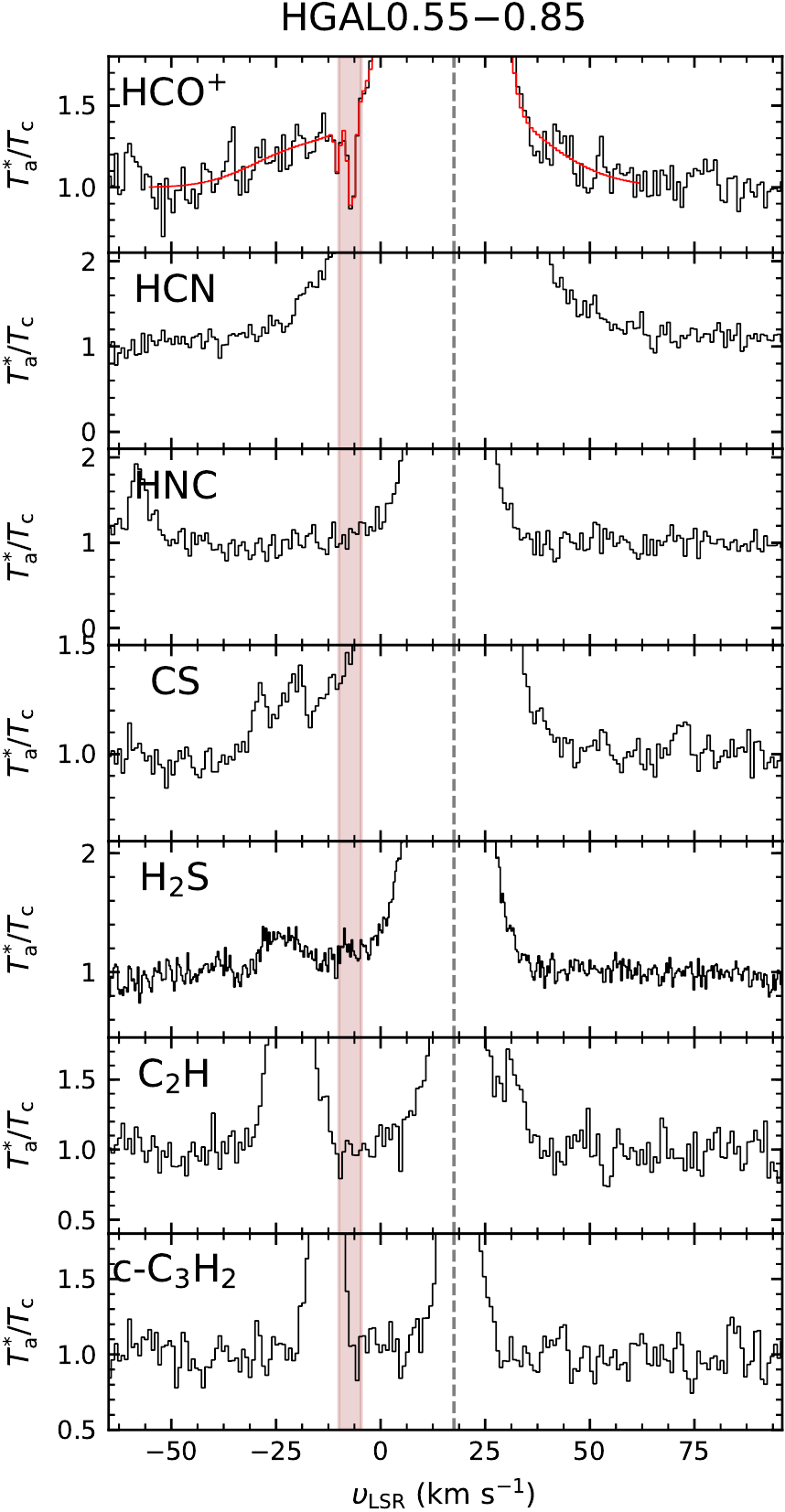}    
    \includegraphics[width=0.315\textwidth]{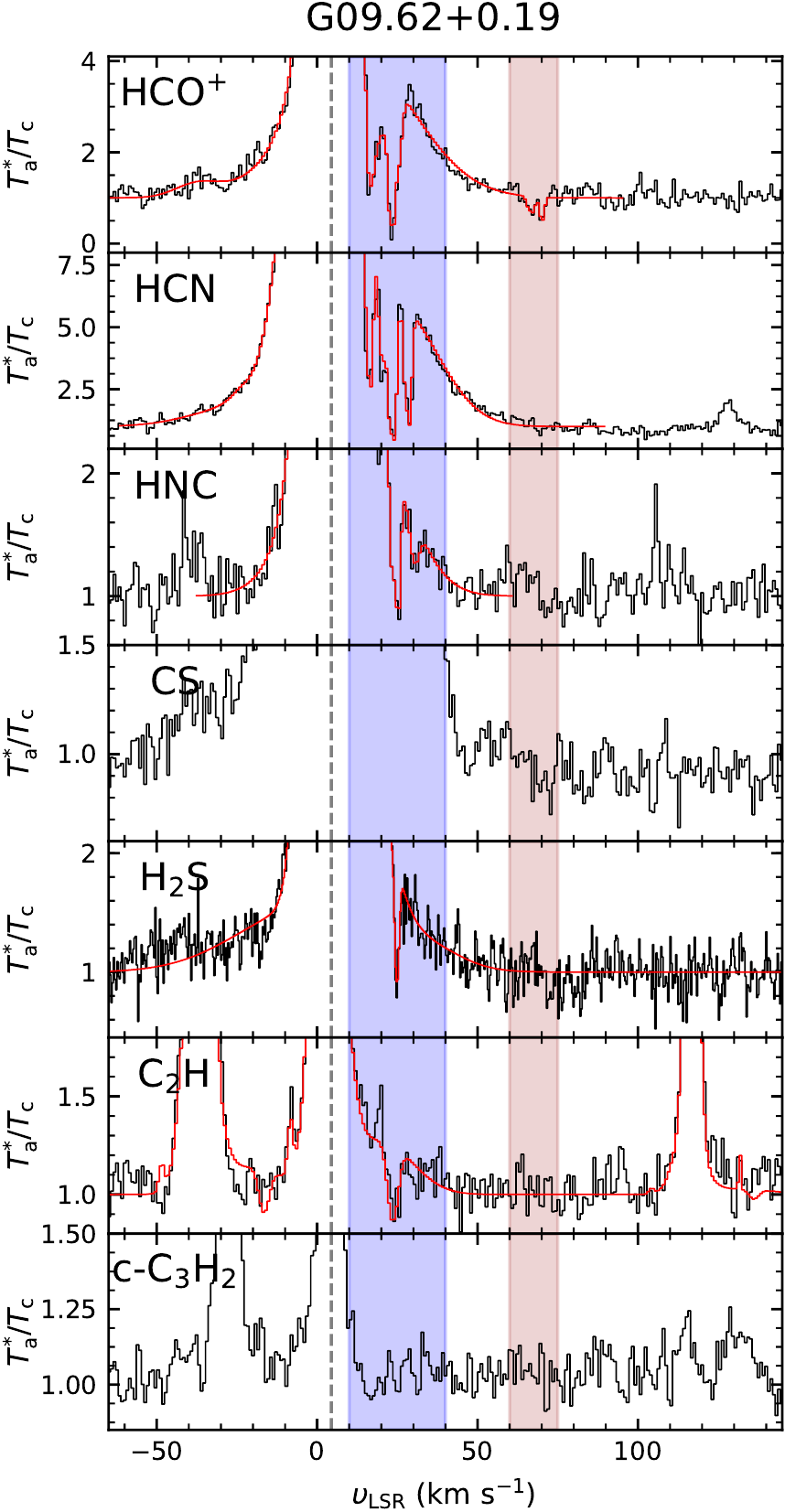}    
     \caption{Same as Fig.\,\ref{fig:abs_spec_w3} but toward HGAL0.55$-$0.85 and G09.62$+$0.19. The brown and blue shaded areas indicate velocity intervals corresponding to the molecular cloud envelope layers or unidentified foreground clouds, and the Scutum-Centaurus arm, respectively.}
    \label{fig:abs_spec_hgal0p55_g09}   
    \end{figure}        
\subsubsection{HGAL0.55$-$0.85 and G09.62$+$0.19}
HGAL0.55$-$0.85 is located in the Galactic Center region and shows broad emission wings in lines from the \ce{HCO+}, HCN, and CS molecules. As seen in the upper panel of Fig.\,\ref{fig:abs_spec_hgal0p55_g09}, only two narrow absorption components are found in the \ce{HCO+} spectrum at $-$7 and $-$10\,\kms, respectively, in the blue-shifted \ce{HCO+} emission wing. 

G09.62$+$0.19 lies in the Norma spiral arm near the expanding 3\,kpc arm. In addition, the sightline crosses the Scutum-Centaurus and Sagittarius arms over a range of velocities from $+$10 to $+$40\,\kms. Our mm band survey detected three distinct \ce{HCO+} absorption features at $+$16, $+$23, and $+$69\,\kms\ as shown in the lower panel of Fig.\,\ref{fig:abs_spec_hgal0p55_g09} toward G09.62$+$0.19. The first two features have a large velocity shift from the systemic velocity of the embedded \hii\ regions ($-$5--0\,\kms), which is close to velocities of dense molecular cores \citep[$+2$ to $+$5\kms,][]{liu2017_g09p62_co_abs,liu2020_g09p62_hcop_abs} belonging to the Scutum-Centaurus arm. For all other transitions except for CS and \ce{c-C3H2} which we fail to fit because of the broad line emission from the background source, only a single absorption component is detected at $+$23\,\kms. The weak absorption feature at $+$69\,\kms\ only appears in \ce{HCO+} and does not match any identified spiral arm. 

\begin{figure}[h!]
    \centering
    \includegraphics[width=0.315\textwidth]{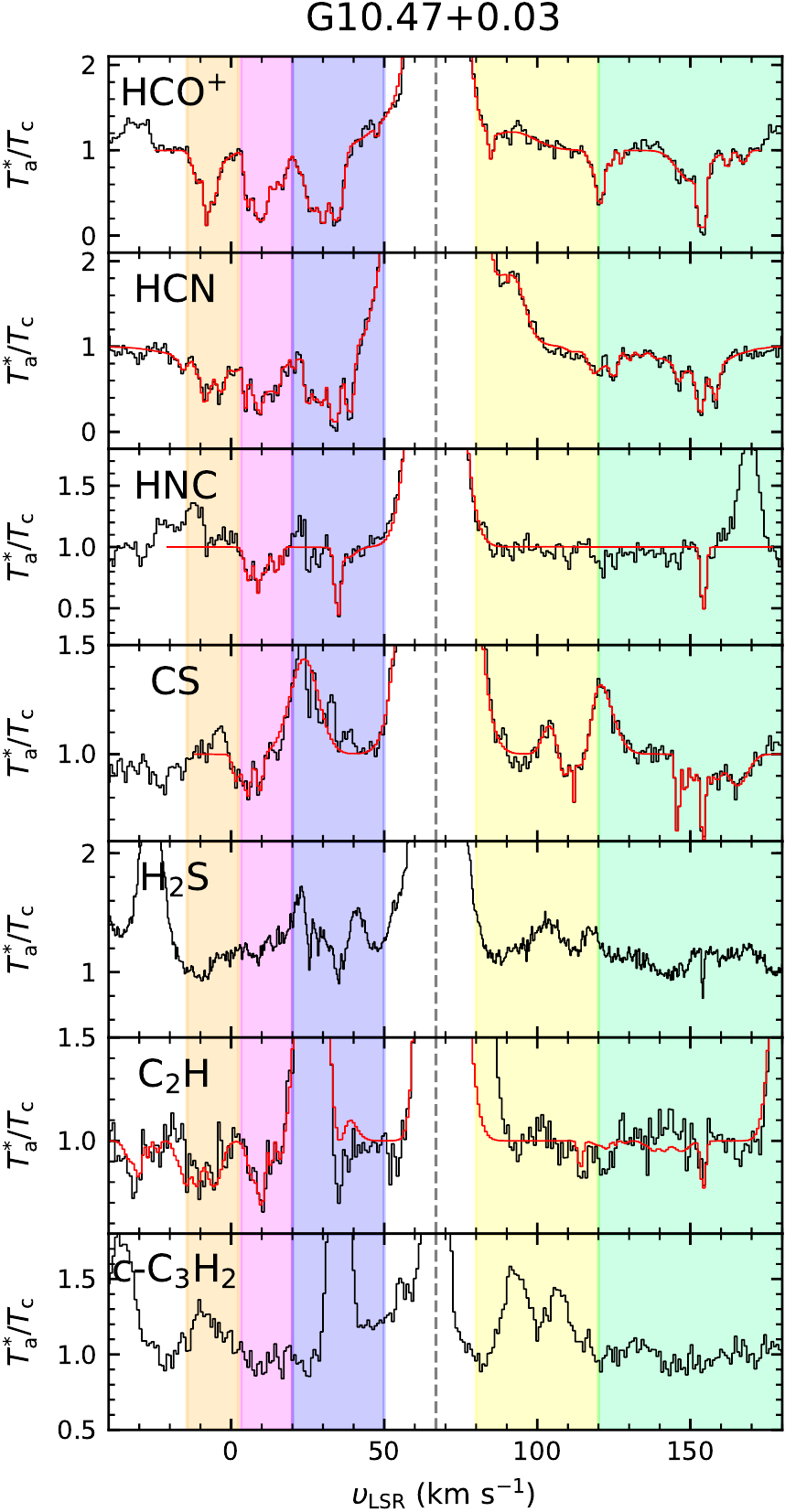}
    \caption{Same as Fig.\,\ref{fig:abs_spec_w3} but toward G10.47$+$0.03. The shaded areas indicate the velocity intervals of specific spiral arms (Sagittarius-Carina arm in magenta, Scutum-Centaurus arm in blue, inter-arm gas in orange, the GC/bar in yellow, and the 135\,\kms\ arm located beyond the GC in blue-green).}
    \label{fig:abs_spec_g10p47}
\end{figure}
\begin{figure}[h!]
    \centering
    \includegraphics[width=0.315\textwidth]{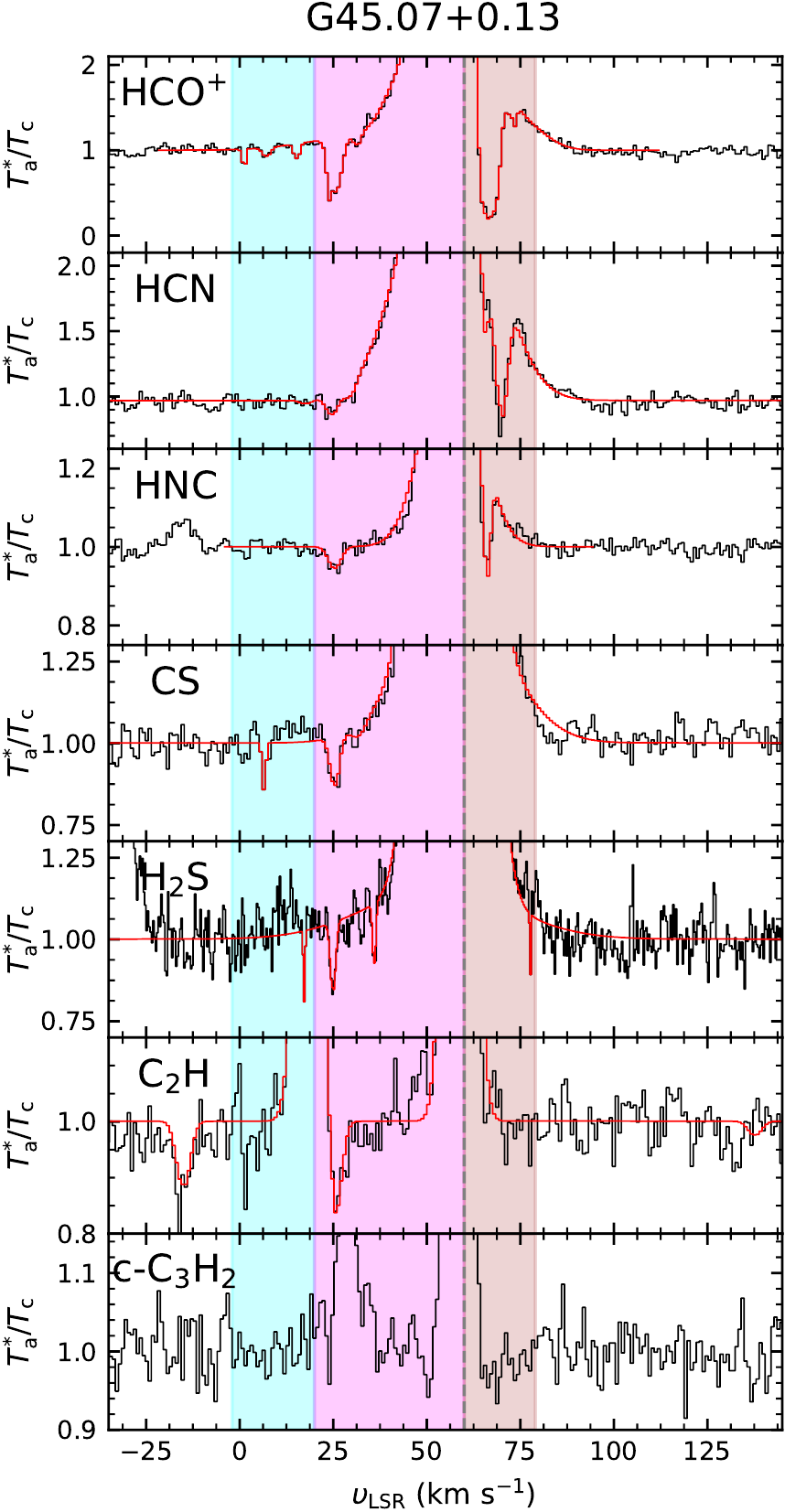} 
    \caption{Same as Fig.\,\ref{fig:abs_spec_w3} but toward G45.07$+$0.13. The shaded areas indicate the velocity intervals of specific spiral arms (local arm in cyan and Sagittarius-Carina arm in magenta), and the molecular cloud envelope or unidentified foreground clouds in brown.}
    \label{fig:abs_spec_g45p07}
\end{figure}

\begin{figure*}[h!]
    \centering
    \includegraphics[width=0.315\textwidth]{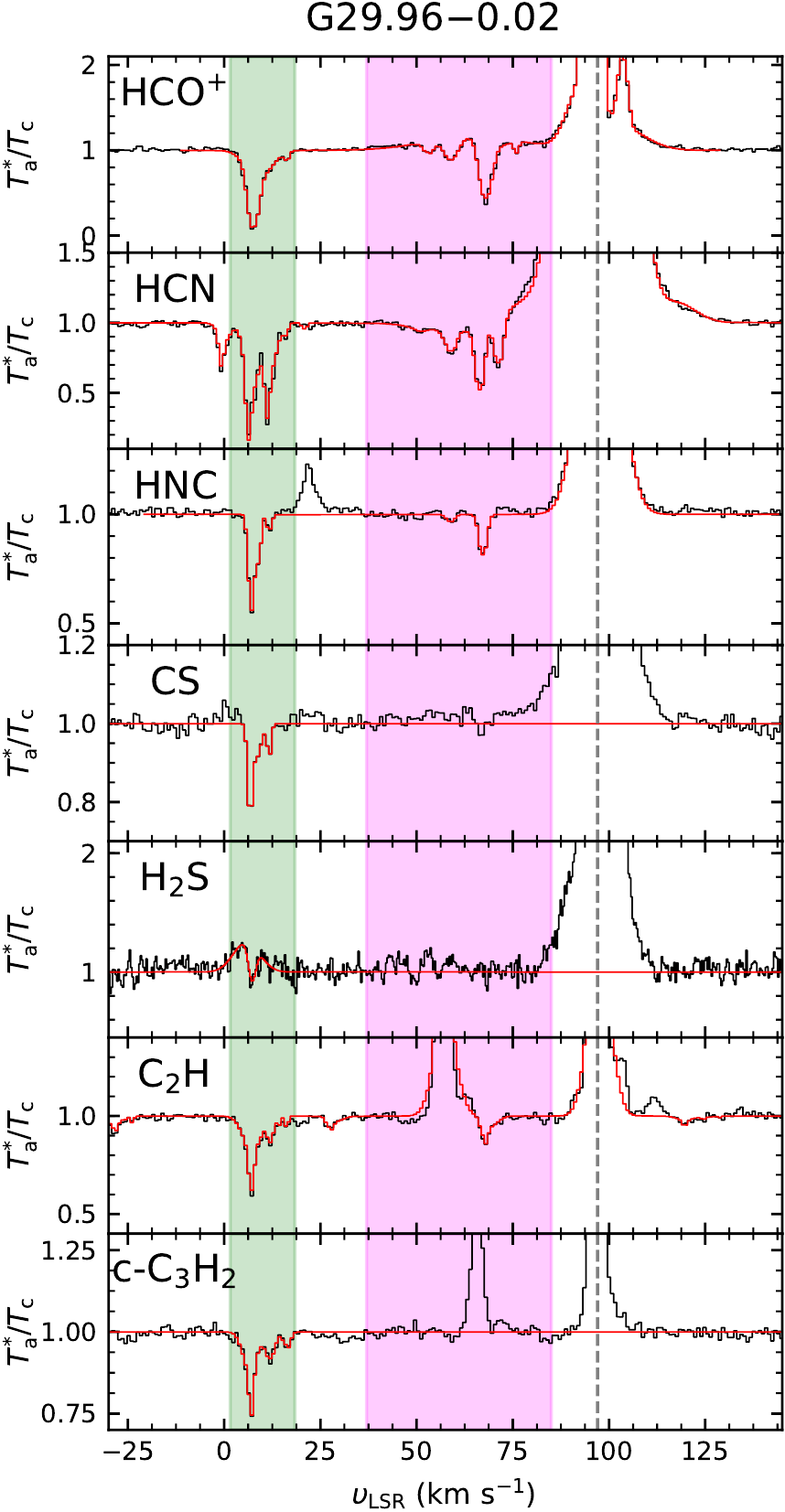}
    \includegraphics[width=0.315\textwidth]{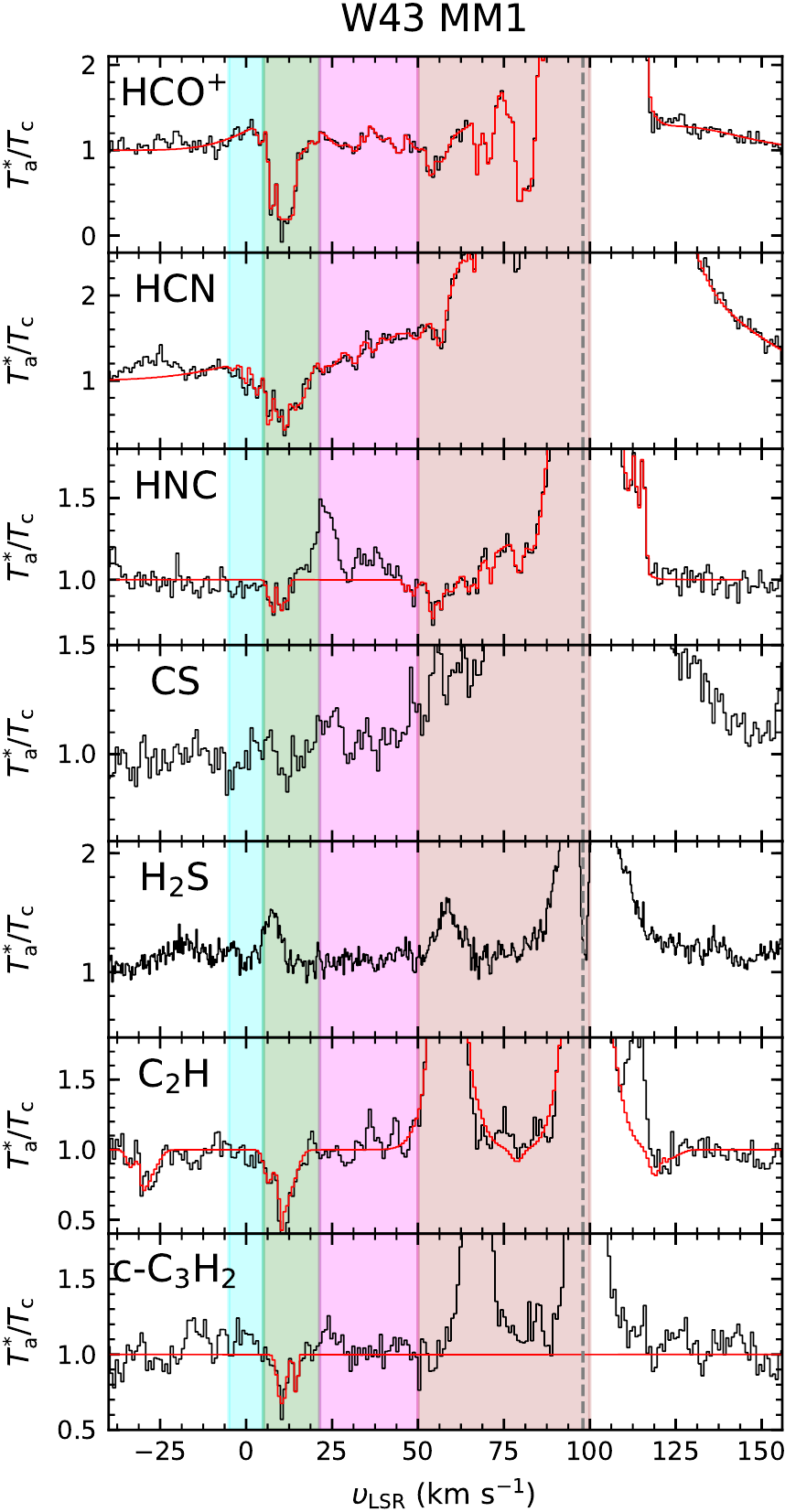} 
    \includegraphics[width=0.315\textwidth]{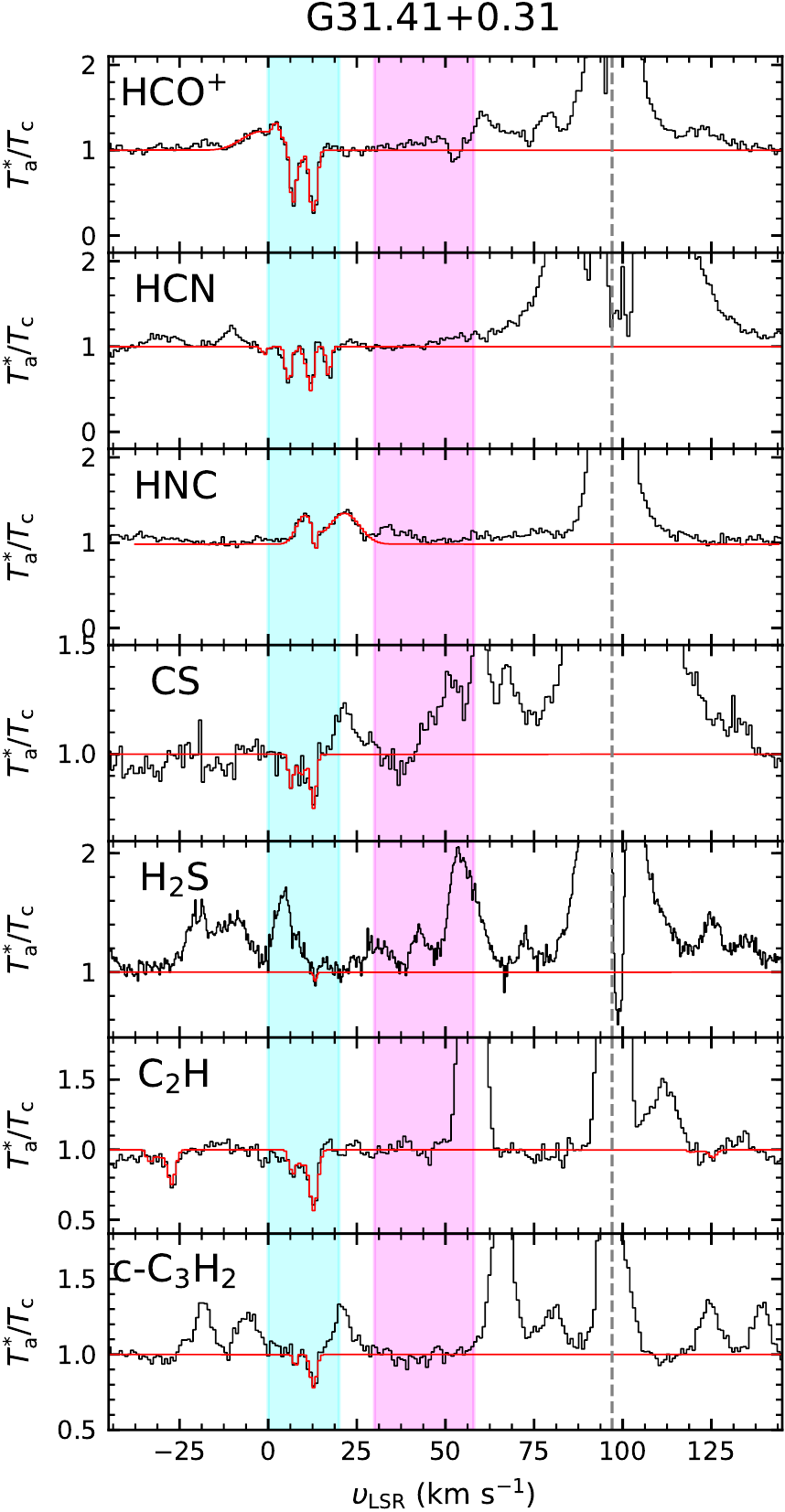}
    \caption{Same as Fig.\,\ref{fig:abs_spec_w3} but toward G29.96$-$0.02, W43 MM1, and G31.41$+$0.31. The shaded areas indicate the velocity intervals of specific spiral arms (local arm in cyan, Sagittarius-Carina arm in magenta, Aquila-Rift in green), and the molecular cloud envelope or unidentified foreground clouds in brown.}
    \label{fig:abs_spec_g29_w43_g31}
\end{figure*}

\subsubsection{G10.47$+$0.03}
This background source lies in the inner Galaxy, behind the GC and the Galactic bar. Thus, several spiral arms pass through its sightline and they cause \ce{HCO+} absorption features spanning a wide velocity range from $-$28 to $+$179\,\kms, while background emission ranges between $+$52 and $+$83\,\kms, as shown in Fig.\,\ref{fig:abs_spec_g10p47}. Such widespread absorption line features are also detected in the hydrides (\ce{ArH+}, \ce{o-H2O+}, \ce{OH+}, and CH; \citealt{jacob2020}), but the hydride absorption lines are broadly continuous, unlike what is seen in the \ce{HCO+} absorption line profiles, which shows well separated components in distinct velocity intervals. The millimeter absorption lines in the velocity interval between $-$28 and $+$52\,\kms\ originate from foreground clouds in inter-arm gas ($-28$ to $+2.4$\,\kms), the near-side crossing of the Sagittarius arm ($+3.2$ to $+20$\,\kms), and the Scutum-Centaurus arm ($+20$ to $+52$\,\kms). Surprisingly, the inter-arm \ce{HCO+} absorption feature shows a similar absorption depth, $T_{\rm a}^*/T_{\rm c}$, when compared to absorption lines arising in the spiral arms. However, the \ce{HCO+} absorption lines belonging to the Scutum-Centaurus arm and the near-side of the Sagittarius arm display flat absorption dip profiles close to being saturated. Such broad features are also seen in the HCN spectra, while the spectra of the other species have mostly narrow absorption features. We notice that these \ce{HCO+} absorption line features cover similar velocities as the \ce{OH+}line at 1033\,GHz line \citep{jacob2020} but have much narrower widths. The absorption around 120\,\kms\ is likely related to diffuse gas in the 3\,kpc arm and the Galactic bar ($\varv > +80$\,\kms), whereas the components at velocities greater than 120\,\kms\ are believed to be associated with the $+$135\,\kms\ spiral arm \citep{Sormani2015MNRAS_135kms_arm} located beyond the GC. Between these two velocity features, the \ce{HCO+} 154\,\kms\ feature is deeper than that belonging to the 3\,kpc arm, while the deepest absorption component of \ce{OH+} is at 120\,\kms\ \citep{jacob2020}. The 154\,\kms\ component identified in \ce{HCO+} is also detected in the spectra of HCN, HNC, \ce{C2H}, CS, and \ce{H2S}. However, the 120\,\kms\ feature is barely seen in HCN, CS, and \ce{C2H} partly due to emission line contamination. As a well known hot core source, G10.47$+$0.03 is rich in complex molecules \citep{widicus_weaver2017_hotcore}. For the \ce{H2S} and \ce{c-C3H2} spectra, we detect multiple, strong emission components that further complicates the assignment and fitting of absorption line features.

\subsubsection{G45.07$+$0.13}
This background source lies in the Sagittarius arm and has an outflow, causing a broad blue-shifted wing \citep{hunter1997_g45p07_cs_abs} in the emission profiles of lines from the \ce{HCO+}, HCN, and CS molecules. The absorption feature seen at 24\,\kms\ in the millimeter transitions, displayed in Fig.\,\ref{fig:abs_spec_g45p07} is believed to belong to its host spiral arm spanning a velocity interval between 20 and 60\,\kms. The conspicuous absorption component at 67\,\kms\ being red-shifted from the $\varv_{\rm sys}$ (60\,\kms) in \ce{HCO+}, HCN, and HNC likely traces infalling foreground gas toward the continuum source as it is similar to that of the CS (2$\leftarrow$1) absorption feature \citep{hunter1997_g45p07_cs_abs} near 65$-$67.3\,\kms\ obtained from interferometric observations revealing infalling gas motion. In addition, we also detect weak absorption lines related to the local arm (0 to 20\,\kms) at 1, 8, and 15\,\kms\ in the \ce{HCO+} spectrum. 
\begin{figure*}[]
    \centering
    \includegraphics[width=0.316\textwidth]{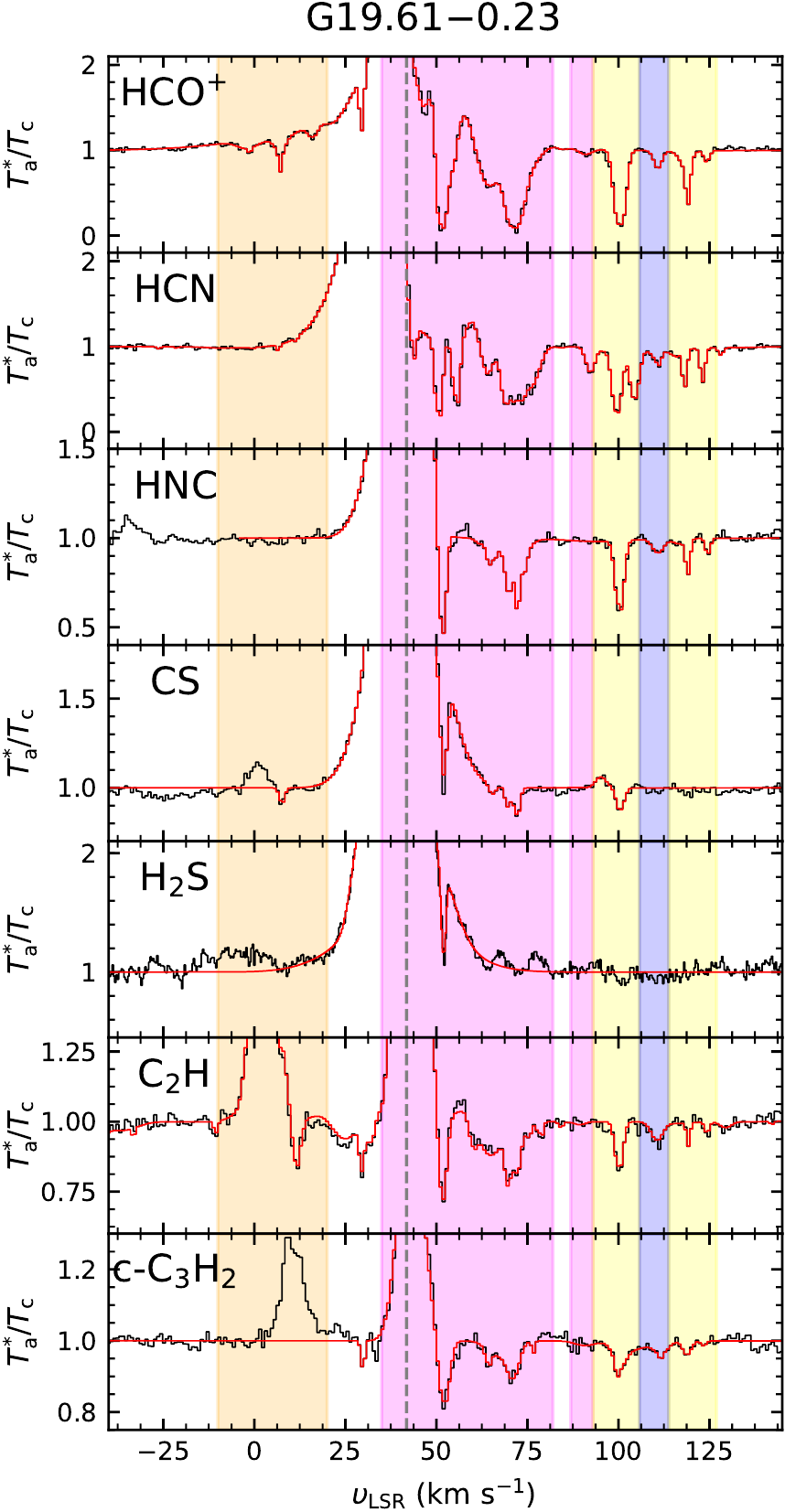}
    \hskip 1 cm
    \includegraphics[width=0.316\textwidth]{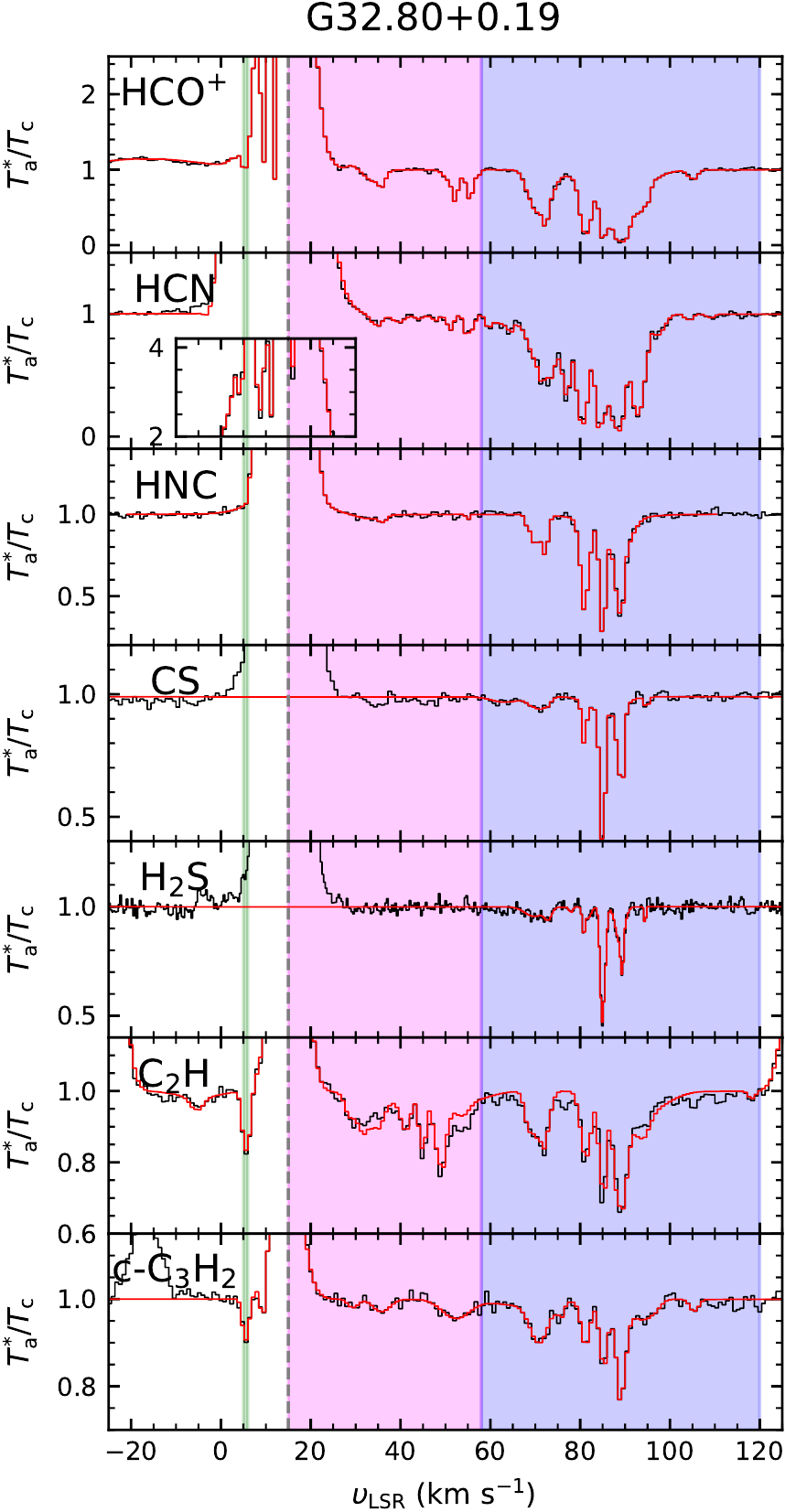}
    \caption{Same as Fig.\,\ref{fig:abs_spec_w3} but toward G19.61$-$0.12 and G32.80$+$0.19. The shaded areas indicate velocity intervals of specific spiral arms (Aquila-Rift in green, Sagittarius-Carina arm in magenta, Scutum-Centaurus arm in blue, inter-arm gas in orange, the GC/bar in yellow), and the molecular cloud envelope or unidentified foreground clouds in brown. Toward G32.80$+$0.19, the insert in the HCN panel zooms in on the HCN absorption features on top of the emission within a velocity window between $-$10\,\kms\ and 30\,\kms.}
    \label{fig:abs_spec_g19_g32}
\end{figure*}

\subsubsection{G29.96$-$0.02, W43 MM1, and G31.41$+$0.31} 
The sightlines toward these three background sources, located in the Scutum-Centaurus arm, are in a similar direction. Along the sightline toward G29.96$-$0.02 (left panel of Fig.\,\ref{fig:abs_spec_g29_w43_g31}), we detected broad, deep absorption features from $-0.6$ to $+$23\,\kms\ in the \ce{HCO+} spectrum, related to the Aquila Rift (Paper I), while multiple absorption components in the range from $+$49 to $+$79\,\kms\ correspond to the near-side crossings of the Sagittarius spiral arm. All of these components are found in other millimeter transitions, except those originating in the Sagittarius arm, which are not detected in the \ce{H2S}, and \ce{c-C3H2} spectra at rms levels, 12.5 and 7.5\,mK, respectively. The \ce{HCO+} absorption line at 8\,\kms\ has a significantly broader profile, while the absorption profiles of HNC, CS, and the two small hydrocarbons show multiple dips within the velocity interval of the Aquila-Rift region. Such multiple dips are also found in \hi\ absorption spectrum \citep{fish2003_g29p96_hi_abs} between $-0.6$ and $+$23\,\kms, showing at least four well separate absorption components. All detected absorption features in these millimeter transitions have counterparts of absorption components in the spectrum of \ce{o-H2Cl+} \citep{neufeld2015}.

W43 MM1 and G31.41$+$0.31 are situated in the Scutum-Centaurus spiral arm and classified as hot cores \citep{nony2018_w43mm1_hotcore, widicus_weaver2017_hotcore} which give rise to a variety of emission line features at millimeter wavelengths. Due to forests of emission lines blending into the absorption features, we only fit certain absorption features, causing that the model fits to some of the absorption features have greater uncertainties others. \ce{HCO+} absorption features toward W43 MM1 (the middle panel of Fig.\,\ref{fig:abs_spec_g29_w43_g31}) appear intermittently from $+$1 to $+$85\,\kms\ and among them, the absorption components with a dip at 11\,\kms\ traces diffuse gas from the Aquila Rift. Some of the \ce{HCO+} absorption features within this velocity range partly correspond to the near-side crossing of the Sagittarius arm (roughly $+$24 to $+$50\,\kms) and are possibly related to local diffuse gas from its host spiral arm, the Scutum-Centaurus arm ($+$50 to $+$100\,\kms) with $\varv_{\rm sys}=+$98\,\kms. It is, however, not clear whether all the absorption features arise in the Scutum-Centaurus arm or partly related to inter-arm gas or to some other region. 

Several absorption lines in the observed transitions are detected along the sightline toward G31.41+0.31 (right panel of Fig.\,\ref{fig:abs_spec_g29_w43_g31}), but we have only fit components associated with the local arm (0 to 20\,\kms) because its velocity range is less affected by emission contamination. In the \ce{HCO+} spectra, both absorption components at 7 and 13\,kms\ have a similar depth, but for other molecular species, the features at 7\kms\ are much weaker than those at 13\,\kms.

\begin{figure*}[h!]
    \centering
    \includegraphics[width=0.33\textwidth]{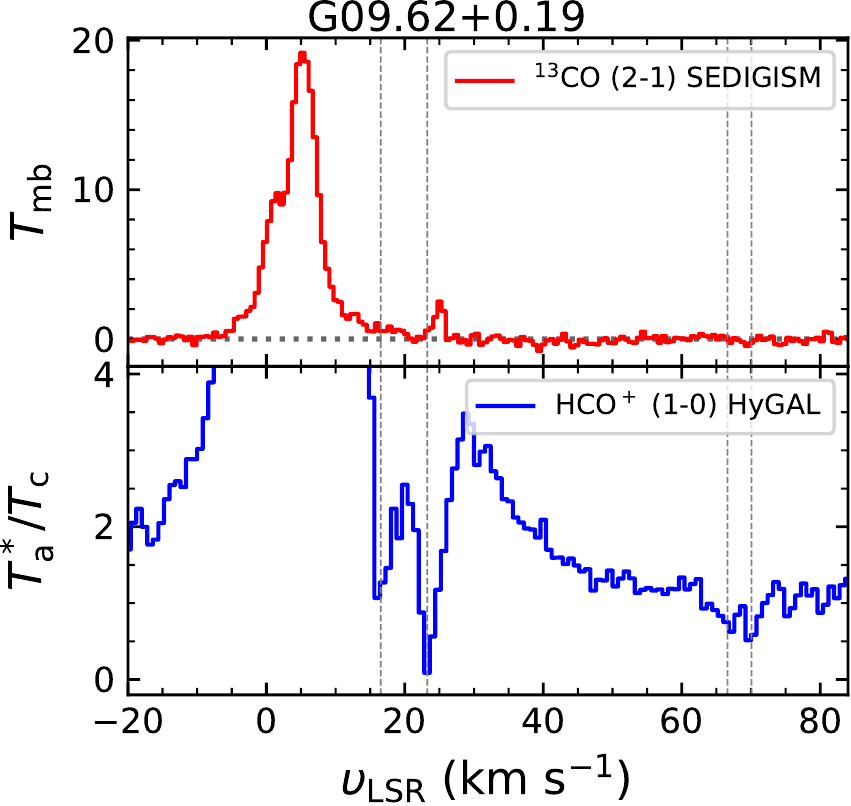}   
    \includegraphics[width=0.33\textwidth]{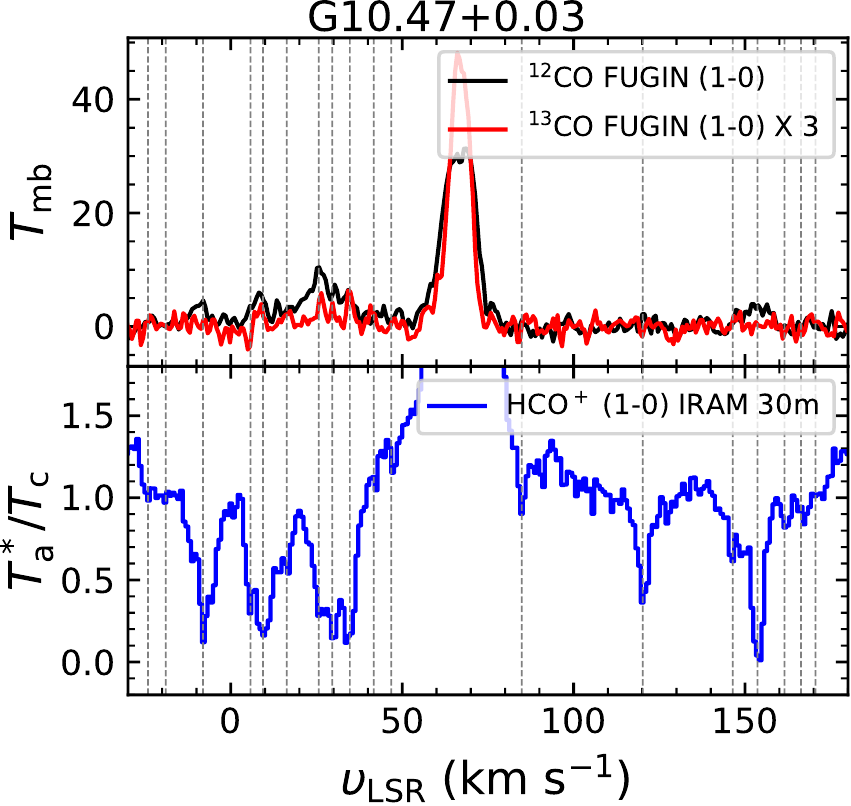}
    \includegraphics[width=0.33\textwidth]{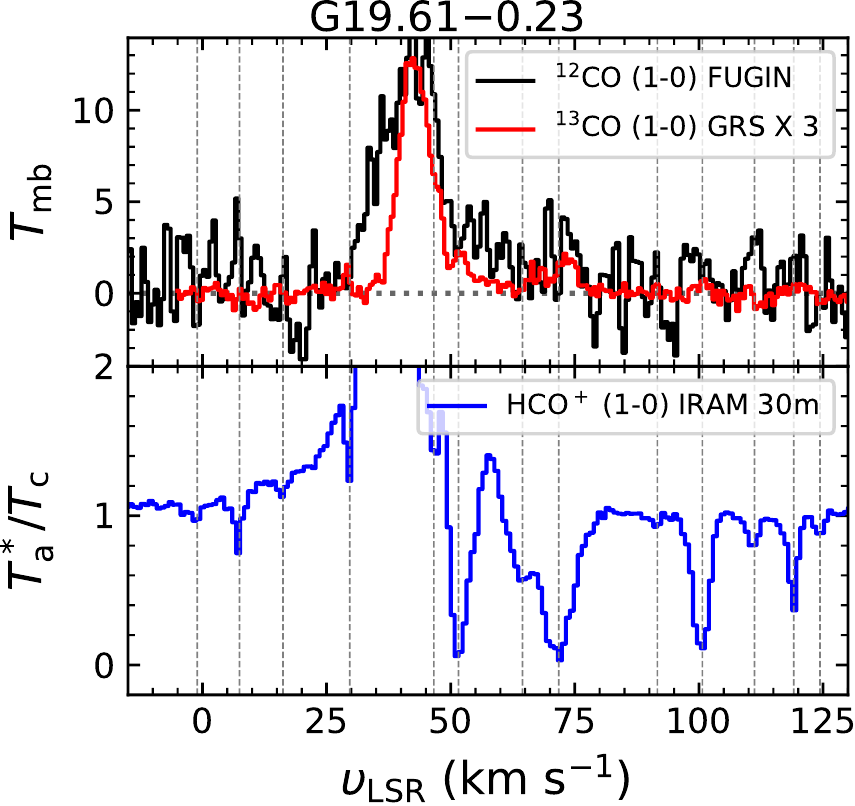}
    \vskip 0.3cm
    \includegraphics[width=0.325\textwidth]{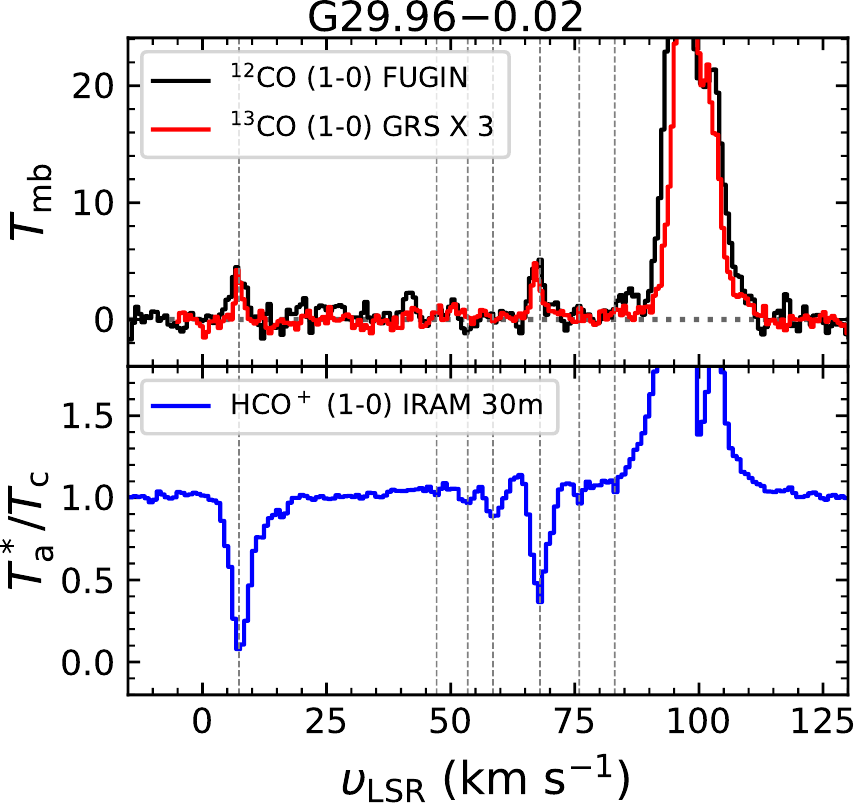}
    \includegraphics[width=0.33\textwidth]{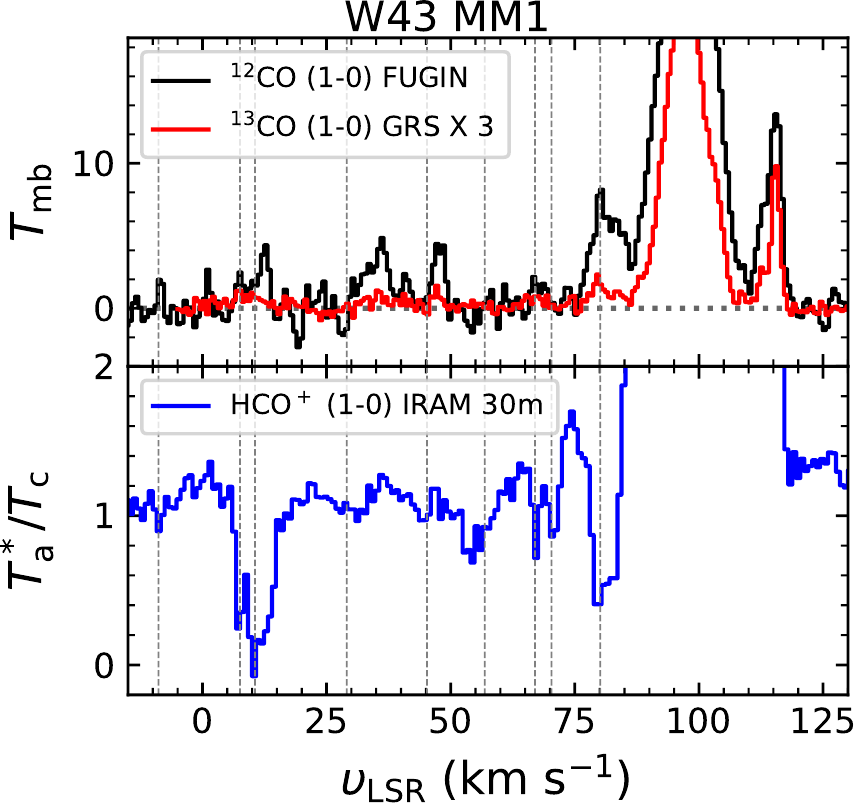}
    \includegraphics[width=0.33\textwidth]{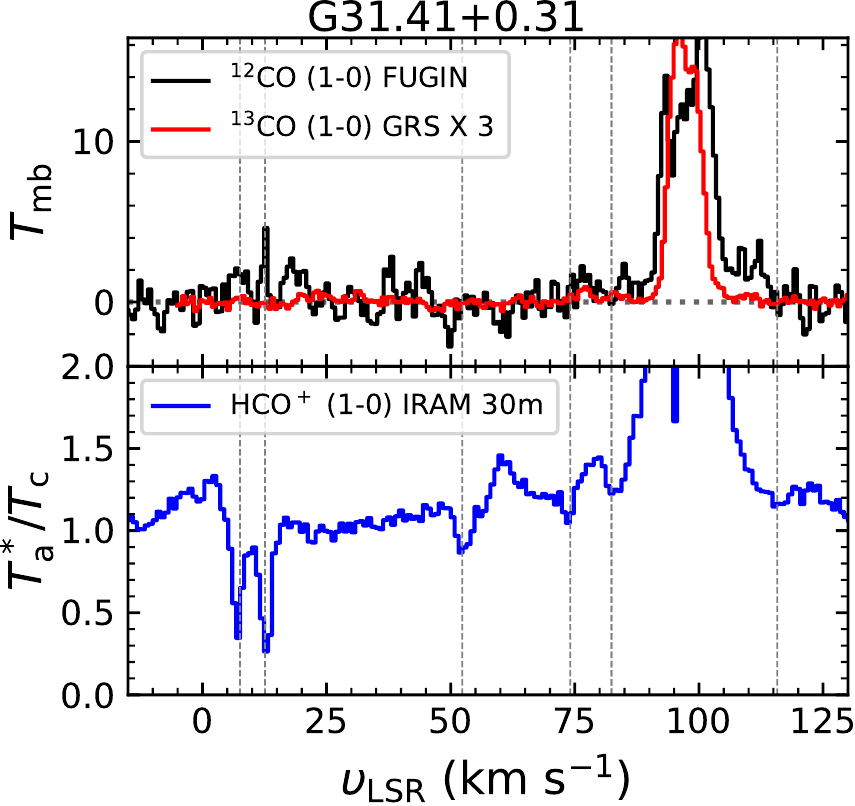}
    \vskip 0.3cm
    \includegraphics[width=0.33\textwidth]{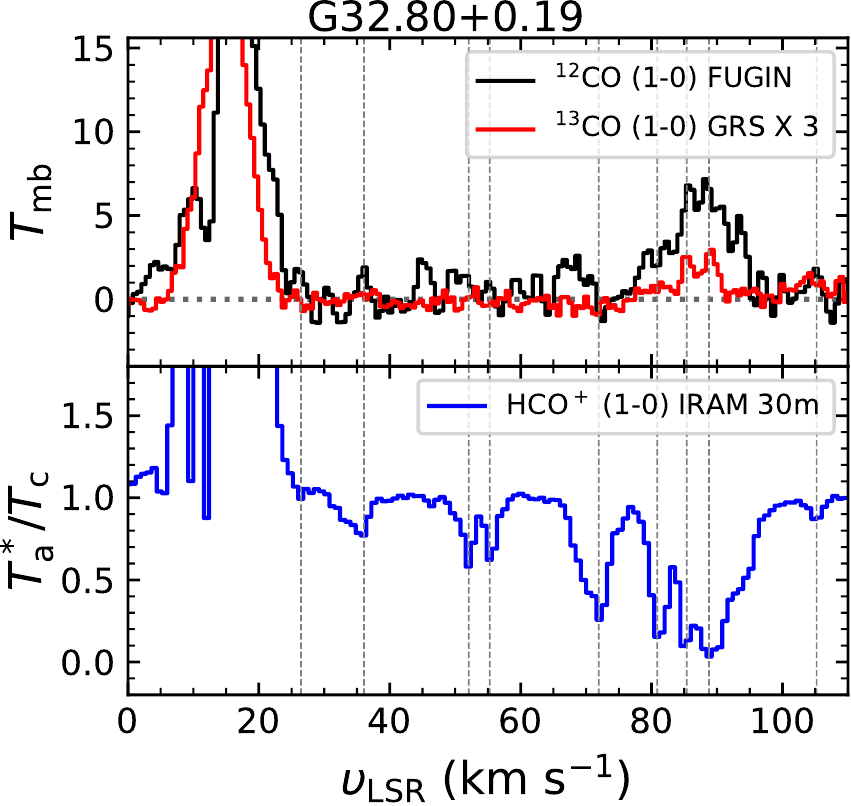}
    \includegraphics[width=0.33\textwidth]{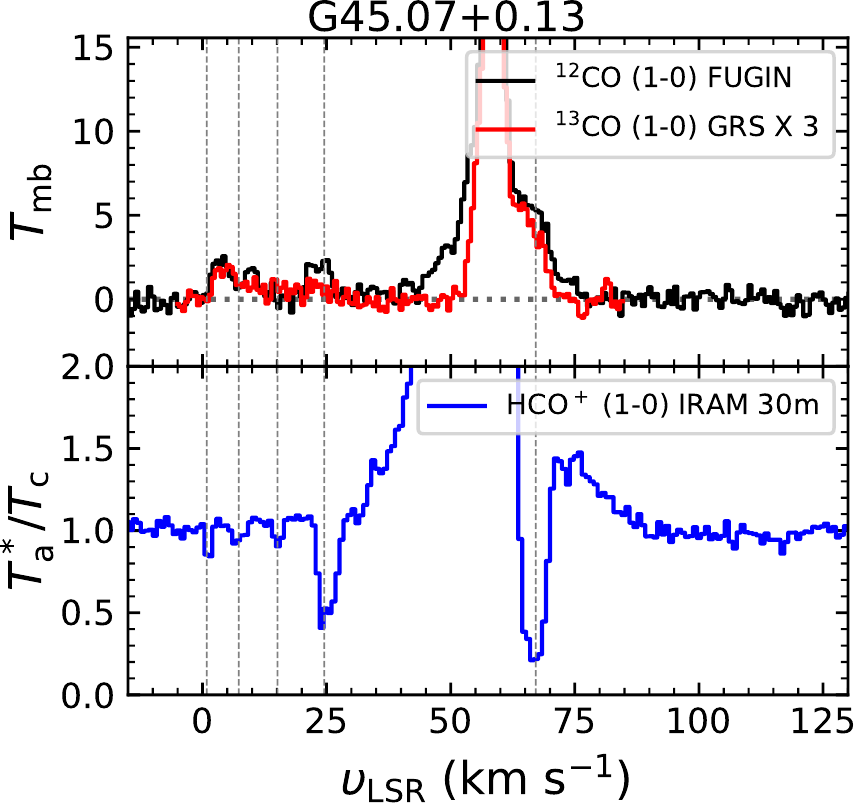}
    \caption{Spectra lines of CO and \ce{HCO+} toward the eight sightlines. The upper panel for each plot shows \ce{^{12}CO} emission lines ($1\rightarrow0$) in black, obtained from the FUGIN survey \citep{umemoto2017}, and \ce{^{13}CO} ($1\rightarrow0$) emission lines in red from the GRS survey \citep{jackson2006}. For only G09.62$+$0.19, the \ce{^{12}CO} (2$\rightarrow$1) is taken from the SEDIGISM survey \citep{schuller2021}. The lower panel for each plot shows absorption and emission lines in \ce{HCO+} spectra in blue from this survey. In the both panels, the vertical dotted lines indicate significant absorption features.}
    \label{fig:co_emission}
\end{figure*}

\subsubsection{G19.61$-$0.23 and G32.80$+$0.19} 
These two sources are inferred to lie in the first quadrant of the Milky Way in the Perseus and Sagittarius-Carina spiral arms at distances of 12.6\,kpc and 9.7\,kpc respectively. Such large distances allow several spiral arms to be intercepted by the sightlines, resulting in multiple absorption components in the molecular spectra observed toward these sources. Along the sightline toward G19.61$-$0.23 (upper panel of Fig.\,\ref{fig:abs_spec_g19_g32}), the \ce{HCO+} spectrum shows absorption features spanning a wide velocity range from $-$7\,\kms\ to 130\,\kms. Several features absorb parts of the emission line arising from the molecular gas related to the background continuum source, which drives an outflows that produces broad red- and blue-shifted emission wings for optically thick transitions \citep{furuya2011_g19p91_hcop_co}. The \ce{HCO+} absorption dip at 46\,\kms\ appears at the velocity of an absorption component of \ce{C^{18}O} (3$\leftarrow$2) \citep{furuya2011_g19p91_hcop_co}, observed with the Submillimeter Array (SMA), implying infalling gas in the low-density outer layers of the central core, which may be interpreted as self-absorption in the background source. On the other hand, another prominent feature at 51\,\kms\ in the absorption spectrum of \ce{HCO+} is absent in the \ce{^{13}CO} (3$\leftarrow$2) absorption spectrum, unlike that at 46\,\kms. Prominent absorption components at 51\,\kms\ and also at higher velocities, 71, 100, and 119\,\kms, have corresponding \hi\ absorption features \citep{kolpak2003_g19p91_hi}. Absorption features at these four velocities appear in millimeter transitions and, therefore, seem to be arising in intervening clouds along the sightline. The first two features at 51 and 71\,\kms\ are associated with the near- and far-side crossings of the Sagittarius spiral arm (spanning from $\sim\,$35 to 74\,\kms), and the latter, at 100 and 119\,\kms, are believed to be located on the 3\,kpc arm. We also find red-shifted absorption features with respect to the $\varv_{\rm sys}$ of 41.8\,\kms\ for this source and assume that these absorption lines are unrelated-foreground clouds as \hi\ absorption line components are also detected in this velocity range \citep{kolpak2003_g19p91_hi}. The HCN, HNC, \ce{C2H}, and \ce{c-C3H2} lines show similar absorption features as the \ce{HCO+} spectrum. In the CS and \ce{H2S} spectra fewer absorption features are identified, for \ce{H2S} spectrum this is caused by its ragged baseline, possibly the result of a number of emission lines.

Toward G32.80+0.19 (lower panel of Fig.\,\ref{fig:abs_spec_g19_g32}), for all seven species, we detect widespread absorption features spanning a broad range of velocities from 2 to 110\,\kms\ with the $\varv_{\rm sys}$ of 15\,\kms\ and emission spanning 1$-$ 28\,\kms\ from the background source. Apart from the red-shifted components with respect to $\varv_{\rm sys}$, the blue-shifted absorption features in the \ce{HCO+}, HCN, \ce{C2H}, and \ce{c-C3H2} spectra are detected in the velocity interval between 4 and 15\,\kms\ which is close to the velocity range of the Aquila-Rift. These absorption lines are possibly associated with foreground cloud components. As seen in the \ce{HCO+} spectra toward the other background continuum sources, the \ce{HCO+} spectrum toward G32.80$+$0.19 clearly shows absorption features that are broader than those of the other six species. Absorption features in the velocity interval between 58 and 120\,\kms\ trace the Scutum-Centaurus spiral arm and show four prominent dips at 70, 81, 85, and 89\,\kms. The deepest dip for \ce{HCO+}, HCN, HNC, and the two small hydrocarbons are at 89\,\kms. On the other hand, CS and \ce{H2S} show the deepest dips at 85\,\kms. Absorption in the $N_{\rm k_a, k_c}=1_{1,0}\leftarrow1_{1,1}$ line of \ce{H2CO} at a frequency of 4.83GHz (wavelength 6.2 cm)) is observed by \citep{araya2002_g32p80_h2co_abs} against the radio continuum emission of the same background source at similar velocities as the CS and \ce{H2S} line with its deepest absorption appearing at $\sim$ 85\,\kms. The Sagittarius arm and the Aquila Rift also cross this sightline and correspond to velocity ranges of 15$-$57\,\kms\ and 5$-$6\,\kms, respectively. The red-shifted absorption features with respect to the source's systemic velocity at 15\,\kms\ could arise from low-density diffuse gas associated with the molecular clouds of the continuum source as the \ce{H2CO} absorption features \citep{araya2002_g32p80_h2co_abs} correspond to the features seen in the \ce{HCO+}, \ce{C2H}, and \ce{c-C3H2} spectra. However, as its velocities correspond to Aquila Rift as well. 

\begin{figure*}[h!]
    \centering
    \includegraphics[width=0.48\textwidth]{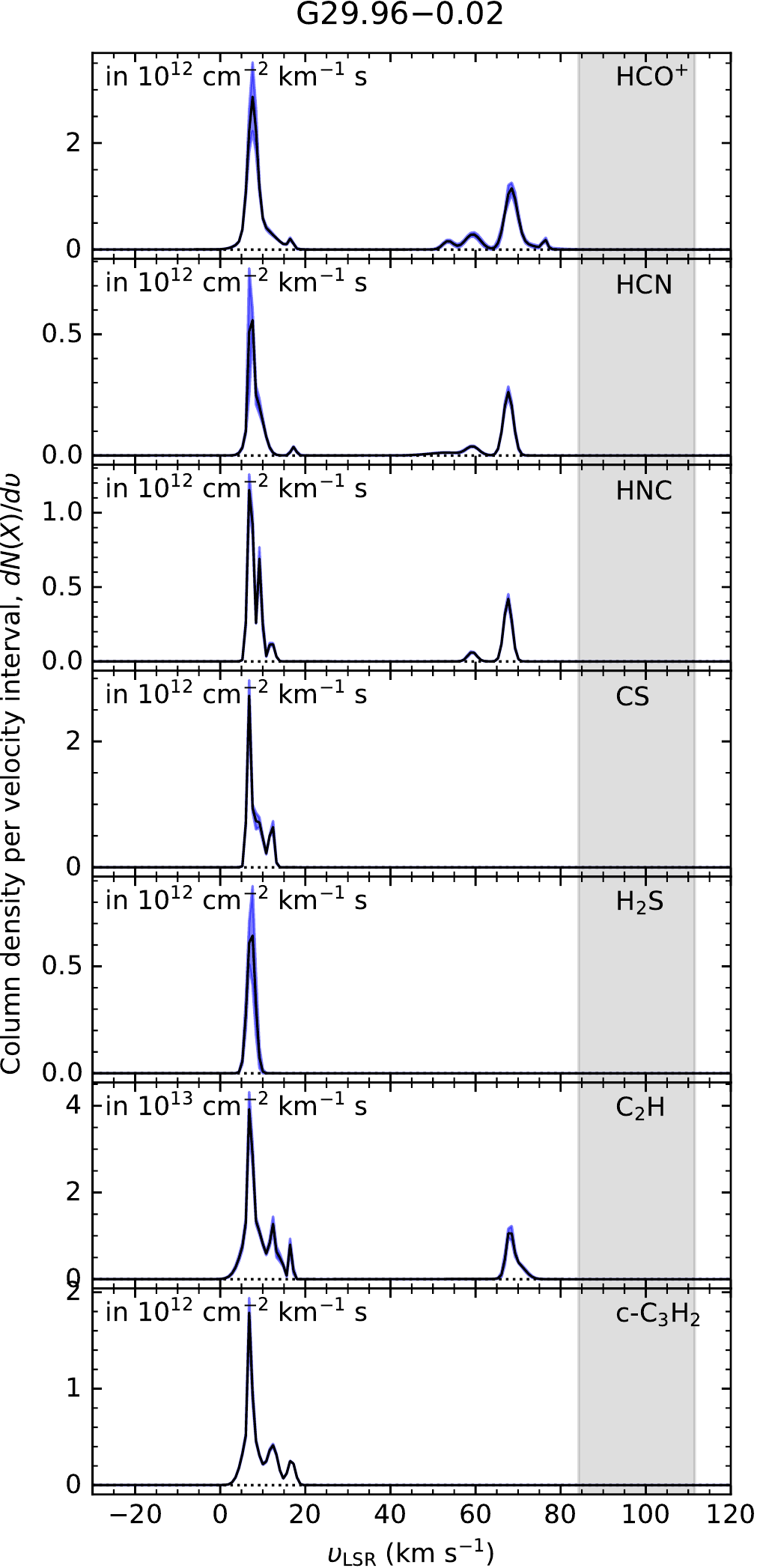}
    \includegraphics[width=0.465\textwidth]{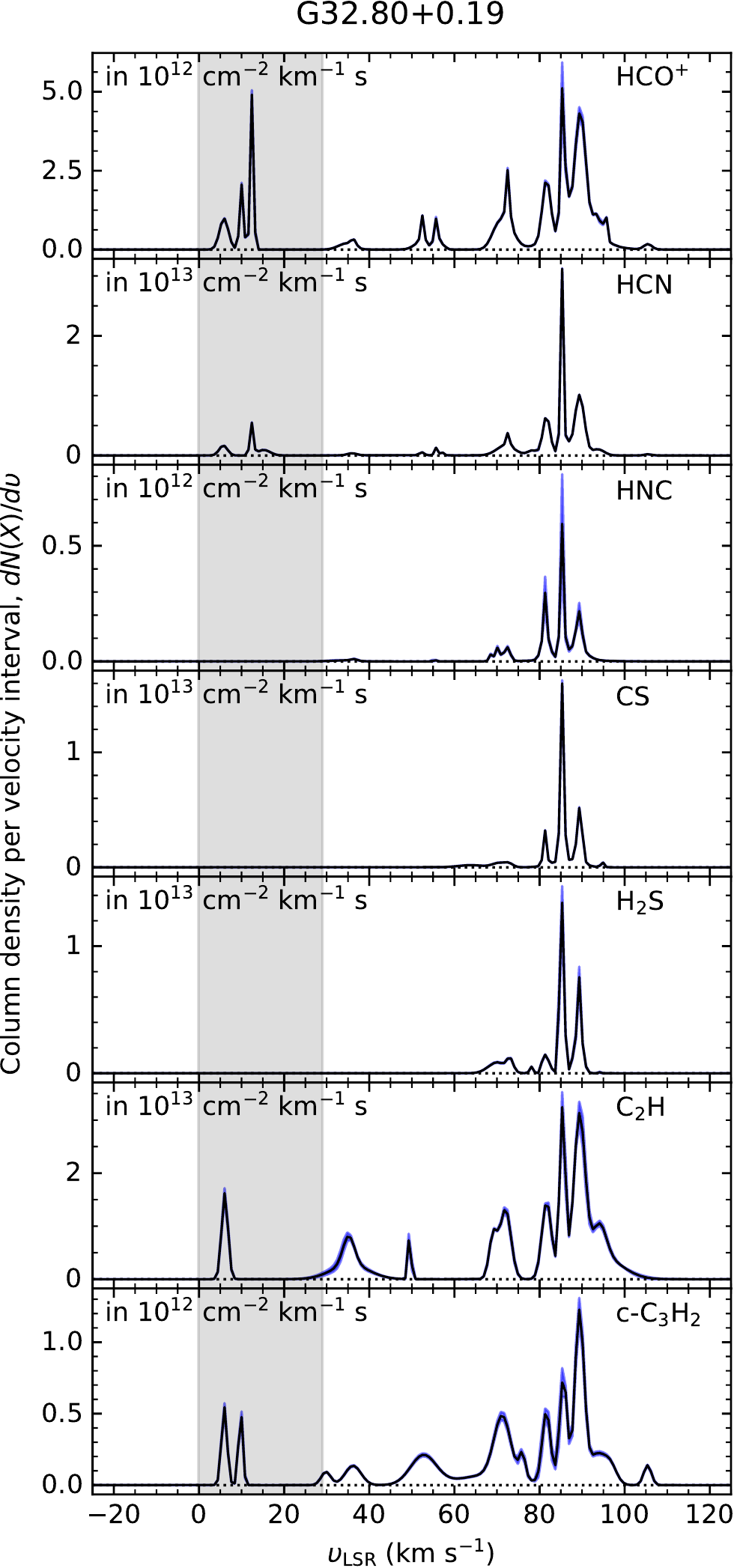}
    \caption{Channel-wise column density spectra ($dN/d\varv$) as a function of velocity toward G29.96$-$0.02 and G32.80$+$0.19. The $dN/d\varv$ in each panel shows \ce{HCO+}, HCN, HNC, CS, \ce{H2S}, \ce{C2H}, and \ce{c-C3H2} from top to bottom. The gray shaded area indicates velocity range of the \ce{HCO+} emission lines. The blue shaded regions represent 2$\sigma$ confidence intervals for the derived column densities (per velocity interval). }
    \label{fig:ntot}    
\end{figure*}

\subsection{CO emission counterparts}
As described in the previous section, the millimeter absorption components of simple molecules detected in this survey generally have narrower line features than those those of the hydride molecular ions (e.g., \ce{OH+} or \ce{ArH+}, \citealt{schilke2014,jacob2020}), whose submillimeter absorption spectra have continuous broad line profiles over wide velocity ranges. Such narrow spectral line shapes might imply that the millimeter absorption lines of neutral species and \ce{HCO+} trace denser gas than that traced by hydride molecular ions. To investigate whether any of the CO emission from dense molecular clouds is detected at the velocities of \ce{HCO+} absorption, we utilized archival data from available CO emission surveys: for \ce{^{13}CO} (2$\rightarrow$1) data mapped with an angular resolution of 30$''$, the Structure, Excitation and Dynamics of the Inner Galactic Interstellar Medium (SEDIGSM, \citealt{schuller2021}) program obtained with the SHeFI single-pixel instrument at the 12~m diameter Atacama Pathfinder Experiment submillimeter telescope; for \ce{^{13}CO} (1$\rightarrow$0) data obtained with an angular resolution of 46$''$ in the inner Galaxy, the Boston University-FCRAO Galactic Ring Survey (GRS, \citealt{jackson2006}) that used the SEcond QUabbin Optical Imaging Array (SEQUOIA) at the Five College Radio Astronomy Observatory 14~m telescope; and for \ce{^{12}CO} (1$\rightarrow$0) and \ce{^{13}CO} (1$\rightarrow$0) observed with an angular resolution of 20$''$ the FOREST Unbiased Galactic plane Imaging survey performed with the Nobeyama 45~m telescope (FUGIN, \citealt{umemoto2017}) and the telescope's four-beam, dual-polarization, sideband-separating SIS receiver. Figure\,\ref{fig:co_emission} displays the CO emission spectra toward eight of our sources obtained from the surveys mentioned above; they are G09.62$+$0.19, G10.47$+$0.03, G19.61$-$0.23, G29.96$-$0.02, W43 MM1, G31.41$+$0.31, G32.80$+$0.19, and G45.07$+$0.13. The upper panels show the \ce{^{12}CO} (1$\rightarrow$0) emission (in black) and \ce{^{13}CO} (1$\rightarrow$0) or (2$\rightarrow$1) (in red) on the main-beam temperature ($T_{\rm mb}$) scale, and the lower panels show \ce{HCO+} absorption spectra (in blue) normalized relative to the continuum. The vertical dotted lines indicate representative absorption dips. For G09.62$+$0.19, only the \ce{^{13}CO} (2$\rightarrow$1) data from the SEDIGIM survey is available, and G10.47$+$0.03 is observed only in the FUGIN survey, but not the GRS survey. Since the rms noise level of the FUGIN \ce{^{13}CO} (1$\rightarrow$0) data is higher than that of the GRS survey, we mainly use the \ce{^{13}CO} (1$\rightarrow$0) data of the GRS survey for the remaining sightlines observed in our survey. After extracting the CO spectrum within an area covered by a 20\,$''$ radius, we resampled the spectral data to 0.6\,\kms\ velocity bins that are 2.4 $-$ 6 times broader than the original spectral resolutions of the SEDIGIM and GRS surveys. For the FUGIN data, we smoothed the spectral channel resolution to 1.3\,\kms\ to improve S/N levels.

In Fig.\,\ref{fig:co_emission}, the most prominent absorption features have CO emission counterparts while weak absorption lines tend be not to have a detection of CO emission except for the absorption features around $0-20$\,\kms\ toward G45.07$+$0.13. In addition, not all deep absorption features have clear CO emission lines at their velocity ranges. For some velocity components toward some sources, CO emission features have different intensities, although the \ce{HCO+} absorption line depths are comparable. For example, toward G10.47$+$0.03, the depth of the absorption feature at around 154\,\kms\ is similar to that of the absorption components spanning $0-50$\,\kms, but its CO emission counterpart is weaker. Frequently, compared to \ce{HCO+} with narrow line profiles, broad \ce{HCO+} absorption features tend to have CO emission counterparts.

\begin{figure*}[h!]
    \centering
    \includegraphics[width=0.34\textwidth]{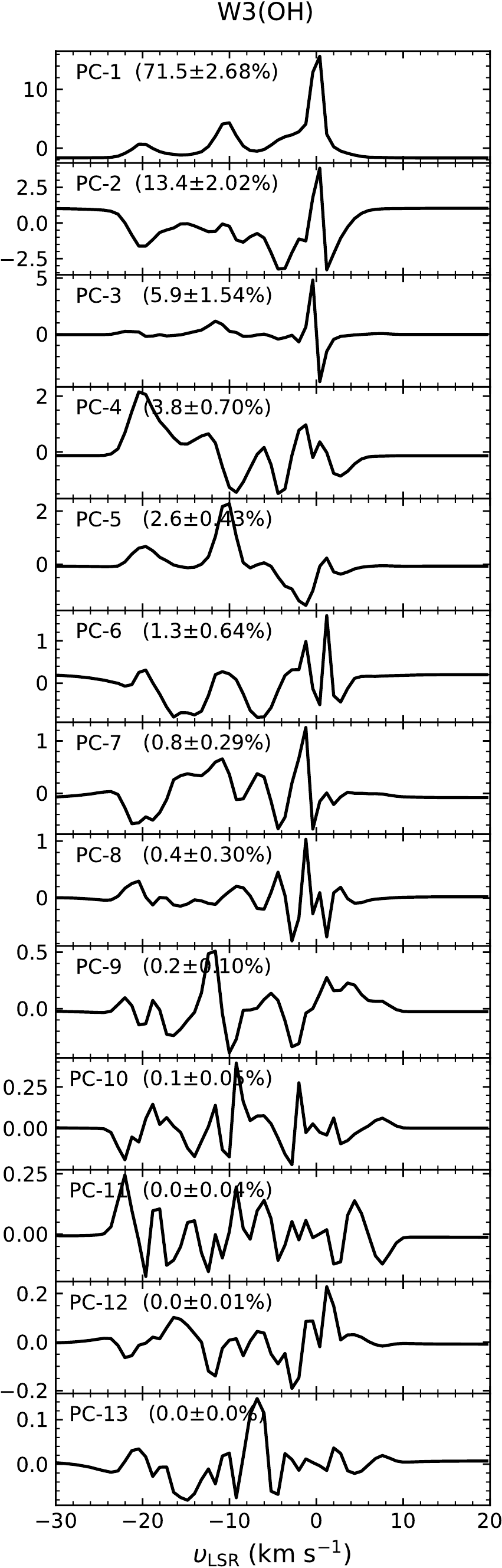}
    \hskip 2cm
    \includegraphics[width=0.346\textwidth]{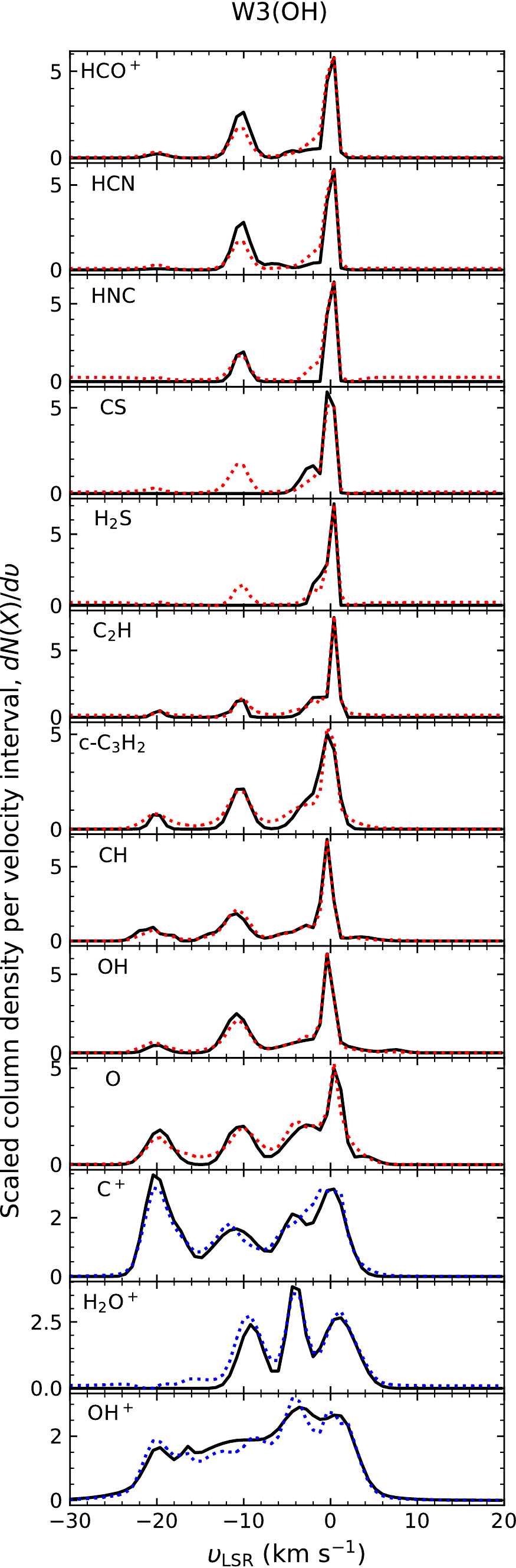}
    \caption{\textit{Left}: Eigen spectra of thirteen principal components with percentage of variances as well as uncertainties (2$\sigma$), toward W3(OH). \textit{Right}: Scaled channel-wise column density spectra for the thirteen species (\ce{HCO+}, HCN, HNC, CS, \ce{H2S}, \ce{C2H}, \ce{c-C3H2}, CH, OH, O, \ce{C+}, \ce{H2O+}, and \ce{OH+}) as a function of $\varv_{\rm LSR}$. Black curves represent the observational data, and red dotted curves are the spectra reproduced with only the first three principal components (PC-1, PC-2, and PC-3). For \ce{C+}, \ce{H2O+}, and \ce{OH+}, the reproduced spectra indicated by blue dotted curves include one or two more PCs, i.e., PC-4 and PC-5.}
    \label{fig:pca_fit}
\end{figure*}

\section{Analysis}\label{sec:analysis}

\subsection{Line parameter determination with XCLASS}
The spectral lines observed were modeled using the eXtended CASA Line Analysis Software Suite (XCLASS\footnote{https://xclass.astro.uni-koeln.de}, \citealt{moeller2017}) by solving the 1-dimensional radiative transfer equation with the assumption of local thermal equilibrium (LTE) for an isothermal source. The \texttt{myXCLASSFit} program computes synthetic spectra, by fitting the spectral lines with Gaussian profiles, thereby ensuring that opacity effects on the line shapes are properly taken into account. The fit is performed using optimization package \texttt{MAGIX} \citep{Moeller2013}, which provides an interface between the input codes and an iterating engine. It minimizes the deviation between the modeled outputs from the observational data. Molecular properties (e.g., Einstein $A$ coefficients, partition functions, etc.) are taken from an embedded SQLite database containing entries from the Cologne Database for Molecular Spectroscopy (CDMS, \citealt{mueller2001,mueller2005}) and the Jet Propulsion Laboratory database (JPL, \citealt{pickett1998}) in its Virtual Atomic and Molecular Data Center (VAMDC, \citealt{endres2016}) implementation, along with an extended set of partition function calculations. The parameter set fitted for each absorption line component consists of the excitation temperature ($T_{\rm ex}$), the total column density ($N_{\rm tot}$), the line width ($\Delta \varv$), and the velocity offset ($\varv_{\rm off}$) from the systemic velocity. The column density per velocity interval, $dN/d\varv$ for each velocity channel, $i$, is related to the optical depth, $\tau_i$, (Paper I) by
\begin{equation}
\begin{aligned}
\label{eq:optically_thin}
\left(\frac{dN}{d\varv}\right)_i = \frac{8\pi\nu^{3}}{c^{3}} \frac{Q(T_{\rm ex})}{g_{u}A_{ul}} \frac{{\rm exp} ({E_{u}}/{kT_{\rm ex}} )}{{\rm exp}(h\nu/{kT_{\rm ex}}) -1} \tau_i,
\end{aligned}
\end{equation}
where $E_{\rm u}$ is the energy of the upper state for the selected transition, $g_{\rm u}$ is the upper state degeneracy, and $\nu$ is the frequency of a molecular transition. The quantity $A_{\rm ul}$ is the Einstein coefficient for spontaneous emission, and $Q$($T_{\rm ex}$) and $k$ are the partition function and Boltzmann constant, respectively. For the absorption components, we adopt a fixed excitation temperature for all observed molecular transitions equivalent to the cosmic microwave background temperature of 2.73\,K. This is motivated by the high dipole moments of most of the species, which lead to critical densities for the observed transitions that lie a few orders of magnitude above the typical number density of hydrogen molecules ($n_{\rm \ce{H2}}$) in translucent clouds of a few hundred cm$^{-3}$ (e.g., \citealt{snow2006,gerin2016}). Thus, the excitation temperature of these transitions are determined by the 2.73\,K cosmic background radiation field. XCLASS offers various algorithms to find the best-fit parameters by minimizing the $\chi^2$ value, and in this work, we used the Levenberg-Marquardt algorithm. In addition, in the fitting procedure, all the hfs splitting transitions of HCN and \ce{C2H} are taken into account, and the fitting deconvolves the hfs of these molecular transitions. We note that we only consider the ortho transitions of \ce{H2S} and \ce{c-C3H2}, and thus the derived column densities of these molecules only apply to their ortho species, not to the sum of their ortho and para species. The spectral line profiles observed are the combination of emission and absorption features. We have performed simultaneous fitting of the emission and absorption lines to determine the background for the absorption lines and derive accurate physical properties for the absorption components. However, since the focus of this paper is the absorption system (rather than deriving reliable physical parameters of the emission lines), we do not present the fit parameters of the emission lines. 
For uncertainties of column densities derived here, we perform the error estimation by sampling a posterior distribution of the optical depths ($\propto$~ln($T_{\rm a}^*/T_{\rm c}$)). The sampling was performed using XCLASS fits to multiple realizations of sample spectra that were generated by adding independent pseudorandom noise to the observed spectra. For the XCLASS fits, as done to the main fitting for the original data sets, we also fit emission and absorption components simultaneously on the noise-added spectra in order to take into account variability of the posterior distributions caused by emission components.

Figures\,\ref{fig:ntot} and \ref{appen:ncol1}--\ref{appen:ncol5} 
display the column density per velocity interval, $dN$(X)/$d\varv$, determined for all seven species studied and toward all sources, with uncertainties indicated as blue shaded areas. The gray parts indicate the velocity ranges of the emission lines that partly overlap with the absorption lines. As a result, the column densities of the absorption components within the gray velocity ranges might have higher uncertainties than those implied by the blue shaded areas because of additional errors that may arise when emission components are simultaneously fitted with the absorption components. 

\begin{figure*}[!h]
    \centering
    \includegraphics[width=0.45\textwidth]{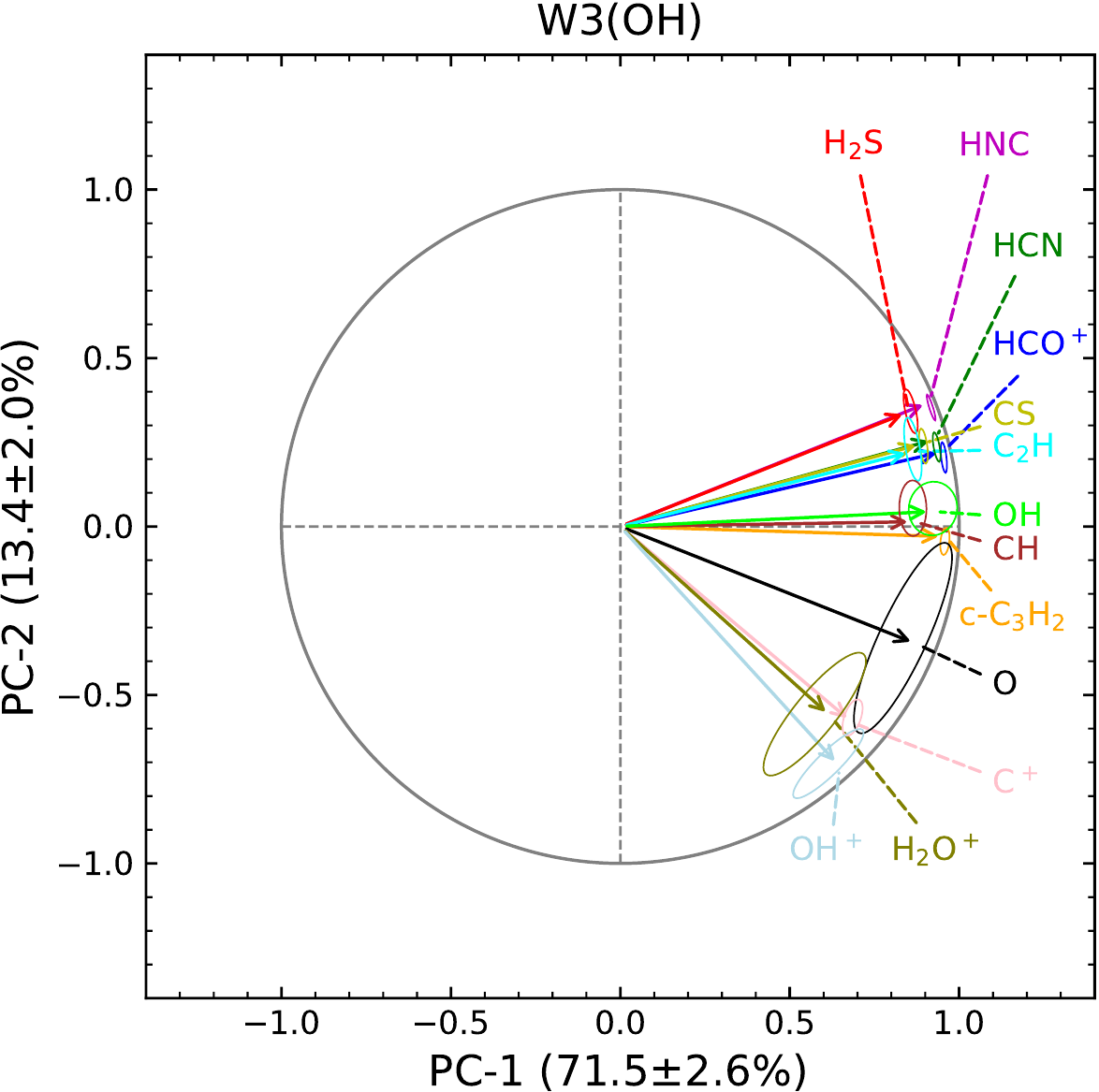}
    \includegraphics[width=0.45\textwidth]{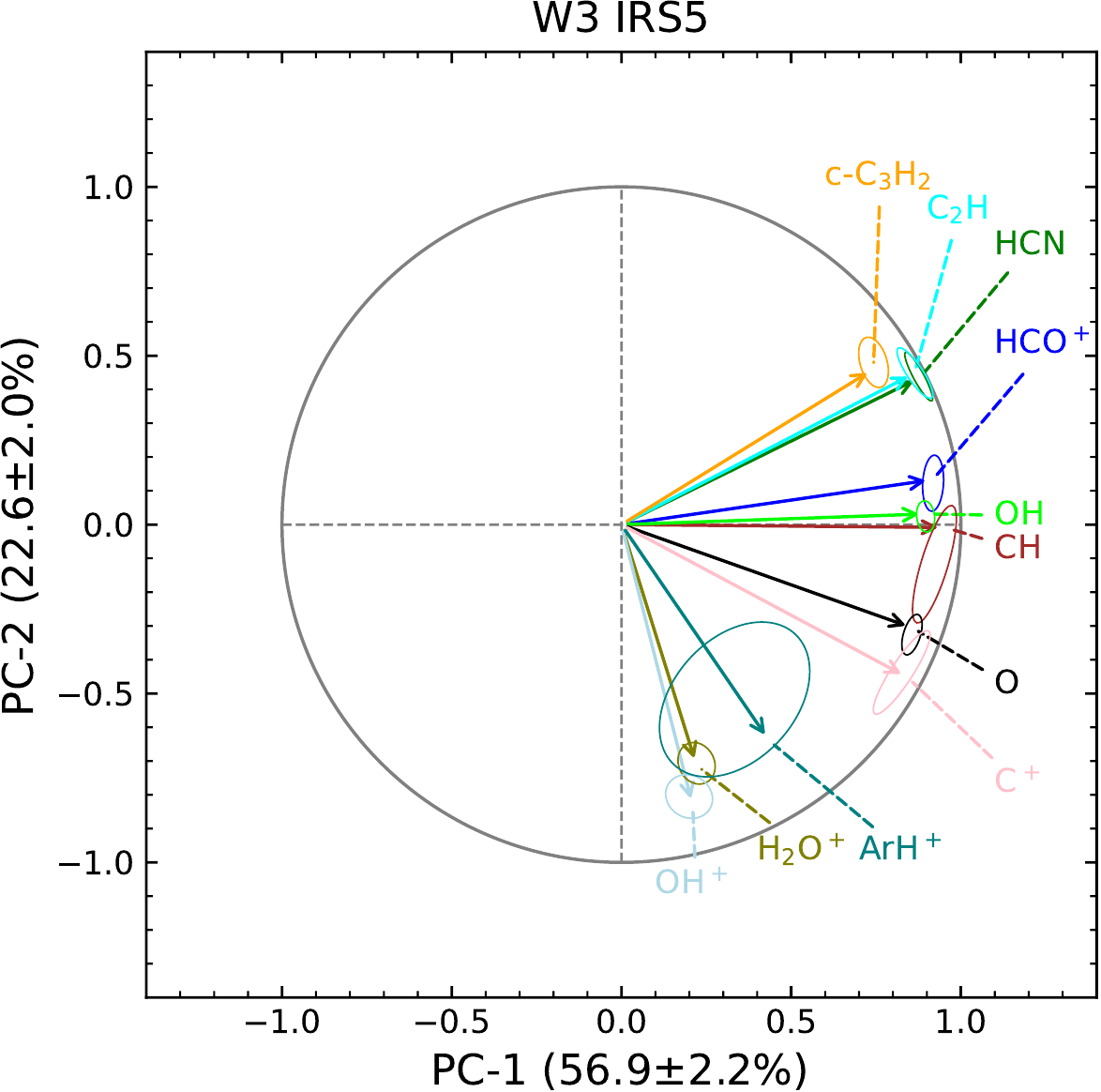}
    \caption{Variable correlation circle plots for the first and second PCs toward W3(OH) in the left panel and W3 IRS5 in the right panel. The solid straight lines close to the unit circle imply PC-1 and PC-2 contain the most information about the variables. Different colors are used to denote the different species analyzed. Each colored ellipse is a 2$\sigma$ confidence interval for the uncertainty of the PCA results for the corresponding species. 
    }
    \label{fig:pc1pc2_circle}
\end{figure*}

\begin{figure*}[!h]
    \centering
    \includegraphics[width=0.45\textwidth]{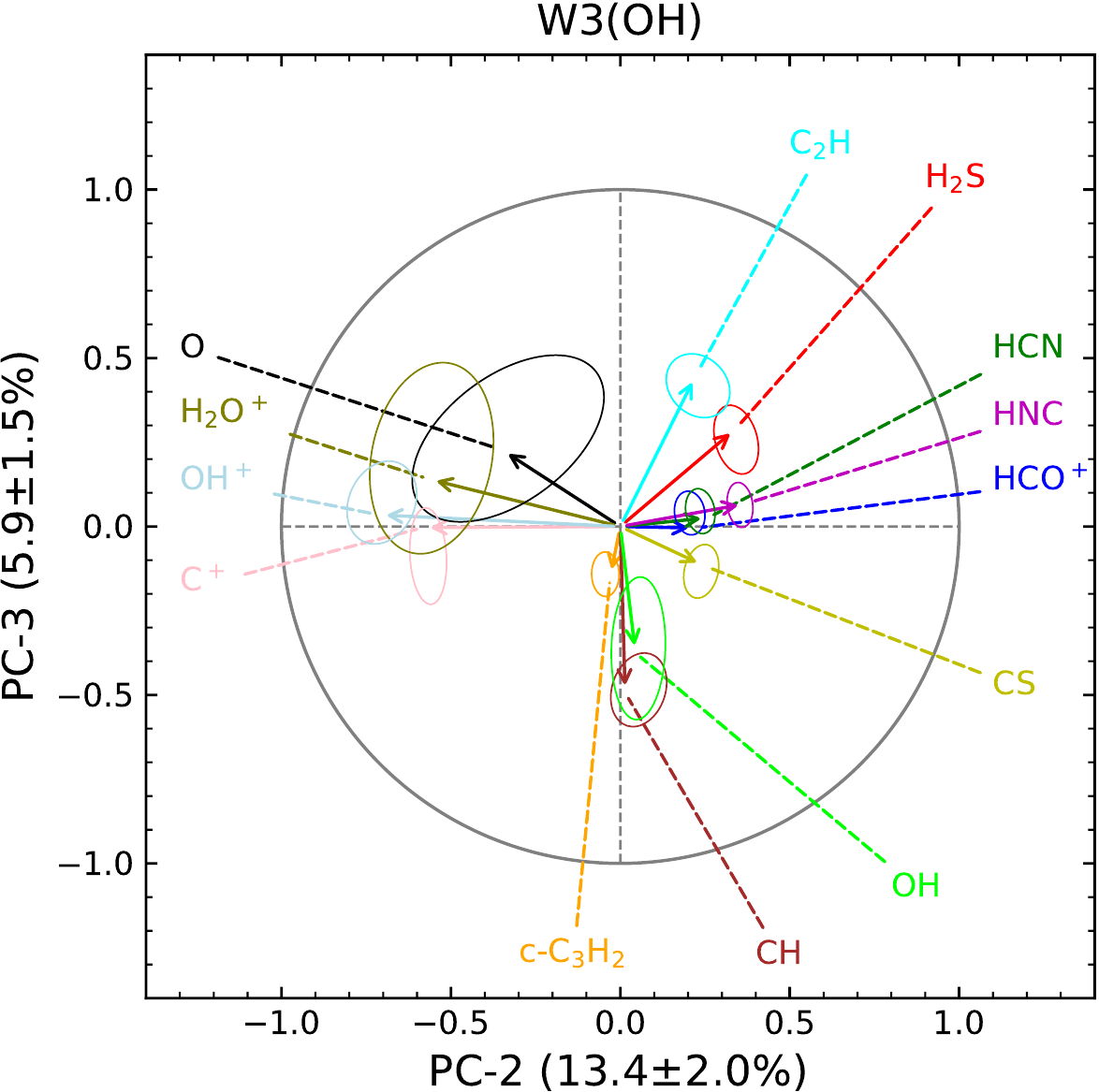}
    \includegraphics[width=0.45\textwidth]{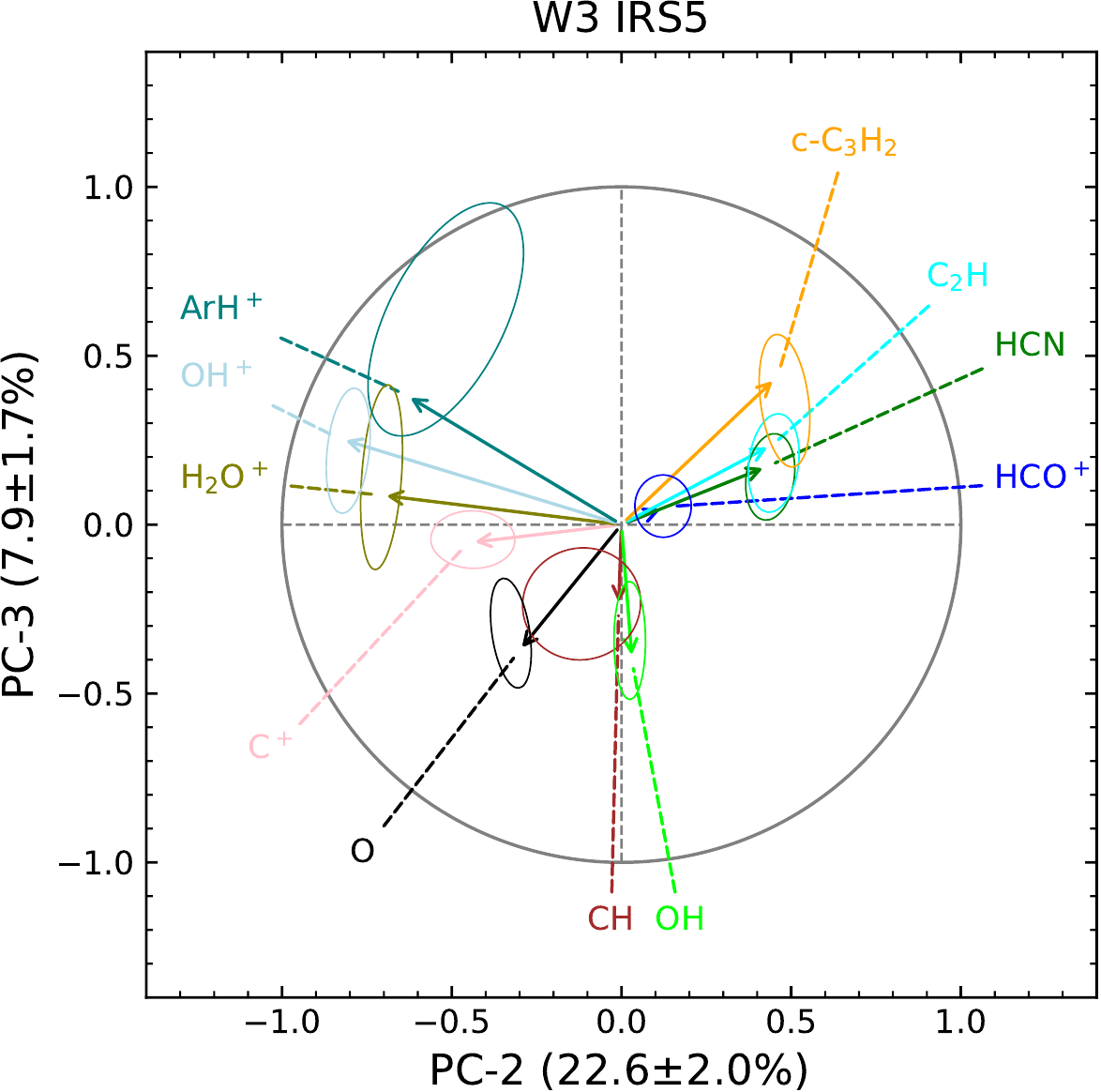}    
    \caption{Same as Fig.\,\ref{fig:pc1pc2_circle} but for the second and third PCs (PC-2 and PC-3) toward the same sources, from left to right.}
    \label{fig:pc2pc3_circle}
\end{figure*}

\subsection{Principal component analysis}
Principal component analysis (PCA) has been used to investigate differences and similarities in the distributions of multiple species detected in absorption spectra (e.g., Sect. 4.2 in \citealt{neufeld2015}).
Here, we carried out PCA of our W3(OH) data for 13 species (\ce{HCO+}, HCN, HNC, CS, \ce{H2S}, \ce{C2H}, \ce{c-C3H2}, CH, OH, O, \ce{C+}, \ce{H2O+}, and \ce{OH+}) and for W3 IRS5 for 11 species (same as for W3(OH) except with the addition of \ce{ArH+} and the omission of HNC, CS, and \ce{H2S}). The lattermost six species listed for W3(OH) were observed with SOFIA (Paper I). The SOFIA results published to date provide spectra of these species toward three sightlines, W3(OH), W3 IRS5, and NGC 7538 IRS1. Since only \ce{HCO+} was detected toward NGC 7538 IRS1 in the present work, this sightline was not considered in our PCA analysis. Moreover, the SH absorption features detected with SOFIA toward these sightlines are all associated with the background continuum source; toward them foreground absorption is detected. Therefore, SH is not included in this PCA analysis. Instead of using transmission ($T_{\rm a}^*/T_{\rm c}$) or optical depth spectra, we used column densities, $dN/d\varv$, as the input to the PCA, thereby minimizing contributions from the emission features and avoiding complications caused by hfs. All the column density spectra were computed with a common channel width of 0.8\,\kms. As discussed further below, we also accounted for uncertainties in the derived column densities, which can significantly affect the results of the PCA.

Prior to performing the PCA analysis, standardization is imperative: this involves rescaling the spectra so that their variations have a mean of zero and a variance of unity. PCA involves writing the column density for each species, $i$, as a linear combination of orthogonal eigenfunctions (a.k.a. principal components), $f_{j}(\varv)$
\begin{equation}
    N_{i}(\varv) = a_{i}+ b_{i}\sum_{j=1}^{k} C_{ij}{f_{j}(\varv)}, 
\end{equation}
where $a_{i}$ and $b_{i}$ and $C_{ij}$ are coefficients, and $k$ is the number of species included (13 for W3(OH) or 11 for W3 IRS5). The $C_{ij}$ obey a normalization constraint, $\sum_{j=1}^{k} C^2_{ij} = 1$. To examine the reliability of the PCA results, uncertainty estimates are essential. Therefore, we made use of 1000 realizations obtained by random sampling within the column density probability distribution function and then for each realization of the 1000 realizations, we created a data set containing PCA results for all used species from the each PCA performance toward W3(OH) and W3 IRS5, respectively. From these multiple realizations, we obtained a distribution of the PCA results for each species. Using the posterior distributions for all considered species from the PCA performances for the 1000 realization, we obtained 2$\sigma$ confidence intervals for all the PCA results; these are included in the results presented in the following section.

In Figures\,\ref{fig:pca_fit} and \ref{append:pca_fit}, the left panels show eigen spectra of principle components with percentage of variances and uncertainties (2$\sigma$), toward W3(OH) and W3 IRS5, respectively. As the first component ($f_{1}(\varv)$, labeled PC-1) accounts for more than half of the variance for each source, its profile resembles the column density distributions of most of the species included in the PCA. However, there are significant differences between the W3(OH) (Fig.\ref{fig:pca_fit}) and W3 IRS5 (Fig.\,\ref{append:pca_fit}) sightlines in regard to the fraction of the variance accounted for by each component. For W3(OH), PC-1 and PC-2 account respectively for 71.5$\pm$2.6\,\% and 13.4$\pm$2.0\,\% of the total variance, Toward W3 IRS5, PC-2 accounts for a more significant fraction of the total variance, 22.6$\pm$2.0\,\%, and PC-1 accounts for only slightly more than half (56.9$\pm$2.2\,\%) of the total. The right panel of Figs.\,\ref{fig:pca_fit} and \ref{append:pca_fit} represent scaled channel-wise column density spectra as black curves for the observed species (for W3(OH), \ce{HCO+}, HCN, HNC, CS, \ce{H2S}, \ce{C2H}, \ce{c-C3H2}, CH, OH, O, \ce{C+}, \ce{H2O+}, and \ce{OH+}; for W3 IRS5, \ce{HCO+}, HCN, \ce{C2H}, \ce{c-C3H2}, CH, OH, O, \ce{C+}, \ce{H2O+}, \ce{OH+}, and \ce{ArH+}). The column density spectra reproduced (red and blue dotted lines) by summing the first three terms in the expansion, for most of the species but for \ce{C+}, \ce{H2O+}, \ce{OH+}, and \ce{ArH+} by summing up to the first five terms : $C_{i1}f_{1}(\varv)$, $C_{i2}f_{2}(\varv)$ and $C_{i3}f_{3}(\varv)$, and additionally $C_{i4}f_{4}(\varv)$ and $C_{i5}f_{5}(\varv)$ for the mentioned species above. For both sources, the first three principal components account for $\sim 90\%$ of the total variance, and the sum of their contributions provides a reasonable approximation to the column densities of the observed species.

\begin{figure*}[]
    \centering
    \includegraphics[width=0.7\textwidth]{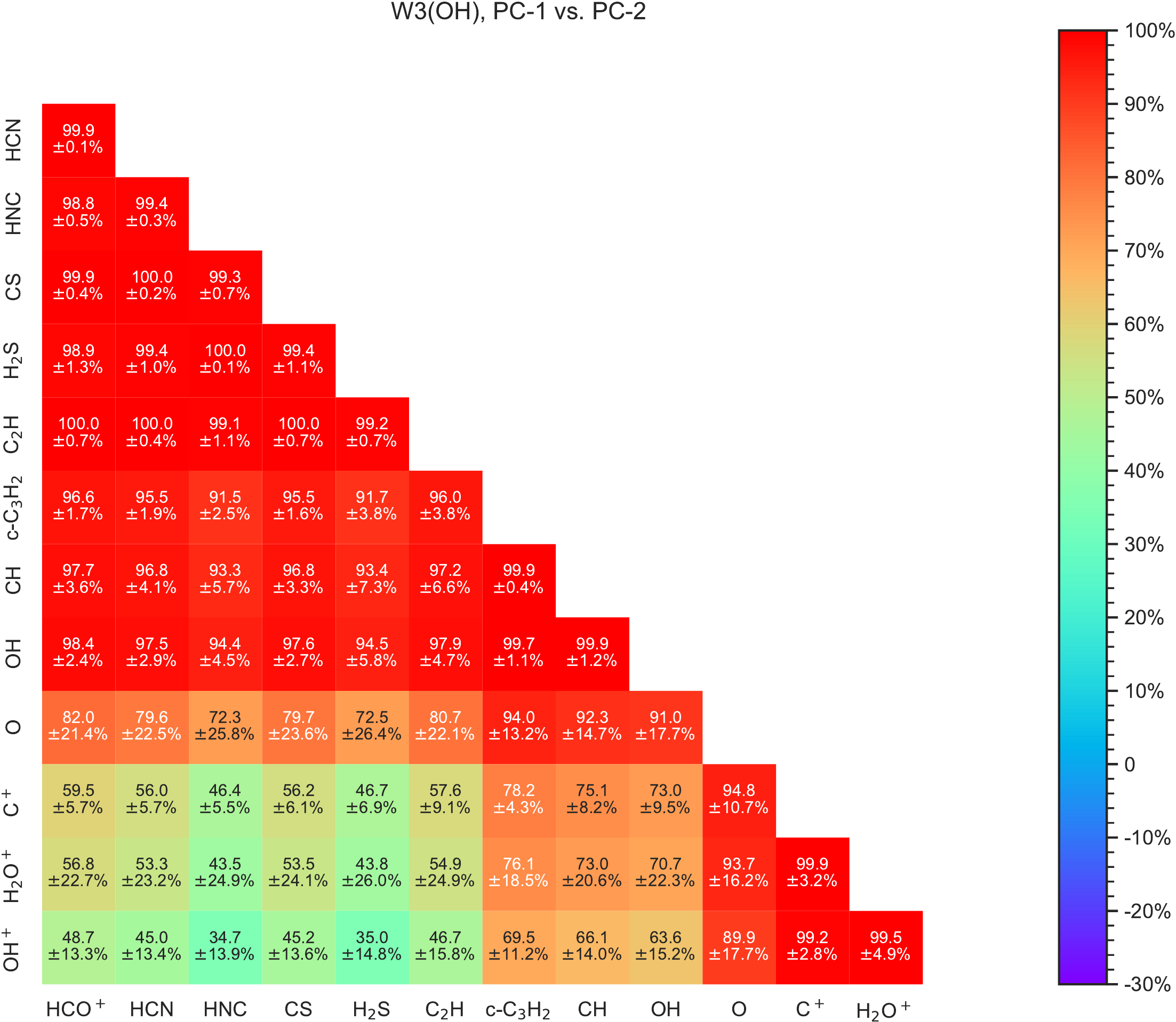}
    \vskip 0.53 cm
    \includegraphics[width=0.7\textwidth]{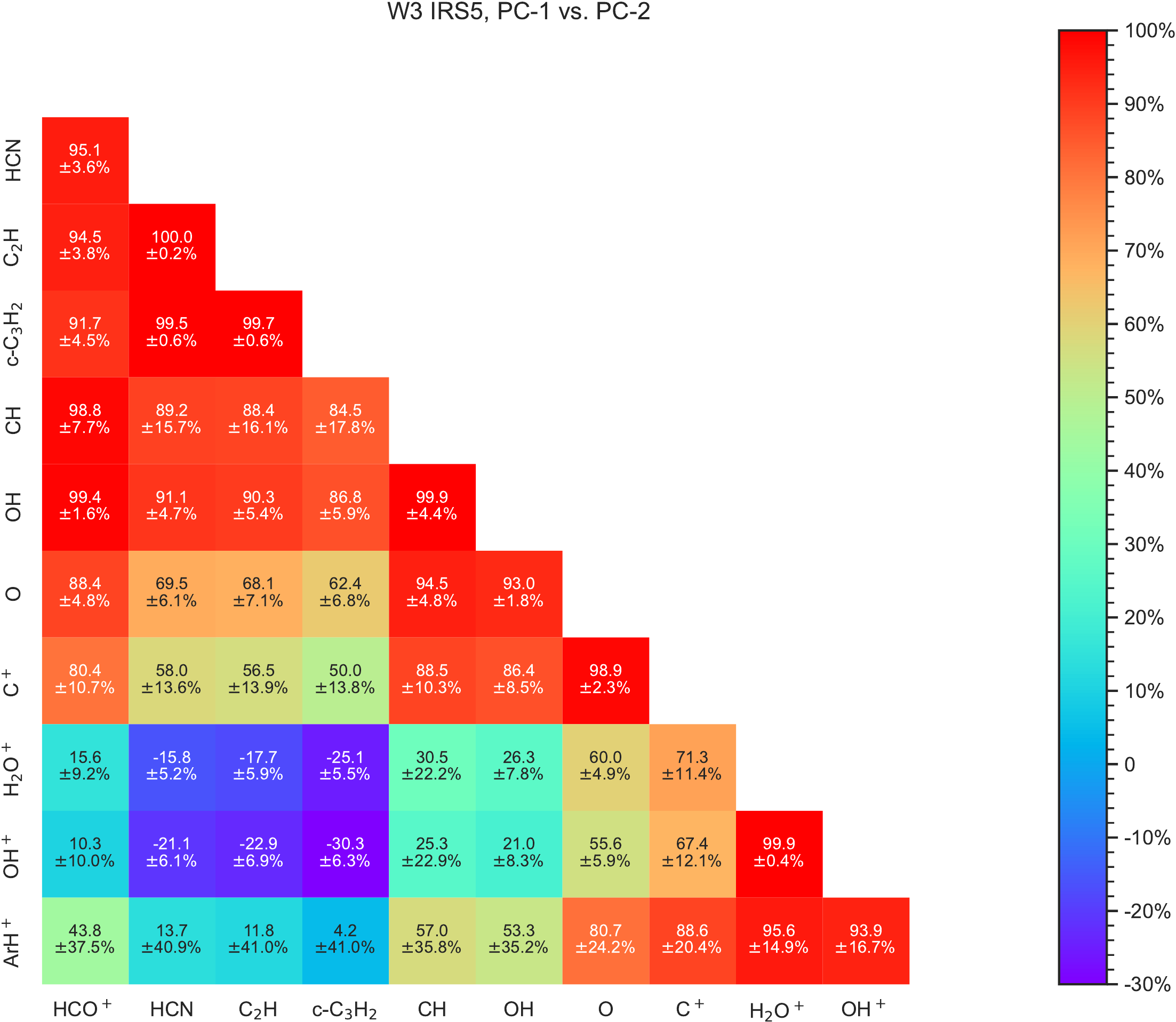}
    \caption{Cosine similarities showing correlation between two pairs of species in the circle plots for PC-1 versus PC-2 toward W3(OH) (top) and W3 IRS5 (bottom). }
    \label{fig:corr_heatmaps_pc1pc2}
\end{figure*}

\begin{figure*}[]
    \centering
    \includegraphics[width=0.7\textwidth]{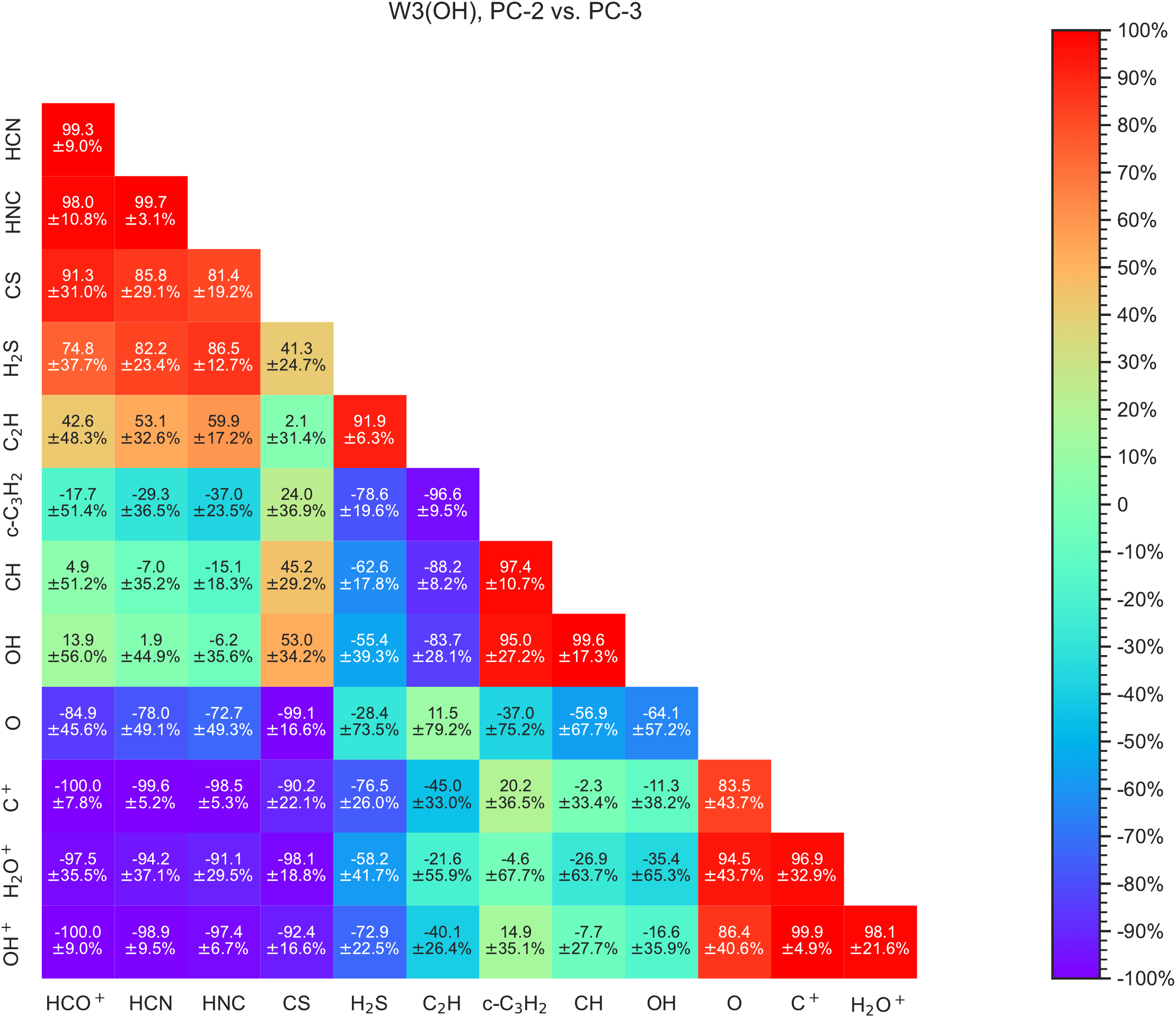}  
    \vskip 0.535 cm
    \includegraphics[width=0.7\textwidth]{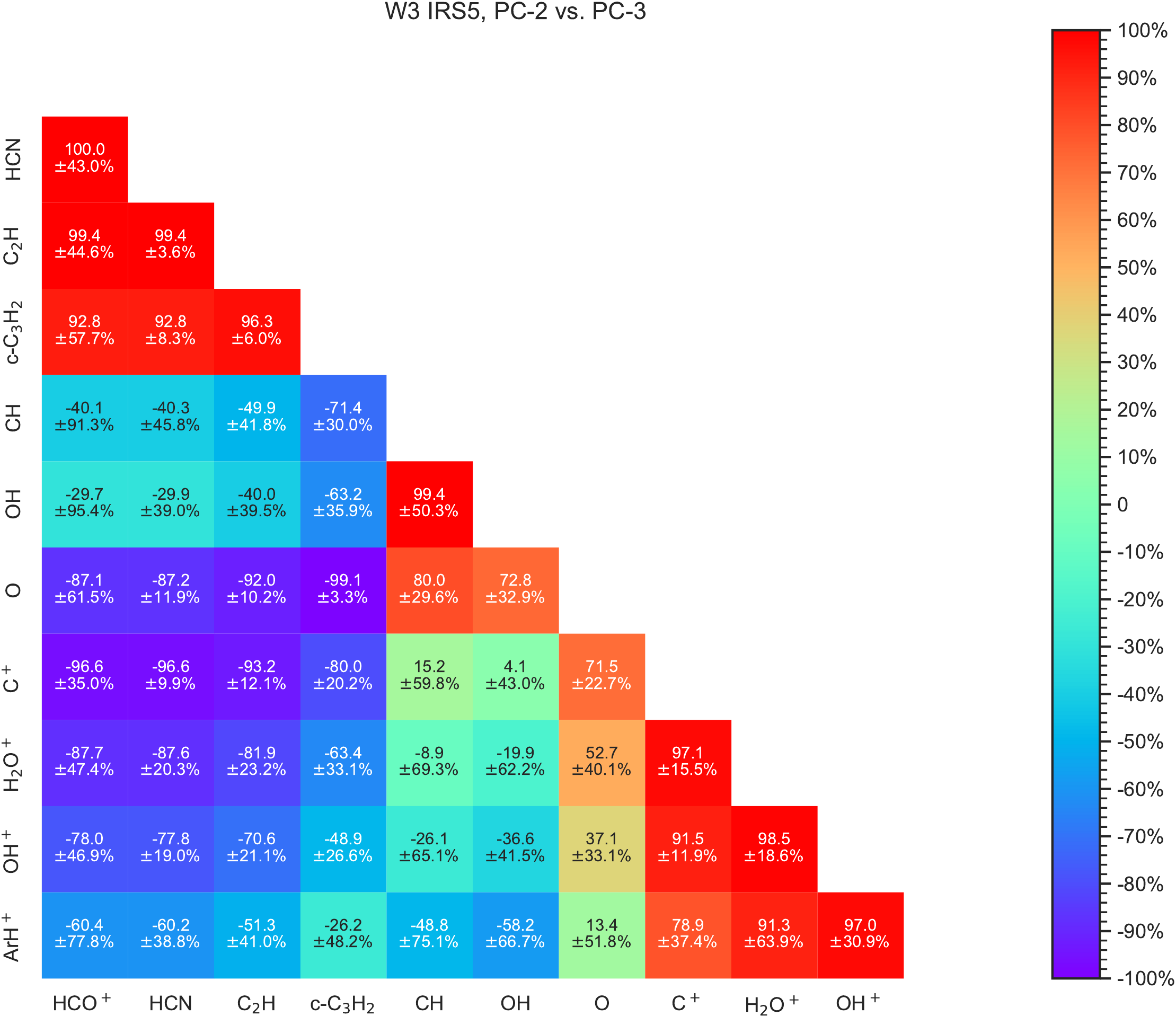}  
    \caption{Cosine similarities showing correlation between two pairs of species in the circle plots for PC-2 versus PC-3 toward W3(OH) (top) and W3 IRS5 (bottom).}
    \label{fig:corr_heatmaps_pc2pc3}
\end{figure*}

\begin{figure*}[ht!]
    \centering
    \includegraphics[width=0.33\textwidth]{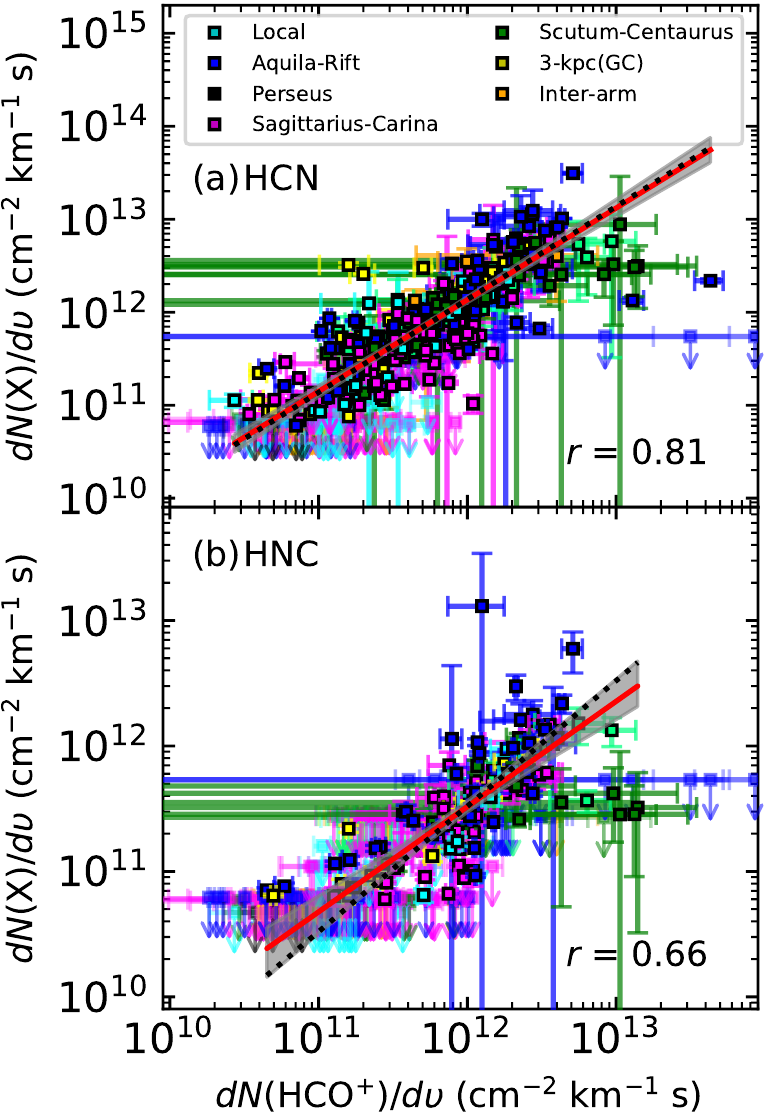}
    \includegraphics[width=0.33\textwidth]{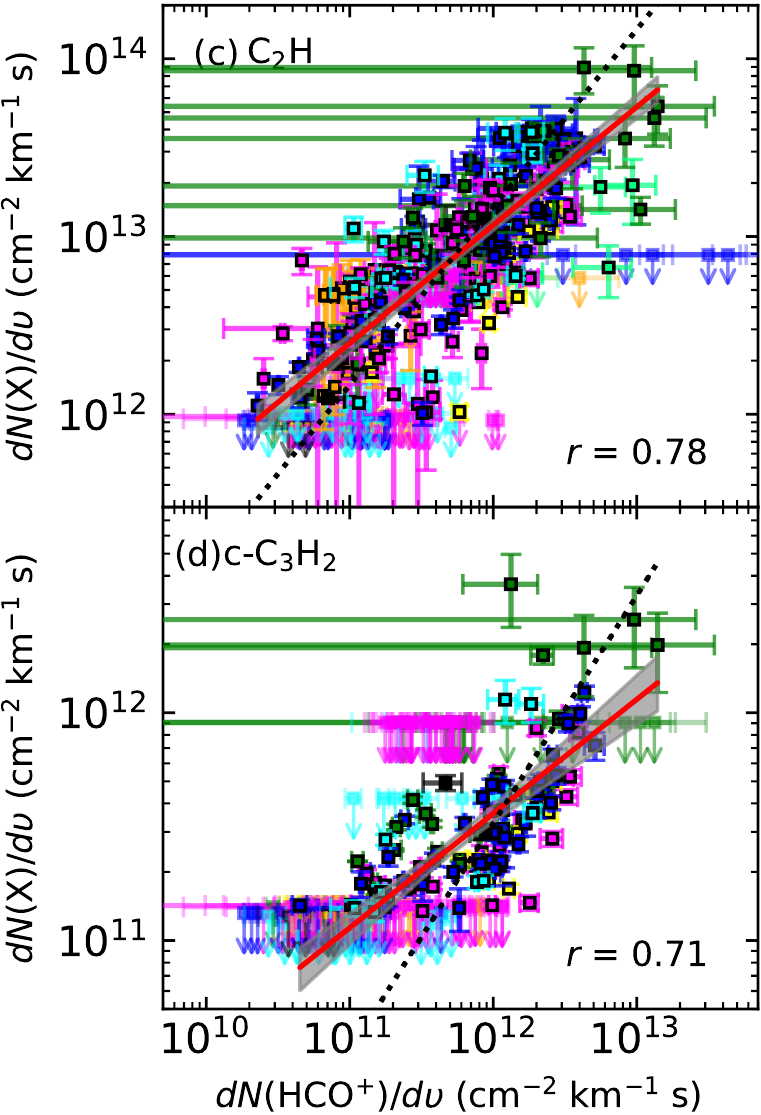}
    \includegraphics[width=0.33\textwidth]{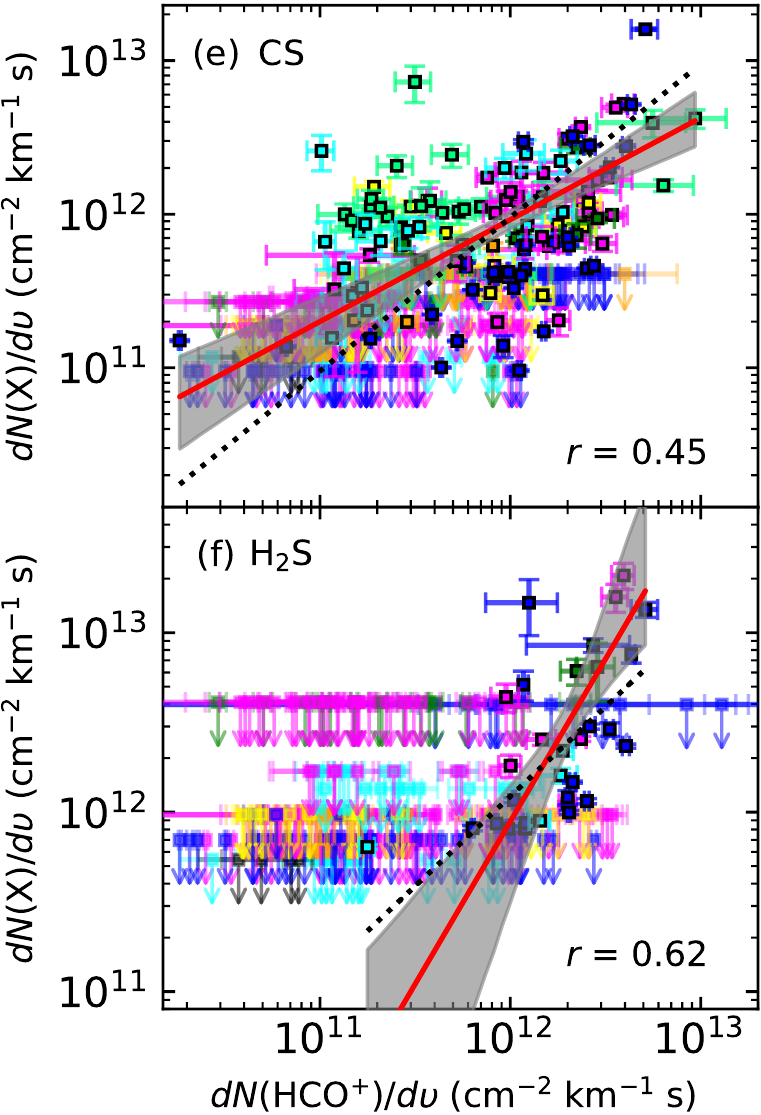}
    \caption{Comparisons of column densities in 0.8\,\kms\ velocity channel bins for seven molecular species; (a) HCN versus \ce{HCO+}, (b) HNC versus \ce{HCO+}, (c) \ce{C2H} versus \ce{HCO+}, (d) \ce{c-C3H2} versus \ce{HCO+}, (e) CS versus \ce{HCO+}, and (f) \ce{H2S} versus \ce{HCO+}. The color codes correspond to specific spiral arms (local arm in cyan, Perseus arm in black, Aquila-Rift in blue, Sagittarius-Carina arm in magenta, Scutum-Centaurus arm in green, inter-arm regions in orange). All symbols without black edges indicate upper limits for nondetections, while the data points above the detection threshold are marked with black edges. In addition, the blue-green colored points indicate clouds that belong to the 135\,\kms\ arm located beyond the Galactic center, toward G10.47$+$0.03. The black dotted lines represent the median value of the column density ratio in each panel. Uncertainties in the derived channel-wise column densities are presented as error bars. The red lines present a best-fit taking into account uncertainties of column densities. The gray areas are 2$\sigma$ confidence interval for the best fit (red lines) with errors. }
    \label{fig:ncol_channel}
\end{figure*}

\begin{figure*}[ht!]
    \centering
    \includegraphics[width=0.33\textwidth]{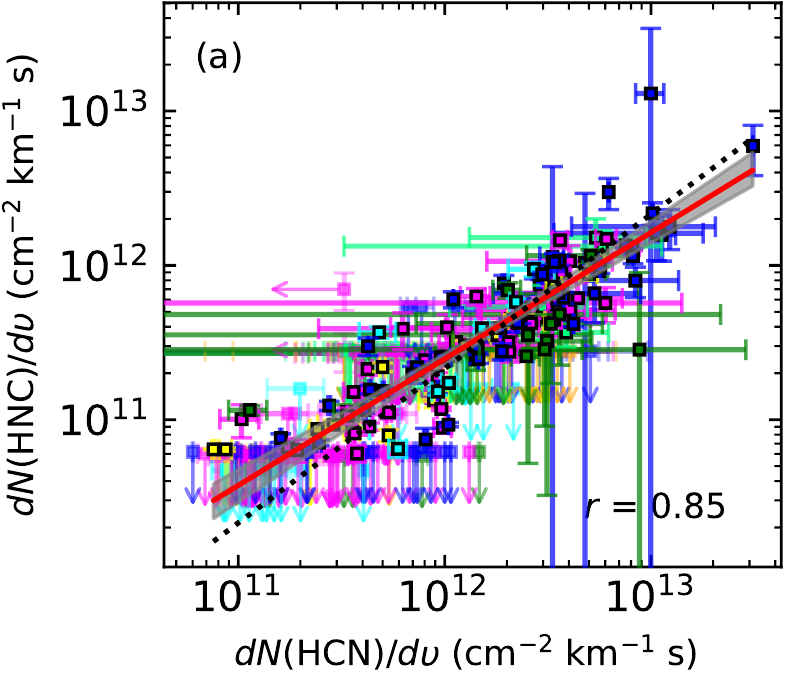}
    \includegraphics[width=0.33\textwidth]{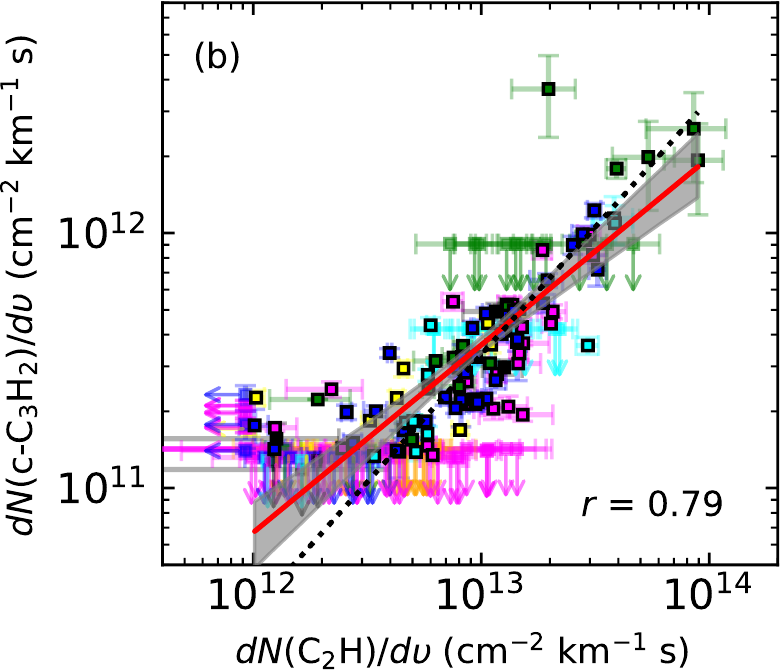}
    \includegraphics[width=0.33\textwidth]{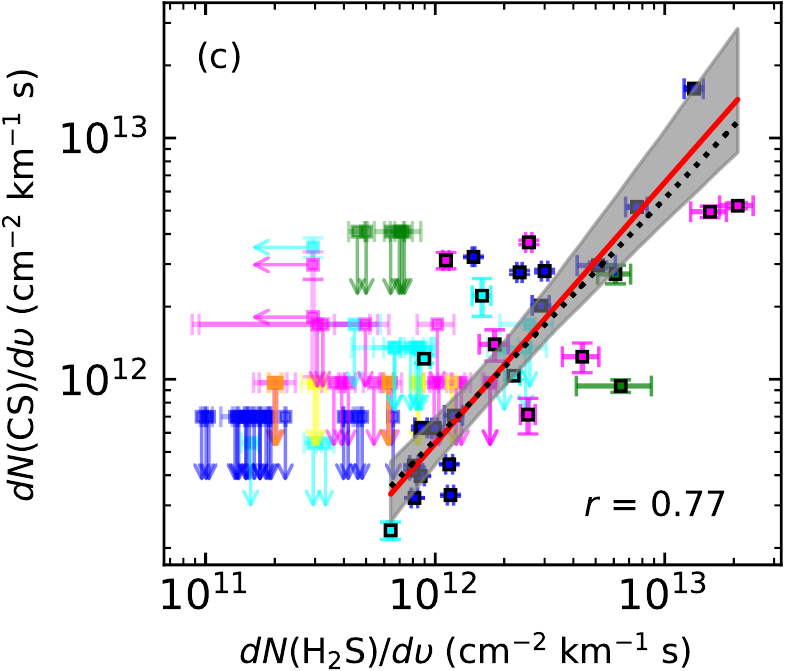}  
    \caption{Same as Fig.\,\ref{fig:ncol_channel} but for (a) HCN versus HNC, (b) \ce{C2H} versus \ce{c-C3H2}, and (c) CS versus \ce{H2S}. }
    \label{fig:ncol_channel_pair}
\end{figure*}

\section{Discussion}\label{sec:discussion}
\subsection{Correlations between species for W3(OH) and W3 IRS5}

Figures\,\ref{fig:pc1pc2_circle} and \ref{fig:pc2pc3_circle} show the coefficients $C_{ij}$ in the PCA expansion for each observed species. These are sometimes referred to as h-plots (e.g., \citealt{ungerechts1997}). Figure\,\ref{fig:pc1pc2_circle} shows the coefficients, $C_{i1}$ and $C_{i2}$, for the first and second components, PC-1 and PC-2, obtained for W3(OH) (left panel) and W3 IRS5 (right panel), while Fig. \ref{fig:pc2pc3_circle} shows the coefficients for the second and third terms in the expansion. Each colored vector corresponds to a different species as indicated, and the normalization condition for the $C_{ij}$ coefficients implies that all the plotted vectors must lie within the unit circle (black). Colored ellipses indicate the uncertainties (corresponding to 2$\sigma$ confidence intervals) for each vector. 

The angle ($\theta$) between any two vectors (corresponding to two different species) indicates how well the two species of a given pair are correlated with each other. The similarity between any two vectors may be characterized by the cosine of the angle between the vectors (known as cosine similarity); $\cos \theta = \bf{A} \cdot \bf{B}/ \lVert A\rVert\,\lVert B \rVert$. Figures\,\ref{fig:corr_heatmaps_pc1pc2} and \ref{fig:corr_heatmaps_pc2pc3} present similarity coefficients estimated for each pair of species, with $\cos \theta$ represented as percentages; acute angles in the h-plots indicate a strong similarity ($\sim 100$\,\% as noted in Figs.\,\ref{fig:corr_heatmaps_pc1pc2} and \ref{fig:corr_heatmaps_pc2pc3} ), orthogonal (90$^{\circ}$) angles mean no similarity (0\,\%, i.e., lack of correlation), and opposite vectors with an angle of 180$^{\circ}$ represent negative similarity ($\sim -$100\,\%, i.e., anticorrelation). Although the sets of species detected are not exactly identical toward these sightlines, the relative positions of the vectors appear to be similar for most of the species, as shown in Fig.\,\ref{fig:pc1pc2_circle}. In addition, we note that when we recompute the vector positions for the same set of species (\ce{HCO+}, HCN, \ce{C2H}, \ce{c-C3H2}, CH, OH, O, \ce{C+}, \ce{H2O+}, and \ce{OH+}) for both sightlines, the relative positions stay almost the similar as shown as in Figs.\,\ref{fig:pc1pc2_circle} and \ref{fig:pc2pc3_circle}, and the PCA h-plots for the recomputed results are presented in \ref{append:pca_fit_subsample}. From these plots, which present the PCA results, we observe three notable trends summarized below. 

\begin{figure*}[h!]
    \centering
    \includegraphics[width=0.33\textwidth]{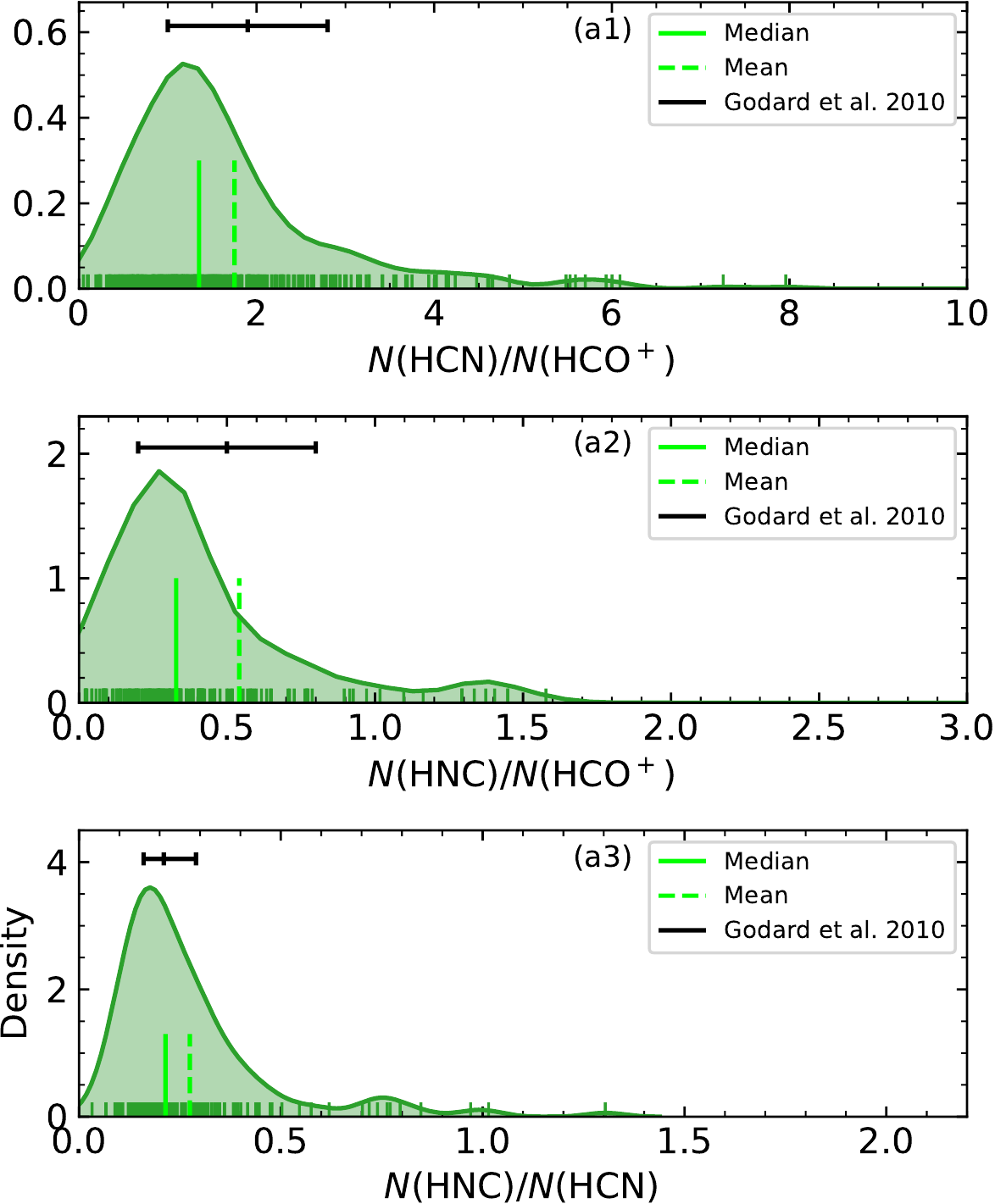}
    \includegraphics[width=0.33\textwidth]{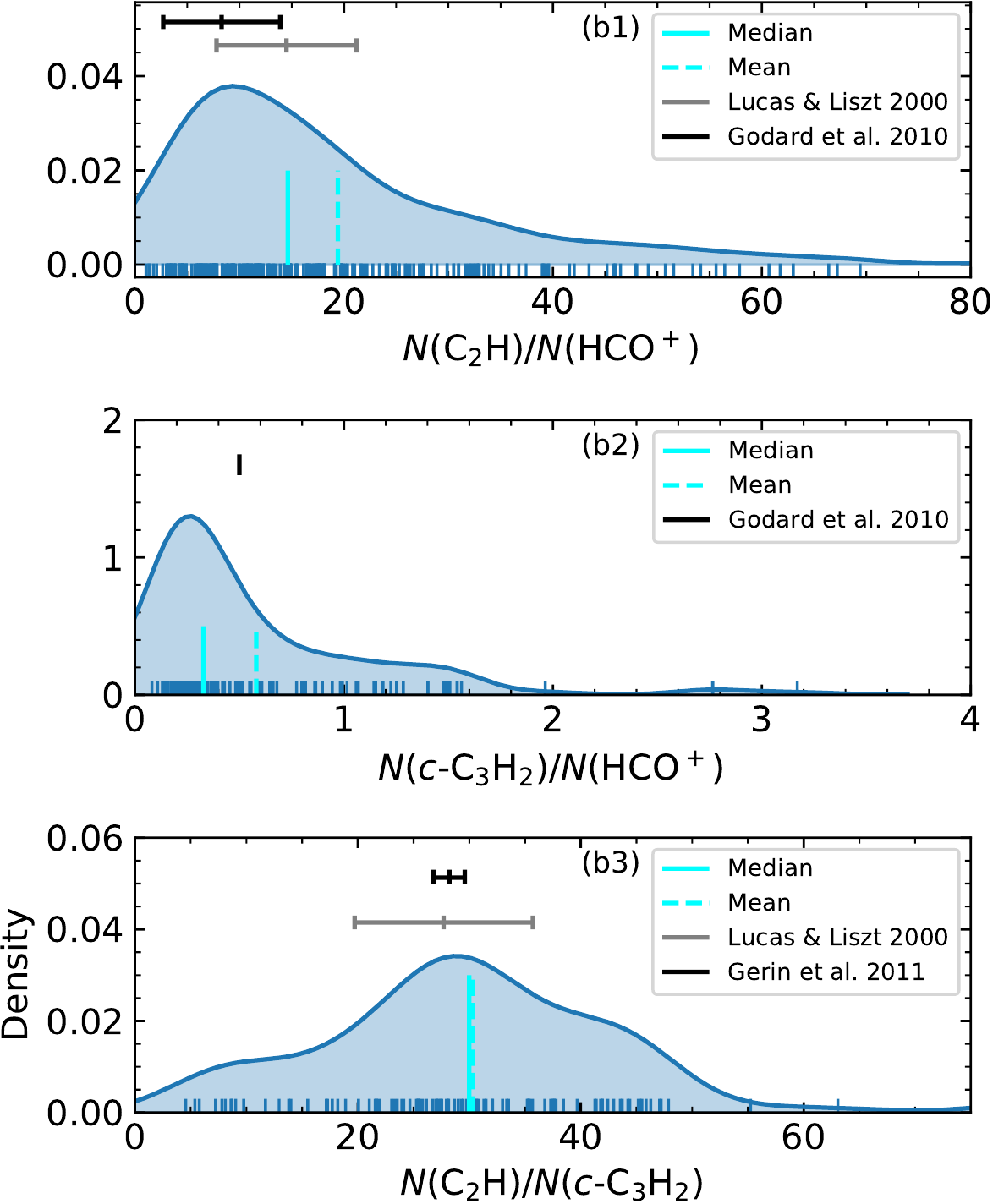}
    \includegraphics[width=0.33\textwidth]{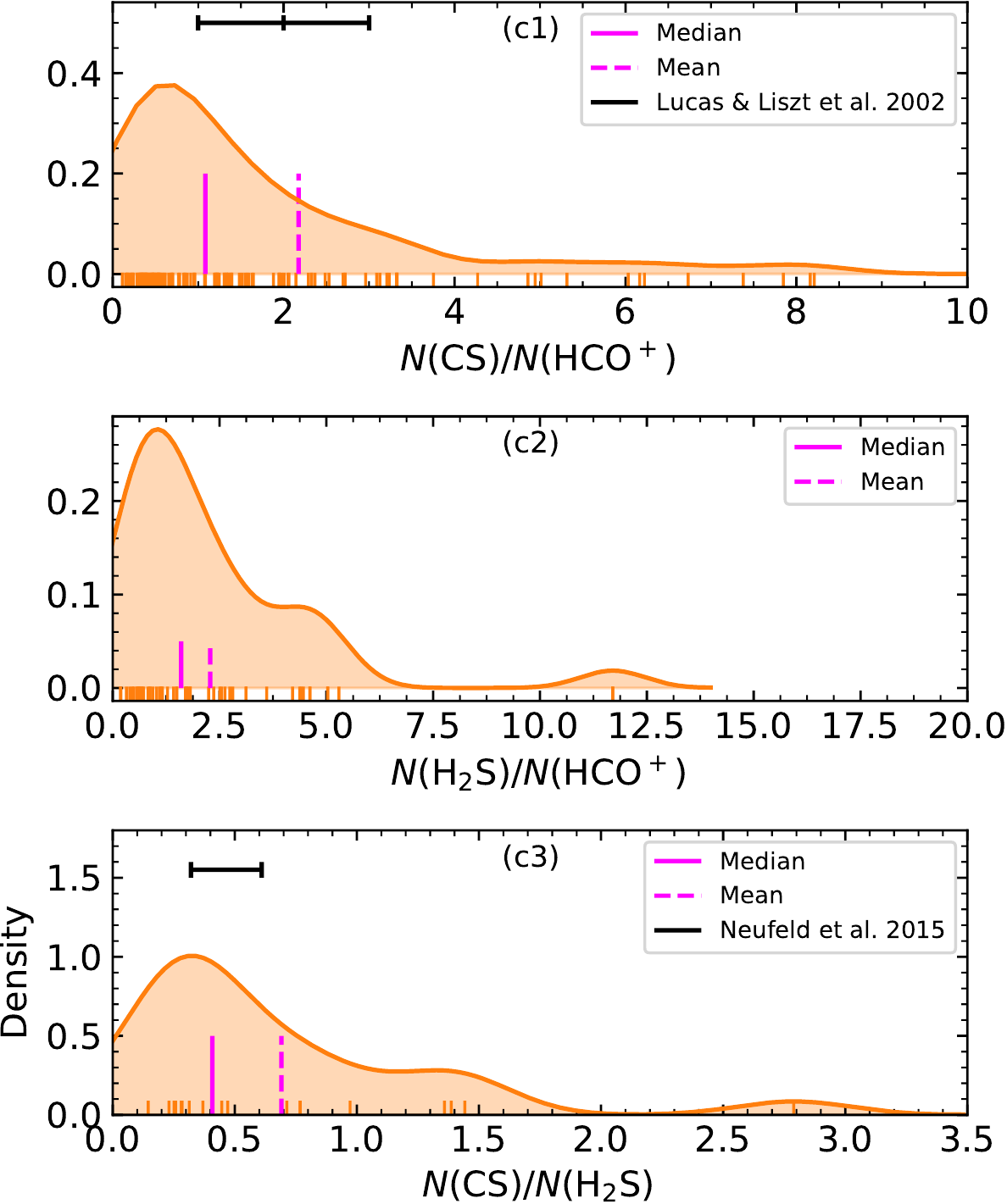}
    \caption{Kernel density estimate (KDE) distributions of the channel-wise column density ratios. Black and gray bars indicate unweighted mean ratio values measured from previous studies of absorption lines \citep{lucas2000_small_hydrocarbons, lucas2002_sulfur,godard2010_3mmabs,gerin2011_smallhydro, neufeld2015}. Median and mean values of the KDE distributions are marked by solid and dashed vertical lines (in lime, cyan, and fuchsia).}
    \label{fig:kde_col_ratio}
\end{figure*}

First, the molecular ions \ce{OH+}, \ce{H2O+}, and \ce{ArH+} and \ce{C+} lie away from neutral molecular species HCN, HNC, CS, \ce{H2S}, \ce{C2H}, and \ce{c-C3H2} and \ce{HCO+} in the h-plots. The first three mentioned molecular ions and \ce{C+} are considered to be tracers of gas with small molecular fractions, especially \ce{ArH+}, which traces gas with a molecular fraction $\lesssim10^{-3}$ \citep{schilke2014, jacob2020}. In Figs.\,\ref{fig:corr_heatmaps_pc1pc2} and \ref{fig:corr_heatmaps_pc2pc3}, the similarity coefficients between the group of hydride ions and the group of neutral molecules show that these two groups are poorly correlated (cooler colors). On the other hand, all the species {\it within} each group have very similar distributions, as shown by the warm colors in the heat maps. Clearly, \ce{HCO+} traces the same gas as that traced by neutral molecules, showing correlation coefficients higher than 92\,\% in the PC-1 versus PC-2 heat map. 

Second, the sulfur-bearing species, CS and \ce{H2S}, lie close to the neutral molecular species rather than to the molecular ions in the h-plots, and exhibit significant similarities as shown in Figs.\,\ref{fig:corr_heatmaps_pc1pc2} and \ref{fig:corr_heatmaps_pc2pc3}. Such a pattern is also found for other sightlines, W31C and W49N \citep{neufeld2015}. In agreement with \cite{neufeld2015}, we find the abundances of these two sulfur-bearing species to peak in regions of larger molecular fraction rather than in regions traced by \ce{C+} and the hydride ions \ce{OH+}, \ce{H2O+}, and \ce{ArH+}. 

Third, the vectors for CH and OH lie very close to each other in all the h-plots shown in Figs.\,\ref{fig:pc1pc2_circle} and \ref{fig:pc2pc3_circle}. Their correlation coefficients exceed 99.9\,\% (Figs.\,\ref{fig:corr_heatmaps_pc1pc2} and \ref{fig:corr_heatmaps_pc2pc3}). In addition, these molecules lie reasonably close to oxygen in the h-plots. The vectors corresponding to these three species lie between the neutral and ion molecules in both the h-plots, with the vectors for CH and OH located closer to those for the other neutral species and the vector for O located closer to those for the ionized species. CH is considered a good tracer for \ce{H2} because it is present in gas dominated by both atoms and molecules \citep{gerin2010_ch,wiesemeyer2018_ch,jacob2019_ch}. On the other hand, the vector for O tends to lie closer to those of the molecular ions that are found in gas with small molecular fractions. 

\subsection{Column densities and abundances}
In Fig.\,\ref{fig:ncol_channel}, we compare channel-wise column densities for all sightlines -- except for NGC 7538 IRS1 and HGAL0.55$-$0.85, toward which only \ce{HCO+} is detected -- to investigate whether the column densities show a linear relationship and whether there is any noticeable variation associated with specific spiral- or inter-arm regions. Due to high noise levels and contamination by emission lines, we have fewer detections for \ce{H2S} and CS. To obtain a reasonable number of data points required for reliable statistical results, we constrain the detection thresholds of their column densities to the 1.5\,rms and 2\,rms levels for \ce{H2S} and CS, respectively, while for the remaining five molecules, we adopt a 2.5\,rms threshold to mitigate uncertainties caused by poor sensitivity. 

The comparison of d$N$(HCN)/d$\varv$ and d$N$(\ce{HCO+})/d$\varv$ in Fig.\,\ref{fig:ncol_channel} (a) shows the best correlation, with a Pearson correlation coefficient ($r$) of 0.81. d$N$(\ce{C2H})/d$\varv$ in Fig.\,\ref{fig:ncol_channel} (c) has the second-best correlation with d$N$(\ce{HCO+})/d$\varv$, $r=$ 0.78: both these values differ from zero at a very high level of statistical significance. Along the sightlines toward extragalactic sources, both HCN and \ce{C2H} also show such good correlations and are considered to be proxies for the total column density of \ce{H2} \citep{liszt2001cn-bearing,lucas2000_small_hydrocarbons}. In Figs.\,\ref{fig:ncol_channel} (b) and (d), d$N$(HNC)/d$\varv$ and d$N$(\ce{c-C3H2})/d$\varv$ show moderate correlations ($r=$ 0.66 -- 0.71) with d$N$(\ce{HCO+})/d$\varv$. In contrast, the two sulfur-bearing molecules appear to be less correlated ($r=$ 0.45 -- 0.62) with d$N$(\ce{HCO+})/d$\varv$, as shown in Figs.\,\ref{fig:ncol_channel} (e) and (f). On the other hand, all the species show better correlations between the same chemical families as seen in the subplots of Fig.\,\ref{fig:ncol_channel_pair}, with correlation coefficients which are 0.85 for (a) $dN$(HCN)/$d\varv$ versus $dN$(HNC)/$d\varv$, 0.79 for (b) $dN$(\ce{C2H})/$d\varv$ versus $dN$(\ce{c-C3H2})/$d\varv$, and 0.77 for (c) $dN$(CS)/$d\varv$ versus $dN$(\ce{H2S})/$d\varv$. Such strong correlations between the same chemical families (HCN and HNC, \ce{C2H} and \ce{c-C3H2}, as well as CS and \ce{H2S}, in Fig\,\ref{fig:ncol_channel_pair}) have also been found in other diffuse and translucent clouds in the galactic plane and the solar neighborhood \citep{lucas2000_small_hydrocarbons,liszt2001cn-bearing,lucas2002_sulfur,gerin2011_smallhydro,godard2010_3mmabs,neufeld2015}.

\begin{table}[h!]
    \centering
    \caption{Median, mean and standard deviation of the column density ratios for the seven species studied in this work.}
    \begin{tabular}{l c c c}
    \hline \hline
    Ratio & Median & Mean & SD$^{\dagger}$  \\
    \hline 
    $N$(HCN)/$N$(\ce{HCO+})  & 1.37 & 1.78 & 1.68 \\
    $N$(HNC)/$N$(\ce{HCO+})  & 0.33 & 0.49 & 0.89 \\
    $N$(HNC)/$N$(HCN)        & 0.21 & 0.27 & 0.20 \\
    \hline
    $N$(\ce{C2H})/$N$(\ce{HCO+})     &  14.06  & 19.81 & 18.28 \\    
    $N$(\ce{c-C3H2})/$N$(\ce{HCO+})  &  0.33   & 0.57  & 0.53 \\
    $N$(\ce{C2H})/$N$(\ce{c-C3H2})   &  29.99  & 30.26 & 13.57 \\
    \hline
    $N$(CS)/$N$(\ce{HCO+})        & 0.95  & 1.99 & 3.33 \\
    $N$(\ce{H2S})/$N$(\ce{HCO+})  & 1.22   & 2.07 & 2.13 \\   
    $N$(CS)/$N$(\ce{H2S})         & 0.56   & 0.73 & 0.58 \\       
    \hline
    \end{tabular}
    \label{tab:ncol_ratios}
    \tablefoot{$^{\dagger}$ SD is the standard deviation. For \ce{H2S} and CS, the detection threshold lowers to 1.5$\times$\,rms, and 2$\times$\,rms levels, respectively, while ratios of the other species are obtained from data restricted by $>$ 2.5$\times$\,rms level. }
\end{table}

\begin{table*}[h!]
\centering
\tiny
\caption{Derived column densities over specific velocity intervals. }
\begin{tabular}{l c c c c c c c c c }
\hline
\hline
Source & $\varv_{\rm LSR}$ range & $N$(\ce{H2})& $N$(\ce{HCO+}) & $N$(HCN) & $N$(HNC) & $N$(CS) & $N$(\ce{H2S}) & $N$(\ce{C2H}) & $N$(\ce{c-C3H2}) \\
& (\kms) & ($10^{21}$cm$^{-2}$) & ($10^{12}$cm$^{-2}$) & ($10^{12}$cm$^{-2}$) & ($10^{12}$cm$^{-2}$) & ($10^{12}$cm$^{-2}$) & ($10^{13}$cm$^{-2}$) & ($10^{13}$cm$^{-2}$) & ($10^{12}$cm$^{-2}$) \\
 \hline
W3 IRS5 & $-$24, $-$18 & 0.37$^{+0.10}_{-0.10}$ & 1.11$^{+0.31}_{-0.30}$ & 1.67$^{+0.50}_{-0.49}$ & $\cdots$ & $\cdots$ & $\cdots$ & 2.17$^{+0.80}_{-0.69}$ & 0.79$^{+0.08}_{-0.08}$\\
 & $-$6, $-$1 & 0.08$^{+0.01}_{-0.01}$ & 0.24$^{+0.03}_{-0.03}$ & $\cdots$ & $\cdots$ & $\cdots$ & $\cdots$ & $\cdots$ & $\cdots$\\
 & $-$1, 5 & 0.27$^{+0.06}_{-0.06}$ & 0.82$^{+0.18}_{-0.18}$ & $\cdots$ & $\cdots$ & $\cdots$ & $\cdots$ & $\cdots$ & $\cdots$\\
\hline
W3(OH) & $-$23, $-$15 & 0.07$^{+0.03}_{-0.03}$ & 0.22$^{+0.09}_{-0.08}$ & 0.08$^{+0.06}_{-0.08}$ & $\cdots$ & $\cdots$ & $\cdots$ & 0.30$^{+0.01}_{-0.01}$ & 0.13$^{+0.01}_{-0.13}$\\
 & $-$14, $-$7 & 0.60$^{+0.05}_{-0.05}$ & 2.24$^{+0.20}_{-0.20}$ & 2.71$^{+0.22}_{-0.22}$ & 0.36$^{+0.03}_{-0.03}$ & $\cdots$ & $\cdots$ & 1.01$^{+0.02}_{-0.02}$ & 0.52$^{+0.03}_{-0.03}$\\
 & $-$7, 3 & 0.87$^{+0.10}_{-0.10}$ & 3.44$^{+0.48}_{-0.47}$ & 3.72$^{+0.41}_{-0.36}$ & 0.78$^{+0.04}_{-0.04}$ & 2.67$^{+0.15}_{-0.15}$ & 0.34$^{+0.02}_{-0.02}$ & 4.57$^{+0.41}_{-0.41}$ & 1.36$^{+0.10}_{-0.10}$\\
\hline
NGC 6334 I & 0, 12$^{a}$ & 4.79$^{+2.32}_{-2.56}$ & 25.84$^{+15.39}_{-15.39}$ & 9.17$^{+0.53}_{-0.53}$ & 1.35$^{+0.15}_{-0.14}$ & $\cdots$ & $\cdots$ & 1.98$^{+0.43}_{-0.40}$ & 0.72$^{+0.25}_{-0.23}$\\
\hline
HGAL0.55$-$0.85 & $-$14, $-$4$^{a}$ & 0.56$^{+0.28}_{-0.29}$ & 2.04$^{+1.26}_{-1.18}$ & $\cdots$ & $\cdots$ & $\cdots$ & $\cdots$ & $\cdots$ & $\cdots$\\
\hline
G09.62$+$0.19 & 9, 20 & 46.14$^{+45.78}_{-32.94}$ & 922.76$^{+915.66}_{-658.84}$ & 13.4$^{+0.56}_{-0.56}$ & 4.20$^{+0.64}_{-0.64}$ & $\cdots$ & 5.66$^{+17.06}_{-4.99}$ & $\cdots$ & $\cdots$\\
 & 20, 29 & 2.29$^{+0.65}_{-0.69}$ & 10.82$^{+3.72}_{-3.72}$ & 24.80$^{+3.93}_{-3.93}$ & 14.46$^{+23.68}_{-13.18}$ & $\cdots$ & 2.25$^{+0.66}_{-0.66}$ & 13.34$^{+7.83}_{-7.83}$ & $\cdots$\\
 & 62, 73 & 0.82$^{+0.62}_{-0.59}$ & 3.23$^{+3.06}_{-2.50}$ & $\cdots$ & $\cdots$ & $\cdots$ & $\cdots$ & $\cdots$ & $\cdots$\\
\hline
G10.47$+$0.03 & $-$18, $-$9 & 0.54$^{+0.14}_{-0.14}$ & 1.61$^{+0.42}_{-0.42}$ & 3.59$^{+1.13}_{-1.13}$ & $\cdots$ & $\cdots$ & $\cdots$ & $\cdots$ & $\cdots$\\
 & $-$9, 3 & 2.00$^{+0.82}_{-0.89}$ & 9.21$^{+4.63}_{-4.63}$ & 15.95$^{+7.50}_{-7.50}$ & $\cdots$ & 2.65$^{+0.32}_{-0.32}$ & $\cdots$ & $\cdots$ & $\cdots$\\
 & 3, 7 & 0.65$^{+0.15}_{-0.16}$ & 2.45$^{+0.69}_{-0.69}$ & 7.18$^{+7.31}_{-5.63}$ & 1.17$^{+0.32}_{-0.32}$ & 4.46$^{+0.82}_{-0.82}$ & $\cdots$ & 1.85$^{+0.27}_{-0.27}$ & $\cdots$\\
 & 7, 14 & 2.64$^{+0.95}_{-1.01}$ & 12.83$^{+5.58}_{-5.58}$ & 20.99$^{+12.26}_{-12.26}$ & 3.38$^{+0.82}_{-0.82}$ & 5.44$^{+0.81}_{-0.81}$ & $\cdots$ & 12.51$^{+4.56}_{-4.47}$ & $\cdots$\\
 & 14, 19 & 0.95$^{+0.20}_{-0.21}$ & 3.83$^{+0.96}_{-0.96}$ & 3.62$^{+4.78}_{-2.53}$ & 1.54$^{+0.39}_{-0.39}$ & $\cdots$ & $\cdots$ & 2.49$^{+1.45}_{-1.24}$ & $\cdots$\\
 & 19, 28 & 1.52$^{+0.38}_{-0.40}$ & 6.70$^{+2.01}_{-2.01}$ & 9.57$^{+7.96}_{-7.81}$ & $\cdots$ & $\cdots$ & $\cdots$ & 17.59$^{+2.06}_{-1.82}$ & $\cdots$\\
 & 28, 32 & 1.71$^{+0.75}_{-0.82}$ & 7.67$^{+4.11}_{-4.11}$ & 7.92$^{+3.76}_{-3.76}$ & $\cdots$ & $\cdots$ & $\cdots$ & 11.05$^{+1.20}_{-1.20}$ & $\cdots$\\
 & 32, 44 & 2.53$^{+0.74}_{-0.78}$ & 12.20$^{+4.31}_{-4.31}$ & 36.84$^{+22.83}_{-22.79}$ & 4.87$^{+1.38}_{-1.38}$ & $\cdots$ & $\cdots$ & 22.93$^{+8.24}_{-5.39}$ & $\cdots$\\
 & 44, 51 & 0.06$^{+0.03}_{-0.06}$ & 0.18$^{+0.08}_{-0.18}$ & $\cdots$ & $\cdots$ & $\cdots$ & $\cdots$ & $\cdots$ & $\cdots$\\
 & 82, 94 & 0.23$^{+0.07}_{-0.05}$ $^{b}$ & 0.70$^{+0.20}_{-0.14}$ & 22.05$^{+2.23}_{-2.23}$ & $\cdots$ & $\cdots$ & $\cdots$ & $\cdots$ & $\cdots$\\
 & 114, 133 & 1.26$^{+0.27}_{-0.28}$ $^{b}$ & 5.36$^{+1.38}_{-1.38}$ & 9.75$^{+3.56}_{-3.56}$ & $\cdots$ & 9.58$^{+0.96}_{-0.96}$ & $\cdots$ & $\cdots$ & $\cdots$\\
 & 137, 151 & 0.85$^{+0.15}_{-0.16}$ & 3.37$^{+0.73}_{-0.73}$ & 7.05$^{+2.11}_{-2.11}$ & $\cdots$ & 13.44$^{+2.94}_{-2.93}$ & $\cdots$ & $\cdots$ & $\cdots$\\
 & 151, 159 & 4.02$^{+1.47}_{-1.57}$ & 21.02$^{+9.33}_{-9.33}$ & 19.50$^{+13.7}_{-13.7}$ & 3.08$^{+0.92}_{-0.92}$ & 13.78$^{+2.30}_{-2.30}$ & $\cdots$ & 4.23$^{+1.43}_{-1.41}$ & $\cdots$\\
 & 159, 164 & 0.24$^{+0.06}_{-0.06}$ & 0.71$^{+0.18}_{-0.18}$ & 1.81$^{+0.47}_{-0.47}$ & $\cdots$ & 4.49$^{+0.54}_{-0.54}$ & $\cdots$ & $\cdots$ & $\cdots$\\
 & 164, 175 & 0.22$^{+0.04}_{-0.04}$ & 0.65$^{+0.13}_{-0.13}$ & 1.15$^{+0.81}_{-0.54}$ & $\cdots$ & 7.44$^{+1.02}_{-1.02}$ & $\cdots$ & $\cdots$ & $\cdots$\\
\hline
\end{tabular}
\label{tab:nmol1}
\tablefoot{$N$(\ce{H2}) is derived using $N$(\ce{HCO+}); $N$(\ce{H2}) $= N$(\ce{HCO+})/ $3\times10^{-9}$ for a case with $N$(\ce{HCO+})~$< 2.8\times10^{12}$\,cm$^{-2}$, $N$(\ce{H2}) $= N$(\ce{HCO+})/$2\times10^{-8}$ for a case with $N$(\ce{HCO+})~$> 9.4\times10^{13}$\,cm$^{-2}$, and an interpolated value between these two ranges of $N$(\ce{HCO+}), log($N$(\ce{H2})) = log($N$(\ce{HCO+}))$\times${\bf 1.64}. The column densities derived for each species toward the remaining sources is tabulated in Appendix\,\ref{tab-appendix:ntot}. The velocity intervals are defined, based on \ce{HCO+} absorption line profiles. \\
$^{a}$ refers to the velocity intervals corresponding to molecular gas components associated with the background continuum sources, and thus the derived column densities of these velocity intervals indicate lower limits. \\ $^{b}$ indicates the $N$(\ce{H2}) for the velocity intervals belonging to the GC region where likely have higher $X$(\ce{HCO+}) ($> 3.9\times10^{-9}$ for the central molecular zone \citep{riquelme2018_cmz_hcop}) than the Galactic disk regions.}
\end{table*}

In Figs.\,\ref{fig:ncol_channel} and \ref{fig:ncol_channel_pair} the black dotted straight lines have a slope that represents the average column density ratio between any two given species. The red solid lines are the best fits to the data (only including the data points above the detection thresholds), including column density uncertainties, shown as error bars on the plots. The gray areas represent the 2$\sigma$ confidence interval of the best-fit uncertainties. These uncertainties are obtained by resampling the data points and then iteratively fitting the resampled data sets. Comparing $dN$(\ce{HCO+})/$d\varv$ versus $dN$(HCN)/$d\varv$, we find that the median value and the best-fit agree fairly well. For the other cases, comparing different chemical groups, the best fit lines are not as steep as the lines representing the median. Such discrepancies become significant when comparing sulfur-containing molecules (CS and \ce{H2S}) with \ce{HCO+}, shown in plots (e) and (f) of Fig.\,\ref{fig:ncol_channel} since their relationships are weakly correlated. In addition, each color represents a different spiral-arm or intermediate arm region, or an unassigned foreground clouds, as tabulated in Table\,\ref{tab:arm_info}. For HCN, HNC, and the two small hydrocarbons, there is no apparent difference in column densities between clouds associated with the different environments. However, in panel (e) of Fig.\,\ref{fig:ncol_channel} for CS, the cloud components belonging to the 135\,\kms\ arm, marked in blue-green color, have higher $dN$(CS)/$d\varv$ compared to the other spiral arms that have $dN$(\ce{HCO+})/$d\varv$~$\approx 1-9\times10^{11}$~cm$^{-2}$~km$^{-1}$~s. 

In panel (b) of Fig.\,\ref{fig:ncol_channel_pair}, the two small hydrocarbons, have a nonlinear relationship for values of $dN$(\ce{C2H})/$d\varv < 1\times10^{13}$~cm$^{-2}$~km$^{-1}$~s, which plateaus as $dN$(\ce{c-C3H2})/$d\varv$ increases ($\approx 1-4\times10^{11}$~cm$^{-2}$~km$^{-1}$~s). As $dN$(\ce{C2H})/$d\varv$ increases beyond the inflection point from ~$\approx 1\times10^{13}$\,cm$^{-2}$~km$^{-1}$~s to higher \ce{C2H} column densities, the column densities of these small hydrocarbons follow a linear relationship. Taking into account fitting uncertainties, the flat slope between the column densities of the two small hydrocarbons at lower values of $N$(\ce{C2H}) remains constant.

\begin{figure*}[htpb]
    \centering
    \includegraphics[width=0.33\textwidth]{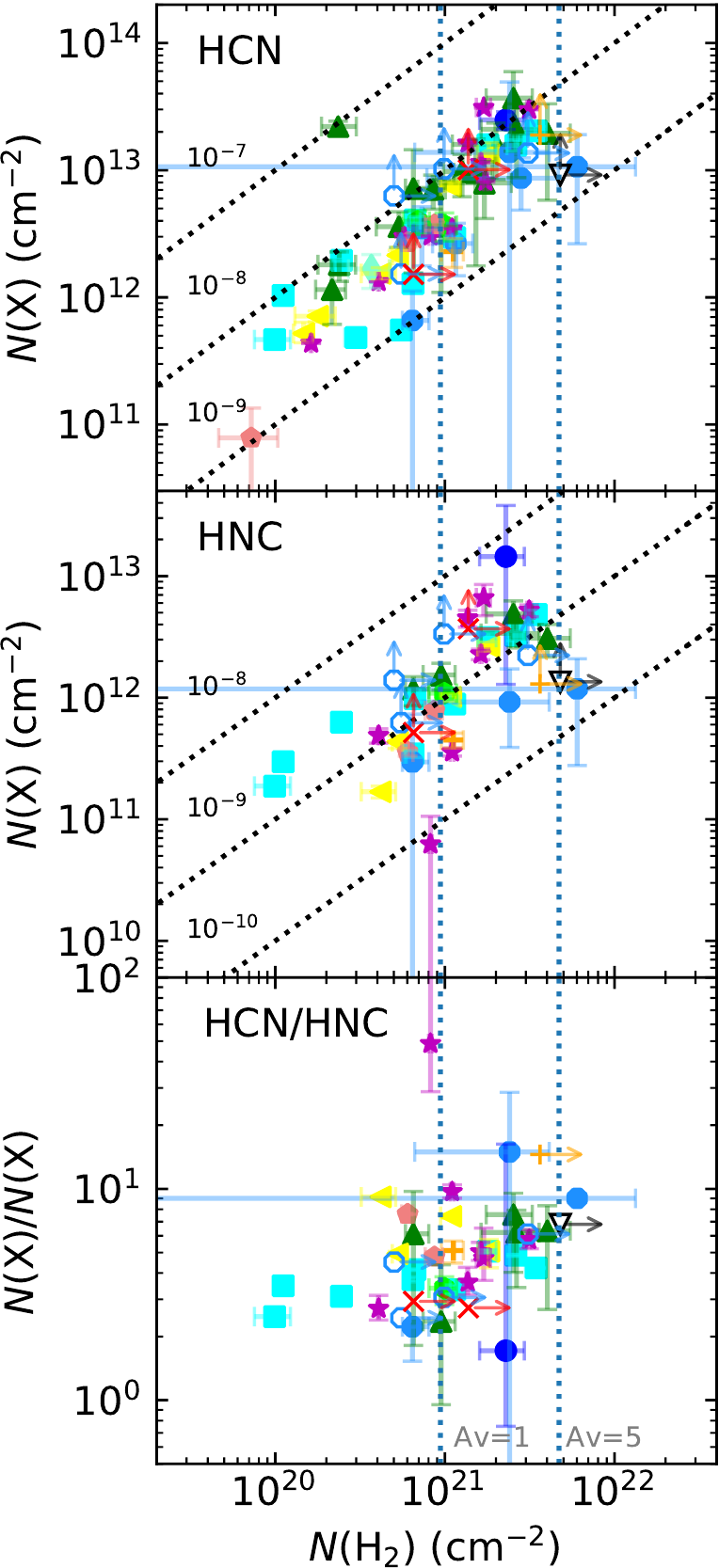}
    \includegraphics[width=0.33\textwidth]{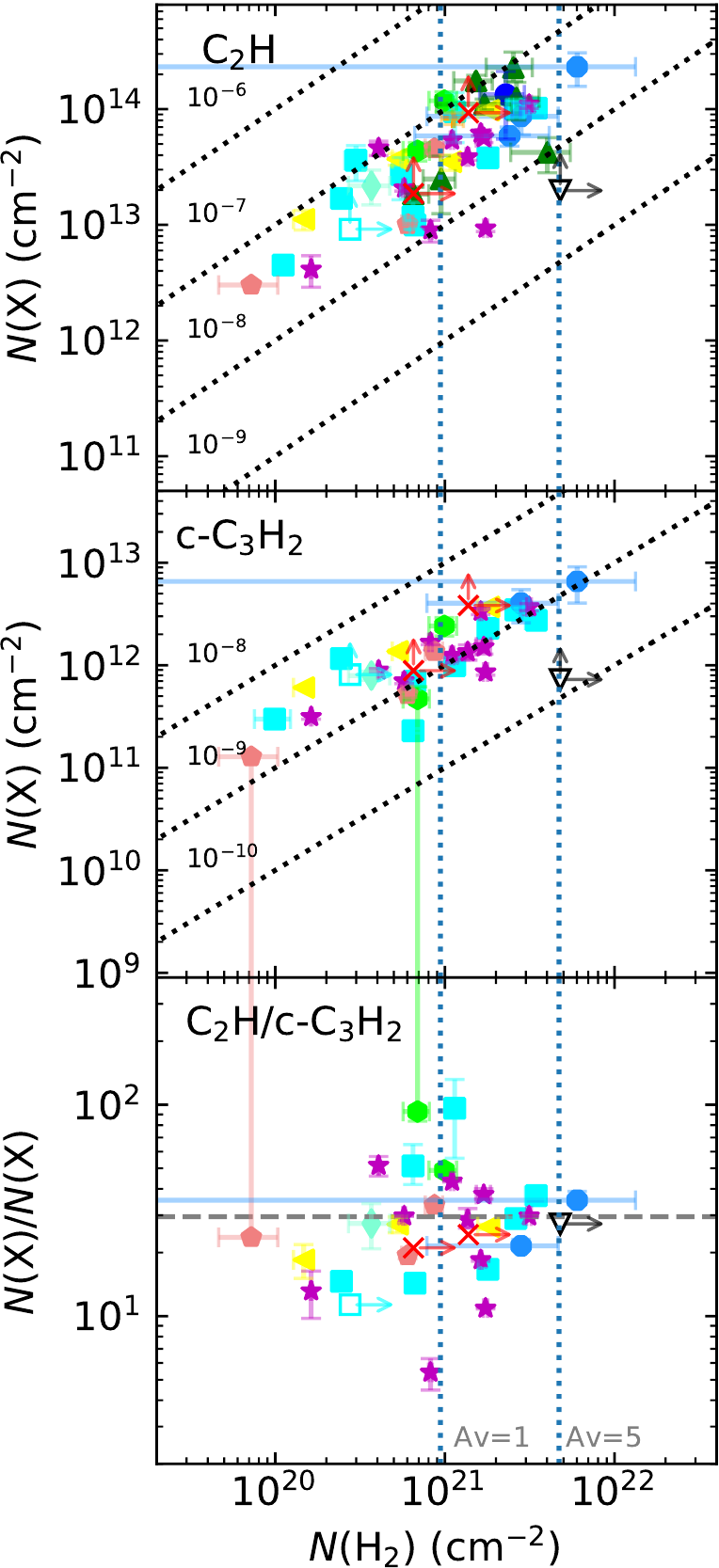}
    \includegraphics[width=0.33\textwidth]{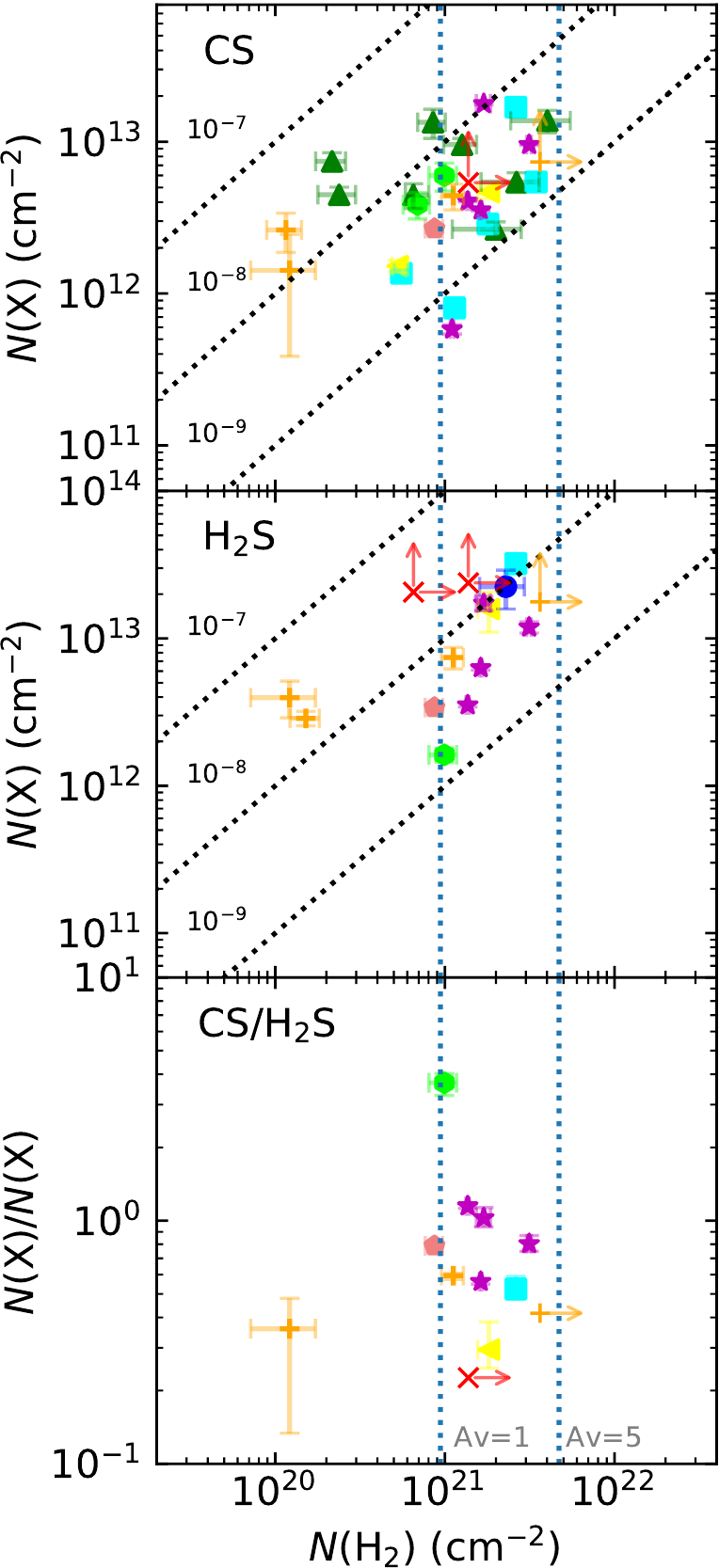}
    \caption[]{Comparison of the column density of the presented molecular species and $N$(\ce{H2}). Each colored symbol represents different individual clouds in the sightline toward the observed background continuum sources (DR21: Red X, G09.62$+$0.19: Blue circle, G10.47$+$0.03: Green up-triangle, G19.61$-$0.23: Cyan square, G29.96$-$0.02: Yellow left-triangle, G31.41$+$0.31: Lime hexagon, G32.80$+$0.19: Magenta star, G45.07$+$0.13: Orange cross, NGC 6334 I: Black down-triangle, W3 IRS5: Aquamarine thin diamond, W3(OH): Light-coral pentagon, W43 MM1: Dodger-blue octagon). Filled and unfilled markers indicate foreground components and molecular envelope components, respectively. For the molecular envelope components, only lower limits are marked by arrows instead of fitting uncertainty bars. Black diagonal dotted lines represent abundances of given molecules on the y-axis relative to $N$(\ce{H2}) with their values labeled. In each panel, two vertical blue dotted lines indicate the visual extinction of 1 and 5, respectively. In the plot of $N$(\ce{C2H})/$N$(\ce{c-C3H2}), the gray horizontal dashed line indicates a mean value of 30.26 in the column density ratio.}
    \label{fig:int_ntot}
\end{figure*}

Figure\,\ref{fig:kde_col_ratio} shows kernel density estimate (KDE) plots to examine distributions of channel-wise column density ratios. With the KDE method the probability density function of sample data can be estimated based on kernels as weights. KDE provides smooth profiles of the density, unlike histograms. Here we used a Gaussian kernel placed on each ratio data point as marked by the short vertical lines (in green, blue, and orange colors). The KDEs, shown as solid curves, are the sum of all kernels at each column density ratio. The solid and dashed vertical lines are median and mean values, respectively, of the distributions. The values for the median and mean as well as the standard deviations (SDs) are listed in Table\,\ref{tab:ncol_ratios}. We compare the values with ratios obtained from previous absorption line studies \citep{lucas2000_small_hydrocarbons,lucas2002_sulfur,godard2010_3mmabs,gerin2011_smallhydro, neufeld2015}, which are marked by black and gray bars.
 From this work we get mean ratios of 1.78, 0.49, and 0.27 for $N$(HCN)/$N$(\ce{HCO+}), $N$(HNC)/$N$(\ce{HCO+}) and $N$(HNC)/$N$(HCN), respectively. The mean ratios of $N$(HCN)/$N$(\ce{HCO+}) (Fig.\,\ref{fig:kde_col_ratio} (a1)) and $N$(HNC)/$N$(\ce{HCO+}) (Fig.\,\ref{fig:kde_col_ratio} (a2)) are very similar to values from the previous studies, while the mean ratio of $N$(HNC)/$N$(HCN) (Fig.\,\ref{fig:kde_col_ratio} (a3)) is slightly higher than that derived by \citet{liszt2001cn-bearing} and \citet{godard2010_3mmabs} of (0.21$\pm$0.71) but within the error range. The median $N$(HNC)/$N$(HCN) ratio of 0.22 from this survey agrees well with the other studies. According to \citealt{liszt2001cn-bearing}, a small $N$(HNC)/$N$(HCN) ratio such as 0.2 is expected to indicate warmer gas at higher $A_{\rm v}$ or denser gas conditions. In addition, this ratio distribution is very similar to the distribution of HNC/HCN, between 0.01 and 1.0, toward star-forming regions (e.g., \citealt{graninger2014}) and Orion-KL/the OMC regions (e.g., \citet{schilke1992} and \citealt{hacar2020}; HNC/HCN $\sim$ 0.013 -- 0.2). Toward some sightlines, $N$(HNC)/$N$(HCN) ratios are between 0.2 and 1.0. This range might imply that the observed HCN and HNC trace warmer regions and/or inner parts of diffuse molecular clouds since the high ratio of these two molecules are found toward TMC-1 (ratio of 1, \citealt{ohishi1992}).
 
The mean ratios of the column densities of \ce{C2H} and \ce{c-C3H2} to $N$(\ce{HCO+}) (Figs.\,\ref{fig:kde_col_ratio} (b1) and (b2)) and $N$(\ce{C2H})/$N$(\ce{c-C3H2}) (Fig.\,\ref{fig:kde_col_ratio} (b3)) are 19.81, 0.57, and 30.26. The mean value of the column density ratios of the two small hydrocarbons agrees well with previously measured values of 27.7$\pm$0.8 and 28.2$\pm$1.4 from \citet{lucas2000_small_hydrocarbons} and \citet{gerin2011_smallhydro}, as shown in Fig.\,\ref{fig:kde_col_ratio} (b3). For $N$(CS) versus $N$(\ce{HCO+}), $N$(\ce{H2S}) versus $N$(\ce{HCO+}) and $N$(CS) versus $N$(\ce{H2S}), the mean ratios are 1.99, 2.07, and 0.73, respectively. Like other species, these two sulfur-containing molecules have mean ratios greater than their median values (Figs.\,\ref{fig:kde_col_ratio} c1, c2, and c3, and Table\,\ref{tab:ncol_ratios}). The previously measured mean ratios of 2$\pm$1 for $N$(CS)/$N$(\ce{HCO+}) \citep{lucas2002_sulfur} and 0.32$-$0.61 for $N$(CS)/$N$(H2S) \citep{neufeld2015} come very close to the measurements obtained here.

\begin{table*}[h!]
    \centering
    \caption{Median value of abundances relative to \ce{H2} and $N$(\ce{H2}) of each targeted species X observed toward different cloud components.}
    \begin{tabular}{l c c c c c c c c c c c }
    \hline \hline
    Species &\multicolumn{3}{c}{All clouds$^{a}$} & &\multicolumn{3}{c}{Diffuse cloud} & & \multicolumn{3}{c}{Translucent cloud}  \\ \cline {2-4} \cline{6-8} \cline{10-12}
     & \# &$N$(X)/$N$(\ce{H2}) & $N$(\ce{H2}) (cm$^{-2}$) & & \# & $N$(X)/$N$(\ce{H2}) & $N$(\ce{H2}) (cm$^{-2}$) & & \#& $N$(X)/$N$(\ce{H2}) & $N$(\ce{H2}) (cm$^{-2}$)  \\
    \hline
    \ce{HCN} & 64 & $4.80\times10^{-9}$ & $9.69\times10^{20}$ & & 31 & $4.49\times10^{-9}$ & $5.26\times10^{20}$ & & 30 & $6.13\times10^{-9}$ & $1.72\times10^{21}$  \\    
    \ce{HNC} & 42 & $1.17\times10^{-9}$ & $1.11\times10^{21}$ &  & 16 & $1.02\times10^{-9}$ & $5.77\times10^{20}$ & & 23 &  $1.40\times10^{-9}$ & $1.80\times10^{21}$  \\
    \hline
    \ce{C2H} & 46 & $3.84\times10^{-8}$ & $1.09\times10^{21}$ & & 20 & $4.11\times10^{-8}$ & $5.40\times10^{20}$ & & 24 & $3.82\times10^{-8}$ & $1.72\times10^{21}$  \\
    \ce{c-C3H2} & 33 & $1.36\times10^{-9}$ & $8.68\times10^{20}$ & & 17 & $1.93\times10^{-9}$ & $5.26\times10^{20}$ & &  14  & $1.23\times10^{-9}$ & $1.72\times10^{21}$ \\       
    \hline
    \ce{CS} & 28 & $3.26\times10^{-9}$ & $1.20\times10^{21}$ & & 10 & $9.28\times10^{-9}$ & $5.40\times10^{20}$ & & 18 & $2.77\times10^{-9}$ &  $1.74\times10^{21}$  \\
    \ce{H2S} & 16 & $7.67\times10^{-9}$ & $1.50\times10^{21}$ & & 4  & $2.53\times10^{-8}$ &  $4.02\times10^{20}$ & & 11 & $6.62\times10^{-9}$ & $1.69\times10^{21}$  \\
    \hline
    \end{tabular}
    \label{tab:X_nh2}
    \tablefoot{$^{a}$ includes all cloud types (dense, translucent, and diffuse). The \# columns indicate the number of foreground components used for the statistics. All features associated with background sources are excluded in these calculations because the column densities of these features only represent lower limits.}
\end{table*}

The detected absorption features of simple molecular species studied in this work have different velocity components, and some of them have CO emission counterparts, shown in Fig.\,\ref{fig:co_emission}. Therefore, to characterize the properties of individual diffuse and translucent foreground clouds, we integrated column densities over given velocity intervals to identify individual cloud components and list the average column densities for all 14 line-of-sight sources in Tables\,\ref{tab:nmol1} and \ref{tab:nmol2}. Figure\,\ref{fig:int_ntot} shows the column densities (upper and middle image) of HCN, HNC, \ce{C2H}, \ce{c-C3H2}, CS, and \ce{H2S} and column density ratios (lower panel) as a function of $N$(\ce{H2}). The \ce{H2} column density is derived from $N$(\ce{HCO+}) using a \ce{HCO+} abundance of $3\times10^{-9}$ relative to \ce{H2} ($N$(\ce{H2}) $= N$(\ce{HCO+})/$3\times10^{-9}$) \citep{lucas2000_small_hydrocarbons,liszt2010} for $N$(\ce{HCO+}) $< 2.8\times10^{12}$\,cm$^{-2}$, $2\times10^{-8}$ ($N$(\ce{H2}) $= N$(\ce{HCO+})/$2\times10^{-8}$) for $ N$(\ce{HCO+}) $> 9.4\times10^{13}$\,cm$^{-2}$, and an interpolated value between these two ranges of $N$(\ce{HCO+}), log($N$(\ce{H2})) = log($N$(\ce{HCO+}))$\times$1.64, because $N$(\ce{HCO+}) and $N$(\ce{H2}) does not follow a linear correlation in regions with higher \ce{H2} column density \citep{thiel2019}. Visual extinctions of 1 and 5 are marked with vertical blue dotted lines and estimated using $A_{\rm v}$ (mag) $=N$(\ce{H2})/($9.4\times10^{20}$ cm$^{-2}$) \citep{thiel2019}. In Fig.\,\ref{fig:int_ntot}, symbols of the same color indicate foreground clouds in the same direction, and the black diagonal lines represent lines of constant abundance for each individual molecular species presented. We exclude the cloud component in the velocity range between 9.0 and 20.2\,\kms\ for G09.62$+$0.19 as its extremely large $N$(\ce{HCO+}) value of 9.2$\times10^{14}$\,cm$^{-2}$ indicates that the component does not originate from either diffuse or translucent clouds.

On the left-hand side of Fig.\,\ref{fig:int_ntot}, $N$(HCN) and $N$(HNC) increase as $N$(\ce{H2}) ($A_{\rm v}$) increases, but slightly nonlinearly. In particular, the HCN abundance increases briefly relative to \ce{H2} at $N$(\ce{H2}) $\gtrsim 10^{21}$\,cm$^{-2}$, regions considered to be translucent clouds. \citealt{liszt2001cn-bearing} also reported such nonlinear relationships of the column densities of CN-bearing molecules. They show a rapid increase in abundances of both the CN-bearing molecules detected toward sightlines with $N$(\ce{HCO+}) $\gtrsim 10^{12}$ cm$^{-2}$ equivalent to $N$(\ce{H2}) $\sim 0.3-0.5\times10^{21}$\,cm$^{-2}$ by assuming an abundance of $3\times10^{-9}$ relative to \ce{H2}, against extra-galactic continuum sources. The lower-left panel shows the column density ratios of HCN and HNC as a function of $N$(\ce{H2}), and these ratios tend to increase slightly toward higher molecular hydrogen column densities, tracing higher visual extinction regions of molecular clouds.

\begin{figure*}
    \centering
    \includegraphics[width=0.65\textwidth]{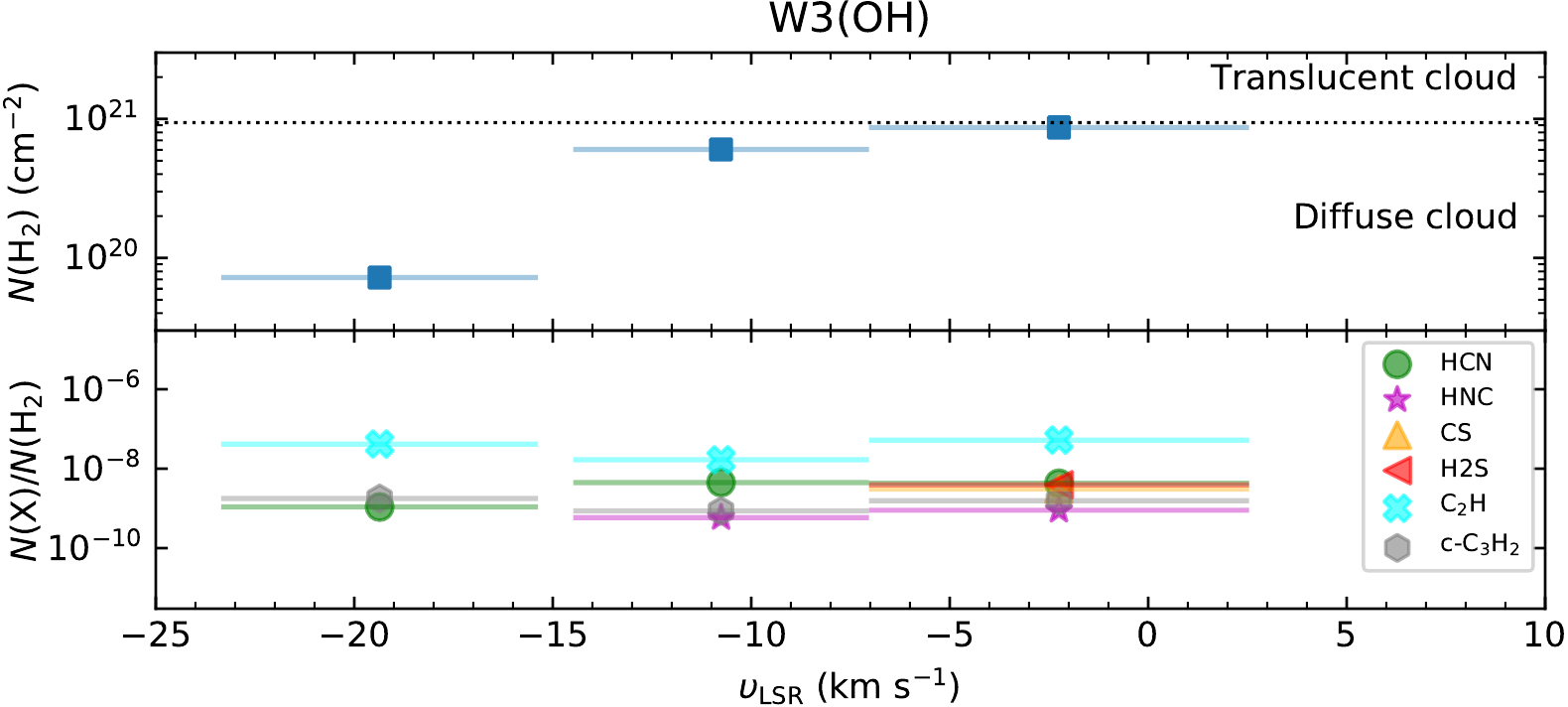}
    \includegraphics[width=0.65\textwidth]{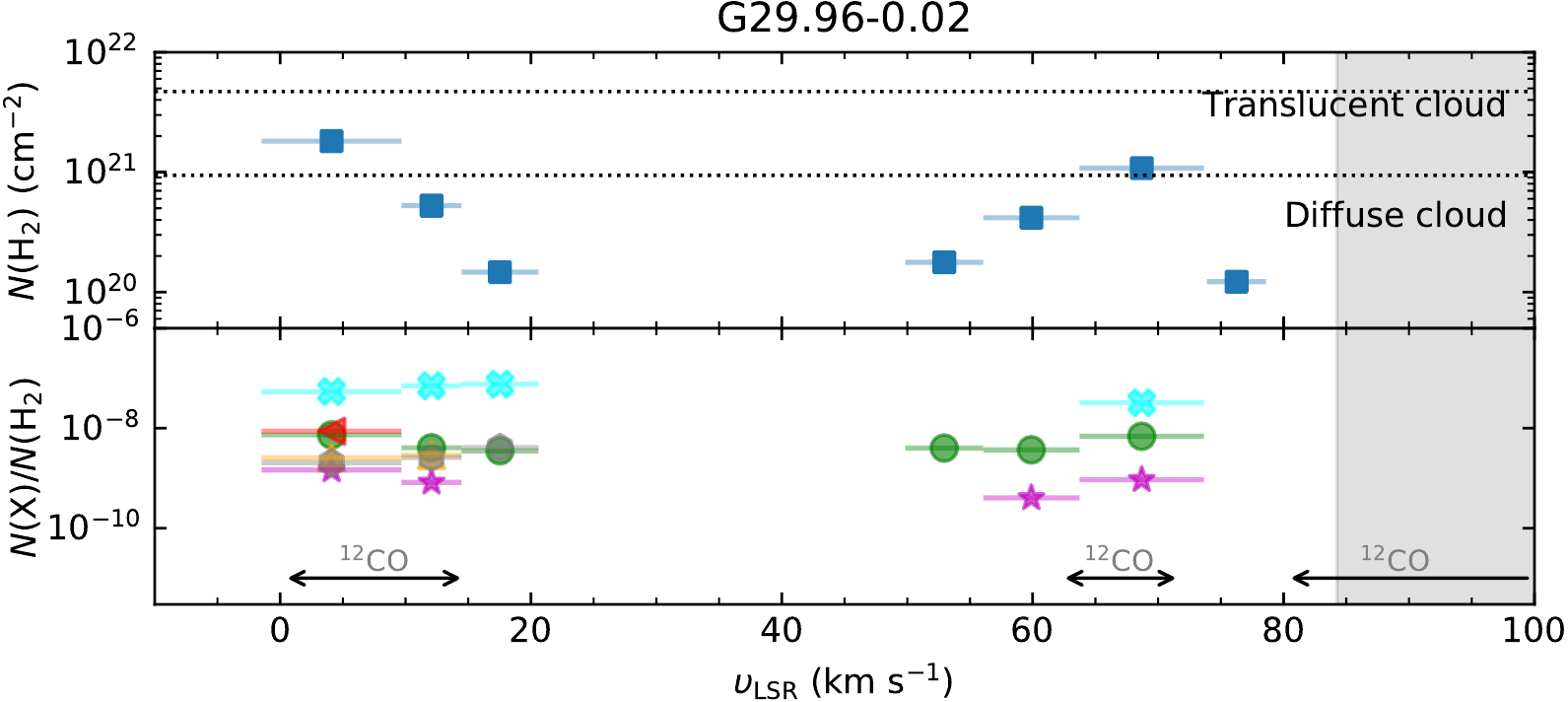}    
    \includegraphics[width=0.65\textwidth]{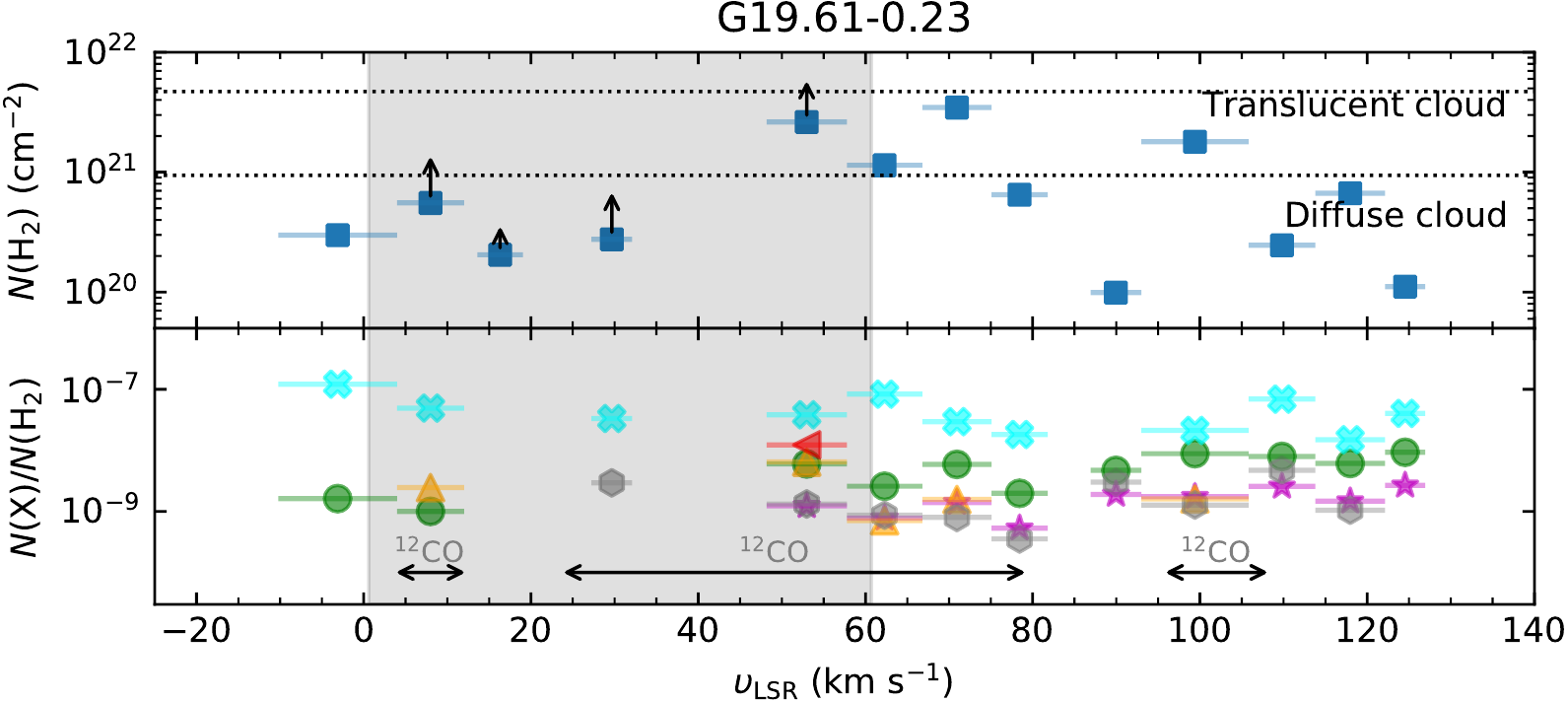}
    \includegraphics[width=0.65\textwidth]{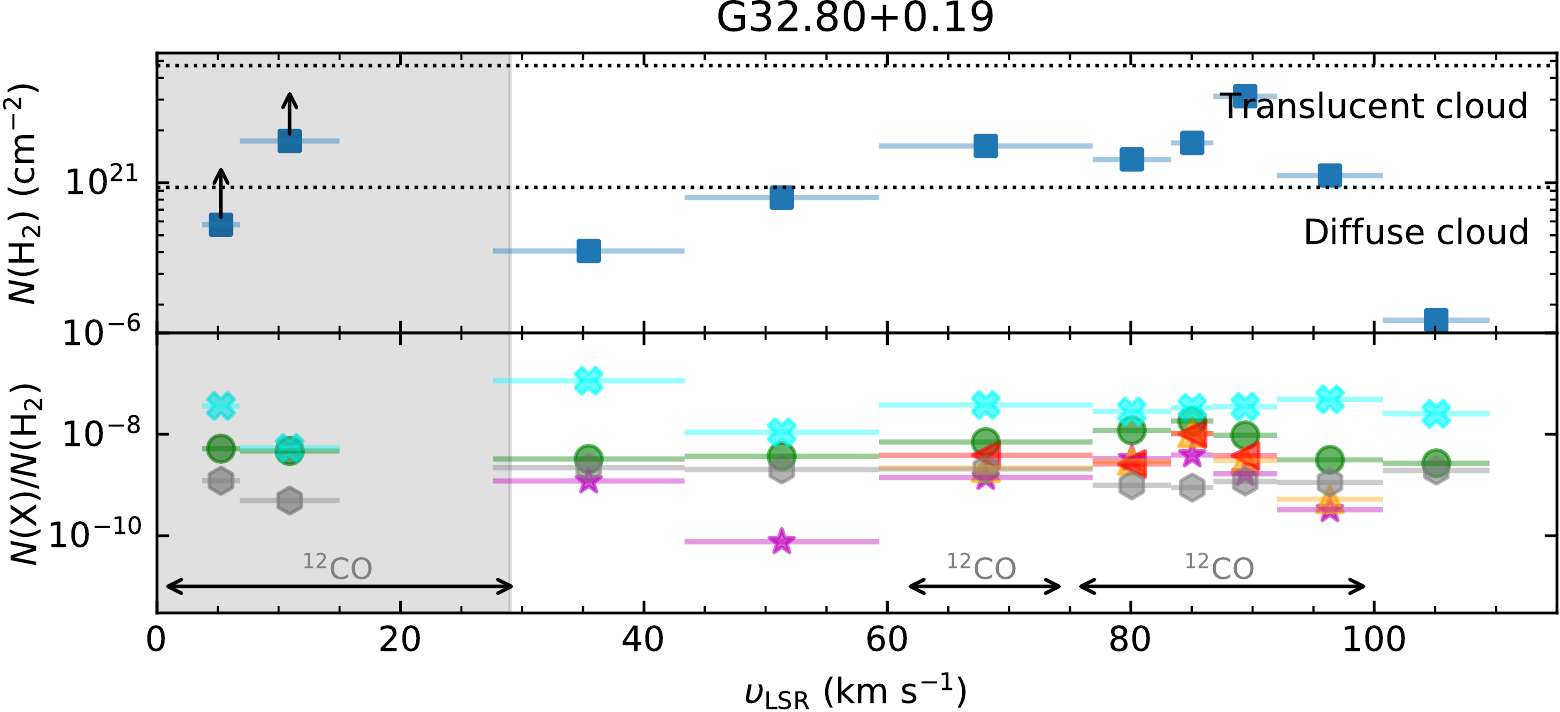}   
    \caption{$N$(\ce{H2}) and $N$(X)/$N$(\ce{H2}) ratio integrated over specific velocity intervals from top to bottom toward W3(OH), G29.96$-$0.02, G19.61$-$0.23, and G32.80$+$0.19. Different species are labeled with different symbols and colors as shown in the legend displayed in the right-hand corner of the top panel. The gray areas indicate the velocity ranges of the emission lines. In the upper panels, the dotted horizontal lines correspond to $A_{\rm v}$ of 1 and 5, respectively. In the lower panels, the black horizontal arrows indicate the velocity ranges in which CO emission features have been detected from archival data at its given sensitivity. All the other sources are displayed in Figs.\,\ref{fig:x_vel_add1} and \ref{fig:x_vel_add2}.}
    \label{fig:x_vel}
\end{figure*}

As expected, in the middle plots of Fig.\,\ref{fig:int_ntot}, the column densities of the two small hydrocarbons (\ce{C2H} and \ce{c-C3H2}) show tight linear correlations with $N$(\ce{H2}), and their abundance varies only a little across different environments from diffuse ($A_{\rm v}<1$) gas to dense molecular gas ($A_{\rm v}>5$), through translucent molecular gas regions in between. The \ce{C2H} abundance measured here lies in an approximate range from $10^{-8}$ to $10^{-7}$, while that of \ce{c-C3H2} is pretty much within $10^{-9}-10^{-8}$. These abundances relative to \ce{H2} agree well with the $X$(\ce{C2H})$~\sim 4-7\times10^{-8}$ \citep{lucas2000_small_hydrocarbons, gerin2011_smallhydro} and $X$(\ce{c-C3H2})$~\sim\,2-3\times10^{-9}$ \citep{liszt2012_c3h2} in diffuse gas, measured in previous absorption line studies. As seen in the bottom plot, the $N$(\ce{C2H})/$N$(\ce{c-C3H2}) ratios in individual cloud components of our targets are constant over the values of $N$(\ce{H2}) probed, with an average column density ratio of $N$(\ce{C2H})/$N$(\ce{c-C3H2})~$\sim$\,30.26 indicated by the gray horizontal dashed line. This average ratio agrees well with the values obtained by \citealt{lucas2000_small_hydrocarbons}.

As already seen in Fig.\,\ref{fig:ncol_channel}, $N$(CS) and $N$(\ce{H2S}) also show poor correlations with $N$(\ce{H2}) in Fig.\,\ref{fig:int_ntot}, in contrast to the column densities of HCN, HNC, \ce{C2H} and \ce{c-C3H2}, although $N$(CS) has some degree of correlation with $N$(\ce{H2}). The abundances of CS and \ce{H2S} derived in diffuse clouds are higher $>10^{-8}$ than those in translucent clouds ($10^{-9}<$ abundance $<10^{-8}$). However, apart from the fitting uncertainties, the derived column densities of CS and \ce{H2S} probably contain higher uncertainties than the lines of the other species, for example, due to poor sensitivities, as seen in Fig\,\ref{fig:abs_spec_g29_w43_g31}. Therefore, it is difficult to conclude that the abundance of these sulfur-bearing species is higher toward diffuse clouds than toward translucent clouds. We need more observational data points with higher sensitivity toward more diffuse clouds to confirm this hypothesis since we only have a small sample (10 for CS and 4 for \ce{H2S}). The CS abundances toward the Galactic translucent clouds are in good agreement with $X$(CS)~$\sim~4\times10^{-9}$ determined for high latitude diffuse clouds in the solar neighborhood \citep{lucas2002_sulfur} in spite of the sulfur chemistry being different in these environments. Likewise, the abundances of \ce{H2S} are similar with $X$(\ce{H2S}) determined from diffuse clouds in different parts of the Galactic plane \citep{neufeld2015}. In the plot of $N$(CS)/$N$(\ce{H2S}), a majority of the foreground clouds with detections for both these molecules are identified as translucent clouds but still associated with relatively less dense regions ($A_{\rm v}\sim1-3$ mag). In addition, the column density ratios do not show any correlation with the total hydrogen column density in the sightlines studied here.

\begin{figure*}[h!]
    \centering
    \includegraphics[width=0.23\textwidth]{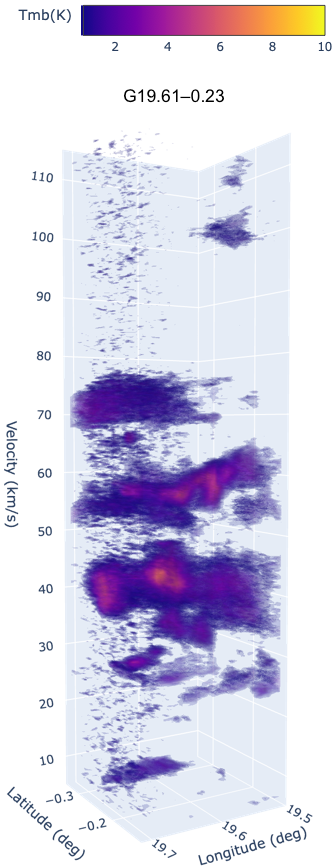}
    \includegraphics[width=0.234\textwidth]{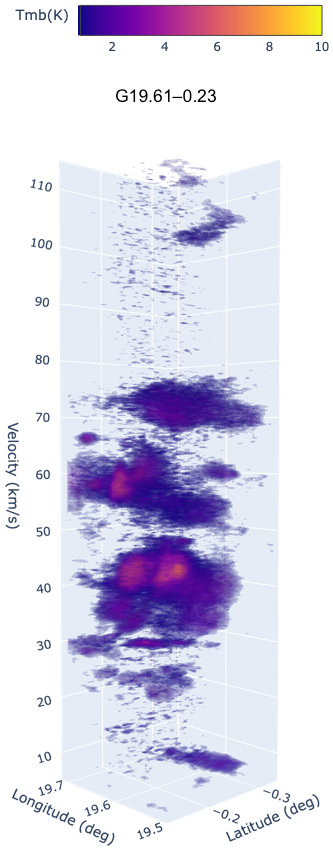}
    \includegraphics[width=0.23\textwidth]{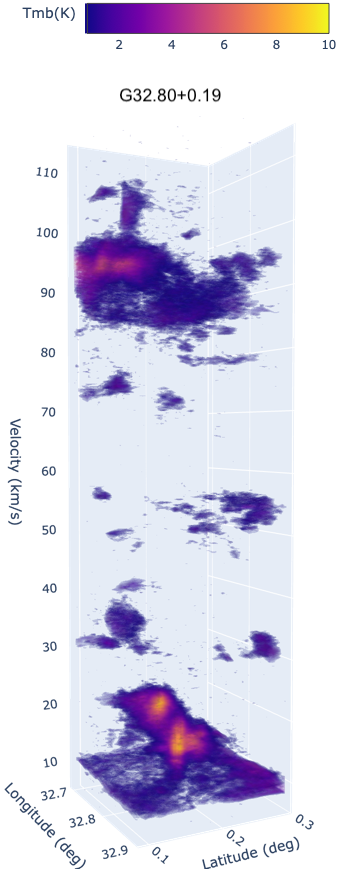}        
    \includegraphics[width=0.233\textwidth]{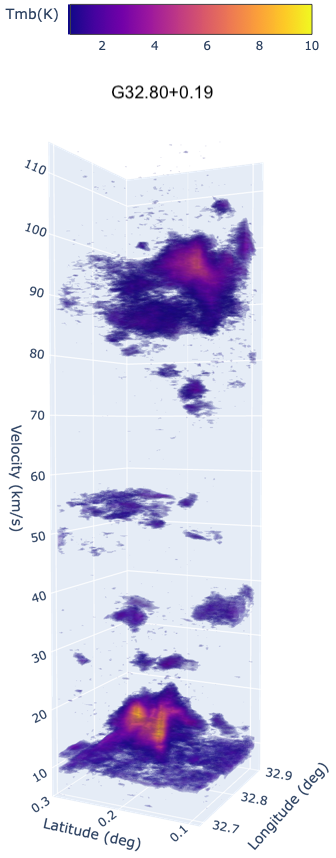} 
    \caption{Position-position-velocity images of \ce{^{13}CO} emission using data taken from the GRS survey toward two selected sightlines; G19.61$-$0.23 (approximately 90$^{\circ}$ angle difference between left two images) and G32.80$+$0.19 (approximately 90$^{\circ}$ angle difference between right two images). }
    \label{fig:co_3dmaps}
\end{figure*}

\subsection{Variation of abundances along the sightline}
Figures\,\ref{fig:x_vel}, \ref{fig:x_vel_add1} and \ref{fig:x_vel_add2} show variations in molecular abundance relative to $N$(\ce{H2}) in the lower panel and $N$(\ce{H2}) in the upper panel as a function of velocity along 12 sightlines. Two sightlines, HGAL0.55$-$0.85 and NGC 7538 IRS1, are not plotted here because only the \ce{HCO+} transition is seen in absorption toward these two sources. The column density of \ce{H2} and the abundance of the molecules relative to \ce{H2} are average values over given velocity ranges (see tables\,\ref{tab:nmol1} and \ref{tab:nmol2}). In the upper panel, visible extinctions of 1 and 5 are indicated with horizontal dotted lines separating the three types of molecular gas clouds; diffuse clouds below $A_{\rm v} < 1.0$, translucent clouds in $1.0 < A_{\rm v} < 5.0$, while regions with $A_{\rm v} > 5$ are considered to be dense molecular clouds. The black arrows imply that the derived column densities and abundances may be underestimated within the velocity ranges marked with gray shaded areas that are affected by spectral emission features. Abundances of the considered species are shown in the lower panels with different colored markers; green circles for HCN, purple asterisks for HNC, orange up triangles for CS, red left triangles for \ce{H2S}, cyan cross for \ce{C2H}, and gray hexagons for \ce{c-C3H2}.

Overall, some variations in abundance along the selected sightlines are found. These variations are not attributed to variable rms levels since the variability of abundances appears in all the sightlines, nor fitting uncertainties. Despite potential uncertainties, the abundances of \ce{C2H} and \ce{c-C3H2} change little and are instead consistent in both velocity ranges or between diffuse and translucent clouds. In Table\,\ref{tab:X_nh2} the mean abundances of these small hydrocarbons are surprisingly similar, while $N$(\ce{H2}) of diffuse clouds is smaller than that of translucent clouds by a factor of 4. We note that all calculations for the median values do not include the absorption features associated with background continuum sources. The median abundances of HCN and HNC do not show significant changes from diffuse to translucent clouds. In contrast, we note that $X$(\ce{H2S}) (with a median of $2.5\times10^{-8}$) and $X$(CS) (with a median of $9.3\times10^{-9}$) in diffuse clouds are significantly higher than in translucent clouds; median values of $X$(\ce{H2S}) and $X$(CS) are $2.8\times10^{-9}$ and $6.6\times10^{-9}$, respectively. These species' higher abundances in diffuse clouds are already noticeable in Figs.\,\ref{fig:int_ntot}, \ref{fig:x_vel}, \ref{fig:x_vel_add1}, and \ref{fig:x_vel_add2}. Overall, the mean or median abundances of the discussed species fit well within the ranges of values derived for high-altitude clouds or other clouds in the Galactic plane, considering the uncertainties. The abundance variations toward individual clouds may signify the influences of their environments which might be affected by, for example, pristine infalling gas in low metallicity \citep{de_cia2021_large_metal}, the high UV fields ($\sim5\times10^5$ in Habing units, \citealt{rizzo2003}) from embedded massive star-forming regions, or outflows, or the feedback from young high mass stars at the late stage of molecular clouds (e.g., influence of turbulence and shocks caused by expanding H{\sc ii} regions and stellar winds). All these effects could, to different degrees, mix gas from star-forming regions with diffuse or/and translucent gas, and result in significant time dependent chemistry (e.g., \citealt{pilleri2013}).

To understand the abundance variations along the sightlines in relation to molecular gas structures, we plot position-position-velocity images of two selected sightlines (G19.61$-$0.23 and G32.80$+$0.19). Compared with other sightlines, these two sightlines have absorption features well separated from the background emission lines. As a result, absorption lines of all seven transitions are clearly detected (see Fig.\,\ref{fig:abs_spec_g19_g32}). Figure\,\ref{fig:co_3dmaps} displays position-position-velocity images of \ce{^{13}CO} emission extracted over an area of $12'\times12'$ from the GRS survey with the original spectral resolution (0.2\,\kms). Toward both sightlines, we notice complex structures of dense molecular gas rather than simple shapes like spheres or cylinders. We can clearly see dense and clumpy molecular gas structures at the systemic velocities of these sightlines (for G19.61$-$0.23, $\varv_{\rm sys}=$~41.8\,\kms, and for G32.80$+$0.19, $\varv_{\rm sys}=$~15\,\kms). Some foreground clouds (velocity ranges of 28 -- 40\,\kms\ and 58.5 -- 70.8\,\kms\ for G19.61$-$0.23 and 90 -- 110\,\kms for G32.80$+$0.19) show an open ring structure. We need to study the association and interaction between diffuse material filling voids of dense molecular gas and the dense material by combining absorption and emission spectral analyses to address the abundance variations found in this work.

\subsection{UV-dominated chemistry}
We used the Meudon PDR code\footnote{\url{https://ism.obspm.fr/pdr.html}} \citep{le_petit2006_meudon,goicoechea2007_meudon,gonzalez_garcia2008_meudon} to study the physical conditions of diffuse and translucent clouds in which the molecules are studied in this investigation and are produced. The Meudon PDR code simulates a plane-parallel slab cloud illuminated on both sides by assuming that the cloud is in a static state. Throughout the model cloud, the thermal and chemical profiles are computed by assuming thermal and chemical equilibrium for every location. All models simulated here are isobaric models that consider constant thermal pressure. The calculated temperature and density profiles show that the temperature in the cloud decreases with increasing density. Self and mutual shielding for H, \ce{H2}, CO, isotopologues of CO (i.e., \ce{^{13}CO} and \ce{^{18}CO}), and HD becomes vital for models of diffuse clouds. Thus, toward the foreground clouds studied in this work, the H/\ce{H2} transition should be considered in PDR models in detail with regard to the shielding processes. Therefore, we have applied an exact radiative transfer method described in \citep{goicoechea2007_meudon} to the PDR models rather than the analytical approximation to the self-shielding process in a line, described by \cite{Federman1979} (FGK approximation). In the exact method, the spherical harmonics method \citep{goicoechea2007_meudon} is implemented to solve for the far-ultraviolet (FUV) radiation fields in externally and internally illuminated clouds, considering gas absorption and scattering properties by dust grains.

\begin{table}[h!]
    \centering
    \small
    \caption{Input parameters of the Meudon PDR isobaric models.}
    \label{tab:meudon_para}
    \begin{tabular}{l | c}
    \hline \hline
        \multicolumn{2}{c}{Fixed parameters}\\
        \hline 
        Oxygen abundance & $3.2 \times 10^{-4}$\\
        Carbon abundance & $1.3 \times 10^{-4}$\\
        Nitrogen abundance & $7.5 \times 10^{-5}$ \\
        Sulfur abundance & $1.9 \times 10^{-5}$ \\
        $R_{\rm v} = A_{\rm v}/E(B-V)$ & 3.1 \\
        Gas to Dust ratio & 100 \\
        Grain size distribution index & 3.5 \\
        Minimum grain radius ($\mu$m) & 0.003 \\
        Maximum grain radius ($\mu$m) & 0.3 \\
        \hline 
        \multicolumn{2}{c}{Free parameters}\\
        \hline
        Radiation field $G_{\rm 0}$ & 1 \& 100 \\
        Cosmic ray ionization rate $\zeta$ (s$^{-1}$)  & $5\times10^{-17}$, $10^{-16}$ \& $10^{-15}$ \\
        $A_{\rm v, max}$ (mag) & 0.1, 1, 5  \& 10  \\
        Pressure (K cm$^{-3}$) & $10^{3}$, $3\times10^{3}$, $5\times10^{3}$ \& $10^{4}$ \\
    \hline
    \end{tabular}
    \tablefoot{The carbon and sulfur atoms are in their ionized form in the chemical file. The UV interstellar radiation field is in Mathis unit; $G_{\rm 0} = 1$ corresponds to a flux of $3.03 \times 10^{-4}$ ergs cm$^{-2}$s$^{-1}$.  }
\end{table}

Table\,\ref{tab:meudon_para} shows the fixed and free parameters used in the simulated PDR models. We used default values for initial abundances of elements, extinction curve ($R_{\rm v}$), and dust properties and considered them to be fixed parameters. However, we have changed the values for the interstellar radiation field ($G_{\rm 0}$ in Mathis units), which for the solar neighborhood is $G_{\rm 0} = 1$ and for FUV illuminated PDRs such as the Horsehead Nebula is $G_{\rm 0} = $100 \citep{pety2005}. The value of $G_{\rm 0}$ only changes on the observer side and is fixed to 1 for the backside of the cloud in all models. Since the observed galactic sightlines pass over different parts of the Milky Way, we assume that diffuse clouds along the sightlines are controlled by various cosmic ray ionization rate ($\zeta$) values where $\zeta$ is the total ionization rate by cosmic rays per \ce{H2} molecule. Therefore, we varied $\zeta$ to $5\times10^{-17}$, $1\times10^{-16}$ and $1\times10^{-15}$ s$^{-1}$ in our models. In addition, we have assumed typical values of interstellar pressure in diffuse and translucent clouds of 1000, 3000, 5000, and 10\,000\,K\,cm$^{-3}$ and chosen total extinctions along the plane-parallel slab of $A_{\rm v} =$ 0.1, 1, 5, and 10 mag. We note that in the Meudon PDR isobaric models, each velocity-integrated component listed in Tables\,\ref{tab:nmol1} and \ref{tab:nmol2} is a separate cloud (diffuse or translucent clouds) with a uniform pressure.

\begin{figure}
    \centering
    \includegraphics[width=0.50\textwidth]{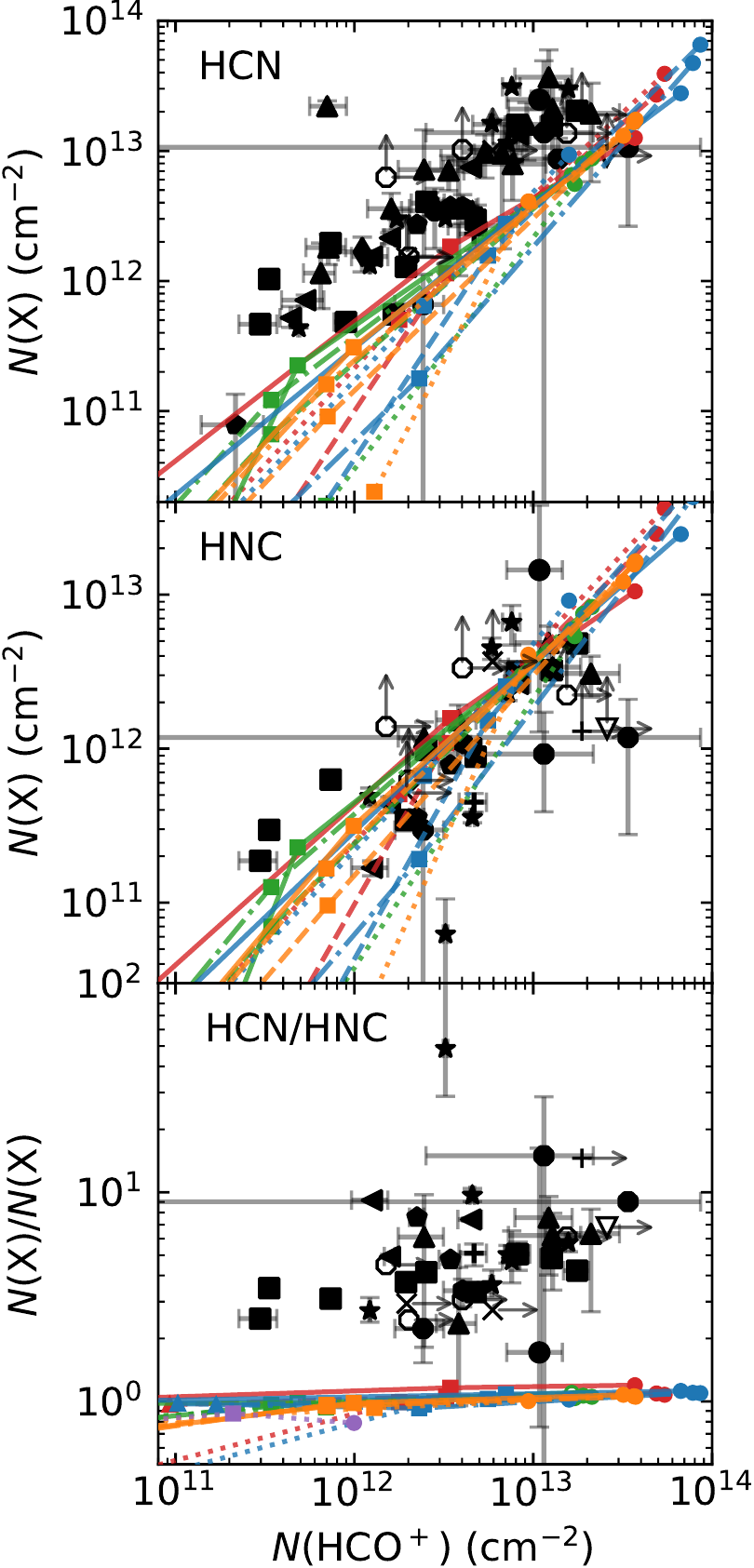}
    \caption[l]{Column densities of HCN and HNC and their column density ratio as a function of $N$(\ce{HCO+}), with predictions from isobaric PDR models overlaid. Filled and unfilled black markers indicate foreground components and molecular envelope components observed, respectively. The arrows refer to lower limits of $N$(X), and the markers are as same as Fig\,\ref{fig:int_ntot}. Each color represents different radiation field strengths and cosmic-ray ionization rates; red: G$_0 = 1$ and $\zeta = 5\times10^{-17}$\,s$^{-1}$, green: G$_0 = 100$ and $\zeta = 5\times10^{-17}$\,s$^{-1}$, blue: G$_0$ = 1 and $\zeta =1\times10^{-16}$\,s$^{-1}$, orange: G$_0$ = 100 and $\zeta = 1\times10^{-16}$\,s$^{-1}$, and purple: G$_0$ = 100 and $\zeta = 1\times10^{-15}$\,s$^{-1}$. Initial pressures of 1000, 3000, 5000, 10\,000 K cm$^{-3}$ are indicated with dotted-, dashed-, dashed-dotted-, and solid-lines, respectively. $A_{\rm v} = $1, 5, and 10~mag are marked with colored triangles, squares, and circles, respectively. }
    \label{fig:pdr_model_hcn_hnc}
\end{figure}

\begin{figure}
    \centering
    \includegraphics[width=0.50\textwidth]{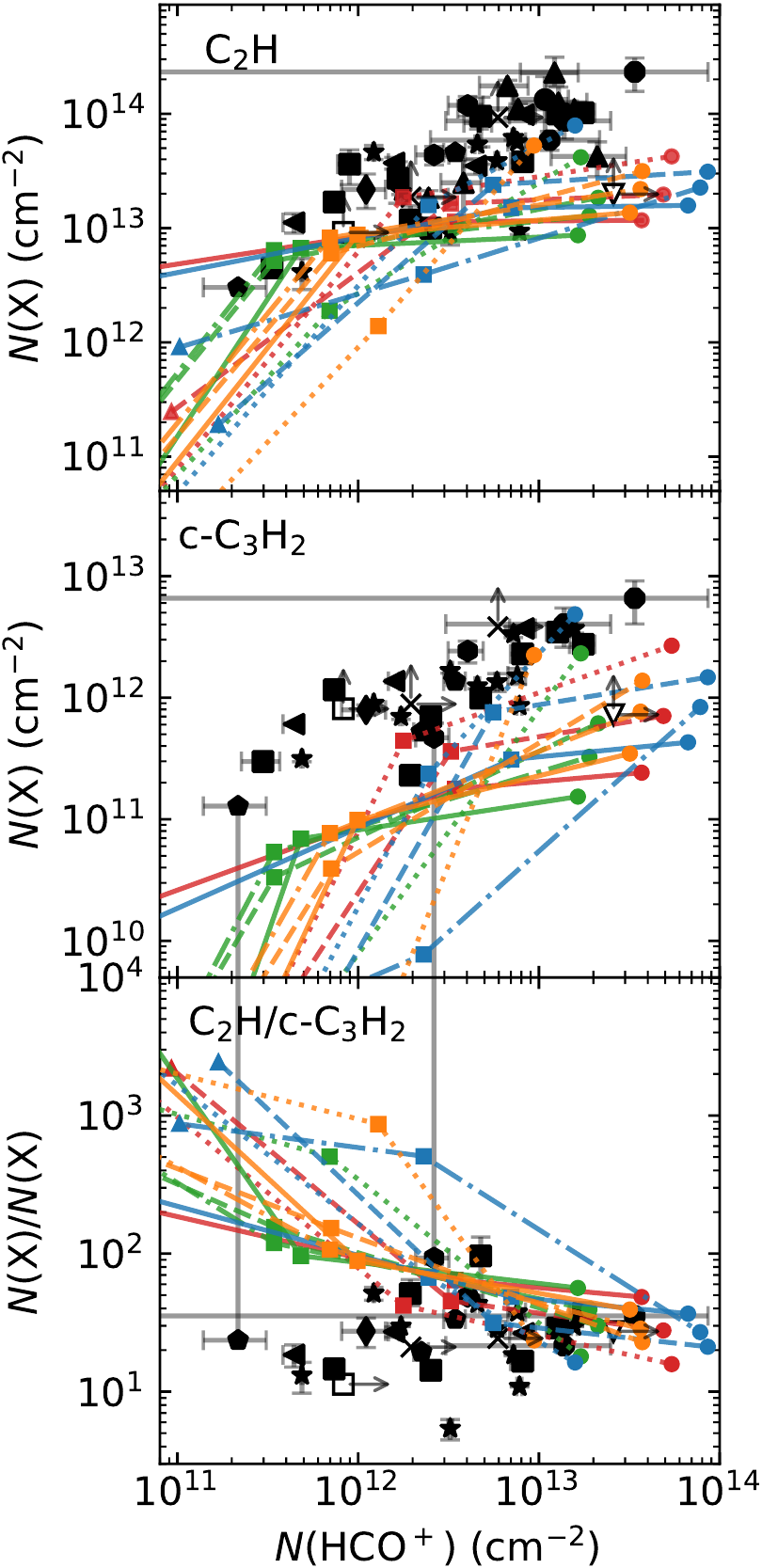}
    \caption[l]{Column densities of \ce{C2H} and \ce{c-C3H2} and their column density ratio as a function of $N$(\ce{HCO+}), with predictions from isobaric PDR models overlaid. All symbols and lines used to represent the models are the same as Fig.\,\ref{fig:pdr_model_hcn_hnc}.} 
    \label{fig:pdr_model_cch_c3h2}
\end{figure}

Figures\,\ref{fig:pdr_model_hcn_hnc}, \ref{fig:pdr_model_cch_c3h2} and \ref{fig:pdr_model_cs_h2s} show comparisons of $N$(X) for each species X studied and column density ratios with $N$(\ce{HCO+}), from top to bottom, and the predicted column densities and ratios from isobaric PDR models are marked with colored symbols and lines. For the predicted values from the models, different colors indicate various radiation field strengths and cosmic-ray ionization rates; G$_0$ = 1 and $\zeta =5\times10^{-17}$\,s$^{-1}$ in red, G$_0$ = 100 and $\zeta =5\times10^{-17}$\,s$^{-1}$ in green, G$_0$ = 1 and $\zeta =1\times10^{-16}$\,s$^{-1}$ in blue, G$_0$ = 100 and $\zeta =1\times10^{-16}$\,s$^{-1}$ in orange, and G$_0$ = 100 and $\zeta =1\times10^{-15}$\,s$^{-1}$ in purple. Initial pressures of 1000, 3000, 5000, and 10\,000\,K\,cm$^{-3}$ are indicated with dotted-, dashed-, dashed-dotted-, and solid-lines, respectively. In addition, the results with different visual extinctions are represented by colored triangles ($A_{\rm v} =1$~mag), colored squares ($A_{\rm v} =5$~mag), and colored circles ($A_{\rm v} =10$~mag). We note that all the predicted column densities at $A_{\rm v} =0.1$~mag fall outside the range of column densities shown in Figs\,\ref{fig:pdr_model_hcn_hnc}, \ref{fig:pdr_model_cch_c3h2} and \ref{fig:pdr_model_cs_h2s}. 

\subsubsection{\ce{HCO+}, HCN, and HNC}
The top and middle plots of Fig.\,\ref{fig:pdr_model_hcn_hnc} show comparisons of two CN-bearing molecules (i.e., HCN and HNC) with $N$(\ce{HCO+}) as well as the column densities predicted from the isobaric PDR models. For HCN, most of the models underestimate the observed quantities, only those models with $\zeta\,=\, 5\times10^{-17} - 1\times10^{-16}$\,s$^{-1}$ and an initial pressure of 10\,000 K\,cm$^{-3}$ result in values close to the lower limit of the observed column densities by a factor of a few higher. The modeled gas conditions are at $n_{\rm H}\,\sim\,1400 - 2000$\,cm$^{-3}$ and decrease in gas temperature at $A_{\rm v}\,\sim\,1$~mag. For HNC, all models fit reasonably within the observed column density ranges. In addition, models with $G_0 = 1$ (red and blue colors) make better predictions than those with $G_0 = 100$. The predicted $N$(HCN), $N$(HNC), and $N$(\ce{HCO+}) from the isobaric models result from regions of higher visual extinction between 1 and 10 mag. In contrast, the estimated \ce{H2} column densities from the observations correspond to lower than $A_{\rm v} \leq 5$ mag. These models provide reasonable predictions of column densities of these species. However, the models underestimate column densities by roughly an order of magnitude, especially for HCN and \ce{HCO+}. The modeled $N$(HCN)/$N$(HNC) against $N$(\ce{HCO+}) as shown in the bottom-left panel of Fig.\,\ref{fig:pdr_model_hcn_hnc}, is lower than the ratios from the observations due to an underestimation of $N$(HCN) in the isobaric PDR models.

In the isobaric PDR models, there are two main formation pathways for \ce{HCO+} and the two CN-bearing molecules (see \citealt{godard2009_TDR, gerin2021_COp} for details). For \ce{HCO+}, the first one involves the chains, from OH $+$ \ce{C+} $\rightarrow$ \ce{CO+} $+$ H to \ce{CO+} $+$ \ce{H2} $\rightarrow$ \ce{HCO+} $+$ H, and the second pathway is via \ce{H2O} $+$ \ce{C+} $\rightarrow$ \ce{HCO+} $+$ H. The latter reaction can also lead to the production of \ce{HOC+} that can produce \ce{HCO+} via hydrogenation. The formation of HCN and HNC involves nitrogen via the reaction NH $+$ \ce{C+} $\rightarrow$ \ce{CN+} $+$ \ce{H2} or H $\rightarrow$ \ce{HCN+} or \ce{HNC+} $+$ \ce{H2} $\rightarrow$ \ce{HCNH+}; \ce{NH2} $+$ \ce{C+} $\rightarrow$ \ce{HCN+} $+$ \ce{H2} $\rightarrow$ \ce{HCNH+}. The dissociative recombination of \ce{HCNH+} forms HCN and HNC. Additionally, HCN can be produced via the neutral-neutral reaction of \ce{CH2} $+$ N. As seen in Fig.\,\ref{fig:pdr_model_hcn_hnc}, the chemistry involving only the UV-dominated chemistry cannot reproduce the observed column densities of these species in diffuse and translucent clouds. In the turbulence-dominated chemistry \citep{godard2009_TDR, godard2010_3mmabs, godard2014_tdr} involving turbulent dissipation rate, relaxation and ion-neutral drift \citep{valdivia2017_CHp}, several crucial endoenergetic reactions can significantly increase the abundances of important precursors of HCN, HNC, and \ce{HCO+}, such as \ce{CH+}, \ce{CH^{+}2}, \ce{CH^{+}3} as well as NH. The TDR models from \cite{godard2009_TDR,godard2014_tdr} show a considerable increase in the formation of these three species via ion-neutral reactions, \ce{CH+} $+$ N $\rightarrow$ \ce{CN+} $+$ \ce{H2} $\rightarrow$ \ce{HCN+} or \ce{HNC+} $+$ \ce{H2} $\rightarrow$ \ce{HCNH+}; \ce{CH^{+}2} $+$ N $\rightarrow$ \ce{HCN+} $+$ \ce{H2} $\rightarrow$ \ce{HCNH+}; \ce{CH^{+}3} $+$ N $\rightarrow$ \ce{HCNH+}; \ce{CH^{+}3} $+$ O $\rightarrow$ \ce{HCO+} $+$ \ce{H2}. Additionally, the reaction of \ce{CH+} with oxygen enhances the abundance of \ce{CO+}, leading to \ce{HCO+}. 

\subsubsection{\ce{C2H} and \ce{c-C3H2}}
Figure.\,\ref{fig:pdr_model_cch_c3h2} shows comparisons between column densities predicted from isobaric PDR models and observations for \ce{C2H} and \ce{c-C3H2}. Similar to the HCN and \ce{HCO+} results, the models underestimate the column densities of these two small hydrocarbons in these foreground cloud components. Nevertheless, their observed column densities are in better agreement with values from the isobaric models corresponding to low-density gas in $n_{\rm H} \approx 100 - 500$\,cm$^{-3}$ up to 1000\,cm$ ^{-3}$. The $N$(\ce{C2H}) toward the diffuse clouds with $N$(\ce{HCO+}) $< 2\times 10^{12}$\,cm$^{-2}$ (corresponding to an observed $A_{\rm v}\sim 1$) is close to the column densities from the isobaric models with moderate pressure values (dashed and dash-dotted green lines) of $3000-5000$\,K\,cm$^{-3}$ giving $n_{\rm H} \approx 500 - 1000$\,cm$^{-3}$ and strong UV radiation of $G_0 = 100$ and $\zeta = 5\times10^{-17}$\,s$ ^{-1} $, for less shielded regions $A_{\rm v} < 5$ (colored square symbols). By contrast, in denser regions having $N$(\ce{HCO+})~$> 2\times10^{12}$\,cm$^{-2}$, the low-density models (dashed lines, $n_{\rm H} \approx$ $100 - 500$\,cm$^{-3}$) with lower $G_{0}$ of 1 (red and blue colors) at high $A_{\rm v} > 5$~mag seem to fit the observed column densities better. For \ce{c-C3H2}, the low-density PDR models with $G_0 =1$ only in the direction of translucent clouds predict results near the low end of the column densities obtained from the observational data, as in the case of \ce{C2H}. The comparison between $N$(\ce{C2H})/$N$(\ce{c-C3H2}) and $N$(\ce{HCO+}) for the models and observations show a better match than those for $N$(HCN)/$N$(HNC) versus $N$(\ce{HCO+}) but still have some discrepancy. Overall, the models with lower density and the solar neighborhood $G_0$ generate column densities of these hydrocarbons close to the observed values, unlike the cases of HCN, HNC, and \ce{HCO+} which are associated better with high density gas materials. Nevertheless, the predicted column densities of the small hydrocarbons are underestimated by a factor of a few. In particular, for \ce{c-C3H2}, toward diffuse clouds with $N$(\ce{c-C3H2})~$<2\times10^{12}$\,cm$^{-2}$, the discrepancy is more than about a factor of 10. In the isobaric PDR models, the predicted column densities of \ce{c-C3H2} are total column densities of ortho and para species without separate spin symmetries (with a spin-weight ratio of 3/1 for ortho/para). Thus, with consideration of this fact, the discrepancies between the observations and the models become larger.

In the PDR chemistry, \ce{C2H} is mainly produced by the recombination of \ce{C2H^{+}2} and \ce{C2H^{+}3} with electrons (e.g., \citealt{godard2009_TDR,mookerjea2012_small_hydrocarbons}). In addition, it is formed by the photodissociation of \ce{C2H2} \citep{lee1984_C2H2_C2H} as well as by the neutral-neutral reaction of \ce{CH2} with atomic carbon \citep{turner2000_hydrocarbons,sakai2010_C2H}. For the formation of \ce{c-C3H2}, once \ce{C3H+} is formed by reactions of \ce{C2H2} with \ce{C+}, \ce{C3H+} produces the linear and cyclic \ce{C3H^{+}3} isomers via reactions with \ce{H2} \citep{maluendes1993JChPh_c3h2p}. The resulting isomers produce linear and cyclic-\ce{C3H2} by dissociative recombination \citep{fosse2001_c3h2}. In addition, the recombination of \ce{C3H+} also contributes to the abundances of \ce{C2H} \citep{guzman2015_hydrocarbons}. 

Similar to HCN, HNC, and \ce{HCO+}, turbulence-dominated chemistry can considerably increase the abundances of \ce{C2H} and \ce{c-C3H2} by enhancing abundances of their precursor, \ce{C2H+} which is formed from \ce{CH2} via hydrogenation reactions. Not only that, but also the endoenergetic reaction of \ce{C^{+}2} $+$ \ce{H2} with $-\Delta E/k = 1260$\,K \citep{godard2009_TDR} increases the abundance of \ce{C2H}. Aside from the contribution of turbulent dissipation, the formation of small hydrocarbons possibly involves the top-down hydrocarbon chemistry that describes polycyclic aromatic hydrocarbons (PAHs) and small carbonaceous grains being broken down by UV erosion into small hydrocarbons that react in the gas phase (e.g., \citealt{pety2005, gerin2011_smallhydro, guzman2015_hydrocarbons}).

\subsubsection{CS and \ce{H2S}}
\begin{figure}
    \centering
    \includegraphics[width=0.50\textwidth]{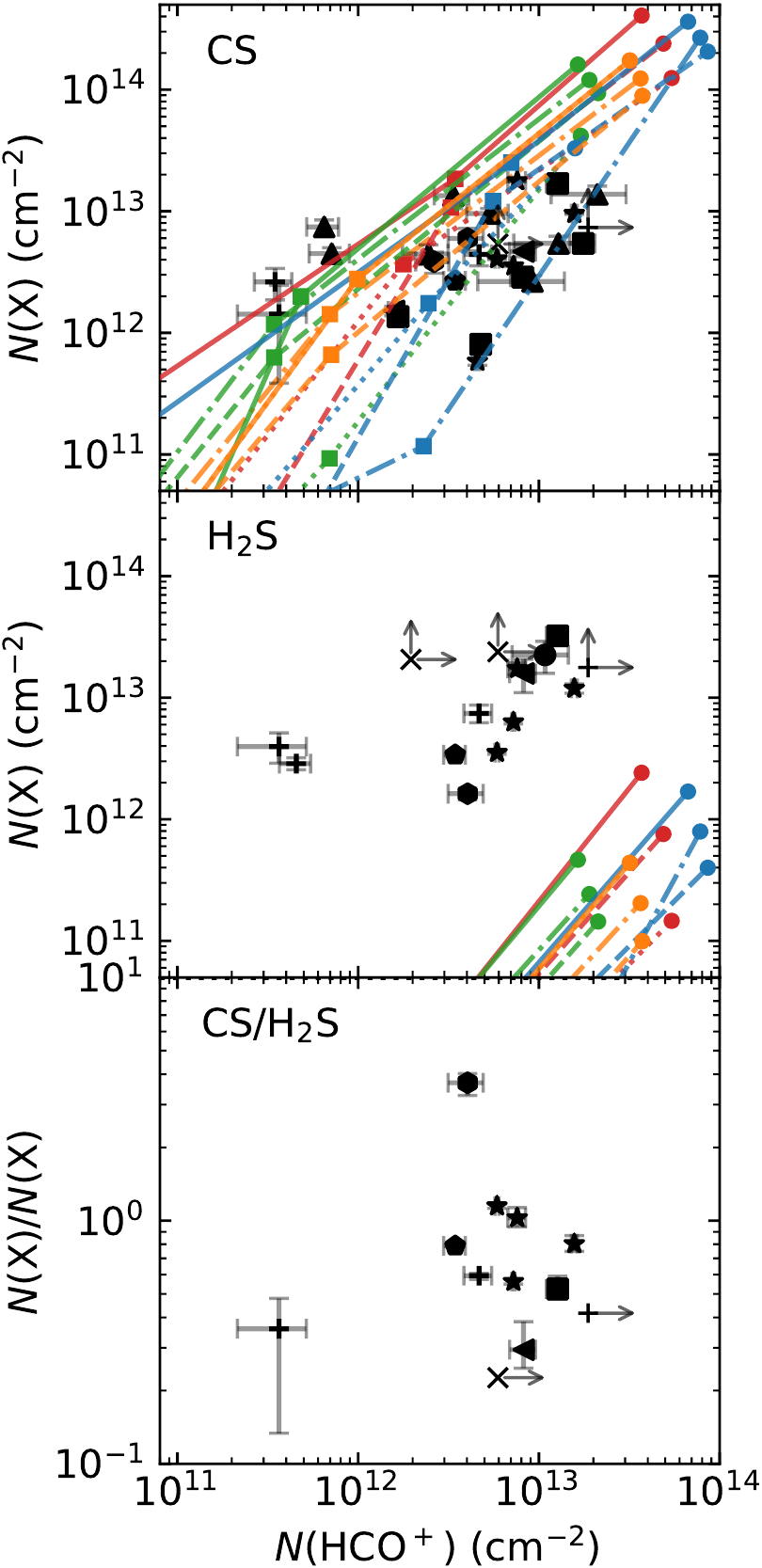}
    \caption[l]{Column densities of CS and \ce{H2S} and their column density ratio as a function of $N$(\ce{HCO+}), with predictions from isobaric PDR models overlaid. All markers and lines for the models are same as Fig.\,\ref{fig:pdr_model_hcn_hnc}.}
    \label{fig:pdr_model_cs_h2s}
\end{figure}

Figure\,\ref{fig:pdr_model_cs_h2s} shows comparisons of the column densities of the two sulfur-bearing species and their column density ratios with $N$(\ce{HCO+}). The top panel shows a good agreement between the predicted and derived column densities of CS. All the models with a wide range of physical parameters ($G_{\rm 0}=$1 and 100, $\zeta = 5\times10^{-17}$ and $1\times10^{-16}$ s$^{-1}$, and gas pressures of 1000--10\,000\,K\,cm$^{-3}$) broadly fit the derived column densities. In particular, toward translucent cloud components, the models with $G_{\rm 0}=1$, $\zeta=1\times10^{-16}$ s$^{-1}$, and pressure of 3000 -- 5000\,K\,cm$^{-3}$ (blue dashed- and dashed-dotted-lines) well fit the derived values from the observations. The primary pathway of CS formation is the dissociative recombination reaction of \ce{HCS+} and electrons, which, however, leads to CS $+$ H in 19\,\% of reactions \citep{montaigne2005_HCSp}. There are two formation routes for \ce{HCS+}, either through CH $+$ \ce{S+}$\,\rightarrow$\,\ce{CS+}; \ce{CS+} $+$ \ce{H2}\,$\rightarrow$\,\ce{HCS+} in the PDR route or through S $+$ \ce{CH^{+}3}\,$\rightarrow$\,\ce{HCS+} in the TDR route \citep{neufeld2015}. We have found that the results of our PDR models are obtained in more shielded gas ($n_{\rm H} \sim 100 - 1000$\,cm$^{-3}$ and $A_{\rm v} > 1$~mag) compared to the more diffuse gas ($n_{\rm H}=50$\,cm$^{-3}$ and $A_{\rm v}=0.4 $ mag) considered in the TDR and shock models presented in \cite{neufeld2015}. The destruction of CS reacting with \ce{H^{+}3} directly contributes to the formation of \ce{HCS+} alongside other CS destruction pathways driven by photoionization and ion-exchange reactions.

The middle panel in Fig.\,\ref{fig:pdr_model_cs_h2s} shows that the isobaric PDR models underestimate the observed column densities of \ce{H2S}. We also note that, as in the case of \ce{c-C3H2}, the \ce{H2S} column densities predicted from the models show the total column densities of both ortho and para species, while the observational measurements refer to the column densities of only ortho-\ce{H2S}. In UV-dominated chemistry, reactions of \ce{S+} $+$ \ce{H2} $\rightarrow$ \ce{H2S+} $+$ \ce{e-} $\rightarrow$ \ce{H2S} is the main pathway to form \ce{H2S}. However, in low A$_{\rm v}$ ($<$\,0.5\,mag) regions, the endothermic reaction of SH $+$ \ce{H2} $\rightarrow$ \ce{H2S} $+$ $h\nu$ (with $-\Delta E/k=6862$\,K) becomes the predominant formation pathway for \ce{H2S} (e.g., \citealt{sternberg1995_sp, neufeld2015}). In addition, the abundances of \ce{H2S+}, which is the precursor of \ce{H2S}, can be enhanced by two other endothermic reactions: \ce{SH+} $+$ \ce{H2} $\rightarrow$ \ce{H2S+} $+$ H ($-\Delta E/k=6380$\,K), and \ce{S+} $+$ \ce{H2} $\rightarrow$ \ce{H2S+} $+$ $h\nu$ ($-\Delta E/k=9860$\,K). Similar to the species discussed in the previous sections, these endothermic reactions activated by turbulent dissipation or shocks \citep{neufeld2015} could significantly enhance the abundances of \ce{H2S} toward the observed sightlines. However, TDR or shock models alone cannot address the high observed column densities \citep{neufeld2015}. This implies that another route for the formation of \ce{H2S} is needed. This route could be related to dust grain-surface chemistry (e.g., \citealt{navarro-almaida2020,cazaux2022_h2s}). Sulfur is depleted onto dust grains \citep{ruffle1999_sulphur_depletion} because sulfur mainly exists as \ce{S+} \citep{sternberg1995_sp} and most dust grains are negatively charged. According to the gas-grain models of \cite{navarro-almaida2020}, the chemical desorption of \ce{H2S} (via solid-H $+$ solid-HS $\rightarrow$ \ce{H2S}) can explain the high \ce{H2S} abundances observed in translucent clouds (cloud edges, n(\ce{H2}) $< 10^4$\,cm$^{-3}$) surrounding the dark clouds TMC-1 and Barnard 1b. Although the clouds in our sample have different physical conditions to these objects, the addition of chemical desorption may also help explain the high \ce{H2S} abundances derived toward the Galactic sightlines in our sample.

\section{Summary and conclusions}\label{sec:summary}
Using the IRAM 30\,m telescope, we have studied seven simple molecular species -- \ce{HCO+}, HCN, HNC, \ce{C2H}, \ce{c-C3H2}, CS, and \ce{H2S} -- in absorption toward 15 submillimeter sources in the Milky Way. We have detected \ce{HCO+} absorption lines in the direction of all 14 sources that show detectable millimeter wave continuum radiation, identifying 78 foreground gas components. A significant number of sightlines also have detectable absorption lines of HCN (12 sightlines), HNC (11 sightlines), \ce{C2H} (12 sightlines), and \ce{c-C3H2} (9 sightlines), CS (8 sightlines) and \ce{H2S} (6 sightlines). Our sensitivity to CS and \ce{H2S} is reduced by the presence of significant emission line contamination near the CS frequency and the poorer S/N ratios achieved at the higher H$_2$S transition frequency.

Using principal component analysis to investigate correlations among the species observed with the IRAM 30\,m telescope and 7 other species observed with SOFIA (Paper I), we find that \ce{C+} and the hydride ions \ce{OH+}, \ce{H2O+}, and \ce{ArH+} are poorly correlated with neutral molecules and \ce{HCO+}. Sulfur-bearing molecules, CS and \ce{H2S}, appear to be better correlated with the neutral molecules than with hydride ions, which are tracers of gas with small molecular fractions. In addition, CH, OH, and O tend to be partially associated with both mostly molecular and predominantly atomic gas.

For the column densities, we find that $N$(HCN) and $N$(HNC) increase nonlinearly with $N$(\ce{H2}). While the abundances of the two molecules increase linearly in diffuse clouds (A$_{\rm v} < 1$), their abundances suddenly increase in regions with $N$(\ce{H2}) $\gtrsim 10^{21}$\,cm$^{-2}$, where gas is further shielded against ultraviolet radiation. In contrast, \ce{C2H} and \ce{c-C3H2} do not increase as rapidly, and their column densities show a close correlation with $N$(\ce{HCO+}). Moreover, the column density ratios of the two small hydrocarbons remain constant over the $N$(\ce{H2}) range probed and lie close to the ratios measured previously toward external galaxies. Interestingly, we find that the abundances of CS and \ce{H2S} in diffuse clouds are 5 -- 10 times higher than in translucent clouds ($A_{\rm v}\,\approx\,1$). This might imply the onset of sulfur depletion at around $A_{\rm v}$ of 1. Variations in the abundances of observed species along sightlines exist, but finding specific patterns is difficult as the abundances are influenced by the surrounding environment.

We reasonably reproduce column densities of HNC and CS using isobaric PDR models, but the column densities of the other molecular species observed here, except for \ce{H2S}, are underestimated by a factor of 2 -- 10. Any of the models failed to reproduce the column densities of \ce{H2S}. As explored in previous studies on diffuse clouds, TDR models, which include pathways involving endoenergetic chemical reactions, could explain the discrepancy between the predicted and observed column densities of HCN, \ce{HCO+}, \ce{C2H}, and \ce{c-C3H2}. For \ce{H2S}, additional formation routes, for example grain-surface production and chemical desorption of \ce{H2S}, may play an important role in increasing the abundance of \ce{H2S} in diffuse and translucent clouds.

\begin{acknowledgements}
We would like to thank the referee for their constructive comments and suggestions that have helped to improve this paper. This work is based on observations made with the Institut de Radioastronomie Millim\'etrique (IRAM) 30~m telescope and used the data from Paper I obtained with the NASA/DLR Stratospheric Observatory for Infrared Astronomy (SOFIA). SOFIA is jointly operated by the Universities Space Research Association, Inc. (USRA), under NASA contract NNA17BF53C, and the Deutsches SOFIA Institut (DSI) under DLR contract 50 OK 0901 to the University of Stuttgart. This work was supported by DLR/Verbundforschung Astronomie und Astrophysik Grant 50 OR 2007. We gratefully acknowledge the excellent support provided by the SOFIA Operations Team, the GREAT Instrument Team, and the staff of the IRAM 30~m telescope.

\end{acknowledgements}

\bibliographystyle{aa}
\bibliography{hygal30m_Won_Ju_Kim}

%
%

\begin{appendix}
\onecolumn
\newpage
\section{Channel-wise column density spectra of the detected molecular species}\label{append:nmol}
\begin{figure*}[h!]
    \centering
    \includegraphics[width=0.46\textwidth]{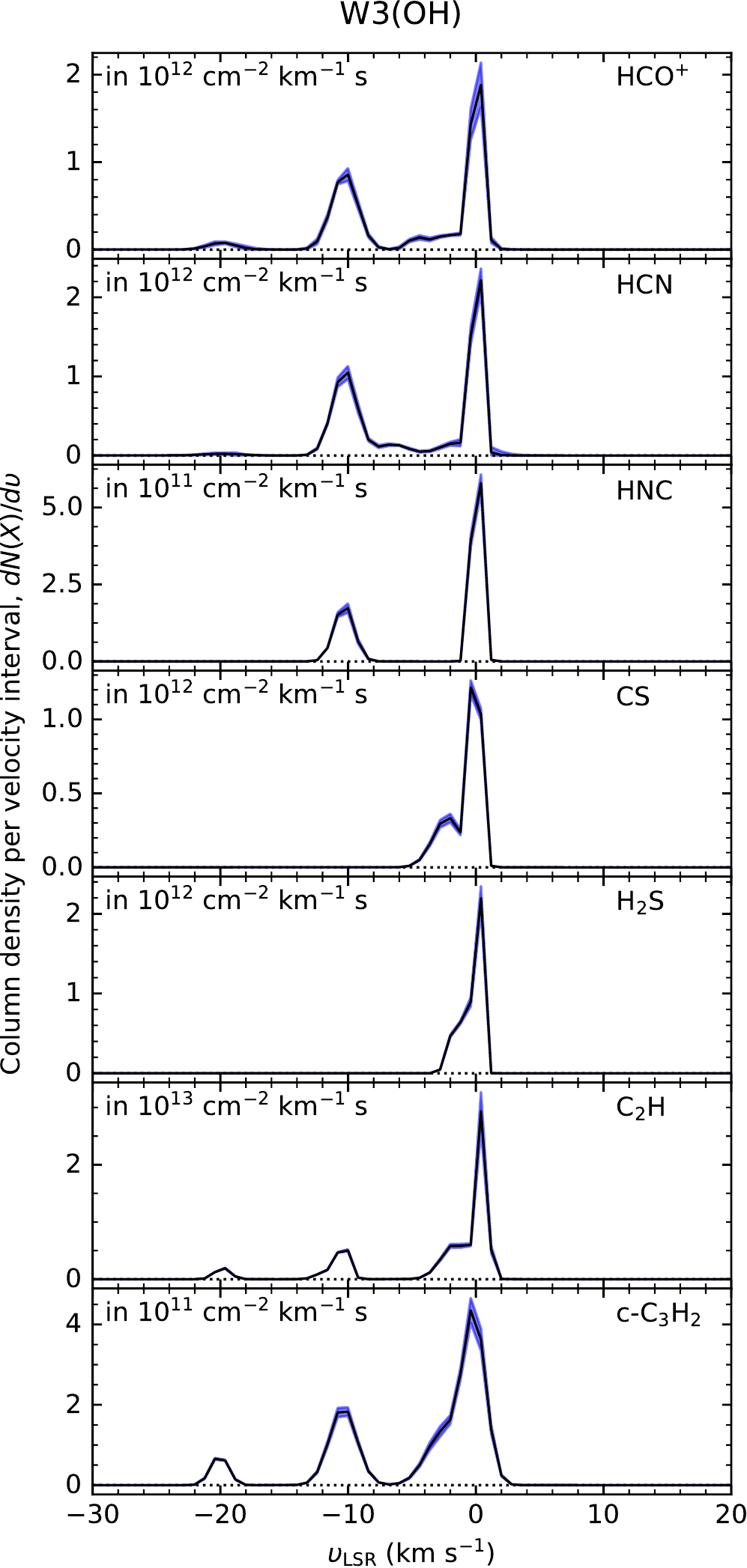}
    \includegraphics[width=0.46\textwidth]{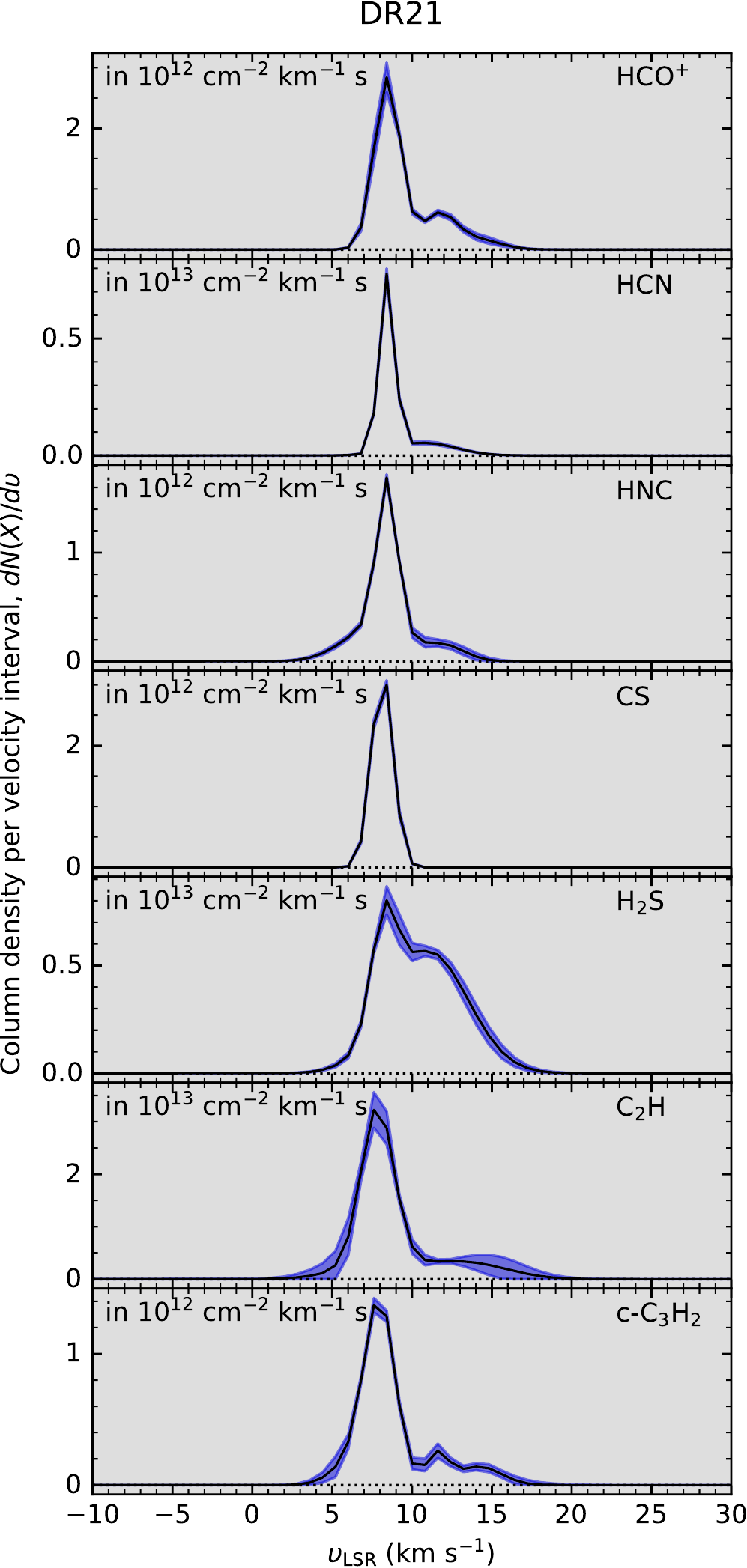}    
    \caption{Channel-wise column density spectra ($dN/d\varv$) as a function of velocity toward W3(OH) and DR21. The $dN/d\varv$ in each panel shows \ce{HCO+}, HCN, HNC, CS, \ce{H2S}, \ce{C2H}, and \ce{c-C3H2} from top to bottom. The gray shaded area indicates dispersion of \ce{HCO+} emission lines from continuum sources, and the blue shaded regions represent 2$\sigma$ confidence intervals for the derived column densities (per velocity interval). }
    \label{appen:ncol1}
\end{figure*}

\begin{figure*}[h!]
    \centering
    \includegraphics[width=0.46\textwidth]{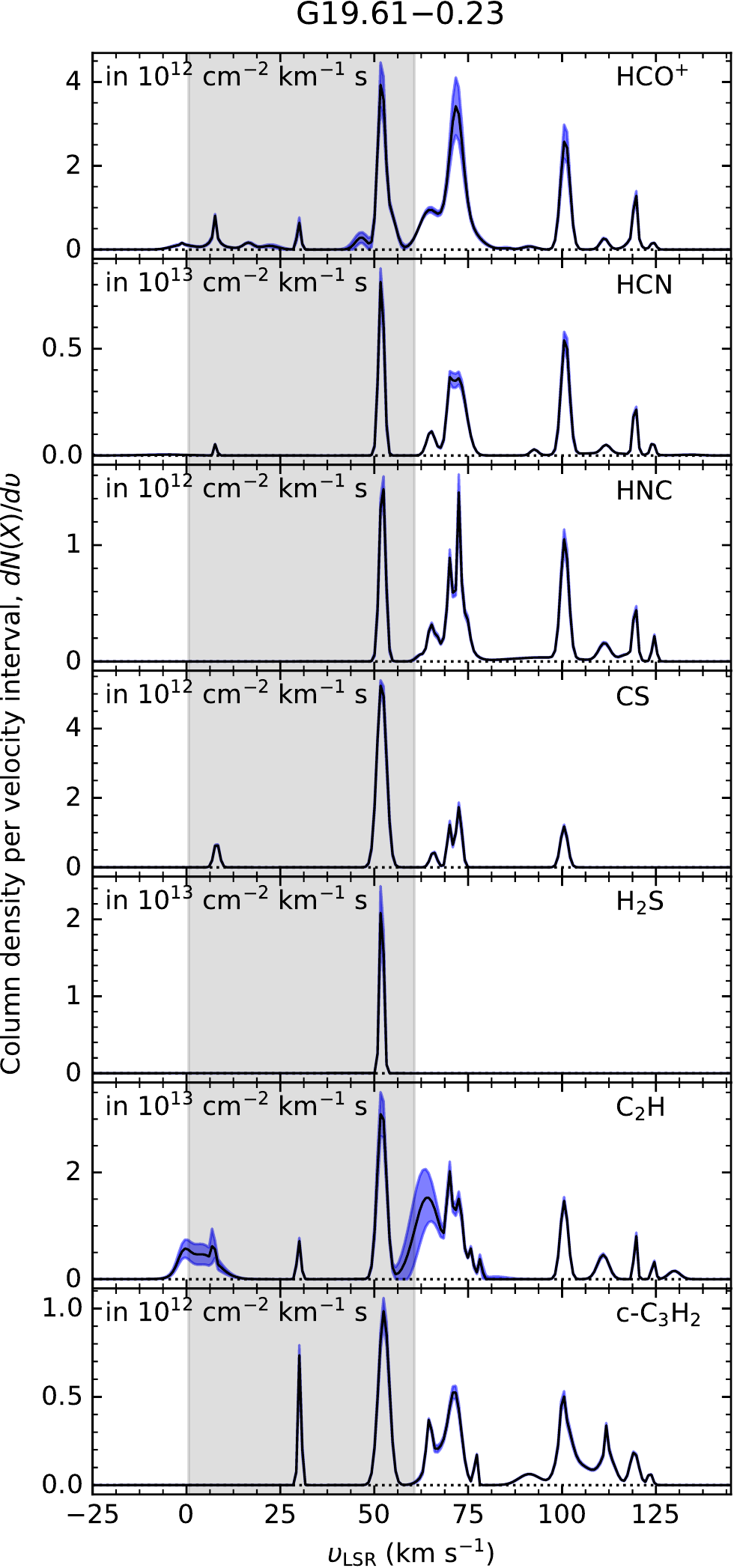}
        \includegraphics[width=0.47\textwidth]{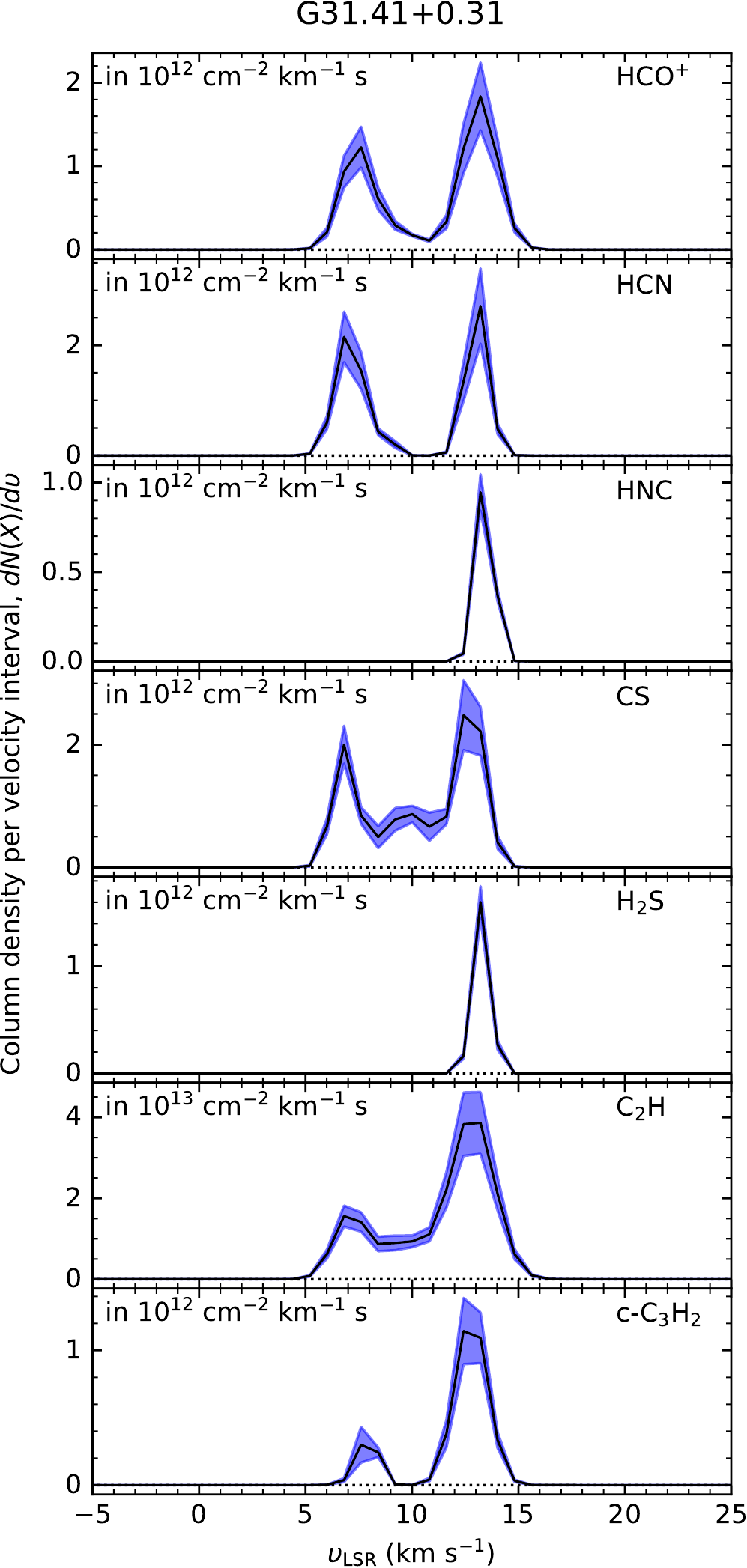}
    \caption{Channel-wise column density spectra ($dN/d\varv$) as a function of velocity toward G19.61$-$0.23 and G31.41$+$0.31 The $dN/d\varv$ in each panel shows \ce{HCO+}, HCN, HNC, CS, \ce{H2S}, \ce{C2H}, and \ce{c-C3H2} from top to bottom. The gray shaded area indicates dispersion of \ce{HCO+} emission lines from continuum sources, and the blue shaded regions represent 2$\sigma$ confidence intervals for the derived column densities (per velocity interval). }
    \label{appen:ncol2}
\end{figure*}

\begin{figure*}[h!]
    \centering
    \includegraphics[width=0.46\textwidth]{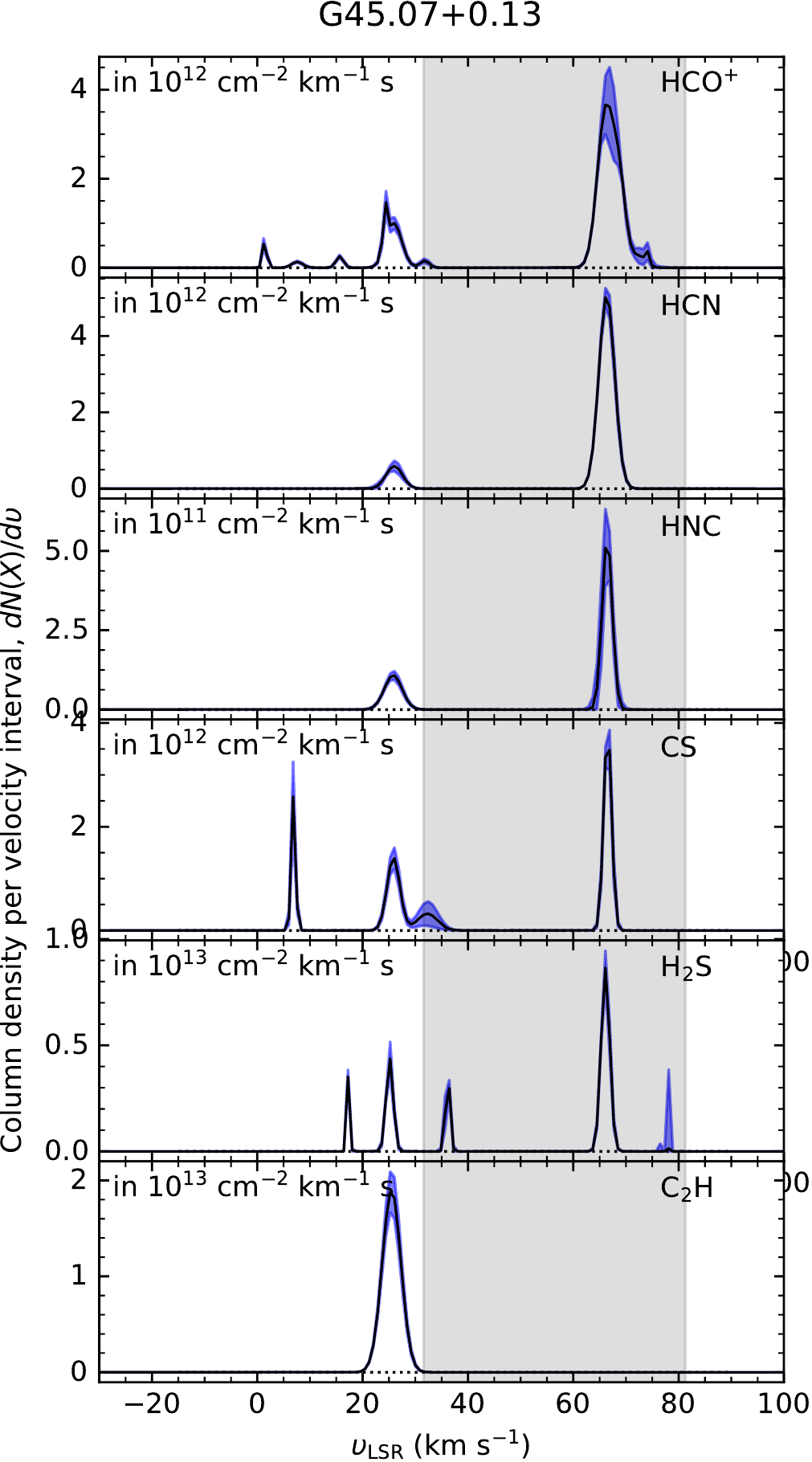}
    \includegraphics[width=0.46\textwidth]{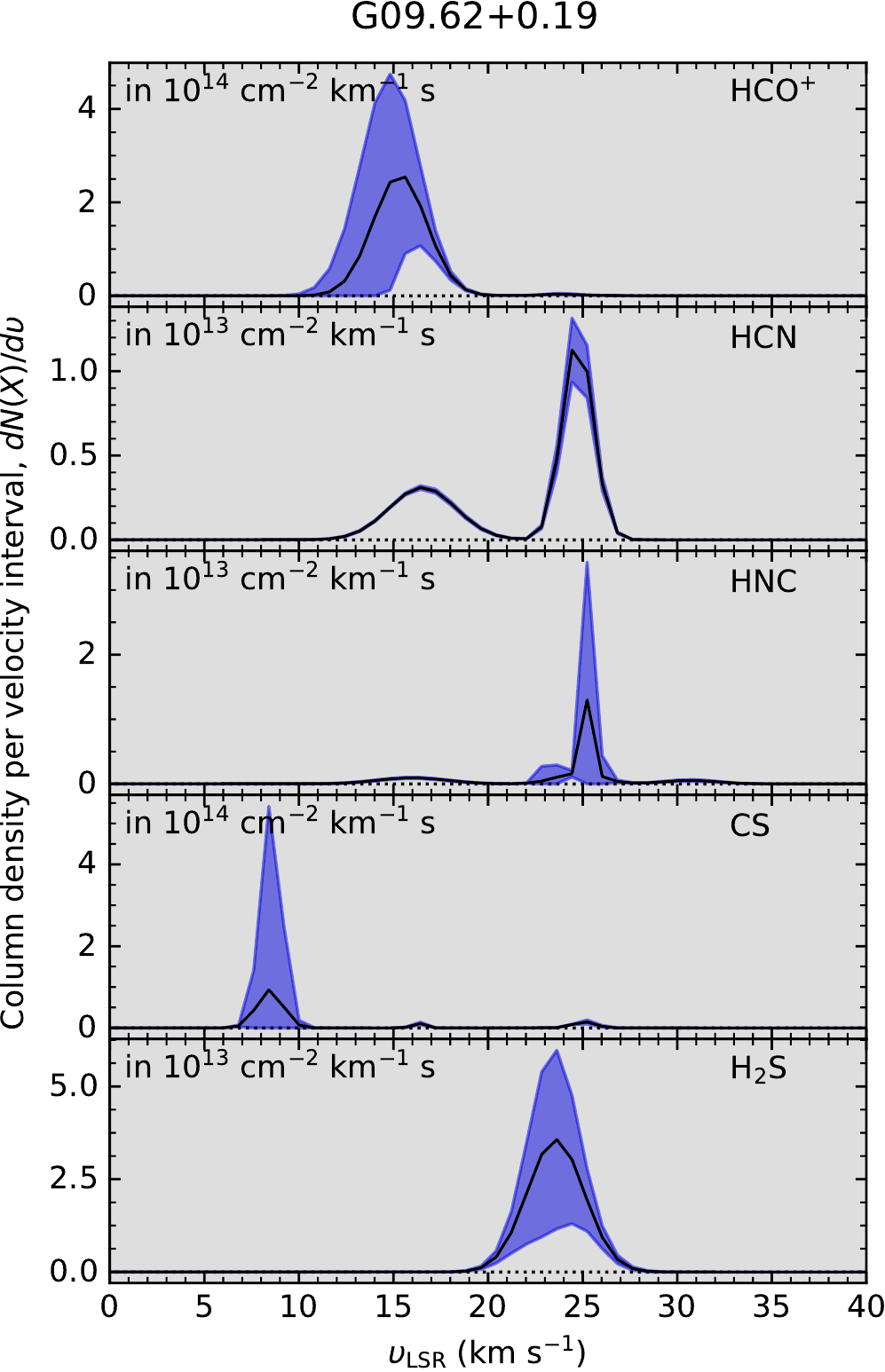}
    \caption{Channel-wise column density spectra ($dN/d\varv$) as a function of velocity toward G45.07$+$0.13 and G09.62$+$0.19. The $dN/d\varv$ in each panel shows \ce{HCO+}, HCN, HNC, CS, \ce{H2S}, and \ce{C2H} from top to bottom, but no \ce{C2H} toward G09.62$+$0.19. The gray shaded area indicates dispersion of \ce{HCO+} emission lines from continuum sources, and the blue shaded regions represent 2$\sigma$ confidence intervals for the derived column densities (per velocity interval). }
    \label{appen:ncol3}
\end{figure*}

\begin{figure}[h!]
    \centering
    \includegraphics[width=0.46\textwidth]{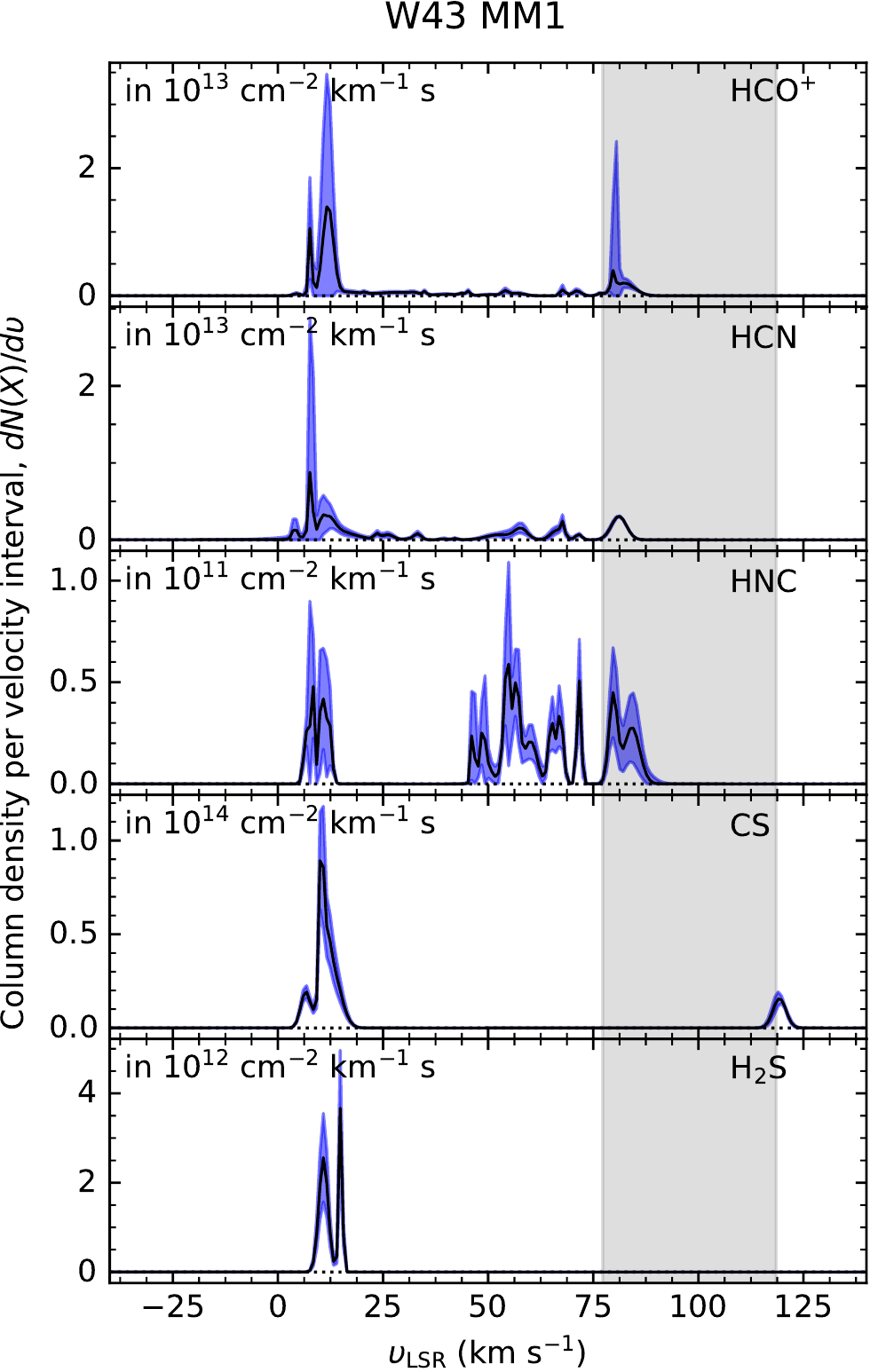}
    \includegraphics[width=0.47\textwidth]{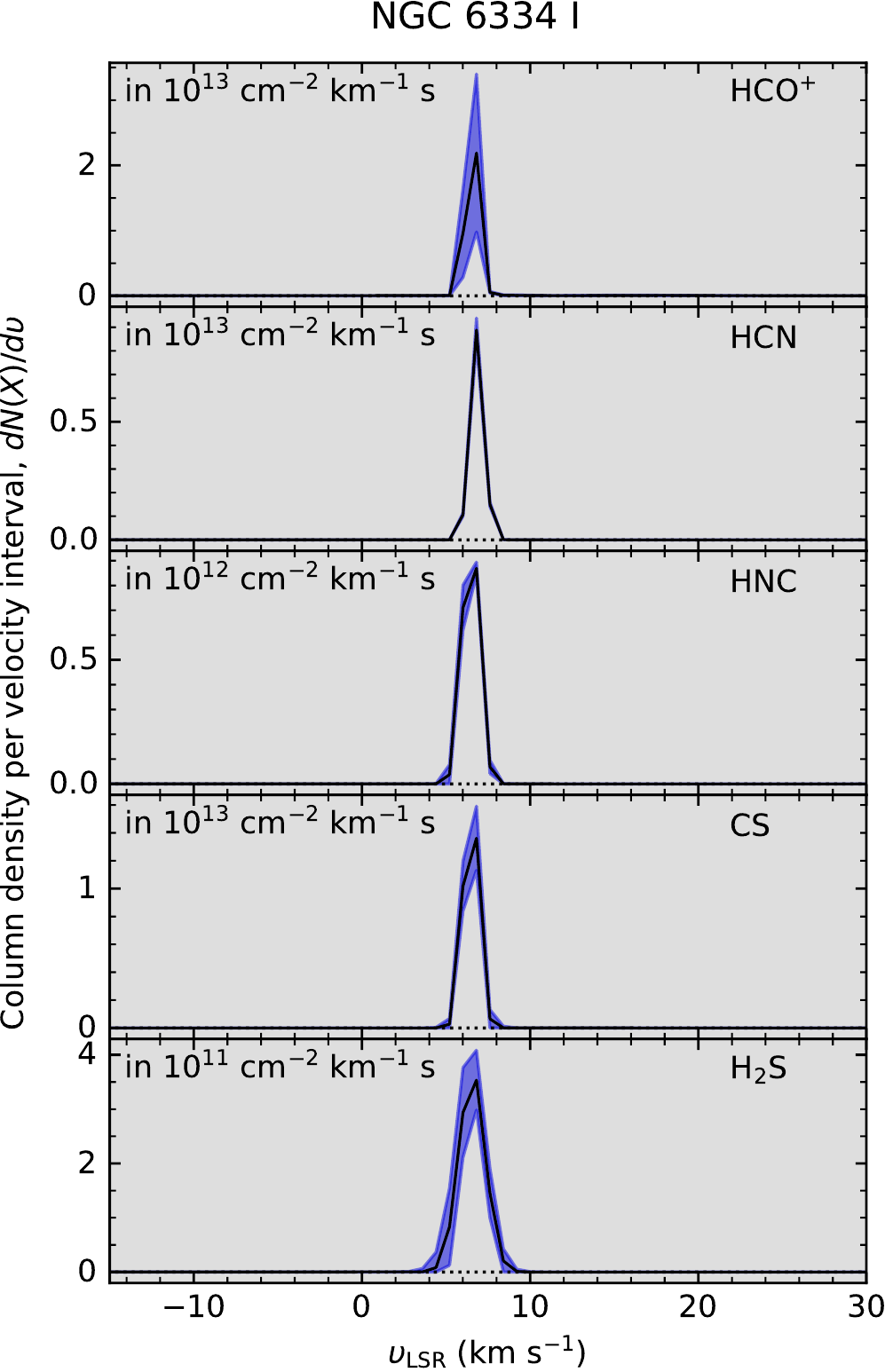}
    \caption{Channel-wise column density spectra ($dN/d\varv$) as a function of velocity toward W43 MM1 and NGC 6334 I. The $dN/d\varv$ in each panel shows \ce{HCO+}, HCN, HNC, CS, and \ce{H2S} from top to bottom. The gray shaded area indicates dispersion of \ce{HCO+} emission lines from continuum sources, and the blue shaded regions represent 2$\sigma$ confidence intervals for the derived column densities (per velocity interval). }
    \label{appen:ncol4}
\end{figure}

\begin{figure*}[h!]
    \centering
    \includegraphics[width=0.46\textwidth]{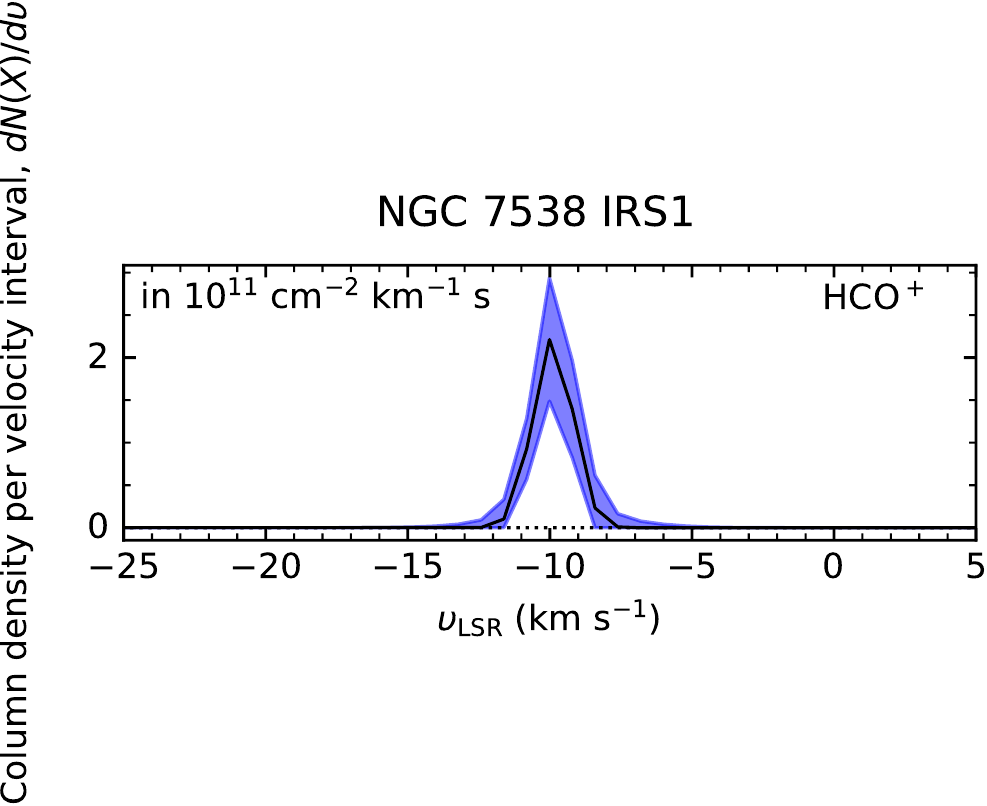}
    \includegraphics[width=0.46\textwidth]{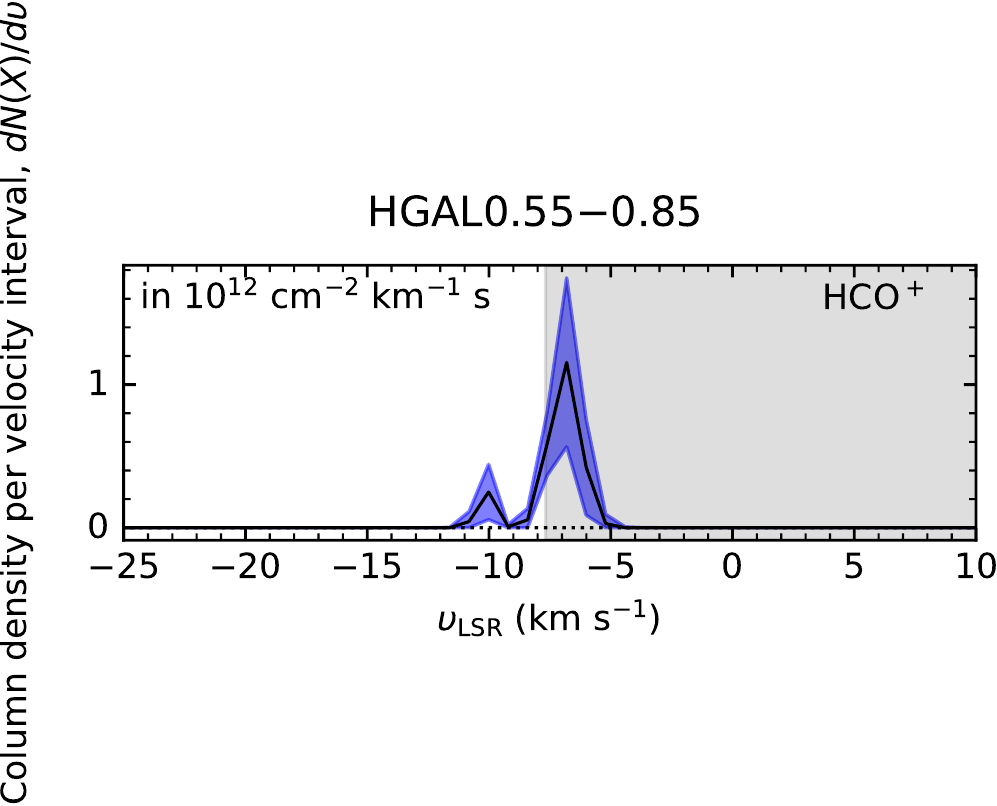}
    \includegraphics[width=0.46\textwidth]{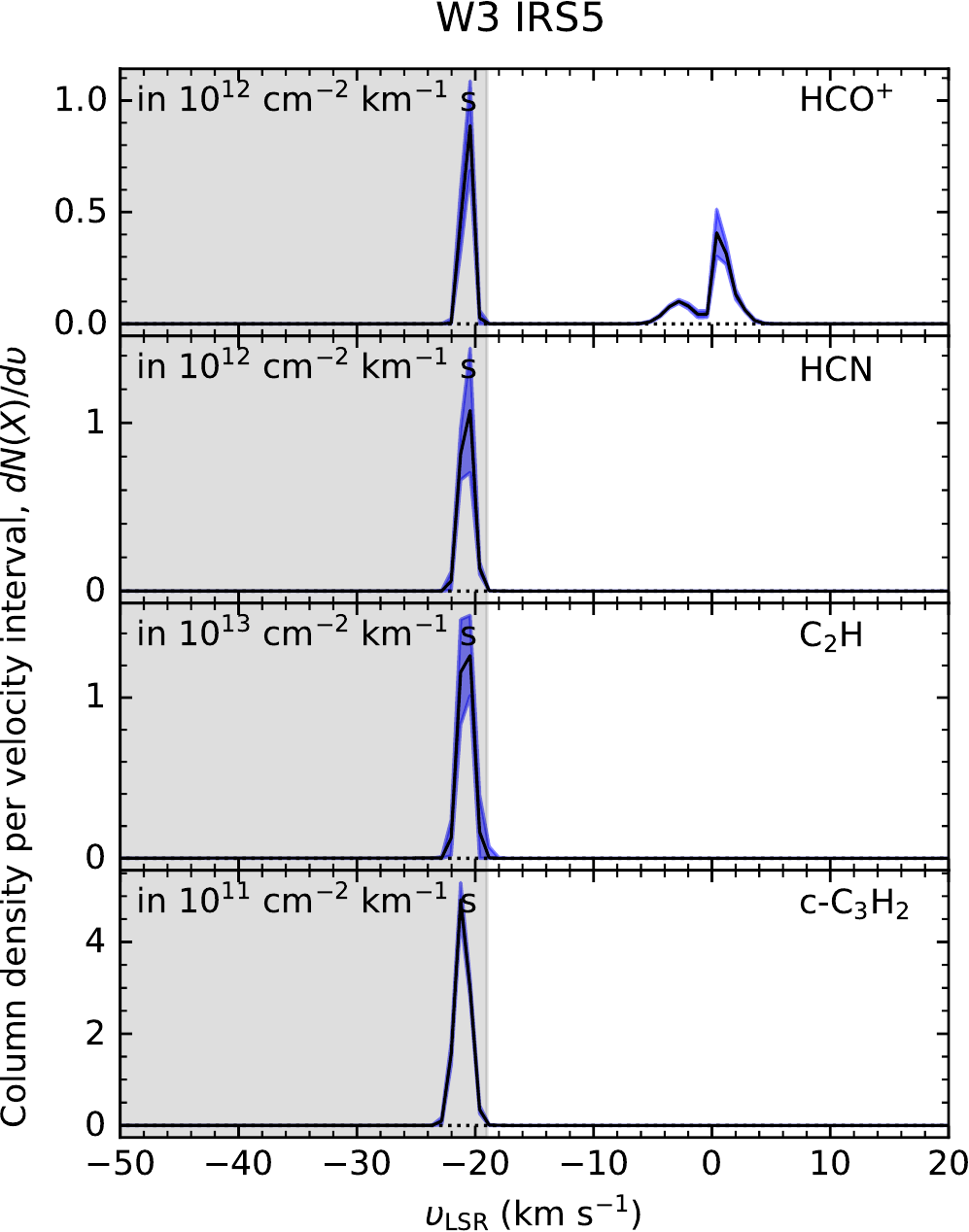}
    \caption{Channel-wise column density spectra ($dN/d\varv$) as a function of velocity toward NGC 7538 IRS1, HGAL0.55$-$0.85, and W3 IRS5. Toward NGC 7538 IRS1 and HGAL0.55$-$0.85, \ce{HCO+} is only detected. For W3 IRS5, the $dN/d\varv$ in each panel shows \ce{HCO+}, HCN, \ce{C2H}, and \ce{c-C3H2} from top to bottom. The gray shaded area indicates dispersion of \ce{HCO+} emission lines from continuum sources, and the blue shaded regions represent 2$\sigma$ confidence intervals for the derived column densities (per velocity interval). }
    \label{appen:ncol5}
\end{figure*}

Figures\,\ref{appen:ncol1}--\ref{appen:ncol5} show the column density per velocity interval for each detected molecular species in each observed source. 

\onecolumn
\newpage
\section{Derived column densities of the detected molecular species}\label{tab-appendix:ntot}
Table\,\ref{tab:nmol2} lists integrated column densities over given velocity interval for each detected molecule in each observed source. 

\begin{table*}[h!]
\centering
\tiny
\caption{\label{tab:nmol2} Continuation of Table\,\ref{tab:nmol1}. }
\begin{tabular}{l c c c c c c c c c }
\hline
\hline
Source & $\varv_{\rm LSR}$ range & $N$(\ce{H2})& $N$(\ce{HCO+}) & $N$(HCN) & $N$(HNC) & $N$(CS) & $N$(\ce{H2S}) & $N$(\ce{C2H}) & $N$(\ce{c-C3H2}) \\
& (\kms) & ($10^{21}$cm$^{-2}$) & ($10^{12}$cm$^{-2}$) & ($10^{12}$cm$^{-2}$) & ($10^{12}$cm$^{-2}$) & ($10^{12}$cm$^{-2}$) & ($10^{13}$cm$^{-2}$) & ($10^{13}$cm$^{-2}$) & ($10^{12}$cm$^{-2}$) \\
\hline
G19.61$-$0.23 & $-$10, 4 & 0.30$^{+0.04}_{-0.04}$ & 0.90$^{+0.13}_{-0.13}$ & 0.48$^{+0.01}_{-0.01}$ & $\cdots$ & $\cdots$ & $\cdots$ & 3.62$^{+1.22}_{-1.22}$ & $\cdots$\\
 & 4, 12 & 0.55$^{+0.05}_{-0.05}$ & 1.66$^{+0.15}_{-0.15}$ & 0.55$^{+0.01}_{-0.01}$ & $\cdots$ & 1.36$^{+0.07}_{-0.07}$ & $\cdots$ & 2.73$^{+1.08}_{-1.08}$ & $\cdots$\\
 & 14, 19 & 0.21$^{+0.03}_{-0.03}$ & 0.62$^{+0.08}_{-0.08}$ & $\cdots$ & $\cdots$ & $\cdots$ & $\cdots$ & $\cdots$ & $\cdots$\\
 & 27, 32$^{a}$ & 0.28$^{+0.06}_{-0.06}$ & 0.83$^{+0.18}_{-0.18}$ & $\cdots$ & $\cdots$ & $\cdots$ & $\cdots$ & 0.92$^{+0.08}_{-0.08}$ & 0.81$^{+0.07}_{-0.07}$\\
 & 48, 58 & 2.62$^{+0.29}_{-0.30}$ & 12.69$^{+1.68}_{-1.68}$ & 15.68$^{+1.56}_{-1.56}$ & 3.23$^{+0.29}_{-0.29}$ & 16.89$^{+0.88}_{-0.88}$ & 3.23$^{+0.51}_{-0.51}$ & 10.03$^{+1.75}_{-1.58}$ & 3.47$^{+0.25}_{-0.25}$\\
 & 58, 67 & 1.14$^{+0.07}_{-0.07}$ & 4.76$^{+0.34}_{-0.34}$ & 2.95$^{+0.07}_{-0.07}$ & 0.89$^{+0.07}_{-0.07}$ & 0.80$^{+0.05}_{-0.05}$ & $\cdots$ & 9.51$^{+4.43}_{-4.37}$ & 0.99$^{+0.07}_{-0.07}$\\
 & 67, 75 & 3.46$^{+0.46}_{-0.47}$ & 17.59$^{+2.81}_{-2.81}$ & 20.36$^{+1.62}_{-1.62}$ & 4.82$^{+0.41}_{-0.41}$ & 5.43$^{+0.51}_{-0.51}$ & $\cdots$ & 10.20$^{+1.03}_{-1.03}$ & 2.75$^{+0.19}_{-0.19}$\\
 & 75, 82 & 0.65$^{+0.08}_{-0.08}$ & 1.95$^{+0.24}_{-0.24}$ & 1.28$^{+0.05}_{-0.05}$ & 0.34$^{+0.02}_{-0.02}$ & $\cdots$ & $\cdots$ & 1.18$^{+0.38}_{-0.26}$ & 0.23$^{+0.01}_{-0.01}$\\
 & 87, 93 & 0.10$^{+0.02}_{-0.02}$ & 0.30$^{+0.07}_{-0.07}$ & 0.47$^{+0.01}_{-0.01}$ & 0.19$^{+0.01}_{-0.01}$ & $\cdots$ & $\cdots$ & $\cdots$ & 0.30$^{+0.01}_{-0.01}$\\
 & 93, 106 & 1.80$^{+0.22}_{-0.22}$ $^{b}$ & 8.12$^{+1.16}_{-1.16}$ & 15.85$^{+1.55}_{-1.55}$ & 3.13$^{+0.25}_{-0.25}$ & 2.87$^{+0.19}_{-0.19}$ & $\cdots$ & 3.80$^{+0.19}_{-0.19}$ & 2.29$^{+0.14}_{-0.14}$\\
 & 106, 114 & 0.25$^{+0.02}_{-0.02}$ & 0.74$^{+0.07}_{-0.07}$ & 1.95$^{+0.03}_{-0.03}$ & 0.63$^{+0.04}_{-0.04}$ & $\cdots$ & $\cdots$ & 1.69$^{+0.11}_{-0.11}$ & 1.16$^{+0.06}_{-0.06}$\\
 & 114, 122 & 0.67$^{+0.05}_{-0.05}$ $^{b}$ & 2.52$^{+0.21}_{-0.21}$ & 4.07$^{+0.24}_{-0.24}$ & 0.98$^{+0.06}_{-0.06}$ & $\cdots$ & $\cdots$ & 1.00$^{+0.09}_{-0.09}$ & 0.70$^{+0.03}_{-0.03}$\\
 & 122, 127 & 0.11$^{+0.01}_{-0.01}$ $^{b}$ & 0.33$^{+0.02}_{-0.02}$ & 1.03$^{+0.03}_{-0.03}$ & 0.3$^{+0.02}_{-0.02}$ & $\cdots$ & $\cdots$ & 0.45$^{+0.05}_{-0.05}$ & $\cdots$\\
\hline
G29.96$-$0.02 & $-$1, 10 & 1.81$^{+0.25}_{-0.25}$ & 8.22$^{+1.34}_{-1.34}$ & 13.34$^{+3.13}_{-3.13}$ & 2.63$^{+0.22}_{-0.22}$ & 4.67$^{+0.44}_{-0.44}$ & 1.58$^{+0.48}_{-0.48}$ & 9.82$^{+0.89}_{-0.89}$ & 3.71$^{+0.23}_{-0.23}$\\
 & 10, 14 & 0.53$^{+0.04}_{-0.04}$ & 1.58$^{+0.11}_{-0.11}$ & 2.14$^{+0.21}_{-0.21}$ & 0.43$^{+0.05}_{-0.05}$ & 1.52$^{+0.13}_{-0.13}$ & $\cdots$ & 3.71$^{+0.43}_{-0.43}$ & 1.37$^{+0.04}_{-0.04}$\\
 & 14, 21 & 0.15$^{+0.02}_{-0.02}$ & 0.44$^{+0.06}_{-0.06}$ & 0.52$^{+0.12}_{-0.08}$ & $\cdots$ & $\cdots$ & $\cdots$ & 1.12$^{+0.22}_{-0.22}$ & 0.61$^{+0.01}_{-0.01}$\\
 & 50, 56 & 0.18$^{+0.05}_{-0.05}$ & 0.53$^{+0.14}_{-0.14}$ & 0.71$^{+0.07}_{-0.07}$ & $\cdots$ & $\cdots$ & $\cdots$ & $\cdots$ & $\cdots$\\
 & 56, 64 & 0.42$^{+0.10}_{-0.10}$ & 1.25$^{+0.29}_{-0.29}$ & 1.54$^{+0.17}_{-0.17}$ & 0.17$^{+0.02}_{-0.02}$ & $\cdots$ & $\cdots$ & $\cdots$ & $\cdots$\\
 & 64, 74 & 1.08$^{+0.14}_{-0.14}$ & 4.47$^{+0.68}_{-0.68}$ & 7.47$^{+0.64}_{-0.64}$ & 1.01$^{+0.07}_{-0.07}$ & $\cdots$ & $\cdots$ & 3.50$^{+0.54}_{-0.54}$ & $\cdots$\\
 & 74, 79 & 0.12$^{+0.04}_{-0.04}$ & 0.37$^{+0.13}_{-0.12}$ & $\cdots$ & $\cdots$ & $\cdots$ & $\cdots$ & $\cdots$ & $\cdots$\\
\hline
W43 MM1 & 1, 9 & 2.40$^{+1.72}_{-1.74}$ & 11.46$^{+10.23}_{-8.95}$ & 13.81$^{+35.52}_{-13.81}$ & 0.92$^{+0.80}_{-0.53}$ & $\cdots$ & $\cdots$ & 5.89$^{+1.10}_{-1.10}$ & $\cdots$\\
 & 9, 13 & 6.02$^{+7.27}_{-6.02}$ & 33.88$^{+52.39}_{-33.88}$ & 10.65$^{+8.42}_{-8.01}$ & 1.18$^{+0.91}_{-0.90}$ & $\cdots$ & $\cdots$ & 23.22$^{+7.49}_{-7.49}$ & 6.57$^{+2.53}_{-2.53}$\\
 & 13, 21 & 2.81$^{+1.84}_{-2.03}$ & 13.79$^{+11.17}_{-10.73}$ & 8.66$^{+3.81}_{-3.81}$ & $\cdots$ & $\cdots$ & $\cdots$ & 8.66$^{+2.66}_{-2.66}$ & 4.03$^{+1.43}_{-1.43}$\\
 & 21, 27 & 0.73$^{+0.21}_{-0.22}$ & 2.81$^{+0.98}_{-0.98}$ & 3.47$^{+1.7}_{-1.7}$ & $\cdots$ & $\cdots$ & $\cdots$ & $\cdots$ & $\cdots$\\
 & 27, 39 & 1.16$^{+0.30}_{-0.31}$ & 4.85$^{+1.51}_{-1.51}$ & 2.65$^{+1.18}_{-1.17}$ & $\cdots$ & $\cdots$ & $\cdots$ & $\cdots$ & $\cdots$\\
 & 39, 47 & 0.64$^{+0.16}_{-0.17}$ & 2.42$^{+0.74}_{-0.74}$ & 0.66$^{+0.46}_{-0.66}$ & 0.30$^{+0.44}_{-0.30}$ & $\cdots$ & $\cdots$ & $\cdots$ & $\cdots$\\
 & 47, 61$^{a}$ & 0.99$^{+0.33}_{-0.35}$ & 4.01$^{+1.64}_{-1.64}$ & 10.24$^{+4.82}_{-4.82}$ & 3.34$^{+1.93}_{-1.84}$ & $\cdots$ & $\cdots$ & $\cdots$ & $\cdots$\\
 & 61, 69$^{a}$ & 0.5$^{+0.33}_{-0.33}$ & 1.50$^{+0.99}_{-0.99}$ & 6.27$^{+2.89}_{-2.82}$ & 1.39$^{+0.59}_{-0.58}$ & $\cdots$ & $\cdots$ & $\cdots$ & $\cdots$\\
 & 69, 75$^{a}$ & 0.55$^{+0.20}_{-0.22}$ & 2.01$^{+0.90}_{-0.90}$ & 1.53$^{+0.65}_{-0.58}$ & 0.62$^{+0.26}_{-0.26}$ & $\cdots$ & $\cdots$ & $\cdots$ & $\cdots$\\
 & 75, 87$^{a}$ & 3.09$^{+4.86}_{-1.68}$ & 15.40$^{+31.62}_{-9.30}$ & 13.64$^{+0.89}_{-0.89}$ & 2.23$^{+1.30}_{-1.30}$ & $\cdots$ & $\cdots$ & $\cdots$ & $\cdots$\\
\hline
G31.41$+$0.31 & 3, 10 & 0.69$^{+0.12}_{-0.12}$ & 2.63$^{+0.55}_{-0.55}$ & 3.97$^{+0.83}_{-0.83}$ & $\cdots$ & 3.85$^{+0.76}_{-0.76}$ & $\cdots$ & 4.36$^{+0.79}_{-0.79}$ & 0.47$^{+0.15}_{-0.47}$\\
 & 10, 18 & 0.99$^{+0.18}_{-0.19}$ & 4.03$^{+0.88}_{-0.88}$ & 3.68$^{+0.92}_{-0.92}$ & 1.09$^{+0.11}_{-0.11}$ & 5.98$^{+1.25}_{-1.25}$ & 0.16$^{+0.02}_{-0.02}$ & 11.83$^{+2.30}_{-2.30}$ & 2.42$^{+0.49}_{-0.49}$\\
\hline
G32.80$+$0.19 & 4, 7 & 0.57$^{+0.01}_{-0.01}$ & 1.72$^{+0.03}_{-0.03}$ & 3.00$^{+0.04}_{-0.04}$ & $\cdots$ & $\cdots$ & $\cdots$ & 2.08$^{+0.12}_{-0.12}$ & 0.70$^{+0.04}_{-0.04}$\\
 & 7, 15 & 1.74$^{+0.05}_{-0.05}$ & 7.81$^{+0.25}_{-0.25}$ & 8.15$^{+0.13}_{-0.13}$ & $\cdots$ & $\cdots$ & $\cdots$ & 0.93$^{+0.06}_{-0.06}$ & 0.86$^{+0.06}_{-0.06}$\\
 & 28, 43 & 0.41$^{+0.01}_{-0.01}$ & 1.22$^{+0.01}_{-0.01}$ & 1.33$^{+0.0}_{-0.0}$ & 0.49$^{+0.07}_{-0.07}$ & $\cdots$ & $\cdots$ & 4.62$^{+0.65}_{-0.65}$ & 0.90$^{+0.03}_{-0.03}$\\
 & 43, 59 & 0.82$^{+0.02}_{-0.02}$ & 3.23$^{+0.10}_{-0.10}$ & 3.04$^{+0.01}_{-0.01}$ & 0.06$^{+0.04}_{-0.06}$ & $\cdots$ & $\cdots$ & 0.90$^{+0.19}_{-0.19}$ & 1.67$^{+0.08}_{-0.08}$\\
 & 59, 77 & 1.63$^{+0.03}_{-0.03}$ & 7.23$^{+0.15}_{-0.15}$ & 11.47$^{+0.04}_{-0.04}$ & 2.28$^{+0.23}_{-0.23}$ & 3.55$^{+0.04}_{-0.04}$ & 0.63$^{+0.02}_{-0.02}$ & 6.19$^{+0.31}_{-0.31}$ & 3.35$^{+0.23}_{-0.23}$\\
 & 77, 83 & 1.36$^{+0.03}_{-0.03}$ & 5.86$^{+0.15}_{-0.15}$ & 16.3$^{+0.16}_{-0.16}$ & 4.51$^{+0.71}_{-0.71}$ & 4.06$^{+0.04}_{-0.04}$ & 0.35$^{+0.01}_{-0.01}$ & 3.86$^{+0.24}_{-0.24}$ & 1.35$^{+0.23}_{-0.23}$\\
 & 83, 87 & 1.69$^{+0.16}_{-0.16}$ & 7.58$^{+0.85}_{-0.85}$ & 31.1$^{+0.30}_{-0.30}$ & 6.62$^{+1.90}_{-1.90}$ & 17.8$^{+0.34}_{-0.34}$ & 1.73$^{+0.19}_{-0.19}$ & 5.66$^{+0.53}_{-0.53}$ & 1.51$^{+0.22}_{-0.22}$\\
 & 87, 92 & 3.14$^{+0.10}_{-0.10}$ & 15.69$^{+0.59}_{-0.59}$ & 29.97$^{+0.22}_{-0.22}$ & 5.22$^{+0.59}_{-0.59}$ & 9.57$^{+0.12}_{-0.12}$ & 1.19$^{+0.10}_{-0.10}$ & 10.99$^{+0.8}_{-0.8}$ & 3.70$^{+0.30}_{-0.30}$\\
 & 92, 101 & 1.10$^{+0.03}_{-0.03}$ & 4.57$^{+0.14}_{-0.14}$ & 3.46$^{+0.01}_{-0.01}$ & 0.36$^{+0.03}_{-0.03}$ & 0.58$^{+0.04}_{-0.04}$ & $\cdots$ & 5.41$^{+0.29}_{-0.29}$ & 1.25$^{+0.06}_{-0.06}$\\
 & 101, 109 & 0.16$^{+0.01}_{-0.01}$ & 0.49$^{+0.01}_{-0.01}$ & 0.44$^{+0.01}_{-0.01}$ & $\cdots$ & $\cdots$ & $\cdots$ & 0.41$^{+0.13}_{-0.12}$ & 0.31$^{+0.02}_{-0.02}$\\
\hline
G45.07$+$0.13 & $-$1, 3 & 0.21$^{+0.05}_{-0.05}$ & 0.62$^{+0.15}_{-0.15}$ & $\cdots$ & $\cdots$ & $\cdots$ & $\cdots$ & $\cdots$ & $\cdots$\\
 & 3, 11 & 0.12$^{+0.03}_{-0.03}$ & 0.35$^{+0.08}_{-0.08}$ & $\cdots$ & $\cdots$ & 2.62$^{+0.77}_{-0.75}$ & $\cdots$ & $\cdots$ & $\cdots$\\
 & 12, 19 & 0.15$^{+0.03}_{-0.03}$ & 0.45$^{+0.09}_{-0.09}$ & $\cdots$ & $\cdots$ & $\cdots$ & 0.29$^{+0.03}_{-0.03}$ & $\cdots$ & $\cdots$\\
 & 21, 30 & 1.12$^{+0.16}_{-0.17}$ & 4.66$^{+0.81}_{-0.81}$ & 2.29$^{+0.60}_{-0.57}$ & 0.45$^{+0.06}_{-0.06}$ & 4.41$^{+0.85}_{-0.85}$ & 0.74$^{+0.12}_{-0.12}$ & 8.33$^{+0.96}_{-0.96}$ & $\cdots$\\
 & 30, 38 & 0.12$^{+0.05}_{-0.05}$ & 0.36$^{+0.15}_{-0.15}$ & $\cdots$ & $\cdots$ & 1.42$^{+1.04}_{-1.04}$ & 0.40$^{+0.12}_{-0.11}$ & $\cdots$ & $\cdots$\\
 & 63, 71$^{a}$ & 3.64$^{+0.48}_{-0.49}$ & 18.71$^{+2.91}_{-2.91}$ & 18.88$^{+1.11}_{-1.11}$ & 1.30$^{+0.47}_{-0.42}$ & 7.37$^{+0.93}_{-0.93}$ & 1.77$^{+0.18}_{-0.18}$ & $\cdots$ & $\cdots$\\
 & 71, 80$^{a}$ & 0.34$^{+0.24}_{-0.20}$ & 1.02$^{+0.73}_{-0.59}$ & $\cdots$ & $\cdots$ & $\cdots$ & $\cdots$ & $\cdots$ & $\cdots$\\
\hline
DR21 & 0, 10$^{a}$ & 1.38$^{+0.10}_{-0.10}$ & 5.93$^{+0.53}_{-0.53}$ & 10.08$^{+0.36}_{-0.36}$ & 3.68$^{+0.17}_{-0.17}$ & 5.38$^{+0.21}_{-0.21}$ & 2.38$^{+0.20}_{-0.20}$ & 9.29$^{+1.71}_{-1.54}$ & 3.83$^{+0.33}_{-0.32}$\\
 & 10, 17$^{a}$ & 0.65$^{+0.09}_{-0.09}$ & 1.96$^{+0.27}_{-0.27}$ & 1.52$^{+0.18}_{-0.18}$ & 0.52$^{+0.17}_{-0.16}$ & $\cdots$ & 2.06$^{+0.21}_{-0.21}$ & 1.86$^{+0.81}_{-0.78}$ & 0.89$^{+0.22}_{-0.22}$\\
\hline
NGC 7538 IRS1 & $-$13, $-$5 & 0.13$^{+0.07}_{-0.05}$ & 0.39$^{+0.22}_{-0.16}$ & $\cdots$ & $\cdots$ & $\cdots$ & $\cdots$ & $\cdots$ & $\cdots$\\
\hline
\end{tabular}
\end{table*}


\onecolumn
\newpage
\section{Principal component analysis toward W3 IRS5}
\begin{figure*}[h!]
    \centering
    \includegraphics[width=0.355\textwidth]{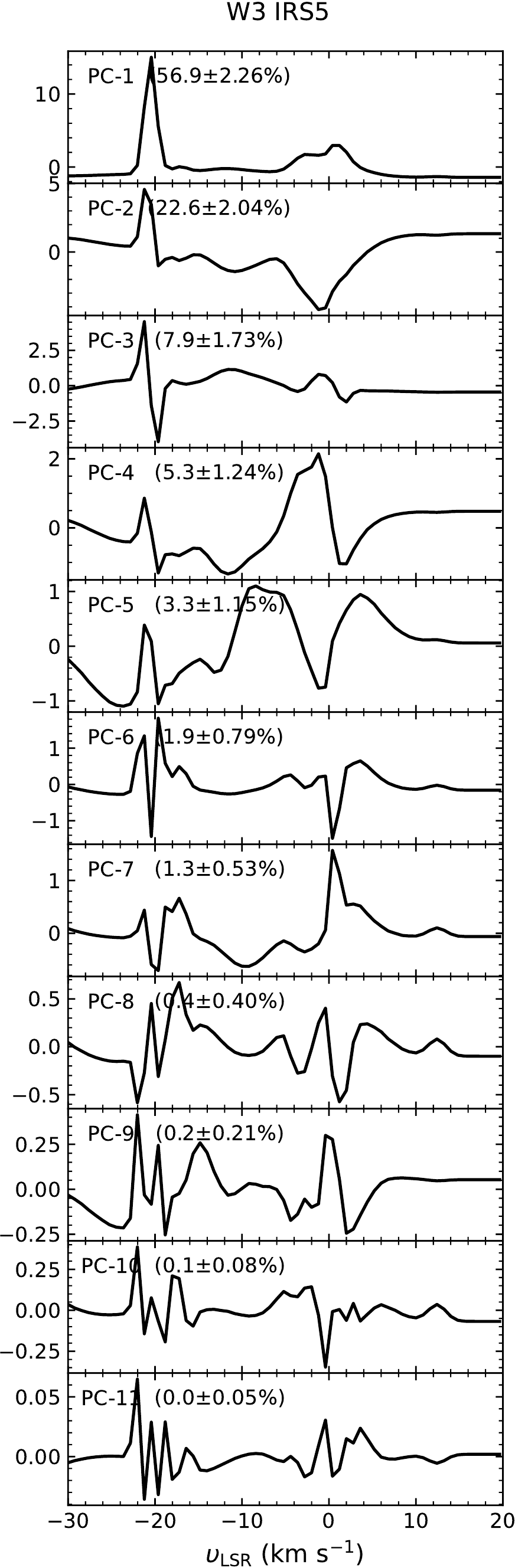}
    \hskip 2 cm 
    \includegraphics[width=0.355\textwidth]{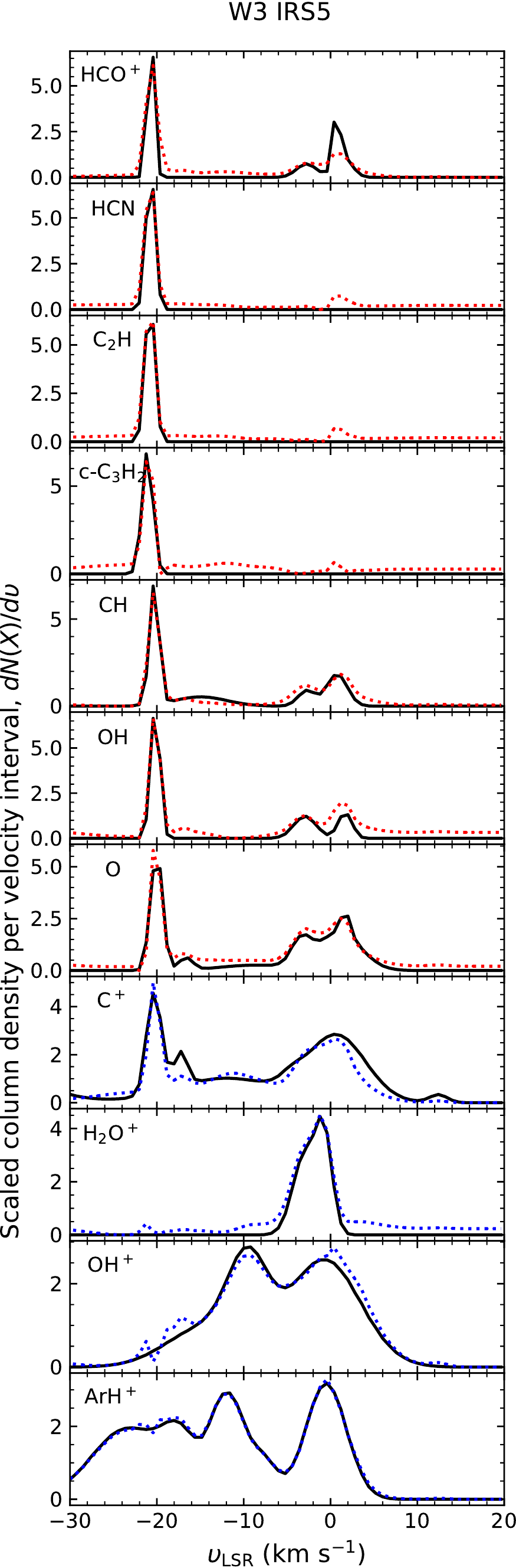}
    \caption{\textit{Left}: Eigen spectra of the eleven principle components with percentage of variances as well as uncertainties, toward W3 IRS5. \textit{Right}: Scaled channel-wise column density spectra of the eleven species (\ce{HCO+}, HCN, \ce{C2H}, \ce{c-C3H2}, CH, OH, O, \ce{C+}, \ce{H2O+}, \ce{OH+}, and \ce{ArH+}) as a function of $\varv_{\rm LSR}$. Black curves represent the observational data, and red dotted curves are the spectra reproduced with only the first three principal components (PC-1, PC-2, and PC-3). For \ce{C+}, \ce{H2O+}, \ce{OH+}, and \ce{ArH+}, the reproduced spectra indicated by blue dotted curves include one or two more PCs, i.e., PC-4 and PC-5.}
    \label{append:pca_fit}
\end{figure*}

Figure\,\ref{append:pca_fit} shows the principal components in the left panel, and in the right panel, the rescaled column density profiles of observed absorption line (black curves) and the reproduced column density profiles (red and blue dotted lines) for each detected species, toward W3 IRS5.

\newpage
\section{H-plots for the ten considered species (\ce{HCO+}, HCN, \ce{C2H}, \ce{c-C3H2}, CH, OH, O, \ce{C+}, \ce{H2O+}, and \ce{OH+}) toward W3(OH) and W3 IRS5}

\begin{figure*}[h!]
    \centering
    \includegraphics[width=0.45\textwidth]{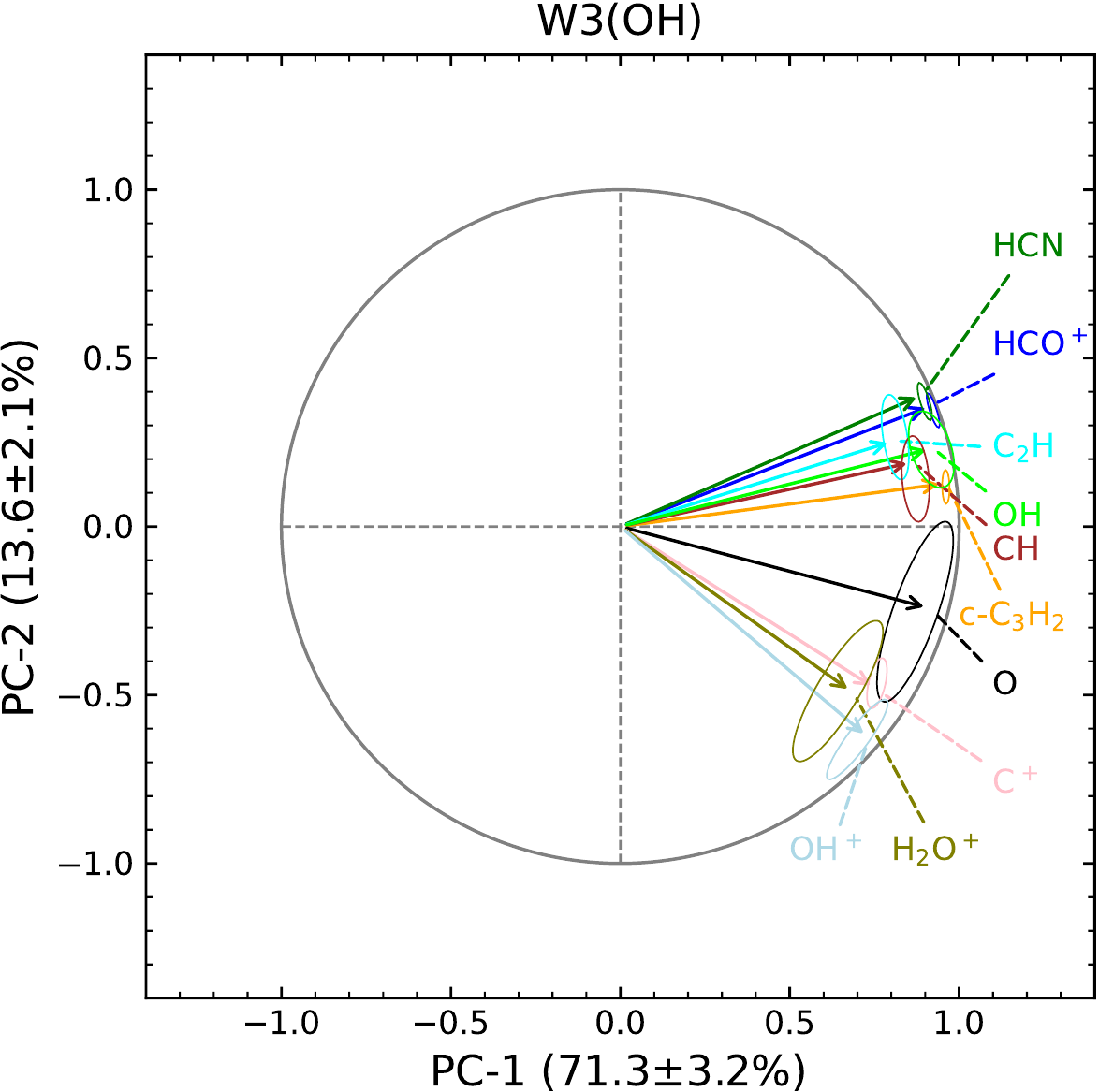}
    \includegraphics[width=0.45\textwidth]{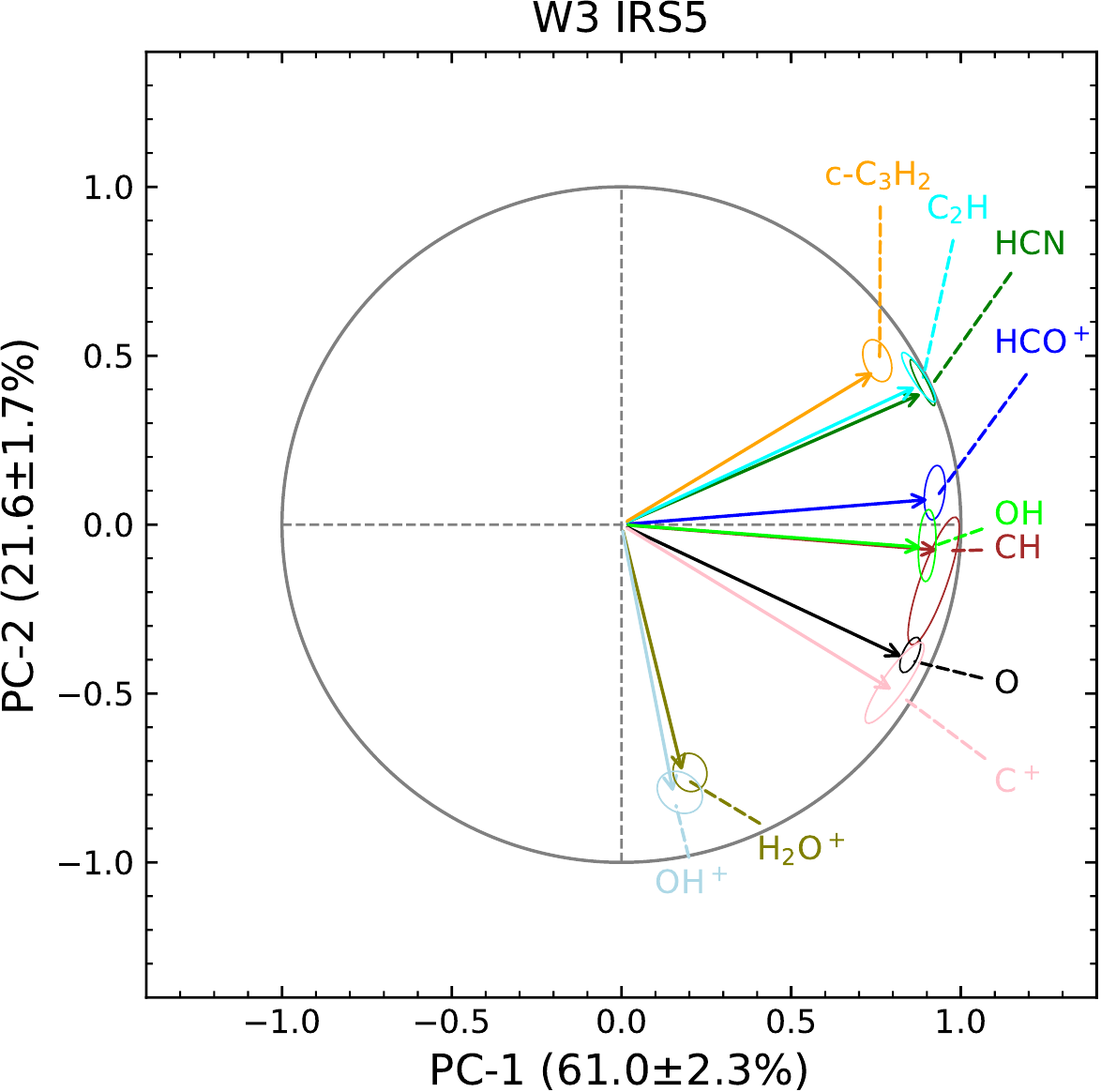}
    \vskip 1 cm 
    \includegraphics[width=0.45\textwidth]{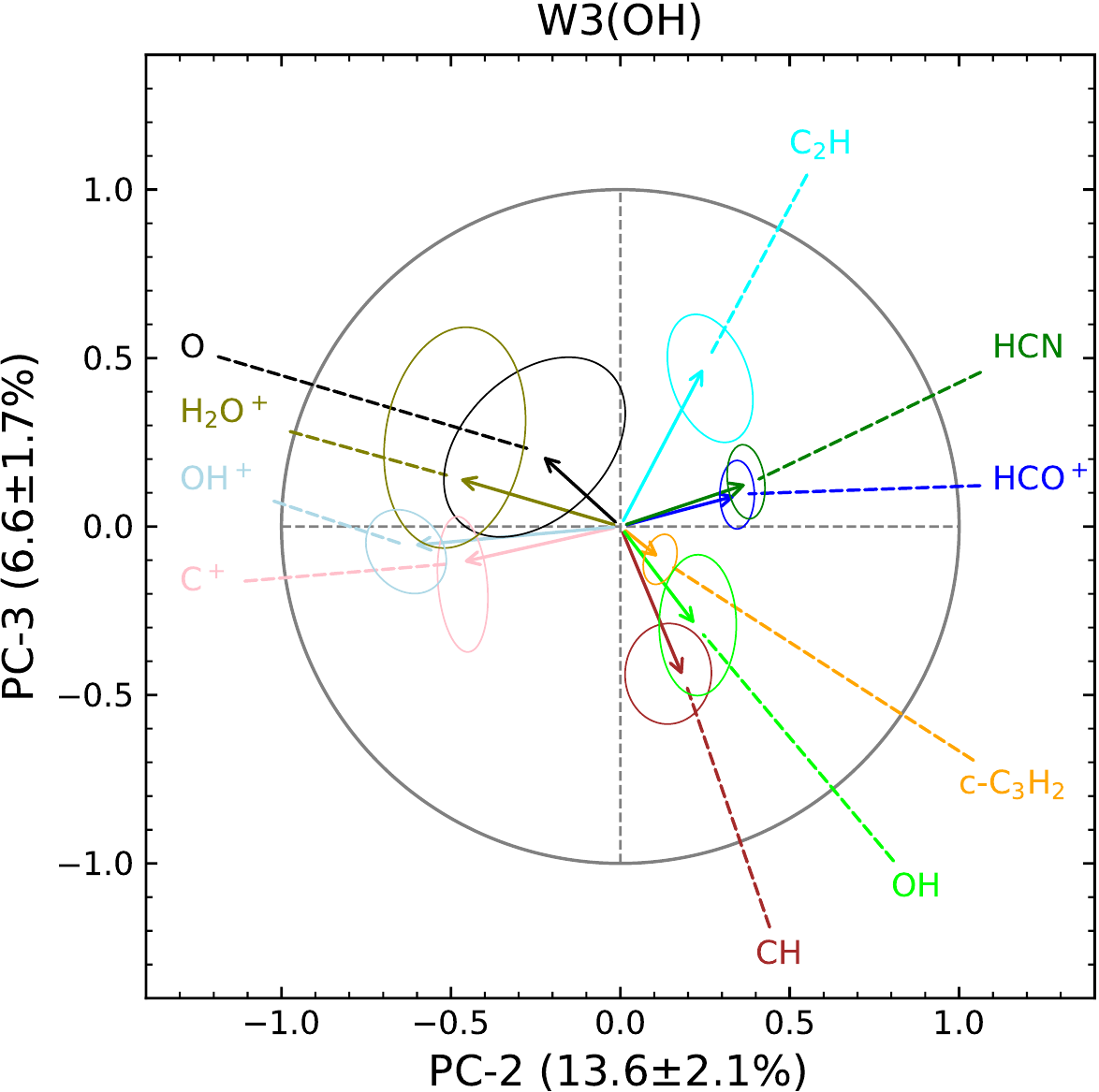}
    \includegraphics[width=0.45\textwidth]{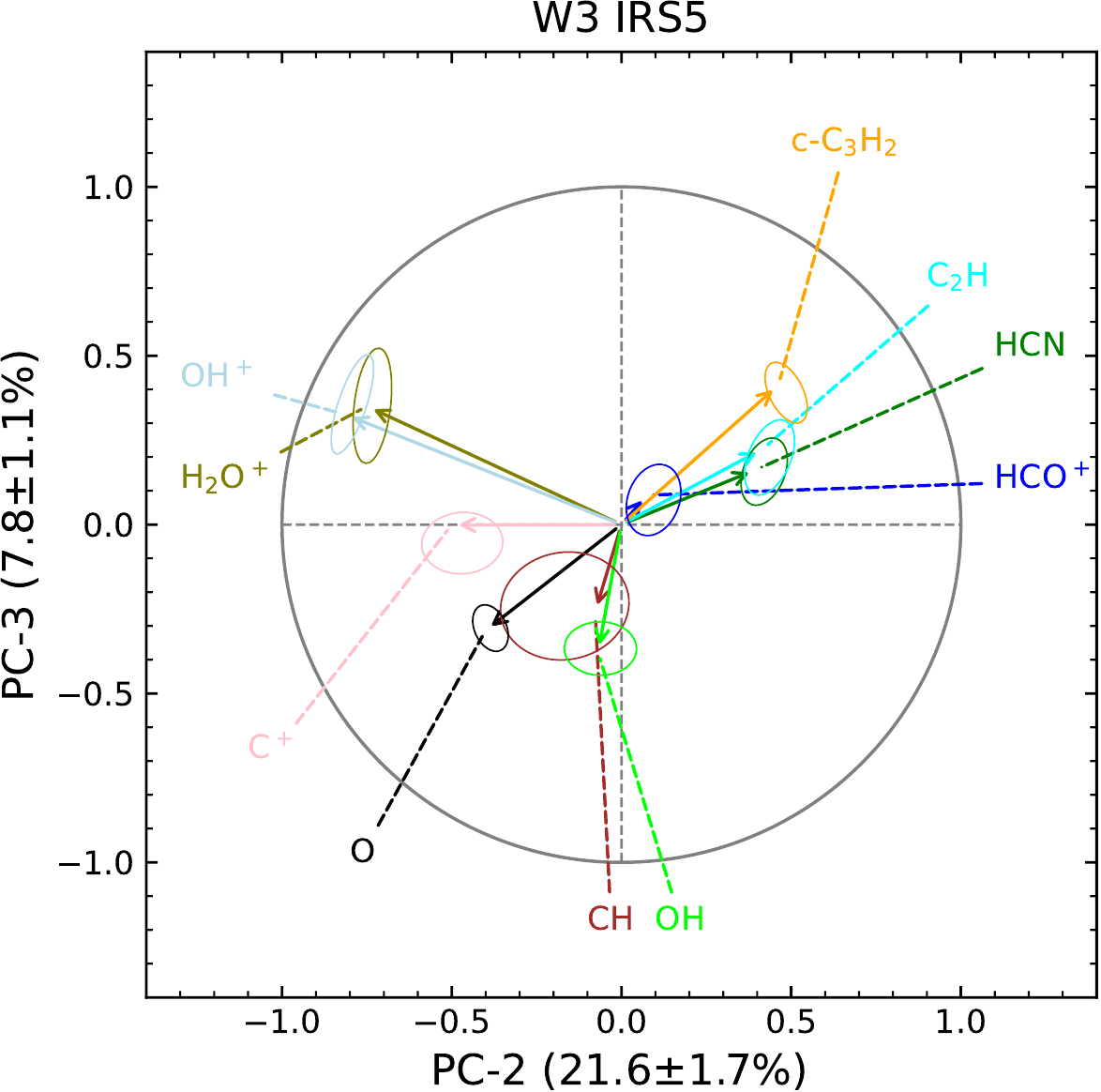}
    \caption{\textit{Upper}: Variable correlation circle plots for the first and second PCs toward W3(OH) in the left panel and W3 IRS5 in the right panel. The solid straight lines close to the unit circle imply PC-1 and PC-2 contain the most information about the variables. Different colors are used to denote the different species analyzed. Each colored ellipse is a 2$\sigma$ confidence interval for the uncertainty of the PCA results for the corresponding species. \textit{Lower}: Same as the upper panel but for the second and third PCs (PC-2 and PC-3) toward the same sources, from left to right. }
    \label{append:pca_fit_subsample}
\end{figure*}

Figure\,\ref{append:pca_fit_subsample} shows the h-plots for ten considered species toward W3(OH) and W3 IRS5 as like as Figs.\,\ref{fig:pc1pc2_circle} and \ref{fig:pc2pc3_circle}, but these h-plots only consider ten species that are commonly detected toward these two sightlines.

\newpage
\section{Variations of $N$(\ce{H2}) and $N$(X)/$N$(\ce{H2}) ratios }
\begin{figure*}[h!]
    \centering
    \includegraphics[width=0.59\textwidth]{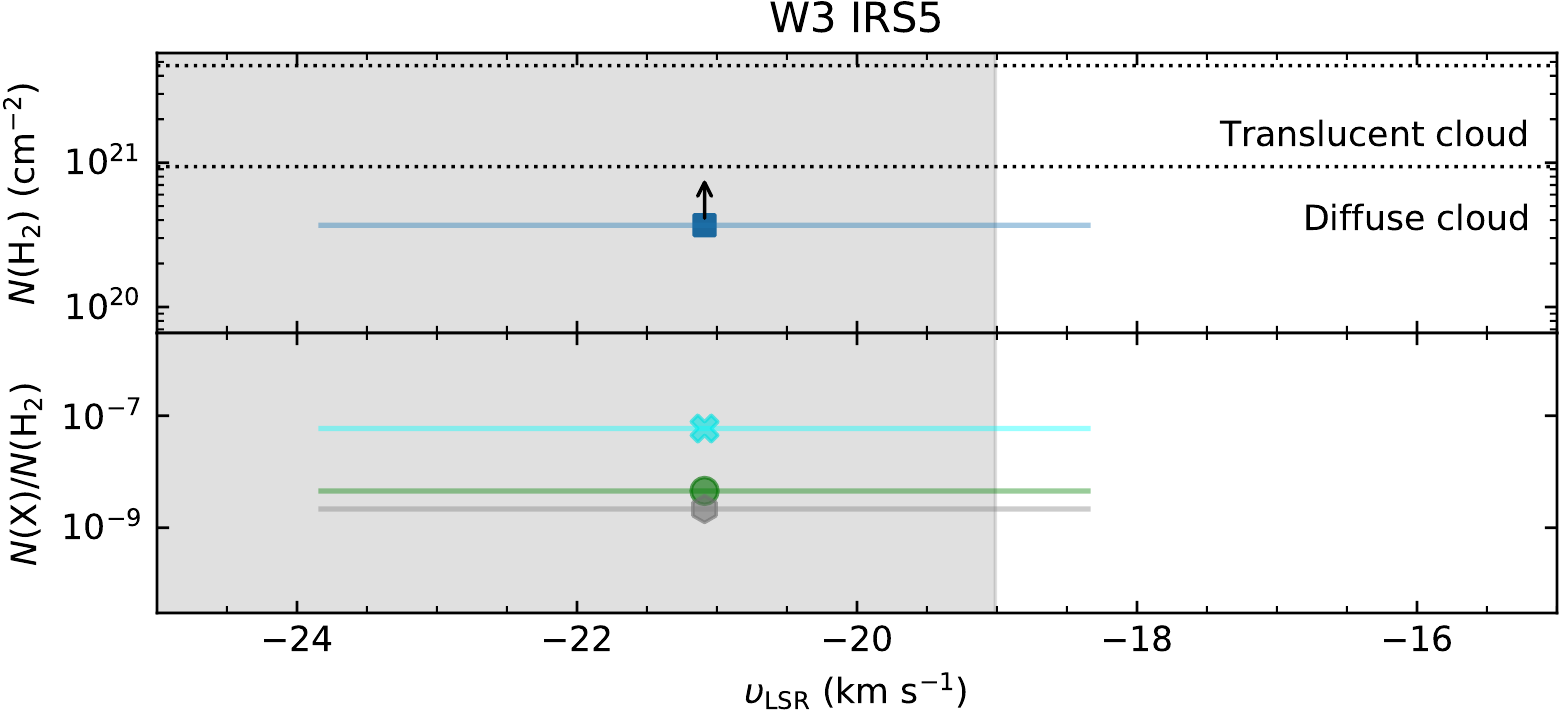} 
    \includegraphics[width=0.59\textwidth]{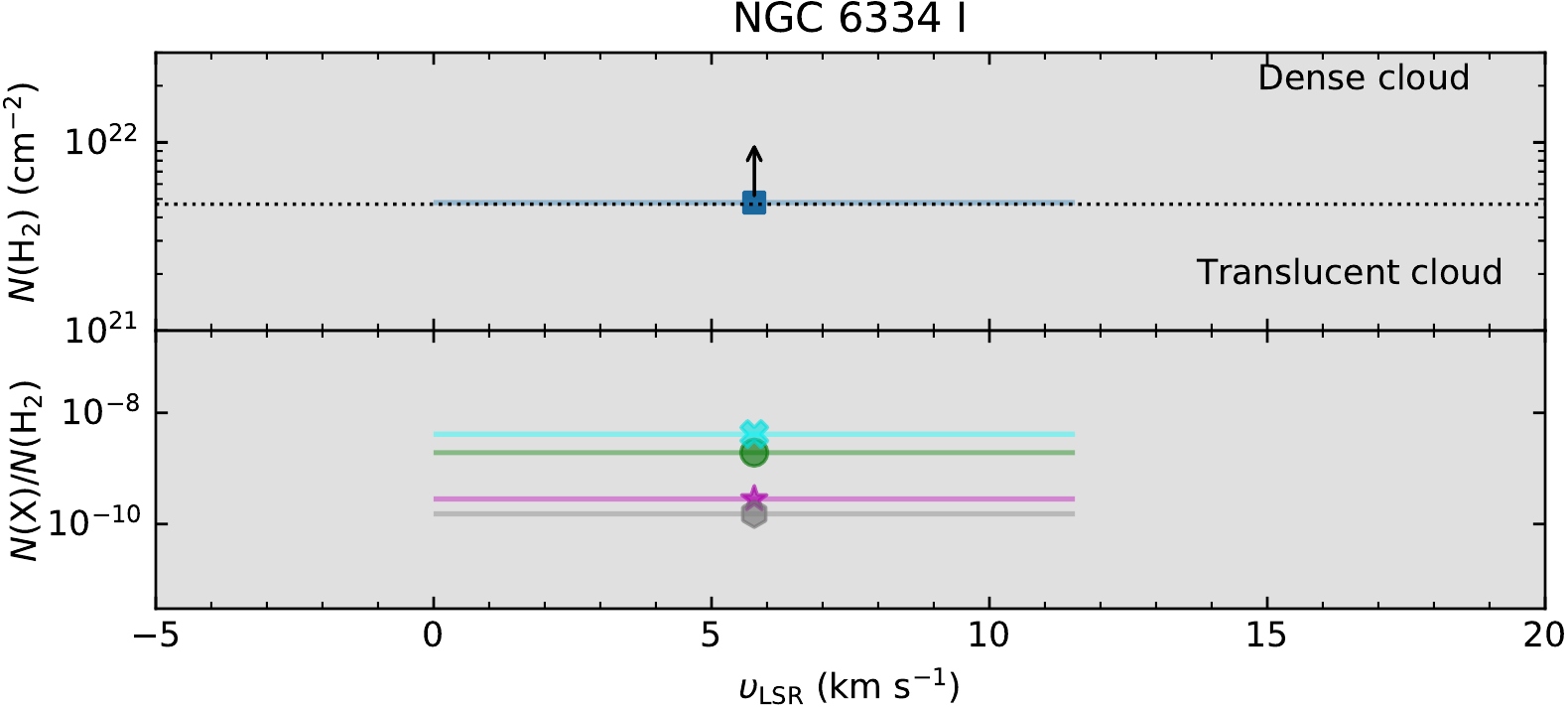}   
    \includegraphics[width=0.595\textwidth]{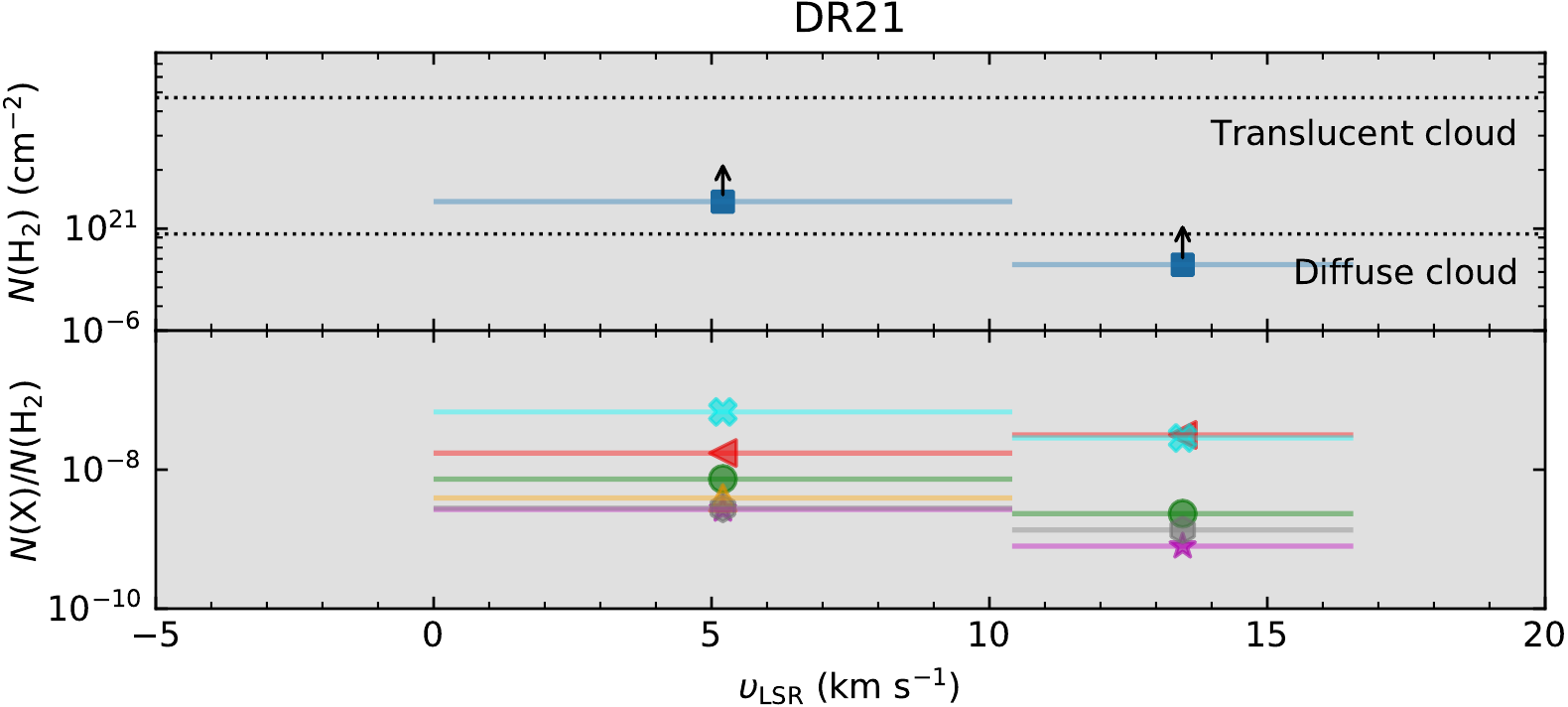}
    \includegraphics[width=0.59\textwidth]{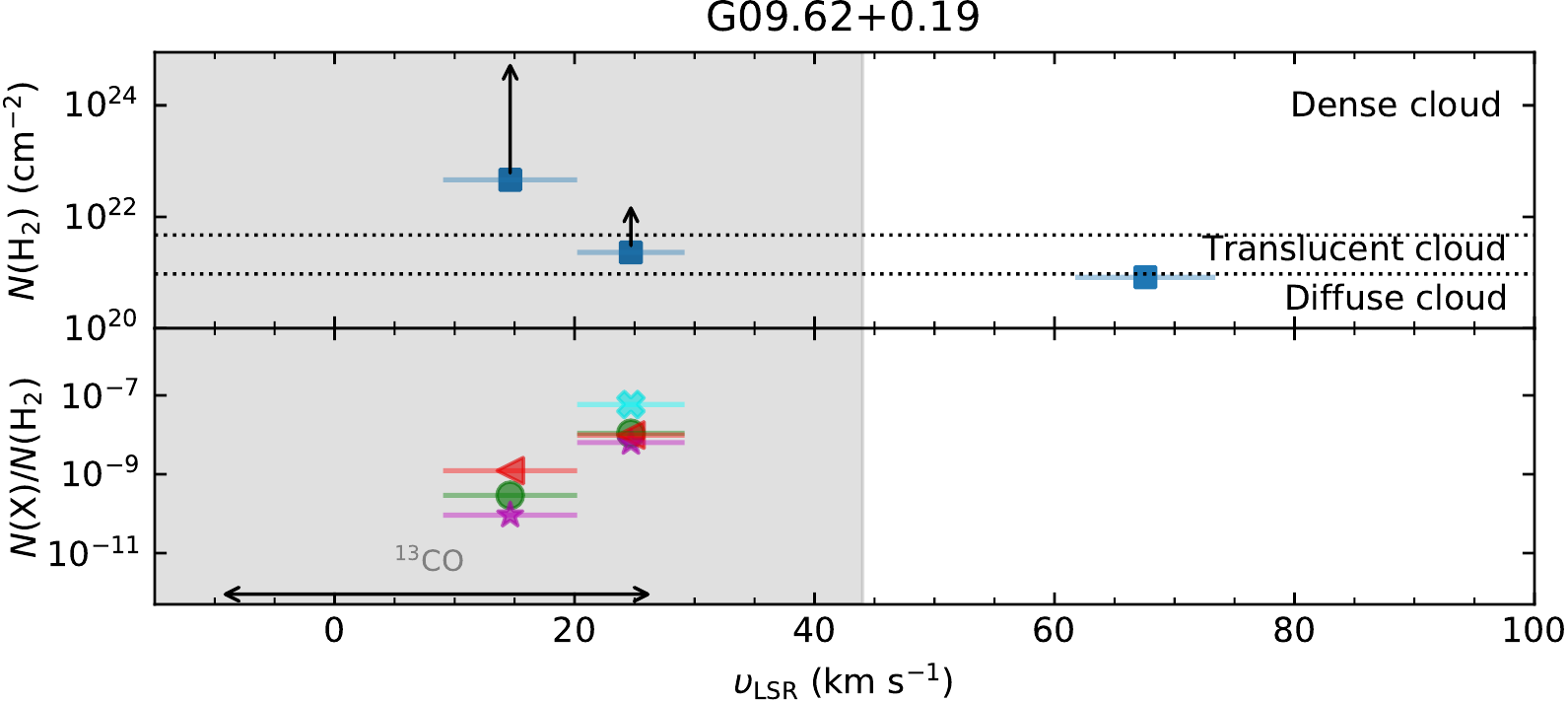}
    \caption{$N$(\ce{H2}) and $N$(X)/$N$(\ce{H2}) ratio integrated over specific velocity intervals from top to bottom toward W3 IRS5, NGC 6334 I, DR21, and G09.62$+$0.19. Different species are labeled with different symbols and colors as shown in the legend displayed in the right corner of the W3(OH) diagram. The gray areas indicate the velocity ranges of the emission lines. In the upper panels, the dotted horizontal lines correspond to $A_{\rm v}$ of 1 and 5, respectively. In the lower panels, the black horizontal arrows indicate the velocity ranges in which CO emission features have been detected from archival data at its given sensitivity.}
    \label{fig:x_vel_add1}
\end{figure*}

\begin{figure*}
    \centering
    \includegraphics[width=0.59\textwidth]{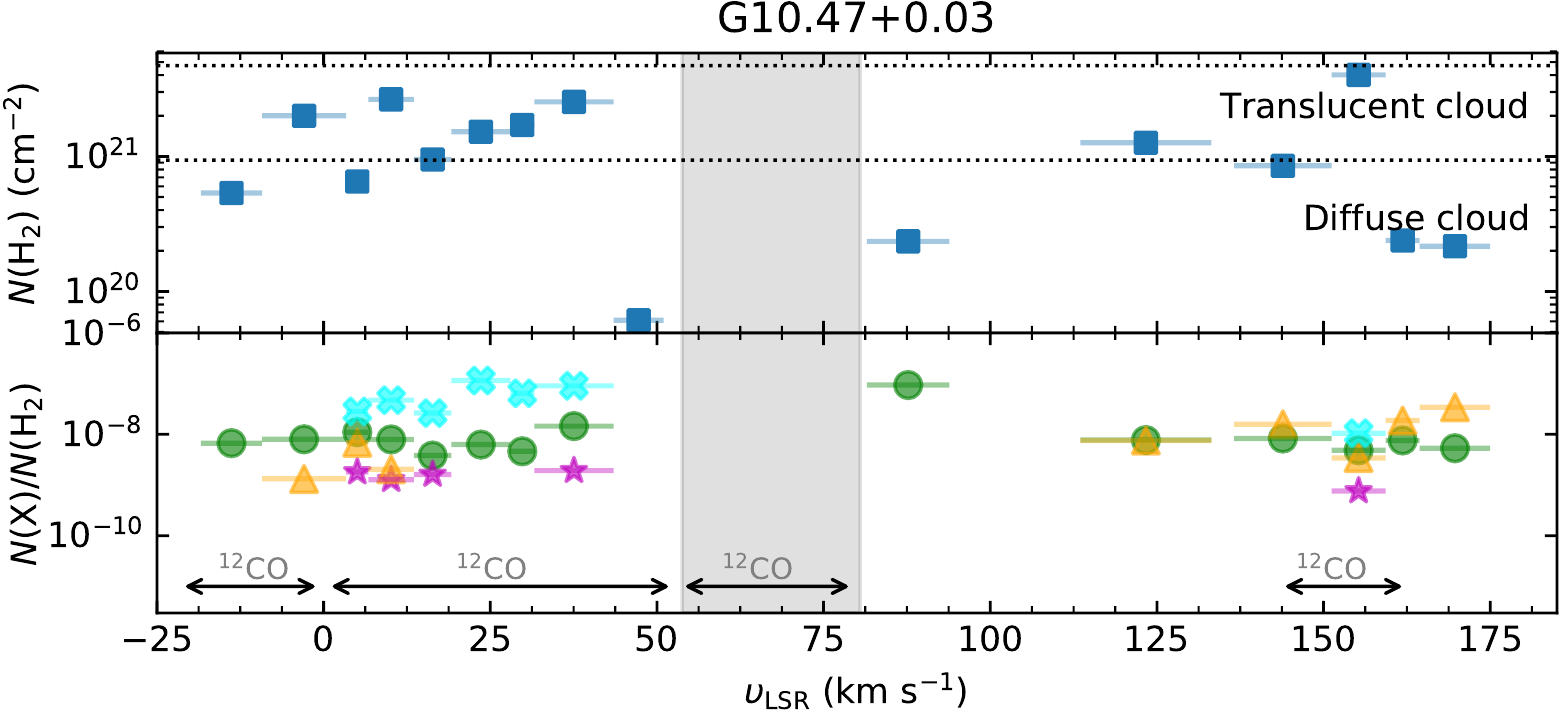}
    \includegraphics[width=0.59\textwidth]{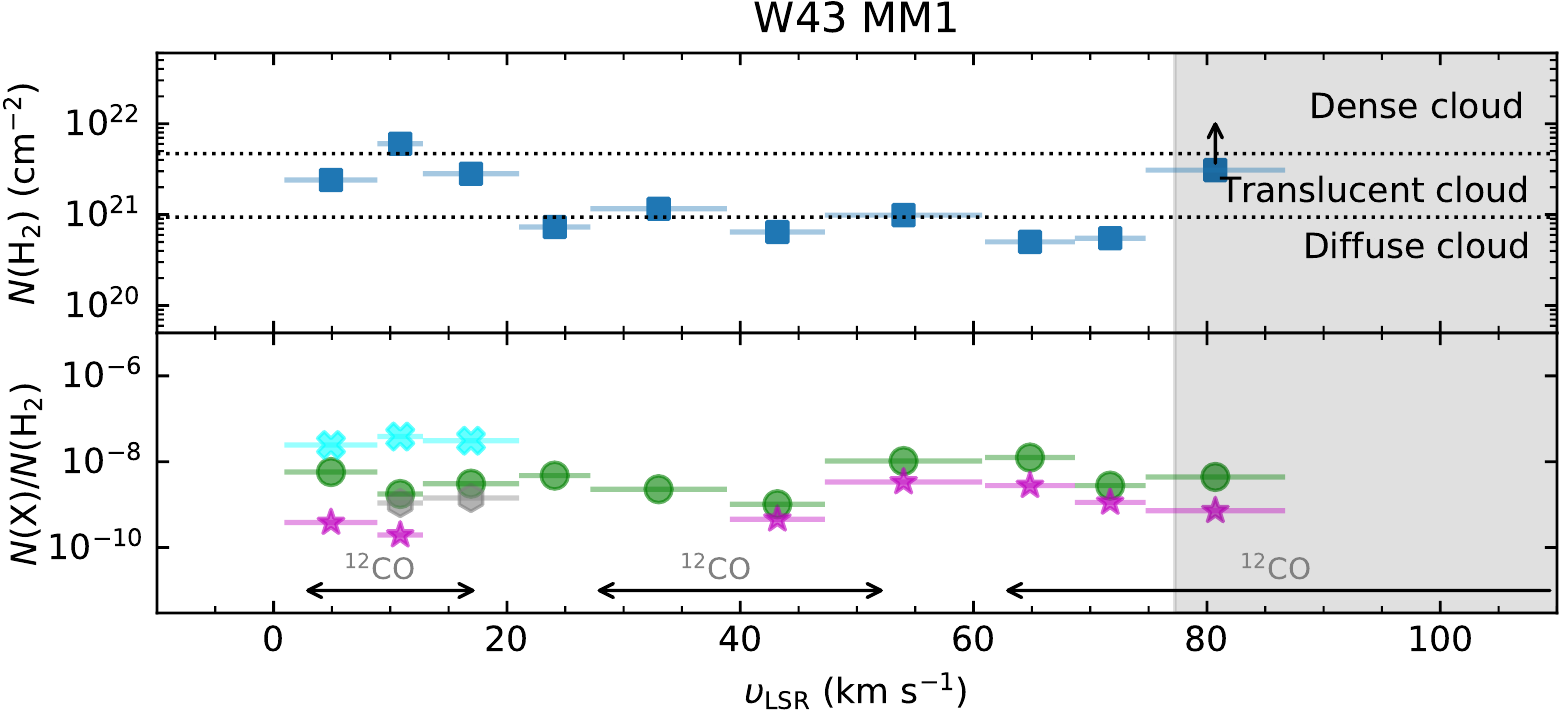} 
    \includegraphics[width=0.59\textwidth]{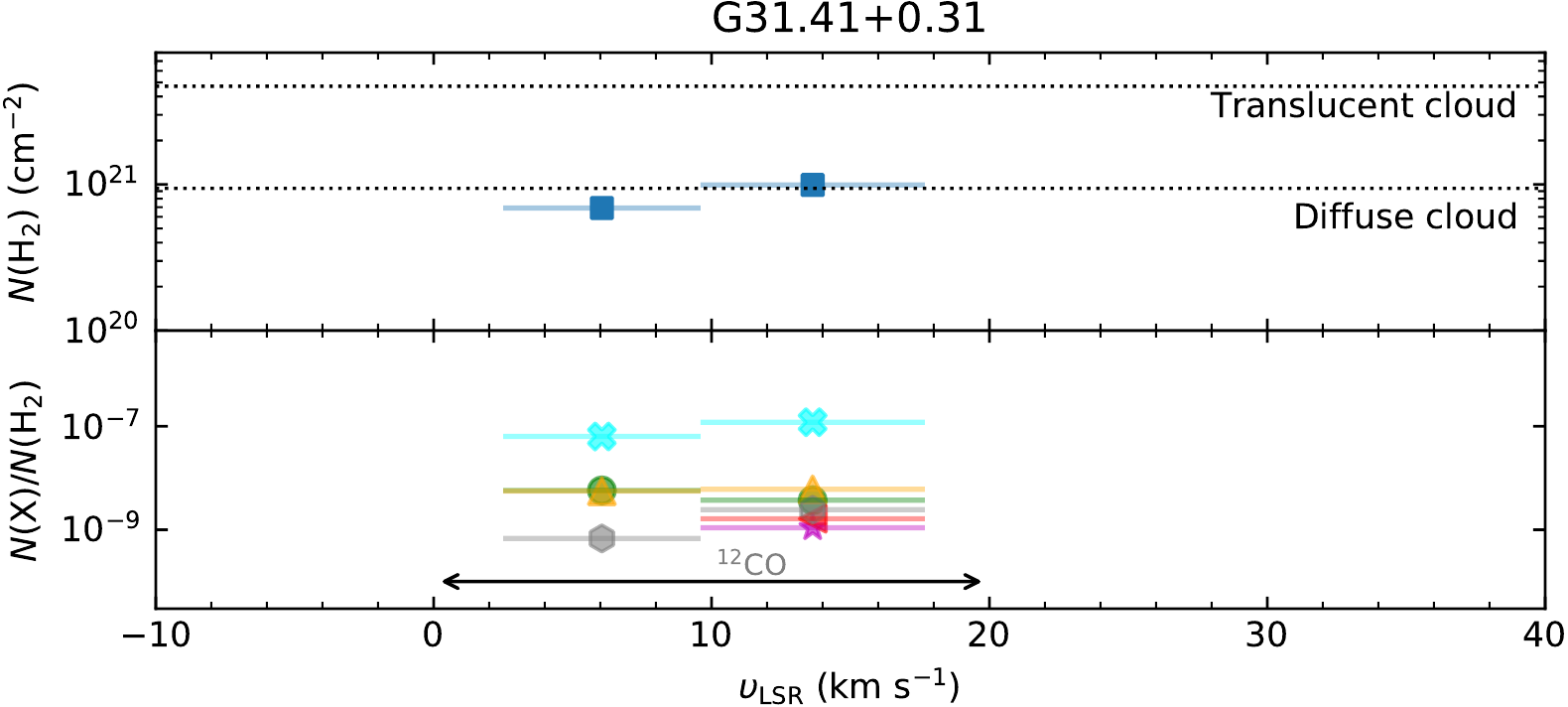}
    \includegraphics[width=0.59\textwidth]{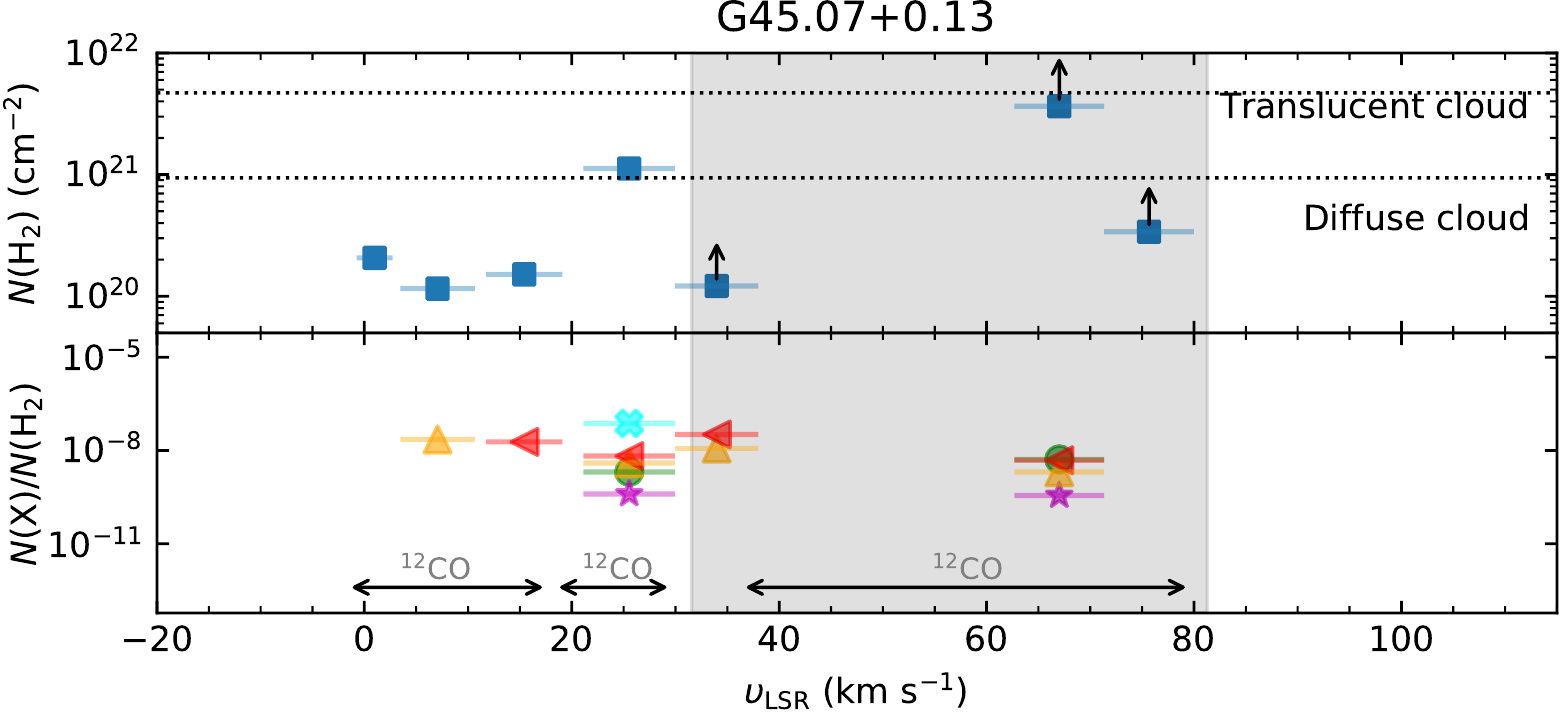}
    \caption{$N$(\ce{H2}) and $N$(X)/$N$(\ce{H2}) ratio integrated over specific velocity intervals from top to bottom, toward G10.47$+$0.03, W43 MM1, G31.41$+$0.31, and G45.07$+$0.13. All symbols and lines are same as Fig.\,\ref{fig:x_vel_add1}.}
    \label{fig:x_vel_add2}
\end{figure*}

Figures\,\ref{fig:x_vel_add1} and \ref{fig:x_vel_add2} show variations in molecular abundance relative to $N$(\ce{H2}) in the lower panel and $N$(\ce{H2}) in the upper panel as a function of velocity in each source. 

\end{appendix}

\end{document}